\documentclass[10pt,a4paper]{book}

\usepackage{amssymb}

\def\O{\Omega}
\def\t{\theta}
\def\F{{\cal F}}
\def\der{\partial}

\def\ve{\varepsilon}
\newcommand{\ft}[2]{{\textstyle\frac{#1}{#2}}}
\newcommand{\be}{\begin{equation}}
\newcommand{\ee}{\end{equation}}
\newcommand{\ov}{\overline}

\newcommand{\la}{\langle}
\newcommand{\ra}{\rangle}

\newcommand{\wh}{\widehat}
\newcommand{\mscr}[1]{\mbox{\scriptsize #1}}

\def\bea{\begin{eqnarray}}
\def\eea{\end{eqnarray}}

\renewcommand{\c}{\gamma}
\renewcommand{\d}{\delta}
\newcommand{\pa}{\partial}
\newcommand{\g}{\gamma}

\newcommand{\e}{\epsilon}

\renewcommand{\L}{\Lambda}
\newcommand{\La}{\Lambda}
\newcommand{\m}{\mu}

\newcommand{\n}{\nu}

\newcommand{\s}{\sigma}

\newcommand{\h}{\eta}

\begin{document}

\begin{titlepage}

\begin{center}
\hfill {\tt hep-th/0007195}\\

\vskip 3cm

{ \LARGE \bf Black Hole Entropy, Special Geometry and Strings}

\vskip .3in

{\bf Thomas Mohaupt\footnote{\mbox{
\tt 
mohaupt@hera1.physik.uni-halle.de}}}
\\

\vskip 1cm

{\em 
\centerline{Martin-Luther-Universit\"at Halle-Wittenberg, 
Fachbereich Physik,
D-06099 Halle, Germany}    }

\vskip .1in
\end{center}

\vskip .2in

\begin{center} {\bf ABSTRACT } \end{center}
We review work done over the last years on the macroscopic and
microscopic entropy of supersymmetric black holes in fourdimensional
${\cal N}=2$ supergravity and in ${\cal N}=2$ compactifications of 
string theory and M-theory.
Particular emphasis is put on the crucial role of 
higher curvature terms and of modifications of the area law
in obtaining agreement between the macroscopic entropy, which 
is a geometric property of black hole solutions and the microscopic
entropy, which is  computed by state counting
in Calabi-Yau compactifications of string or M-theory. 
We also discuss invariance properties of the entropy 
under stringy T-duality and S-duality transformations in 
${\cal N}=2,4$ compactifications in presence of higher curvature terms.

In order to make the paper self-contained we review the
laws of black hole mechanics in higher derivative gravity, the
definition of entropy as a surface charge, the superconformal 
off-shell description of ${\cal N}=2$ supergravity, special geometry,
and ${\cal N}=2$ compactifications of heterotic and type II string
theory and of M-theory.


\vfill

December 1999 / September  2000\\
\end{titlepage}


\newpage

\tableofcontents

\chapter{Introduction and Overview}

\section{Introduction}

Black holes are truly unique objects: For theoretical physicists
they pose various fascinating problems, which may offer a clue
for solving the riddles of quantum gravity. At the same time
black holes are an established part of pop culture. And even more
astonishing it is very likely that black holes really exist beyond
imagination: Currently about 20 stellar binaries are known in our
galaxy which are believed to contain black holes of some solar masses, 
whereas 
supermassive black holes provide the only explanation for the 
processes observed
in the centers of active galaxies \cite{FroNov:1998}. The 
gravitational wave detectors GEO 600 \cite{geo600}, VIRGO \cite{virgo} 
and LIGO \cite{ligo},
which are currently under construction,
aim to directly observe processes involving black holes, including
collisions of black holes, in our cosmic neighbourhood of about
25 Mpc.

In this paper we will consider black holes from the theoretical 
perspective. The most interesting questions of black hole 
physics are related to one of the fundamental problems of
contemporary physics, namely to find a synthesis of the
ideas of dynamical space-time geometry and of quantum physics.
This has nourished the hope that black holes might turn out
to be 'the hydrogen atom of quantum gravity'. Though it remains
to be seen whether this is finally true, tremendous progress on
several aspects of black holes has been made during the last four
years using string theory. String theory 
\cite{GreSchWit:1987,LueThe:1989,Pol:1998}
is the leading candidate
for a unified quantum theory of elementary particles and their
interactions, including gravity. During the 'first
string revolution' of the mid-eighties it became clear that 
string theory is consistent at the level of perturbation
theory.
This was an important step, but left non-perturbative 
problems, such as the study of black holes, out of reach.
The 'second string revolution' of the mid-nineties 
has changed this dramatically. The discovery of string dualities
led to a new picture, where the strong coupling limit of
every string theory is described
by a weakly coupled dual theory, which is either a string 
theory, or eleven-dimensional supergravity. Moreover all
five superstring theories and eleven-dimensional supergravity
seem to be asymptotic expansions of one single underlying theory,
called M-theory. A central role in establishing string dualities
has been played by solitons, and in particular by $p$-branes,
which are higher dimensional analogues of black holes. Conversely
string theory can now be used to investigate problems of black 
hole physics. 
One approach is to interpret black holes and
string excitations as dual descriptions of the same 
object \cite{Sus:1993,HorPol:1996}. This so-called 
string - black hole correpondence passes some qualitative tests:
it predicts the black hole entropy in terms of string states up
to a constant of order unity and it gives an explanation for the
final state of a Schwarzschild black hole: It heats up until it 
reaches the Planck temperature and then converts into a highly
excited string state. 

More precise and quantitative tests are possible for the 
subclass of charged extremal black holes. 
When embedded into theories with extended supersymmetry, extremal
black holes are BPS solitons \cite{Gib:1981}. Since the essential properties
of such objects are determined by supersymmetry, it is possible
to do computations in the perturbative regime and then to extrapolate
them to finite coupling. This became a powerful tool after
the discovery that charged black holes in string theory
have a dual, perturbative description in terms of 
D-branes \cite{Pol:1995}. Four- and five-dimensional extremal 
black holes can be embedded into string theory. From the 
point of view of ten-dimensional string theory or eleven-dimensional
M-theory extremal black holes are bound states of $p$-branes and 
other gravitational solitons. Using the BPS property one can
study these brane configurations in the perturbative regime. In particular
one can count the microstates of a black hole, which means all states of the
brane configuration which lead to the same four- or five-dimensional black 
hole, and one can compute the corresponding 'microscopic' or 
'statistical' entropy.
Since one can independently compute the 'macroscopic' entropy of the 
black hole, which means its Bekenstein-Hawking entropy, the comparison
of the two results provides a test both for the hypothesis that
the Bekenstein-Hawking entropy really is an entropy, and for the 
microscopic model of black holes provided by D-branes. Using this
strategy Strominger and Vafa showed that the two entropies agree 
quantitatively for certain classes of black holes in 
string compactifications \cite{StrVaf:1996}. 
Here 'quantitatively' means that the 
leading terms for both entropies are exactly the same. 
The D-brane method has also been applied to derive
Hawking radiation and to compute 
greybody factors. In addition one gets a resolution of the information
or unitarity problem of quantum gravity: According to the D-brane
picture the time-evolution of a black hole is unitary, and the
Planckian character of Hawking radiation is due to averaging
over initial and summing over final states. We refer to \cite{Mal:1996}
for a review and further references and to \cite{Moh:2000/4} for
an elementary introduction to black holes in supergravity and 
string theory.

In this paper we will focus on black hole entropy. The original
work of Strominger and Vafa studied five-dimensional black holes
in compactifications with ${\cal N}=4,8$ supersymmetry. Of course,
one is more interested in four-dimensional black holes and the
method applies whenever extremal black holes are BPS states.
This means that the most general setting is black holes
in compactifications with ${\cal N}=2$ supersymmetry. It turns out
that the generalization to this case is highly non-trivial. The
reason is that four-dimensional supergravity theories with 8 supercharges
(${\cal N}=2$ in four dimensions) have a much richer dynamics
than theories with 16 or 32 supercharges (${\cal N}=4,8$ in four
dimensions). Whereas the low energy Lagrangian (with terms up to
two derivatives) is uniquely fixed by the spectrum in ${\cal N} \geq 4$
theories, the low energy Lagrangian of ${\cal N} = 2$ supergravity
contains arbitrary functions. This leaves room for both
quantum and stringy corrections, which turn out to be very complicated. 
Fortunately the Lagrangian is still much more restricted than in
theories with ${\cal N}=1$ supersymmetry or in non-supersymmetric
theories.
This makes it possible to keep the theory 
under sufficient control in order to solve
various problems exactly. The relevant sector for black hole
solutions consists of the supergravity multiplet together with
a model-dependent number of vector multiplets. It is uniquely
specified by a single function of the scalar fields, 
the prepotential, which is restricted to be holomorphic and to be 
homogenous of degree two. The related geometry of the scalar
sector, called special geometry, is a very powerful tool in
investigating the theory \cite{deWvHvP:1981,deWLauvP:1985,CreEtAl:1985}.

When comparing the macroscopic entropy of four-dimensional 
${\cal N}=2$ black holes to the microscopic entropy computed in the
brane picture one encounters subtleties that have not been noticed
before. 
The microscopic entropy formula
contains a subleading term, which on the macroscopic side 
corresponds to the presence of a non-minimal higher curvature 
term in the effective Lagrangian.
Thus one is forced to include such higher curvature terms
in the construction of black hole solutions and in the computation
of the macroscopic entropy. This is a difficult technical problem,
which can be overcome by using the superconformal off-shell
formulation of ${\cal N}=2$ supergravity, as we will see. 
A second subtlety
is that in the presence of higher curvature terms
the entropy formula itself has to be modified and is no longer
given by the Bekenstein-Hawking area law \cite{Wald:1993}. 
It is the purpose of this
paper to explain all these points in detail, to put them into
perspective and to review the background material that is needed
to understand them.

\section{Overview}

Let us now give an overview of the following chapters.

In chapter \ref{ChapterGravity} we introduce our basic topics.
We start with Einstein gravity and then discuss higher derivative
gravity. Gravity does not lead to a renormalizable and unitary
perturbative quantum field theory and therefore we embed it into
string theory. Within string theory one can compute a low
energy effective action which contains besides the Einstein-Hilbert
term also higher curvature terms. These terms are generated by both 
quantum corrections and stringy $\alpha'$-corrections. After 
recalling basic properties of black holes we review the four
laws of black hole mechanics in the context of Einstein gravity.
Then we discuss quantum aspects of black holes and explain
why the structural equivalence between the laws of
black hole mechanics and the laws of thermodynamics is expected
to be more than a
formal analogy. Special attention is paid to the problem of 
black hole entropy and we explain what we mean by macroscopic
and microscopic black hole entropy. Then we turn to a more
technical discussion of black hole horizons and surface charges.
This enables us to understand why the laws of black hole
mechanics, which were originally derived in Einstein gravity,
are independent of the precise form of the equations of 
motion and also apply, with some subtle modifications, 
to higher derivative gravity. Particular emphasis is put
on the first law and the definition of black hole entropy
given by Wald. 
Finally we discuss extremal black holes and their interpretation
as supersymmetric solitons and BPS states in extended supergravity.

Chapter \ref{ChapterFourDSuGra} is devoted to a review
of ${\cal N}=2$ supergravity in the superconformal off-shell
formulation. After explaining the concept of gauge equivalence
in a non-supersymmetric example, we present the relevant superconformal
off-shell multiplets and the construction of the superconformal
action for an arbitrary number of vector multiplets. The action
depends on the prepotential, which is a holomorphic and homogenous
function of the scalar fields, but is arbitrary otherwise.
Particular
emphasis is put on the Weyl multiplet and its use in
describing a particular class of higher curvature terms,
called $R^2$-terms in the following. The central concept of symplectic
invariance, which generalizes electric - magnetic duality is explained
in this setting. Then we describe how the 
superconformal theory and the gauge-equivalent super Poincar\'e
theory are related. Special geometry is discussed both in terms
of special and of general coordinates. We explain the relation
between the superconformal off-shell and the geometric on-shell
formulation of ${\cal N}=2$ supergravity in order to make contact
with the majority of the recent literature.

In chapter \ref{ChapterN=2BlackHoles} we study model-independent
aspects of black holes in ${\cal N}=2$ supergravity with 
vector multiplets in presence of $R^2$-terms. We prove that the only
static and spherically symmetric ${\cal N}=2$ vacuum 
besides four-dimensional Minkowski space is the
space $AdS^2 \times S^2$. We show that the metric and all
other fields for such a vacuum can be expressed in terms of
a single quantity $Z$, which is closely related to the central 
charge of the supersymmetry algebra. Since the
$AdS^2 \times S^2$ vacuum is the near horizon geometry of an
extremal black hole, we can use the generalized entropy formula
of Wald to derive a model-independent expression for the entropy, 
which is manifestly covariant under symplectic transformations.
Then we review the supersymmetric attractor mechanism and argue
that it generalizes to the case of $R^2$-terms. This implies a
set of algebraic equations, called stabilization equations,
which determine the entropy in terms 
of the electric and magnetic
charges carried by the black hole. All the results mentioned so
far can be obtained without explicitly knowing the black hole
solution away from the horizon. For completeness we review what
is known about explicit black hole solutions in the case
without $R^2$-terms and we briefly indicate how these can be
included. While preparing this paper for publication, we obtained
important new results \cite{CardWKaeMoh}, which are briefly summarized
in section \ref{AddedSection}.

Up to this point the discussion is model independent and does not make use
of string theory. The only input is ${\cal N}=2$ supergravity
and all results apply to general prepotentials. Now we bring
string theory into the game in order to have a microscopic model
for black holes, where one can identify and count microstates. The 
embedding of supergravity into string theory also provides us with
concrete models. In particular the prepotential and the couplings of
higher curvature terms can be computed in string perturbation theory.
In chapter \ref{ChapterFourDimensionalStringsAndM} 
we review how four-dimensional ${\cal N}=2$
supergravity arises from compactifying type II string theory on a
Calabi-Yau threefold, eleven-dimensional M-theory on a 
Calabi-Yau threefold times a circle and heterotic or type I
string theory on a $K3$ surface times a two-torus. 
The structure of the corresponding prepotentials is explained.
In particular we discuss in each case the role of quantum and
$\alpha'$-corrections.
The same is done for the field-dependent couplings of the higher
curvature terms.
We also review the non-perturbative dualities relating the
three compactifications and the role of T-duality in perturbative
heterotic compactifications. Finally we briefly discuss compactifications
with ${\cal N}=4$ supersymmetry.

In chapter \ref{ChapterStringBHs} we consider the 
entropy of black holes in four-dimensional ${\cal N}=2$ string
compactifications both from the macroscopic and microscopic 
point of view. The macroscopic entropy is computed by solving
the stabilization equations and plugging the result into
the model-independent entropy formula. We show that the entropy
is a series in even powers in the charges, with coefficients 
related to the higher curvature couplings.
Then we specifically discuss 
black holes in type II and M-theory compactifications
and describe three classes of black holes where one can completely
solve the stabilization equations and obtain explicit formulae
for the entropy in terms of the charges. 
The most important case is the formula for
the entropy of black holes in a large volume Calabi-Yau compactification. 
This is the
most general case where also the microscopic entropy has been computed.
Then we turn to black holes in heterotic compactifications and 
derive entropy formulae which are manifestly invariant under T-duality.
As a further application we find the S- and T-duality invariant entropy
formula for black holes in ${\cal N}=4$ compactifications with 
$R^2$-terms. In the second part of the chapter
we review the derivation of the microscopic
entropy for black holes in compactifications in ${\cal N}=2,4,8$
supersymmetry. For ${\cal N}=2$ and ${\cal N}=4$ we find
subleading terms which are related to 
$R^2$-terms on the macroscopic side. The matching of 
macroscopic and microscopic entropy is discussed in detail.
We explain why it is a highly non-trivial result and recall
all the subtleties involved in the matching. 

In the final chapter \ref{ChapterLast}
we list the key results on black hole
entropy in ${\cal N}=2$ supergravity and string compactifications
and we give an outlook on further directions of research.

In the five appendices we specify our conventions and collect
several useful formulae.

\chapter{Gravity, Black Holes and Supersymmetry \label{ChapterGravity}}

In this chapter we introduce our topics:
gravity and the role of higher curvature terms in it, 
string theory and its description by a low energy effective action,
the
laws of black hole mechanics and the definition of black
hole entropy, and last but not least the characterization of
extremal black holes as supersymmetric solitons.

\section{Gravity, Strings and Effective Field Theories}

Gravity, as a classical theory, is described by Einstein 
gravity \cite{MTW,Wald:1984,HawEll:1973}, which
is based on the Einstein-Hilbert action
\be
S = - \frac{1}{2 \kappa^2} \int d^4 x \sqrt{-g} R \;.
\ee
The coupling of matter to gravity is obtained by minimal substitution, i.e. by
replacing the partial derivatives occuring in the matter action
by covariant derivatives with respect to the Christoffel and
spin connections. 
The coupling constant $\kappa$ is related
to Newtons constant $G_N$ by $\kappa^2 = 8 \pi G_N$, in natural
units where $\hbar=c=1$. In these units the Planck length and mass
are $G_N = l_{\mscr{Planck}}^2 = m_{\mscr{Planck}}^{-2}$.
In the following we will mostly use
Planckian units, where one sets in addition $G_N=1$. When
dimensional analysis is required we can easily reconstruct 
it. We work in four space-time dimensions and 
take the metric to
have signature $(-+++)$.

Nowadays the only observational facts that might
be in contradiction with the above minimal version of Einstein
gravity are indications for a very small cosmological constant \cite{Rie:1998}.
At the classical level this is a minor modification of the theory, but
within a quantum theory a vanishing or very small cosmological
constant constitutes a naturalness problem, see \cite{Wei:1988} for a
review. This is an important 
topic in quantum gravity, but since it has no direct importance for 
the study of black hole entropy  we will set the cosmological constant 
to zero throughout the text.

Despite that there is no known fact that cannot be
covered by Einstein gravity (possibly extended to contain a
cosmological constant) we believe that Einstein gravity cannot be
a fundamental theory, because it is conceptually incompatible with
the other cornerstone of physics, quantum theory. Therefore a
central problem of theoretical physics consists in finding a 
theory that encompasses both gravity and quantum theory. One
option is to quantize Einstein gravity. At the perturbative level
Einstein gravity is 
non-renormalizable\cite{tHo:1973,tHoVel:1974,DesvNi:1974}. 
It seems to be possible that quantum gravity 
nevertheless exists non-perturbatively, though one cannot get
the appropriate classical limit, namely Einstein 
gravity \cite{Smo:1998}.\footnote{See this paper for a review and 
references on
the background independent approach to quantum gravity and on recent
attempts to make contact with the string theory approach.}
The most prominent candidate for a quantum theory of gravity
is string theory \cite{GreSchWit:1987,LueThe:1989,Pol:1998}. 
At the perturbative level string theory is defined
by quantizing the relativistic string in a fixed background
geometry (and other classical background fields). It has been
shown that five consistent string theories exist, which all
are supersymmetric and ten-dimensional, in the sense that 
ten-dimensional Minkowski space is their most symmetric ground state.
String theories give a consistent description of quantum gravity
in perturbation theory, i.e. one can compute
loop corrections involving gravitons. At the non-perturbative level
no explicit and in particular no background independent definition
is known. During the last years, however, there has been 
tremendous progress in understanding non-perturbative properties
by studying solitons, instantons and string dualities, see \cite{Pol:1998}
for an overview and references. 
Such dualities allow one to consistently interpret the 
strong coupling behaviour of a string theory in terms of the
weak coupling behaviour of another, dual theory. This has led to
a new picture, where all consistent perturbative string theories
arise as asymptotic expansions of one single underlying theory,
called M-theory. This theory has another vacuum, eleven-dimensional
Minkowski space, where it is described at low energies by
eleven-dimensional supergravity.

Among all possible covariant Lagrangians the
Einstein-Hilbert Lagrangian is singled out by
minimality, in the sense that it only contains terms with up to
two derivatives. If this requirement is relaxed, one can write down more
general covariant Lagrangians which then contain higher order
curvature terms. The theory obtained at the four derivative level
is called $R^2$-gravity because it contains terms quadratic
in the curvature\footnote{The definitions of the curvature tensors are
recalled in appendix \ref{AppSpaceTime}.}
\be
S = \int d^4 x \sqrt{-g} \left( - \frac{1}{2 \kappa^2} R 
+ a R^2 + b R_{\m \n} R^{\m \n} + c R_{\m \n \rho \sigma}
R^{\m \n \rho \sigma} + d \square R \right)
\ee
(we neglected the cosmological constant). The last term is a
total derivative and the same is true for the Gauss-Bonnet
combination
\bea
\mbox{GB} &=& \ft12 \e^{\m \nu \alpha \beta} \e^{\rho \sigma \kappa \omega}
R_{\m \n \rho \sigma} R_{\kappa \omega \alpha \beta} \nonumber \\
 & = & R_{\m \n \alpha \beta} R^{\m \n \alpha \beta} - 4 R_{\m \n}
R^{\m \n} + R^2 \;.\\
\nonumber
\eea
$R^2$-gravity is multiplicatively renormalizable, but since it is non-unitary
it does not seem to provide an alternative root to perturbative
quantum gravity
(see \cite{BucOdiSha:1992} for a review and references).
The non-unitarity is related to the fact that the higher curvature
terms introduce new degrees of freedom, which lead to negative norm
states that cannot be eliminated. 
The conformally invariant part of $R^2$-gravity is given 
by the Weyl action
\be 
S = \int d^4 x \sqrt{-g} \frac{1}{g^2_{\mscr{grav}}}  
C_{\mu \nu \rho \sigma}
C^{\m \n \rho \sigma}\;,
\label{WeylAction}
\ee
where $C_{\m \n \rho \sigma}$ is the Weyl tensor 
and $g_{\mscr{grav}}$ is the corresponding coupling. This theory resembles
Yang-Mills theories in many respects. It is classically conformally 
invariant and multiplicatively renormalizable at one loop. Concerning
unitarity the same remarks apply as for the full $R^2$-theory.

Thus higher derivative gravity actions do not seem to make sense as
starting points for the definition of
fundamental theories. They play, however, an important role as
low energy effective actions of fundamental quantum theories of gravity
such as string theory. In string theory one can show that the
low energy effective action indeed contains higher curvature
terms \cite{GreSchWit:1987,Pol:1998}. The couplings of the
effective theory are functions of the scalar fields.
One finds that $R^2$-terms are generated both through
loop corrections and through stringy $\alpha'$-corrections.
The parameter $\alpha'$, which has dimension length squared
(in natural units) is the fundamental dimensionful parameter
of string theory. Up to a constant, $\alpha'$ is the inverse of
the string tension. One then defines the string length
by $l_{\mscr{String}} = (\alpha')^{1/2}$ and the string mass
$m_{\mscr{String}} = (2 \alpha')^{-1/2}$. String units are
obtained by setting $\alpha' = \ft12$, so that the string mass
equals unity. When computing the low energy effective action
of the massless modes of string theory at tree level one finds
that it contains an Einstein-Hilbert term together with an infinite
series of higher curvature terms that are supressed by 
powers of $\alpha'$, so that they are subleading at low energies.
Thus, string theory deviates from Einstein gravity already at the
classical level. These are the $\alpha'$-corrections mentioned above.
On top of this the effective action gets contributions from
loop corrections. String theory contains a dimensionful coupling
constant $\kappa$, which can be identified with the 
gravitational coupling, i.e. the coefficient of the Einstein-Hilbert term.
This quantity is not independent from
the other dimensionful constant, $\alpha'$. Besides the graviton,
every string theory contains another universal state, a massless
scalar called the dilaton $\phi$. The dependence of string perturbation
theory on this field is such that the $g$-loop contribution
to the effective action is proportional to $(e^{-2 \phi})^{1-g}$.
Therefore there is a relation between $\kappa, \alpha'$ and the
vacuum expectation value $\la \phi \ra$ of the dilaton, which in four 
dimensions reads
\be
\frac{\alpha'}{4} = \frac{\kappa^2}{8 \pi} e^{2 \la  \phi \ra} \;.
\label{alphakappadilaton}
\ee
The vacuum expectation value of the dilaton is not fixed by the
equations of motion and this degeneracy is not lifted in string
perturbation theory.\footnote{We only consider supersymmetric
string vacua in this paper. If string theory describes our world
then supersymmetry must be broken, and the supersymmetry breaking
mechanism presumably generates a potential for all scalar fields and
in particular for the dilaton.} 
Therefore $\la \phi \ra$ is a free parameter,
at least in perturbation theory, and has to be specified as
part of the definition of a string background.

Thus there is one fundamental dimensionful parameter which can
be taken to be $\kappa$ or $\alpha'$ and a continous familiy
of vacua parametrized by $\la \phi \ra$.
Since $e^{\la \phi \ra}$ can serve as loop counting 
parameter one introduces the dimensionless string coupling
\be
g_S = e^{\la \phi \ra} \,.
\ee
In terms of the string and Planck masses, the relation 
(\ref{alphakappadilaton}) can be rewritten as
\be
m_{\mscr{String}} = g_S m_{\mscr{Planck}} \;.
\ee

It should be noted that whether an $R^2$-term is due
to a loop or due to an $\alpha'$-correction is not determined
by the term itself, but is a model (or better background) dependent question,
which has to be determined case by case. The Einstein-Hilbert
term itself is always found at tree level in both $g_S$
and $\alpha'$. Consider now an $R^2$-term, by which we denote
any curvature term with four derivatives. In four dimensions
where both $\kappa^2$ and $\alpha'$ have dimension length squared
such an $R^2$ term can, by dimensional analysis, be of the
form $\kappa^{2m} (\alpha')^n R^2$, where $m+n=0$. Thus it could
be either a string loop effect ($m=0=n$) 
or an $\alpha'$-loop effect ($n=1=-m$). Since the
coupling constants of the effective theory are not just constants
but field dependent objects the actual analysis is more 
complicated, as we will see in chapter 
\ref{ChapterFourDimensionalStringsAndM}.
There we will consider terms of the form
$C^2 T^{2g-2}$, where $C$ is the Weyl tensor and $T$ is a field
strength. In one specific theory (IIA string theory on a Calabi-Yau threefold)
these terms arise at the 
$g$-loop level in $g_S$ whereas in another, dual theory (heterotic
string theory on $K3 \times T^2$, with a choice of gauge bundle)
the same term gets contributions at
tree level, one loop level and non-perturbative level 
in $g_S$. In both cases 
the terms get loop and non-perturbative contributions in $\alpha'$.
Moreover in another dual theory (IIB string theory on the mirror 
Calabi-Yau manifold) the terms are $g$-loop in $g_S$ and 
tree level in $\alpha'$.

In the effective action for 
$R^2$-gravity one usually 
separates the $C^2$-term (the square of the Weyl tensor) because
it is the only conformally invariant term:
\be
8 \pi {\cal L} = - \ft12 R + \ft14 \ft1{g_{\mscr{grav}}^2}
C^2 + \ft14 \Theta R_{\m\n\rho\sigma} {}^{\star}R^{\m \nu \rho \sigma}
+ \ft1{\rho^2} R_{\m\n} R^{\m\n} + \ft1{\sigma^2} R^2 \;.
\ee
Here ${}^{\star} R_{\m\n\rho\sigma} = \ft12 \e_{\m\n\alpha\beta}
R^{\alpha \beta}_{\;\;\;\;\rho\sigma}$ is the Hodge dual of
the Riemann tensor.
The conventional choice of couplings for compactifications of the 
heterotic string is such that at tree level all higher 
derivative couplings are topological,
\be
8 \pi {\cal L} =
- \ft12 R + \ft14 \mbox{Re} \; S \;\;GB + \ft14 \mbox{Im} \;S \;\;
R_{\m\n\rho\sigma} \; {}^{\star}{R}^{\m \nu \rho \sigma} \;,
\ee
where $S=e^{-2 \phi} + ia$ is the complex heterotic dilaton,
$GB$ is the Gauss-Bonnet term,
$R_{\m\n\rho\sigma}\; {}^{\star}{R}^{\m \nu \rho \sigma}$ is 
proportional to the Hirzebruch signature density and  $\mbox{Im}\;S$
is sometimes called the 'gravitational $\Theta$-angle' 
\cite{Zwi:1985,GroSlo:1987,Tse:1986}.

At the loop level one can in particular study the 
$C^2$-coupling $g_{\mscr{grav}}$ which is very similar
to a gauge coupling (see for example \cite{CarLueMoh:1995} and references
therein). We will later study this and
related couplings in the context of compactifications with
${\cal N}=2$ supersymmetry. 

Finally note that the precise relation
of quantum corrected string couplings to their counterparts in the
effective field theory involves questions 
such as the scheme dependence of couplings
and the proper definition of the renormalization scale \cite{Kap:1992}. 
We will not need to discuss this systematically (in the 
case of gauge couplings this is different because one wants
to address the problem of coupling unification)
but the distinction
between 'physical' and 'Wilsonian' couplings \cite{DixKapLou:1991}
will play some role
later in chapters \ref{ChapterFourDimensionalStringsAndM} and 
\ref{ChapterStringBHs}.

\section{Black Holes}

\subsection{Basic Properties of Black Holes}

Intuitively a black hole is a region of space-time from which one
cannot escape. A more precise definition is that a black hole
is an asymptotically flat space-time containing a region which is
not in the backward lightcone of future timelike infinity 
\cite{Wald:1984,HawEll:1973}.
The boundary of such a region is a null hypersurface,
called the {\em future event horizon}, or {\em event horizon} for 
short.\footnote{We will
not consider other types of event horizons.
The technical terms used in this section will be explained
in some more detail in the following sections. A complete account of
background material is provided by textbooks on general relativity, see
in particular \cite{Wald:1984}.} The event
horizon is a surface of infinite redshift and this motivates the name
black hole.

Already one of the most simple solutions of Einstein gravity, the 
Schwarzschild solution, describes a black hole. The Schwarzschild
solution is the unique spherically symmetric, and therefore, 
by Birkhoffs theorem, static solution of the vacuum Einstein
equations. It describes the exterior region of a spherical 
mass (or better energy-momentum) distribution. If the massive
body is sufficiently compact, the solution exhibits an event 
horizon. Since no mechanism is known which can stop the collapse
of a star with a mass above the Oppenheimer-Volkov limit, this
is believed to be a physical solution which describes the classical
final state of a spherical collapse.

There exist more general black hole solutions which in addition
carry charge and angular momentum. In Einstein-Maxwell theory
one can show that the most general stationary black hole is
the Kerr-Newmann black hole, which is uniquely characterized by
its mass, charge and angular momentum. Remember that a 
space-time is called {\em stationary} if it is time-independent, i.e.
if it posesses a timelike Killing vector. Such space-times
represent classical final states of a total gravitational
collapse. Much of the work on black holes focusses on stationary
black holes, because these solutions are analytically tractable.
In the following we will restrict ourselves to the subclass of 
static black holes, with the exception of some remarks in this
introductory chapter.  A stationary
space-time is called {\em static}, if the timelike Killing vector field
is hypersurface orthogonal.\footnote{This notion is defined in 
section \ref{SectionBHhorizons}.}
A metric which is both static and spherically symmetry can be brought
to the form
\be
ds^2 =- e^{2h(r)} dt^2 + e^{2k(r)} dr^2 + r^2 d\Omega^2 
\label{staticsphericnonisotropic}
\ee
by a coordinate transformation. Here
$d\Omega^2 = sin^2 \theta d \phi^2  + d\theta^2$
denotes the standard metric on the unit sphere. The adapted
coordinates $t,r,\phi,\theta$ provide a generalization of spheric 
coordinates in Minkowski space-time.

The most general static black hole of Einstein-Maxwell theory is
given by the Reissner-Nordstr{\o}m solution, which has a metric
of the form (\ref{staticsphericnonisotropic}), with 
\be
e^{2h(r)} = e^{-2k(r)} = 1 - \frac{2M}{r} + \frac{Q^2}{r^2} \;,
\ee
where $M,Q$ are the mass and charge carried by the black hole,
respectively. The mass of an asymptotically flat space time
can be determined by considering the non-relativistic motion
of a test particle in the asymptotic region. Such a particle
sees a Newtonian gravitational potential $V=-M/r$, where
$M$ is related to the $1/r$ deviation of the metric from flat space
by $g_{00} = - ( 1 - \ft{2M}{r} + \cdots)$. A more elaborate
definition of the mass is provided by the ADM mass, see for example 
\cite{Wald:1984}.
Like the mass, the charge can be measured at infinity, and is
defined by
\be
Q = \frac{1}{4\pi} \oint_{S^2_{\infty}} {}^{\star} F\;,
\ee
where $S^2_{\infty}$ is an asymptotic spacelike sphere at infinity
and ${}^{\star}F$ is the dual of the field strength 
two-form $F=\ft12 F_{\m\n} dx^\m \wedge dx^\n$. World indices
corresponding to the adapted coordinates are denoted by
$\m,\n=t,r,\phi,\theta$.
The normalization is such that an electric field carries 
charge $Q$ if its asymptotic form at infinity is
\be
F_{tr} = \frac{Q}{r^2} \;.
\ee
We have restricted ourselves here to electric charge, but it
is straigthforward to include magnetic charge,
\be
P = \frac{1}{4\pi} \oint_{S^2_{\infty}} F\;
\ee
as well. This results in replacing $Q^2$ by $Q^2 + P^2$ in the 
metric.

Besides the mass and the charge, which are quantities
measured at infinity, there are two characteristic quantities
defined on the event horizon. One is the surface gravity
$\kappa_S$, which measures the strength of the gravitational
field on the event horizon, the other is the area of the event horizon.
These quantities have a tentative thermodynamic interpretation,
and we will discuss them in more detail soon.

The interpretation of the Reissner-Nordstr{\o}m solution 
depends on the values of mass and charge. If $M > |Q|$  (if
magnetic charge is present one has to replace $|Q|$ by
$\sqrt{Q^2 + P^2}$) then the solution describes a black hole
with two horizons, an exterior event horizon and an interior
so-called Cauchy 
horizon.\footnote{See \cite{Wald:1984,HawEll:1973} for a detailed
account.} This is the non-extremal Reissner-Nordstr{\o}m black hole.
Its surface gravity is given by
\be
\kappa_S = \frac{ \sqrt{ M^2 - Q^2 } }{ 2 M ( M+ \sqrt{M^2 - Q^2}
)-Q^2  } \;,
\ee
whereas the area of the event horizon is
\be 
A =   4 \pi ( M + \sqrt{ M^2 - Q^2 } )^2 \;.
\ee
If $M=|Q|$, the two horizons coincide and the surface gravity
vanishes. The area is now given in terms of the charge by
\be
A= 4 \pi Q^2 \;.
\ee
The resulting black hole is called the extremal
Reissner-Nordstr{\o}m black hole. If
$M < |Q|$ the event horizon vanishes and one is left with a naked
singularity, which according to the cosmic censorship hypothesis
is considered to be an unphysical solution.  Black hole solutions must
respect  the mass bound $M \geq |Q|$ to ensure the existence of
an event horizon. Extremal black holes saturate the bound.

Finally note that the special
case $M>0, Q=0$ is the Schwarzschild black hole, while 
$M=Q=0$ is flat Minkowski space.

\subsection{The Laws of Black Hole Mechanics}

One of the most remarkable results of black hole physics is
that one can derive a set of laws, called the laws of black hole
mechanics, which have the same structure as the laws
of thermodynamics \cite{BarCarHaw:1973}. 
The black hole laws are a priori not
linked to thermodynamics in any obvious way, because they are 
derived using geometrical properties of event
horizons and general covariance.
We will first describe the black hole laws in the context
of Einstein gravity coupled to matter. Later we will discuss
their generalization to higher derivative gravity.

The zeroth law states that the surface gravity of a 
stationary black hole is constant over the event horizon,
\be
\kappa_S = \mbox{const.}
\ee
This resembles the zeroth law of thermodynamics, which 
says that the temperature is constant in thermodynamic 
equilibrium.
The first law of black hole mechanics is energy conservation.
One considers two infinitesimally close stationary black hole
solutions and expresses the change $\d M$ of the mass in terms
of changes of the area of the event horizon, $\d A$,
of the charge, $\d Q$, and of the angular momentum, $\d J$:
\be
\d M = \frac{\kappa_S}{8 \pi} \d A + \m \d Q + \Omega \d J \;.
\ee
This has the same form as the first law of thermodynamics, and
since $\kappa_S$ is the analogue of temperature the area plays the
role of entropy. This is the fact we are most interested in.
The parameter associated with a variation of the charge is
$\mu = Q/r_+$, where $r_+$ is the location of
the event horizon\footnote{For the Reissner-Nordstr{\o}m 
black hole the outer
horizon is  $r_+ = M + \sqrt{M^2 - Q^2}$ 
in the coordinates used in 
(\ref{staticsphericnonisotropic}).}
and $\Omega$ is the
angular velocity of rotation of the black hole. Formally 
we have the first theorem of thermodynamics 
for a grand canonical ensemble with
$\Omega$ and $J$ playing the roles of pressure and volume 
and $\mu$ and $Q$ playing the roles of chemical potential and
particle number, respectively.
This form of the first law applies to Kerr-Newmann solutions of 
Einstein-Maxwell theory. If a more complicated matter sector
is present, then there are more terms \cite{GibKalKol:1996}.

The analogy of area and entropy is confirmed by the second
law of black hole mechanics, the Hawking area law \cite{Haw:1971}. 
This is a statement about non-stationary processes
in a space-time containing black holes, including collisions
and fusions of black holes.\footnote{A black hole cannot split
into two or more black holes, see for example \cite{Wald:1984}. Note that
the notion of a black hole is not time reflection symmetric even
if the black hole is static, because it is based on the concept
of a future event horizon. One can define analogue space-times
involving past even horizons, which sometimes are called white
holes.} Two assumptions have to be made:
1.) The time evolution of the system must be under sufficient control.
This is implemented by requiring that the space-time is 
'strongly asymptotically predictable'. 2.) The matter, represented
by the stress energy tensor must behave 'reasonable'. This is done by
imposing the null energy condition on the stress energy tensor.
We refer to \cite{Wald:1984} for a more detailed explanation. 
Under these assumptions the second law 
states that the total area of all event horizons is non-decreasing,
\be
\d A \geq 0 \;.
\ee
This is a striking analogue of the entropy law of thermodynamics.

Finally there is a third law of black hole mechanics. Here several
versions exist, and the status of this law does not seem to
be fully understood. We only touch upon this and refer to
\cite{Wald:1984} for a more detailed account.
One version of the law states that the extremal limit
cannot be reached in finite time in any 'physical process'.
Here the obvious problem is to define what a physical process is
and to bring such non-stationary processes under sufficient control. 
Another version, which does not refer to non-stationary properties,
states that black holes of vanishing 
'temperature' (surface gravity) have vanishing entropy.
This is in obvious contradiction to the fact that the area
of an extremal black hole can be non-vanishing. There are however
subtleties at the quantum level, and these have been used as
arguments in favour of the second version of the third law.
We will return to this when discussing quantum aspects of black holes.

\subsection{Quantum Aspects of Black Holes and Black Hole
Thermodynamics \label{SectionQuantumAspects}}

The laws of black hole mechanics have been known for quite some time,
but were mostly considered as a curious formal analogy. The most
obvious reason for not believing in a thermodynamic content is 
that a classical black hole is just black: It cannot radiate
and therefore one should assign temperature zero to it, so that
the interpretation of the surface gravity as temperature
has no physical content.

This changes dramatically when taking into account quantum effects.
One can analyse black holes in the context of quantum field theory
in curved backgrounds, where matter is described by quantum field
theory while gravity enters as a classical background, see for example
\cite{BirDav:1982}. In this
framework it was discoverd that black holes can emit Hawking
radiation \cite{Haw:1975}. The spectrum is (almost\footnote{
When backscattering is taken into account
the spectrum is not really Planckian. 
Nevertheless it makes sense to assign a 
temperature to a black hole, because a black hole can be in thermic
equilibrium with a thermal bath if one puts it into a finite box with
perfectly reflecting walls \cite{BirDav:1982,FroNov:1998,Wald:1994}.}) 
Planckian with a temperature,
the so-called Hawking temperature, which is indeed proportional
to the surface gravity,
\be
T_H = \frac{\kappa_S}{2 \pi} \,.
\ee
This motivates to take the analogy of area and entropy
seriously. Since the Hawking temperature fixes the factor of
proportionality between temperature and surface gravity, one
finds the Bekenstein-Hawking area law,
\be
{\cal S} = \frac{A}{4} \;.
\ee
Before the discovery of Hawking radiation, Bekenstein had
already given an independent argument in favour of assigning
entropy on black holes \cite{Bek:1973,Bek:1974}. 
He pointed out that in a space-time
containing a black hole one could adiabatically transport
matter into it. This reduces the entropy in the observable
world and thus violates the second law of thermodynamics. 
He therefore proposed to assign entropy to black holes,
such that a generalized second law is valid, which states that
the sum of thermodynamic entropy and black hole entropy is
non-decreasing. With the discovery of Hawking radiation one
can give an additional argument in favour of this generalization:
By Hawking radiation a black hole looses mass and shrinks. This is
not in contradiction with the second law of black hole
mechanics, because one can show that the null energy condition
is violated in the near horizon region if the effect of quantum
fields is taken into account. Bekenstein's generalized
second law claims that the loss in black hole
entropy is always (at least) 
compensated by the thermodynamic entropy of the
Hawking radiation, so that the total entropy is non-decreasing.

One example of unusual thermodynamic behaviour of black holes is provided
by the mass dependence of the temperature of uncharged black holes.
For the Schwarzschild black hole one finds 
$\kappa_S = (4M)^{-1}$, which shows that the specific heat is
negative: The black hole heats up while loosing mass. This behaviour
is unusual, but nevertheless not unexpected because gravity is a 
purely attractive force. The fact that uncharged black holes seem 
to fully decay into Hawking radiation leads to the 
information or unitarity problem of quantum gravity, see for example 
\cite{Wald:1994}.
Charged black holes behave differently in that
the Hawking temperature vanishes in the extremal limit. Therefore 
extremal black holes are stable against decay by thermic 
radiation. It is less clear whether they are absolutely 
stable in non-supersymmetric gravity, since they could
decay through charge superradiance \cite{FroNov:1998}.
But in (extended) supergravity one can
argue that they are absolutely stable and provide examples of
solitons, as we will see later.

We already mentioned that one version of the third law states
that extremal black holes have vanishing entropy. This 
statement depends on subtleties of the quantum mechanical 
treatment of such objects \cite{GhoMit:1997,DixKapLou:1991}: 
The entropy can be computed 
in semiclassical quantum gravity, i.e. by quantizing 
gravity around a black hole configuration. One can either use
the Euclidean path integral formulation or the 
Minkowskian canonical framework. The result for the entropy depends on 
whether the extremal limit is taken before or after quantization:
If one quantizes around extremal black holes the entropy
vanishes. But if one quantizes around general charged black hole
configurations one finds an entropy that is non-vanishing when
taking the extremal limit. The second option seems to
be more natural and it is the one supported by string 
theory, as we will see later.

The identification of the area with entropy leads to several
questions. Standard thermodynamics provides a macroscopic
effective description of systems in terms of coarse grained
macroscopic observables like temperature and entropy. At
the fundamental, microscopic level systems are described
by statistical mechanics in terms of microstates which
encode, say, the positions and momenta of all particles that
constitute the system. At the microscopic level one can define
the microscopic or statistical entropy as the quantity 
which characterizes the degenaracy of microstates in a given
macrostate, where the macrostate is characterized by 
specifying the macroscopic observables. Assuming ergodic
behaviour the macroscopic and microscopic entropy agree.\footnote{
It is somewhat difficult to imagine how gravity, which is purely
attractive, can show in some sense ergodic behaviour \cite{Wald:1994}. 
This is one
of the problems in relating thermodynamics and gravity. We will
see that nevertheless one can define and compute the microscopic entropy 
for certain black holes in string theory and that it 
quantitatively agrees with the macroscopic black hole entropy,
as defined by the Bekenstein-Hawking area law and generalizations
thereof.} One should therefore address the question whether there
exists a fundamental, microscopic level of description of black holes,
where one can identify microstates and count how many of them
lead to the same macrostate. The macrostate of a black hole is
characterized by its mass, charge and angular momentum.
Denoting the number
of microstates leading to the same mass $M$, charge $Q$ and
angular momentum $J$ by $N(M,Q,J)$, the statistical or microscopic
black hole entropy is defined by
\be
{\cal S}{\mscr{micro}} = \log N(M,Q,J) \,.
\ee
If the Bekenstein-Hawking entropy is the analogue of
thermodynamic entropy and if stationary black holes 
are the analogue of thermodynamic equilibrium states,
then the Bekenstein-Hawking entropy must coincide with 
the microscopic entropy,
\be
{\cal S} = {\cal S}_{\mscr{micro}} \;.
\ee
We will see that the microscopic picture that string theory
gives us for extremal black holes leads to quantitative
agreement between the two entropies.

One of the astonishing properties of the Bekenstein-Hawking
entropy is its simple and universal behaviour: the entropy 
is just proportional to the area. The fact that the entropy
is proportional to the area and not to the volume has led
to the speculation that quantum gravity is in some sense non-local
and  admits a holographic representation on boundaries of 
space-time \cite{tHo:1993,Sus:1994}.
The D-brane picture of black holes 
\cite{StrVaf:1996,CalMal:1996,HorStr:1996,BreMyePeeVaf:1996,MalStr:1996,JohKhuMye:1996}
(see \cite{Mal:1996} for a review)
and the AdS/CFT conjecture \cite{Mal:1997,GubKlePol:1998,Wit:1998}
(see \cite{AhaEtAl:1999} for a review)
provide steps towards a concrete realization of this proposal.
The amount of information that one can store in a black hole
according to the Bekenstein-Hawking area law is roughly one bit
per Planck length squared. This suggests a picture where
gravity lives on a two-dimensional lattice of Planck length
spacing. It is also remarkable that the Bekenstein-Hawking
entropy is several orders of magnitude bigger than the 
thermodynamic entropy of a star of the same mass \cite{FroNov:1998}. Every 
microscopic picture of black holes has to explain where this
huge number of microstates comes from. In string theory 
four-dimensional extremal black holes are effective descriptions
of complicated bound states of solitons of the underlying ten-
or eleven-dimensional fundamental theory. These configurations have
a huge number of internal excitations which all lead to the
same four-dimensional black hole. This accounts
for the huge entropy.

\subsection{Black Hole Horizons \label{SectionBHhorizons}}

In the next sections we will discuss the generalization of the laws
of black hole mechanics to higher derivative gravity. At the same time
we will explain the definitions of surface gravity and 
macroscopic black hole entropy in some more detail.
We start in this section by reviewing properties of black hole horizons
that apply to both Einstein gravity and higher derivative
gravity. Then we formulate the generalization of the zeroth 
law to higher derivative gravity.

Our discussion of event horizons follows \cite{Wald:1984}.
The event horizon is a hypersurface in space-time.
A space-time hypersurface $\Delta$ can be defined by an equation
$f(x)=0$. Alternatively one can specify the hypersurface
in terms of its normal vector field $n^\m(x)$, where 
$n^\m(x) t_\m(x)=0$ for all tangent vectors $t^\m(x)$ of $\Delta$.
If $\Delta$ is defined in terms of a function $f(x)$, then
$\nabla_\m f$ is automatically normal to $\Delta$. 
According to (a special case of)
the Frobenius theorem a vector field $n^\m$ is the normal vector field
of a smooth family of submanifolds if and only if it is
{\em hypersurface orthogonal},
\be
n_{[\m} \nabla_\n n_{\rho]} =0 \;.
\label{Frobenius}
\ee
A hypersurface is called a null surface if its normal vector 
field is null, $n^\m n_\m =0$. Such surfaces appear as boundaries of
lightcones. Event horizons are per definition null hypersurfaces.
Note that for a null hypersurface a normal vector is at the same
time a tangent vector.

By taking a spacelike cross section $\Sigma$ of $\Delta$ one
obtains a two-dimensional spacelike surface which is 
also called the horizon. We will distinguish both objects by use
of the corresponding symbols. Later on we will make use of the
so-called {\em normal bivector} or {\em binormal} $\ve_{\m\n}$ of $\Sigma$.
In the normal space of $\Sigma$ in space-time one can
construct one linear independent antisymmetric tensor. By imposing
the normalization $\ve_{\m\n} \ve^{\m\n}=-2$ one can make
a unique choice (up to an overall sign which is fixed by choosing 
an orientation). This tensor is the
binormal of $\Sigma$. Since the normal space to $\Sigma$ has
signature $(-+)$ one can  write $\ve_{\m\n}$ as the exterior
product of two null vectors. One can take one of these to be
the normal vector $n_\m$ and chooses a second linear independent
null vector $N^\m$, normalized such that $N^\m n_\m =-1$. Then the
bivector
\be
\ve_{\m\n} = N_\m n_\n - N_\n n_\m
\ee
has its non-vanishing components in the directions normal to
$\Sigma$ and has the required normalization.

If the black hole is static and spherically symmetric, then one
can bring the metric to the form
\be
ds^2 = - e^{2g(r)} dt^2 + e^{2f(r)} ( dr^2 + r^2 d \Omega^2 ) \;,
\ee
where $(t,r, \phi,\theta)$ are isotropic coordinates.\footnote{Isotropic
coordinates have the special property that the spatial part of the 
metric is conformally flat. Note that the above coordinates only cover the
exterior region of the black hole. The event horizon is localized at
$r=0$. The coordinate system (\ref{staticsphericnonisotropic}) used in the
last section also covers the interior of the black hole and $r=0$ is the
postion of the singularity.}
For this metric
the binormal takes the form
\be
\ve_{tr} = - \ve_{rt} = e^{g(r) + f(r)} \;.
\ee

A {\em Killing horizon} is defined to be a null hypersurface which has
a Killing vector field as normal vector field. Note that this
notion is independent from that of an event horizon.
In Einstein gravity all event horizons of stationary black holes
are Killing horizons \cite{HawEll:1973}. In higher derivative gravity one can
show that event horizons are Killing horizons
if the black hole is static or if it is stationary, axisymmetric
and possesses a discrete reflection symmetry, called $t-\phi$ reflection
symmetry\footnote{In terms of coordinates this means
that mixed $t-\phi$ components of the metric can be transformed away.}
\cite{Car:1973,RasWal:1992,RasWal:1995}. 
In the following it is understood that 
event horizons are Killing horizons and in particular
that stationary black holes
in higher derivative gravity are required 
to be in addition axisymmetric and to have $t-\phi$ reflection symmetry.

Remember that Killing vectors are defined by the Killing equation
\be
\nabla_{(\m} \xi_{\n)} = 0 \;.
\label{Killing}
\ee
This has the consequence that the second derivative of a Killing
vector is given by
\be
\nabla_\m \nabla_\n \xi_{\rho}
= R_{\nu \rho \m}^{\;\;\;\;\;\; \sigma} \xi_{\sigma}
\ee
and therefore all higher derivatives of a Killing vector can be
expressed in terms of the Killing vector and its curl 
$\nabla_{[\m} \xi_{\nu]}=\nabla_\m \xi_\n$.

For Killing horizons one can define the surface
gravity $\kappa_S$. If $\xi^\m$ is the Killing vector field
normal to the horizon, or 
horizontal Killing vector field for short, then
one can define $\Delta$ by $f=\xi^\m \xi_\m =0$. Since $\nabla_\m f$
is normal to $\Delta$, it must be proportional to $\xi^\m$ itself.
The coefficient of proportionality defines the surface gravity,
\be
\nabla_\m (\xi^\n \xi_\n ) = -2 \kappa_S \xi_\m \;.
\ee
Using the Frobenius theorem (\ref{Frobenius}) and the Killing equation 
(\ref{Killing}) one can
derive several useful relations. In particular one can show
\be
\kappa_S^2 = - \ft12 ( \nabla^\m \xi^\n ) ( \nabla_\m \xi_\n )
\mbox{   and   }
\xi^\m \der_\m \kappa_S = 0\;.
\ee
The second equation implies that the surface gravity is constant
along the integral lines of the Killing field $\xi^\m$ on
$\Delta$. 
The zeroth law of black hole mechanics states that the surface
gravity is constant over $\Delta$ for static and stationary
black holes \cite{BarCarHaw:1973,RasWal:1992,RasWal:1995}.

For static black holes one can give an alternative definition
of the surface gravity \cite{Wald:1984}. 
One considers the problem of keeping a
test body at rest on the event horizon. Though the local
force on the body is infinite (which is another way of defining
an event horizon) the force on an external observer who
tries to keep the testbody fixed is redshifted and finite.
If one fictionally refers to an observer at infinity then
the redshifted force per mass equals the surface gravity as
defined above in terms of the Killing vector. In this sense
the surface gravity measures the force of the gravitational
field on the event horizon. Note that this definition
does not apply to stationary, rotating black holes, because
in this case it is not possible to keep a body at rest inside
the ergosphere. Nevertheless the surface gravity, as defined
by the Killing vector is an intrinsic property of the horizon
and characterizes, in a less direct way, the gravitational
field.

\subsection{Surface Charges \label{SectionSurfaceCharges}}

In order to define black hole entropy in a way that applies
to higher derivative gravity one can make use of a formalism
which associates surfaces charges to local symmetries of the
Lagrangian. 
If the surface charge is evaluated in an on-shell background 
with a rigid symmetry
then one gets a Noether charge in the usual sense, i.e. the generator
of the rigid symmetry of the background. The black hole entropy
is an example of such a Noether charge, and the residual gauge
symmetry is the isometry generated by the horizontal Killing vector field.

The method of Noether surface charges was proposed by
R. Wald in \cite{Wald:1993}, based on \cite{LeeWald:1990,Wald:1990}
and further developed by him and other authors in
\cite{JacKanMye:1993,IyeWal:1994,JacKanMye:1995}.  
In this section we will review a modified 
algorithm for computing surface Noether charges that was described in 
\cite{CardWMoh:1999/04,deW:1999}.

We will consider a generally covariant Lagrangian that
depends on the Riemann tensor but does not contain derivatives
of the Riemann tensor.\footnote{The algorithm of \cite{IyeWal:1994}
can be used when the Lagrangian depends on covariant derivatives of
the Riemann tensor, whereas our algorithm has so far only been developed
for a restricted class of Lagrangians, which contains
the ${\cal N}=2$ supergravity Lagrangians that will be studied later.}
In addition the Lagrangian depends on a matter 
field $\psi_{\m\n}$ (which is a second rank tensor with no particular
symmetry) and on its first derivative. 
The Lagrangian is not strictly invariant
under general coordinate transformations, but transforms
into a total derivative. Consider now a variation of the fields 
$\phi = ( g_{\m \n}, \psi_{\m \n})$ that takes the form of a 
general coordinate transformation multiplied with a test function
$\e(x)$:
\bea
\delta_{\xi}
 g_{\mu\nu}  &=& - \epsilon(x) \Big(\nabla_{\!\mu\,} \xi_\nu(x) +
\nabla_{\!\nu\,} \xi_\mu(x)\Big)  \,, \nonumber\\
\d_{\xi} \psi_{\m\n} &=& -\epsilon(x) \Big(\nabla_{\!\m\,} \xi^\sigma \,
\psi_{\sigma\n}
+ \nabla_{\!\n\,} \xi^\sigma \,\psi_{\m\sigma}  +
\xi^\sigma\,\nabla_{\!\sigma} \psi_{\m\n}\Big)   \,.  
\eea
Such a transformation is not a symmetry of the Lagrangian,
but if $\e(x)$ and its derivatives satisfy the boundary conditions
required by Hamiltons variational principle, then the variation is
proportional to the equations of motion. Therefore the variation is
a total derivative $\der_\m J^\m$, modulo the equations of motion.
This provides the definition of the Noether current 
associated with the local transformation generated by $\xi_\m(x)$.\footnote{
As discussed in \cite{CardWMoh:1999/04} this definition differs from the 
algorithm of \cite{Wald:1993,JacKanMye:1993,IyeWal:1994}
by improvement terms. This reflects ambiguities
in the construction of the Noether current which, however, 
do not affect the definition of entropy as a Noether charge.} 
Explicitly one finds
\bea
J^\mu &=& \xi^\mu \,{\cal L} - 2 \,{\cal L}^{\mu\nu\rho\sigma}\Big[
R_{\lambda \nu\rho\sigma} \,\xi^\lambda + \nabla_{\!\nu} \nabla_{\!\rho\,}
\xi_\sigma \Big] 
 +4 \,\nabla_\rho  {\cal L}^{\mu\nu\rho\sigma}\, \nabla_{(\nu\,}\xi
_{\sigma)} \nonumber \\
&& - {\cal L}_\psi^{\m,\rho\sigma} \, \Big[ \nabla_{\!\rho\,}\xi^\lambda
\,\psi_{\lambda\sigma}  + 
\nabla_{\!\sigma\,} \xi^\lambda\, \psi_{\rho\lambda}  + \xi^\lambda
\,\nabla_{\!\lambda}  \psi_{\rho\sigma} \Big]  \nonumber\\
&& + \ft{1}{2} (\nabla_{\!\lambda\,} \xi_\rho +
\nabla_{\!\rho\,} \xi_\lambda)
\Big[ {\cal L}_\psi^{\m,\rho\sigma} \psi^{\lambda}{}_{\sigma}
+ {\cal L}_\psi^{\m,\sigma \rho} \psi_{\sigma}{}^{\lambda}
+ {\cal L}_\psi^{\rho,\m\sigma} \psi^{\lambda}{}_{\sigma} \nonumber\\
&& \hspace{35mm} +  {\cal L}_\psi^{\rho,\sigma \m} \psi_{\sigma}{}^{\lambda} 
 -  {\cal L}_\psi^{\rho,\lambda \sigma} \psi^{\mu}{}_{\sigma}
-  {\cal L}_\psi^{\rho,\sigma \lambda} \psi_{\sigma}{}^{\mu} \Big]
\,, \label{gravNcurrent}
\eea
where
\be
{\cal L}^{\mu\nu}= {\pa{\cal L}\over \pa g_{\mu\nu}} \,,\qquad 
{\cal L}^{\mu\nu\rho\sigma} = {\pa{\cal L}\over \pa
R_{\mu\nu\rho\sigma}} \,, \qquad {\cal L}_\psi^{\rho,\mu\nu} = {\pa
{\cal L}\over \pa \nabla_{\!\rho} \psi_{\m\n}}\,,
\ee
are partial derivatives of the Lagrangian\footnote{In this and in the
following sections ${\cal L}$ is a function, not a density:
$S = \int d^4 x \sqrt{-g} {\cal L}$. In the later chapters we follow
supergravity conventions and absorbe the density $\sqrt{-g}$ in
the Lagrangian.}
with respect to the
tensors $g_{\m\n}, R_{\m\n\rho\sigma}$ and $\nabla_{\rho} \psi_{\m\n}$.
Since these tensors have symmetries their components
are not independent and the definition of a partial derivative is
ambiguous. To fix this we impose 
that the derivative has the same symmetry properties as the tensor,
i.e. the partial derivative is determined by the variation of the
Lagrangian as a function of the tensor field,
\be
\d {\cal L} = \frac{\der {\cal L}}{ \der T_{\m_1 \cdots \m_n}}
\d T_{\m_1 \cdots \m_n} + \cdots \;.
\ee
The Noether current is a local function of the fields $\phi$ and
depends on the transformation parameter $\xi_\m(x)$ only through
the variations $\d_{\xi} \phi$.

One can directly verify, using the equations of motion, that the
Noether current is conserved on-shell. 
Moreover one can use the equations of motion 
to write the current in terms of a Noether protential
$Q^{\m \n} = Q^{[\m \n]}$ as $J^\m = \nabla_\n Q^{\m \n}$,
so that current conservation becomes trivial.
The explicit expression for the Noether potential for the case at hand
is
\bea
{Q}^{\mu\n} &=& 
- 2\, {\cal L}^{\mu\nu\rho\sigma} \,\nabla_{\!\rho\,}\xi_\sigma 
+ 4 \,\nabla_{\!\rho\,}{\cal L}^{\mu\nu\rho\sigma} \, \xi_\sigma \nonumber\\
&& +\ft12 \Big[-{\cal L}_\psi^{\m,\n\rho} \,\psi^\sigma{}_{\rho} 
-{\cal L}_\psi^{\m,\rho\n} \,\psi_{\rho}{}^\sigma 
+{\cal L}_\psi^{\m,\sigma\rho} \,\psi^\n{}_{\rho} \nonumber \\
&&\hspace{9mm} 
+{\cal L}_\psi^{\sigma,\m\rho} \,\psi^\n{}_{\rho}
+{\cal L}_\psi^{\m,\rho\sigma} \,\psi_{\rho}{}^\n 
+{\cal L}_\psi^{\sigma,\rho\m} \,\psi_{\rho}{}^\n -
(\mu\leftrightarrow\n)  
 \Big] \,\xi_\sigma  \,.
\label{noetgrav}
\eea
The Noether potential is not uniquely defined because one can
add terms of the form  $\nabla_{\rho} X^{\mu \n \rho}$, where 
$X^{\m \n \rho} = X^{[\m \n \rho]}$.
It is a local function of the fields and of the transformation 
parameter: $Q^{\m\n}= Q^{\m \n}(\phi,\xi)$.

The surface charge associated to the transformation $\xi_{\mu}(x)$
is defined by integrating the Noether potential
over a closed spacelike surface $\Sigma$,
\be
Q = \oint_{\Sigma} Q^{\m \n} d \Sigma_{\m \n} \;.
\ee
The surface charge is a Noether charge in the usual sense if
$\xi_{\mu}(x)$ is a Killing vector. In this case the
transformation parametrized by $\xi_{\mu}(x)$ is a
symmetry of the background.
The Noether potential
is proportional to a field dependent linear combination 
$\xi^\m$ and its curl $\nabla_{[\m} \xi_{\n]}$,
because all higher derivatives of the Killing vector can be expressed 
through these two terms.\footnote{We neglect improvement terms
$\nabla_\rho X^{[\m \n \rho]}$ in the following, because they do not
contribute to the Noether charge.}

\subsection{The First Law in Higher Derivative Gravity}

We now turn to the first law of black hole dynamics in the context
of higher derivative gravity. As we will see the requirement that
the first law is valid determines the expression for the entropy.
In presence of higher curvature terms the validity of the first law
implies that 
one has to deviate from the Bekenstein-Hawking area law. 
The idea for the derivation of the first law 
and of the construction of the entropy can be motivated by analysing 
the content of the first law in Einstein gravity
\be
\d M = \frac{\kappa_S}{2 \pi} \d {\cal S} + \Omega \delta J.
\ee
Note that this is an astonishing statement because it relates
variations of the quantities defined at infinity, the mass $M$
and angular momentum $J$, to quantities defined on the horizon,
$\kappa_S$ and ${\cal S}$, without making reference to the behaviour
of the field configuration in the bulk of space time. This suggests that
the first law relates
the variations
of surface charges associated with infinity to variations at
the event horizon.

In the above version of the first law
we omitted the contributions from electric
and magnetic charges and from scalar fields. The presence of matter
fields enters into the first theorem in two ways. First there is
an explicit contribution to the mass. Since we are interested in
the structure of the entropy term, we do not need to care about this.
Second the presence of matter fields modifies the equations of 
motion and this relevant for our discussion because the first
theorem refers to solutions to the field equations.
But in section \ref{SectionSurfaceCharges} we derived the formula for
surface charges in on-shell backgrounds, and this formula holds
irrespective of the details of the equations of motion. Since 
we will see that the entropy is such a surface charge, we do not
need to consider the possible additional terms in the first law.

We will now review the proof of the first law given in
\cite{Wald:1993}.  
Let us consider a continous family of static or stationary 
black hole solutions.
The variation connecting two infinitesimally close solutions is 
denoted by $\delta$. All solutions in the family have the same
symmetries and $\delta$ relates the corresponding 
Killing vectors and Noether currents. 
The natural surface charge in a black hole
space-time is the one associated with the horizontal Killing 
vector field. For static
black holes this is just the static Killing vector field of
time translation invariance, $\xi = \ft{\der}{\der t}$, 
where as for rotating, stationary black holes it is a linear combination
of the static and the axial Killing vector field,
\be
\xi = \frac{\der}{\der t} + \Omega \frac{\der}{\der \phi} \;.
\ee
In the last section we discussed the construction of surface charges
both for general transformations and for the special case of 
Killing vectors. The corresponding variation of the geometry was
denoted $\d_{\xi}$.\footnote{This should not be confused with the deformation
$\d$ which relates a family of solutions.}

The central observation for deriving the first law is that the
motion along the integral lines of the horizontal Killing vector field
is generated by a Hamiltonian $H$, which has the property that its
variation along the continuous family of solution vanishes,
$\delta H =0$ \cite{Wald:1993}. 
Before displaying the explicit form of $\d H$ we need
a few definitions. First we introduce a spacelike Cauchy surface $C$,
which has two boundary components. One is an asymptotic two-sphere
at infinity, $S^2_{\infty}$, the other is a spacelike cross-section
$\Sigma$ of the horizon hypersurface $\Delta$. Second we define
a vector field $\theta^\m$ by considering the variation of the
action under $\delta$ which takes the form 
\be
\delta S = \int d^4 x \der_\m  \left( \sqrt{-g} \theta^\m \right)
\ee
after imposing the equations of motion.
The variation of the Hamiltonian is \cite{Wald:1993}
\be
\d H = \d \left(  \int_C d\Omega_\m J^\m \right) - 
\int_C d \Omega_\m \nabla_\n ( \xi^\m \theta^\n - \xi^\n
\theta^\m ) \;,
\ee
where $d \Omega_\m $ is the volume element on $C$ and $J^\m$ is the
Nother current associated with the horizontal Killing vector field
$\xi^\m$. Using the equations of motions one can rewrite the Noether current
in terms of the Noether potential and use Stokes theorem,
\be
\d H = \int_{\der C} d \Sigma_{\mu \nu} \left(
\d Q^{\m\n} - \xi^\m \theta^\n + \xi^\n \theta^\m \right) \;.
\ee
Next one uses $\d H = 0$ and 
separates the contributions from the two boundary components
to get a relation between the surface charges at infinity and at the
horizon\footnote{The normal is taken with inward direction relative to $C$
for $\Sigma$ and with outward direction for $S^2_{\infty}$.}
\be
\int_{\Sigma}  d \Sigma_{\mu \nu} \left(
\d Q^{\m\n} - \xi^\m \theta^\n + \xi^\n \theta^\m \right) =
\int_{S^2_{\infty}} d \Sigma_{\mu \nu} \left(
\d Q^{\m\n} - \xi^\m \theta^\n + \xi^\n \theta^\m \right) \,.
\ee
This is a relation which has the structure of the first law and it
remains to find the physical interpretation of the terms involved.
We start with the terms at infinity. When decomposing the horizontal
Killing vector field into the static part $\xi_{(t)}^\mu$
and the axial part $\xi_{(\phi)}^\m$ then the variation at infinity
takes the form $\d M - \Omega \d J$, where
\bea
\d M &=&\int_{S^2_{\infty}} d \Sigma_{\m\n} \left(
\d Q[\xi_{(t)}]^{\m \n} - \xi_{(t)}^\m \theta^\n
+ \xi_{(t)}^\n \theta^\m \right) \;,\nonumber \\
\d J &=& \int_{S^2_{\infty}} d \Sigma_{\m\n} \left(
\d Q[\xi_{(\phi)}]^{\m \n} \right) \\
\nonumber
\eea
and $Q[\xi_{(t)}]$, $Q[\xi_{(\phi)}]$ are the Noether charges associated with
the static and axial Killing vector. Note that there is no
contribution proportional to $\xi_{(\phi)}^{[\m} \theta^{\n]}$ 
in the second
line because we have taken the intergration surface tangential to
the axial Killing vector field. The above expressions suggest to
interpret $\d M$ and $ \d J$ as the variations of the mass and of
the angular momentum of the black hole.
A priori it is  not clear how to define these quantities in
higher derivative gravity. If one can find a vector field $b_\m$
such that
\be
\d \int_{S^2_{\infty}} \xi_{(t)}^\m b^\n -
 \xi_{(t)}^\n b^\m = 
\int_{S^2_{\infty}} \xi_{(t)}^\m \theta^\n -
 \xi_{(t)}^\n \theta^\m \;,
\ee
then one can use
\bea
M &=&\int_{S^2_{\infty}} d \Sigma_{\m\n} \left(
Q[\xi_{(t)}]^{\m \n} - \xi_{(t)}^\m b^\n
+ \xi_{(t)}^\n b^\m \right) \;, \nonumber \\
J &=& \int_{S^2_{\infty}} d \Sigma_{\m\n} 
Q[\xi_{(\phi)}]^{\m \n} \label{KomarExpressions} \\
\nonumber
\eea
as definitions of mass and angular momentum. 
In the case of
Einstein gravity these expressions correspond to the Komar expressions
for mass and angular momentum \cite{Wald:1993}. 
The Komar formulae of general relativity
apply for mass and angular momentum in space-times with Killing vectors
and coincide with the more general ADM expressions which do only
assume asymptotic symmetries, namely asymptotic flatness \cite{Wald:1984}.
It is therefore natural to take (\ref{KomarExpressions}) as the definitions
of mass and angular momentum in higher derivative gravity.

Next we turn our attention to the variation of the charge on the horizon.
If this is the entropy term we are looking at, it must take the form
$\frac{\kappa_S}{2 \pi} \d {\cal S}$. Obviously one should consider
a non-singular, i.e. non-extremal situation, where the surface gravity
$\kappa_S$ is finite. This will define the entropy of non-extremal
black holes and in the spirit
of the discussion given in section \ref{SectionQuantumAspects} 
the entropy of extremal black holes
is defined by taking the extremal limit of the resulting expression.

Now we analyse the Noether potential on the horizon $\Sigma$.
Since the Noether potential depends linearly on $\xi^\m$ and its
derivatives, it takes the general form
\be
Q^{\mu \nu} = B^{[\mu} \xi^{\nu]} + C \nabla^{[\m} \xi^{\nu]}\;, 
\ee
where $ B^{\mu}$ and $C$ are local functions of the 
fields in the Lagrangian. For the special case of a Lagrangian depending
on the Riemann tensor but not on its derivatives we found the
explicit expression in section \ref{SectionSurfaceCharges}. 
For Lagrangians which also depend on covariant derivatives of the
Riemann tensor the explicit form of $Q^{\m\n}$ can be found in
\cite{IyeWal:1994}.
Next we use the Frobenius theorem to decompose the curl of the
Killing vector as
\be
\nabla_{[\m} \xi_{\n]} = \kappa_S \ve_{\m \n} + t_{[\mu} \xi_{\n]} \;,
\ee
where $\ve_{\m \n}$ is the binormal and $t^\m$ is a tangent vector
of $\Sigma$. Note that by the Frobenius theorem the non-vanishing
components of the tensor $\nabla_{[\m} \xi_{\n]}$ must have a
non-tangential part. The normalization of the purely normal piece
proportional to the binormal is fixed by the definition of surface
gravity.

The analysis is simplified by a theorem of \cite{RasWal:1995} that states
that the horizon $\Delta$ of a non-extremal black hole 
can be analytically continued such that it becomes a bifurcate
horizon. Bifurcate horizons contain a bifurcation surface $\Sigma_0$.
This is a spacelike surface on which the horizontal Killing vector field
vanishes. Examples of bifurcate horizons are provided by
the maximally extended Schwarzschild and non-extremal Reissner-Nordstr{\o}m
solutions of Einstein-Maxwell gravity. Note that the analytical 
extension is used as a technical tool. It is not relevant whether the
part of space-time containing $\Sigma_0$ is physical.

The special properties of the bifurcation surface can now be
used to simplify the computation of the black hole entropy. 
Since $\d H=0$ is valid for any choice of the Cauchy surface we can
take $\Sigma_0$ as the interior boundary. The surface charges are
by construction conserved under the evolution generated by $H$ and
therefore the horizontal surface charge is the same on any spatial
section $\Sigma$. 
We first  note that by using $\xi^\m=0$ on $\Sigma_0$ 
we get $\nabla_\m \xi_\n = \kappa_S \ve_{\m \n}$
on $\Sigma_0$. In addition  the contribution
$\xi^{\m} \theta^\n - \xi^{\n} \theta^\m$ to the variation
vanishes. Finally define a rescaled Noether potential and
charge by
\be
\tilde{Q}^{\mu \nu} = \kappa_S^{-1} Q^{\mu \nu} \mbox{   and   }
\tilde{Q} = \int_{\Sigma_0} d \Sigma_{\m\n} \tilde{Q}^{\m\n}\;,
\ee
so that the variation at the horizon takes the form
$\ft{\kappa_S}{2 \pi} 2 \pi \d \int_{\Sigma_0} \tilde{Q}$.
Note that on the bifurcation surface $\Sigma_0$ the
rescaled Noether charge takes the simplified form
$\tilde{Q}^{\m\n} = C \ve^{\m \n}$.

Collecting all results we obtain the first law,
\be
\frac{\kappa_S}{2  \pi} \delta {\cal S} = \delta M - \Omega \d J \;,
\ee
with entropy
\be
{\cal S} = 2 \pi \int_{\Sigma_0} d \Sigma_{\m \n} \tilde{Q}^{\m \n} \;.
\ee

Outside the bifurcation surface the rescaled Noether potential
is more complicated, because $\xi_{\mu}$ does not vanish.
But it
has been shown in \cite{JacKanMye:1993} that the terms 
in $\tilde{Q}^{\m\n}$ that vanish on $\Sigma_0$ do not contribute
to the entropy when the integral is evaluated on a general spatial
cross section $\Sigma$. Thus, the entropy is given by
\be
{\cal S} = 2 \pi  \int_{\Sigma} d \Sigma_{\mu \nu} \tilde{Q}^{\mu \nu} \;,\;\;
\tilde{Q}^{\m\n} = C \ve^{\m\n}\;,
\ee
where $\Sigma$ is an arbitrary spatial cross section of $\Delta$.
Note that we droped the additional non-vanishing terms in
$\widetilde{Q}^{\m\n}$, because they do not contribute to the
entropy.

If the Lagrangian does not depend on derivatives of the Riemann tensor,
then the rescaled Noether potential is given by \cite{JacKanMye:1993}
\be
\tilde{Q}_{\mu  \nu} = - \frac{ \der {\cal L}}{ \der R_{\m \n \rho \sigma}}
\ve^{\rho \sigma} \,,
\ee
up to terms proportional to $\xi$. This result also follows using our
formula (\ref{noetgrav}) and
taking into account a conventional normalization 
factor $2$.\footnote{The relative 
normalizations of the Noether charges in both algorithms is not
known a priori. In the approach of \cite{CardWMoh:1999/04,deW:1999}
the construction of the Hamiltionian $H$ 
and the derivation  the first law remain to be done. Therefore the
normalization of the Noether charge has to be fixed 
by either comparing to the results of 
\cite{JacKanMye:1993,IyeWal:1994}
or by imposing that for the Einstein-Hilbert Lagrangian the
Bekenstein-Hawking area law has the correct normalization.} 
Thus the entropy is given by
\be
{\cal S} = 2 \pi \int_{\Sigma} \frac{ \der {\cal L}}
{ \der R_{\m \n \rho \sigma}} \ve^{\mu \nu} \ve^{\rho \sigma} 
\sqrt{h} d\Omega \;.
\label{WaldEntropy}
\ee
It is instructive to check that one indeed gets the 
Bekenstein-Hawking area law in the case of Einstein gravity.
Then the Lagrangian is $8 \pi {\cal L} = - \ft12 R$ and therefore
$8 \pi \ft{\der {\cal L} }{\der R_{\m \n \rho \sigma}} = - \ft12 g^{\m[\rho}
g^{\sigma] \n}$. Now consider a static black hole 
in isotropic coordinates. The
non-vanishing components of the binormal are $\ve_{tr} = - \ve_{rt}$,
whereas the metric is diagonal. Therefore the contraction with
the binormal yields
$8 \pi \ft{\der {\cal L} }{\der R_{\m \n \rho \sigma}} = - (\ve_{tr})^2
g^{tt} g^{rr}$. Now we use the explicit form of these expressions,
$\ve_{tr} = e^{g+f}$, $g^{tt} = - e^{-2g}$ and $g^{rr} = e^{-2f}$
with the result that
\be
{\cal S} = \frac{2 \pi}{8 \pi} \int_{\Sigma} \sqrt{h} d \Omega
= \frac{A}{4} \;.
\ee

Finally we mention some more results on entropy in higher derivative
gravity. For Lagrangians which depend on derivatives of the Riemann
tensor the entropy formula contains additional terms \cite{IyeWal:1994}.
An alternative way of defining the entropy is provided by 
semiclassical Euclidean methods. These also predict deviations from
the Bekenstein Hawking area law and the results are equivalent to those
of the Minkowskian approach, in all cases where both methods apply 
\cite{IyeWal:1995}.
Finally one can try to extend the definition of entropy to non-stationary
black holes. This way it has been argued that the second law 
should be valid in higher derivative gravity \cite{IyeWal:1994}, 
though there seems to be no complete proof. We refer to 
\cite{Wald:1997,Wald:1999}
for recent reviews and more references 
concerning the laws of black hole mechanics.

\subsection{Extremal Black Holes as Supersymmetric Solitons}

Extremal black holes have very special properties and it turns out
that these can be understood in terms of a symmetry principle,
namely supersymmetry. After embedding gravity into extended 
supergravity, extremal black holes provide examples of supersymmetric
solitons. We will consider the explicit example of the extremal
Reissner-Nordstr{\o}m black hole of Einstein-Maxwell theory,
following the orginal work of \cite{Gib:1981,GibHul:1982}, but
with the notation and conventions that we will use in chapter
\ref{ChapterN=2BlackHoles} to analyse more general black holes.

Let us first collect what is special about 
extremal black holes. We already mentioned that they saturate the mass 
bound defined by the existence of an event horizon.
Therefore the mass is determined by the charge 
carried by the black hole. Next the surface gravity vanishes, and
according to the thermodynamic interpretation this means that they
have zero temperature and  are stable against Hawking radiation.
Note that this property makes them particularly interesting for
finding a microscopic interpretation of the black hole entropy.
Closer inspection reveals further particular properties. Consider
for example the asymptotic geometry at the event horizon. In
isotropic coordinates, where the horizon is located at $r=0$ the
asymptotic metric is
\be
ds^2 = - \frac{r^2}{Q^2}dt^2 + \frac{Q^2}{r^2} 
( dr^2 + r^2 d\Omega^2 ) \;.
\ee
This metric is known as the Bertotti-Robinson 
solution \cite{Ber:1959,Rob:1959}.\footnote{
Actually, this is one particular case of the class of metrics 
studied by Bertotti and Robinson. The class consists of all
solutions of the Einstein-Maxwell theory with a covariantly constant
electromagnetic field strength. The above solution is a special case with
vanishing cosmological constant and without charged matter ('dust').}
By switching to a new radial variable $\rho$, where $Q^2/\rho^2= r^2/Q^2$,
or, alternatively, by computing the Weyl tensor one varifies that this
geometry is conformally flat. Moreover the curvature scalar vanishes,
so that only the traceless part of the Ricci tensor is non-trivial.
By inspection of the metric one sees that space-time
factorizes into a 
two-sphere, parametrized by $(\phi,\theta)$ and a second surface
parametrized by $(t,r)$. This second factor is two-dimensional
Anti-de Sitter space, $AdS^2 = SO(2,1)/SO(1,1)$. Both the sphere and
the $AdS$-space have the same curvature radius and since the $AdS$-space
has negative curvature the total curvature scalar is zero, as mentioned
above.

Another special property of the extremal Reissner-Nordstr{\o}m black hole
is that it has a multi-centered generalization, i.e. there exist 
static configurations of black holes, which can be 
placed at arbitrary positions in space as long as their horizons do 
not overlapp and all their charges have equal sign. Then the gravitational
attraction and electrostatic repulsion cancel precisely irrespective of
the position. The corresponding metric belongs to the class of metrics
which was discovered by Majumdar and Papapetrou \cite{Maj:1947,Pap:1947},
\be
ds^2 = - H^{-2}(\vec{x}) dt^2 + H^2(\vec{x})  d \vec{x}^2 \;,
\ee
where $H(\vec{x})$ is a harmonic function with respect to the
Laplacian $\Delta=\sum_{i=1}^3 \der_{x^i} \der_{x^i}$. The most general
choice of $H$ that does not lead to naked singularities is the 
multi-centered extremal Reissner-Nordstr{\o}m solution \cite{HarHaw:1972}:
\be
H(\vec{x}) = 1 + \sum_{i=1}^N \frac{ M_i }{ | \vec{x} - \vec{x}_i | } \;,
\ee
where $M_i, \vec{x}_i$ are the mass and position (of the event horizon)
of the i-th black hole. Since no net forces act between the
black holes the mass is additive,
\be
M = \sum_{i=1}^N M_i = \sum_{i=1}^{N} \sqrt{Q_i^2 + P_i^2} \;,
\ee
where in the second equality we took a general, dyonic charge
configuration with electric charges $Q_i$ and magnetic charges $P_i$. 
In this case the charges must have 'equal sign'
in the sense that all the complex 
numbers $Q_i + i P_i$ must have the same phase.

We will now review how these special properties of the extremal
Reissner-Nordstr{\o}m black hole can be understood in terms of
supersymmetry. First we should explain our use of the term
'soliton'. Usually a soliton is
a static solution of the classical equations of motion that has finite
energy and is regular everywhere. Sometimes one also includes the 
requirement that the soliton can be argued to be quantum mechanically 
stable. Very often solitons interpolate between two different vacua of the
underlying theory. We refer to \cite{Raj:1982} for an overview of 
solitons and instantons in field theory.

In the context of gravity a less restrictive definition is adequate.
Solitons are still required to be
static finite energy solutions of the classical
equations of motion.\footnote{This generalizes to 
solutions with $p$ non-compact spatial isometries ($p$-branes)
by replacing 
'finite energy' by 'finite energy per world volume'.}
'Regular' is interpreted as 'having no naked singularities', so that
solutions where singularities are covered by event horizons 
are admitted. For extremal black holes of extended
supergravity theories one can give an argument in favour of their
quantum mechanical stability that will be reviewed below. 
Finally, we will see
that extremal black holes indeed interpolate between two distinct 
vacua of extended supergravity.

Next we have to explain what we mean by a 'supersymmetric' soliton.
To do so we have to introduce the notion of a BPS state and review
a bit of the representation theory of the supersymmetry algebra,
see for example \cite{WesBag:1992}. We consider
four-dimensional ${\cal N}=2$ supersymmetry, because we are interested
in four-dimensional theories and in four dimensions this is the
most general case where BPS states can be defined.
The supersymmetry algebra can be
brought to the form 
\bea
\{ Q^i_{\alpha}, Q^{j+}_{\beta} \} &=& 2 P_\m \sigma^\m_{\alpha \beta} 
\d^{ij}\;, \\
 \{ Q^i_{\alpha}, Q^{j}_{\beta} \} &=& 2 Z \ve_{\alpha \beta} 
\ve^{ij} \;,
\eea
where we parametrized the supersymmetry generators by two Weyl 
spinors $Q^i_{\alpha}$. The indices $\alpha,\beta,..=1,2$ are Weyl
spinor indices, whereas $i,j,\ldots = 1,2$ count the supercharges.
We denote the hermitean conjugated charges by $Q^{i+}_{\alpha}$ 
and refrain from using
dotted indices for the opposite chirality representation. As in all
cases of extended supersymmetry, the algebra contains central charges,
i.e. operators that commute with all generators of the algebra.
In our case there is one complex central charge $Z$. It follows 
directly from the algebra that all massive representations 
satisfy a mass bound, which is given by the central charge,
\be 
M \geq |Z| \;.
\ee
Moreover states which saturate this bound,
\be
M = |Z|\,,
\ee
have special properties. 
The structure of unitary
representations is different
for the cases $M>|Z|$ and $M=|Z|$. In the case $M=|Z|$ half of the
supercharges act trivially, and therefore the multiplet is generated
by $2^{\cal N}= 2^2$ creation operators instead of $2^{2 {\cal N}}=
2^4$ in the generic case. This is referred to as multiplet shortening,
and the multiplets are called short and long multiplets, respectively.
For short multiplets the mass is termined by
the central charge. This relation cannot be changed by perturbative
or non-perturbative corrections (assuming that the full theory is
supersymmetric).

So far we have only considered the supersymmetry algebra
and now turn to supersymmetric field theories. 
The fundamental fields in the Lagrangian can belong
to either short or long representations. Two important short 
multiplets occuring in ${\cal N}=2$ Super-Yang-Mills theories 
and in ${\cal N}=2$ string compactifications are the short vector
multiplet and the hypermultiplet. The on-shell degrees of freedom of
a short vector multiplet are a (massive) vector, two Weyl fermions
and a real scalar. A hypermultiplet contains two Weyl fermions and
four real scalars. Whereas there is no 'long' version of the hypermultiplet,
there is a long vector multiplet which contains the combined on-shell
degrees of freedom of a short vector multiplet plus a hypermultiplet.

The short vector multiplet and the 
hypermultiplet have the same degrees of freedom 
as the corresponding massless multiplets.
This is related to the two types of the Higgs mechanism that exist in
${\cal N}=2$ gauge theories, see for example \cite{AlvGauHas:1997}. 
The scalar potential has flat 
directions which correspond to vacuum expectation values for 
either the scalars in short vector multiplets or to those in
hypermultiplets. If one gives vacuum expectation values to
scalars in vector multiplets in a generic way, then all charged
massless vector multiplets become short massive vector multiplets
and the gauge group is broken to its maximal torus. If one gives
vacuum expectation values to scalars in hypermultiplets in a generic
way then pairs of massless vector and hypermultiplets combine
into long vector multiplets and the gauge group is broken completely,
if sufficiently many hypermultiplets are present.
These two generic kinds of flat directions constitute the Coulomb and the
Higgs branch of the theory, respectively. Note that the two mechanisms
tend to exclude one another. Once one is at a generic point 
in the Coulomb branch
(Higgs branch) it is no longer possible to give a generic
vacuum expectation
value to scalars in hypermultiplets (vector multiplets).  Depending
on the precise field content there might be mixed branches where 
non-generic vacuum expectation values can be simultanously turned on
for scalars both in vector and hypermultiplets, see for example
\cite{AlvGauHas:1997} for more details.

We now turn to solitons and start with rigid supersymmetric theories.
Like the fundamental fields in the Lagrangian the  solitons 
of ${\cal N}=2$ Super-Yang-Mills theory can sit in
short multiplets. For example the 't Hooft-Polyakov-monopoles 
of Super-Yang-Mills are in short hypermultiplets.\footnote{Due to the
existence of flat directions one is automatically
in the Prasad-Sommerfield limit and the Bogomol'nyi bound of Yang-Mills
theory can be understood as a supersymmetric mass bound.}
This plays a prominent role in the Seiberg-Witten analysis of
${\cal N}=2$ gauge theories, see again \cite{AlvGauHas:1997} for more
details.  In non-supersymmetric
gauge theories monopoles 
saturating the Bogomol'nyi mass bound are called 
Bogomol'nyi-Prasad-Sommerfield or BPS solitons. The special
properties of such objects can be understood in terms of 
supersymmetry after extending the theory to an ${\cal N}=2$
supersymmetric theory. Therefore the terminology has been 
transferred to supersymmetric theories. In particular the
supersymmetric mass bound is also called the BPS bound and
states saturating it are called BPS states.

We now turn to ${\cal N}=2$ supergravity where
we have to take into account
that supersymmetry is realized as a local symmetry. 
Einstein-Maxwell theory can be naturally embedded into pure
${\cal N}=2$ supergravity by adding two Majorana spin $3/2$ fields,
the gravitini. The gravitini $\psi^i_\m$, 
the graviton (vielbein) $e_\m^{\;\;a}$
and the photon $A_{\m}$ 
form the ${\cal N}=2$ supergravity multiplet. 
($i=1,2$ is the internal ${\cal N}=2$ index, $\m$ is a world index and
$a$ is a tangent space index.)
In this context the
gauge field $A_\m$ is usually called the graviphoton. 
In ${\cal N}=2$ supergravity the central
charge transformations are local $U(1)$ symmetries, with the 
graviphoton as gauge field. As a consequence the central charge carried
by a field configuration is related to its electric and magnetic
charge by $Z=Q+iP$ (modulo a conventional phase) \cite{Tei:1977}. 
Thus the black hole
mass bound $M = \sqrt{Q^2 + P^2}$ 
translates into the supersymmetric one, $M=|Z|$.

Whereas it is obvious that the black hole mass bound should
be interpretable as a supersymmetric mass bound in the supergravity 
theory, it is less obvious what we mean by a supersymmetric state
or supersymmetric soliton in a locally supersymmetric theory. 
It is now useful to recall that local symmetries are just 
reduncancies in the parametrization of a theory whereas 
rigid (or global) symmetries are symmetries in the strict sense of
the word, i.e. they lead to the identification of states, rather then
just providing reparametrizations of the same state. Concerning local
transformations one has to distinguish those which are strictly local
(in the sense that the transformation parameter has compact support or
vanishes fast) from those which asymptotically approach a rigid 
transformation (i.e. the transformation parameter approaches a constant).
Thus local transformations in the broader sense 
include rigid transformation and 
local transformations in the narrow sense.

For a black hole supersymmetry transformations with a  
transformation parameter that approaches a 
constant at infinity are rigid transformations. They
describe the collective modes 
of the corresponding field configuration.\footnote{
We refer to \cite{Raj:1982} for a detailed discussion of 
collective modes of solitons.}
This is obvious in the case of translations or rotations, which 
are, like the supersymmetry transformations, part of the supersymmetry
algebra. Typically a soliton (or any other non-trivial field configuration)
has less symmetry than the vacuum. For instance a black hole
solution will never be translationally invariant. This is sometimes 
phrased in the way that the soliton 'breaks translational invariance',
a somewhat trivial statement that should not be confused with spontanous
symmetry breaking, which refers to properties of the vacuum. If the
underlying theory (i.e. the Lagrangian) is translationally invariant,
this will manifest itself in the degeneracy of all solutions
that are related by translations. Such motions constitute the
collective modes of the soliton. But there exist 
solitons which are more symmetric than a generic field configuration, and
which share some of the symmetries of the vacuum. This is 
most easily visualized when thinking about rotations rather than
translations. As we have seen above the stationary black holes 
of Einstein-Maxwell theory are either spherically symmetric or 
axisymmetric. The corresponding transformations do not
generate new states but are isometries of the metric and  
the transformation parameters are Killing vectors.

The same phenomenon can happen with supersymmetry transformations.
If one can find supersymmetry
transformation parameters $\e(x)$ such that a particular field
configuration is
invariant, one has the fermionic analogue of an isometry. The corresponding
parameters are called Killing spinors. 
The prime example is the extremal
Reissner Nordstr{\o}m solution when embedded into ${\cal N}=2$ supergravity.
The relevant field configuration is purely bosonic: 
Only the vielbein and the graviphoton are non-trivial
whereas
the gravitini are trivial, $\psi^i_\m = 0$. In order to prove the existence of
Killing spinors one has to show that
\be
\left. \delta_{\e(x)} \left( e_{\mu}^{\;\;a} (x), A_{\m}(x), \psi^i_\m(x) 
\right) \right|_{\mscr{extr. RN}} = 0
\ee
for some choice of the transformation parameter $\e(x)$. Note that the
supersymmetry transformation has to be evaluated on the background solution,
since we are asking for an invariance of this particular field configuration.
The invariance of the graviton and of the graviphoton is
trivial, because they transform into fermionic quantities that vanish
in the background. The non-trivial part is the variation of the 
gravitini,
\be
\delta_{\e(x)}  \left. \psi^i(x) 
\right|_{\mscr{extr. RN}} = 0 \;.
\ee
One can consider this as an equation for $\e(x)$ and solve it
with the result that one finds four Killing spinors. For later use
it is instructive to proceed in a more general way, which is standard
in the theory of solitons. The point is that the Killing spinor equation
is first order in the bosonic background and therefore much
easier to solve than the equations of motion. The same applies
to the Bogomol'nyi equation of Yang-Mills-Higgs theory. 
One can therefore try to find supersymmetric solitons by taking
the Killing equation as an equation for the supersymmetric background.
In addition one should of course impose further symmetry
properties in order to have a tractable problem. An
ansatz for finding supersymmetric (multi) solitons is a
general 'conformastatic' metric,
\be
ds^2 = - e^{2g(\vec{x})} dt^2  + e^{2 f(\vec{x})} d \vec{x}^2 \;,
\ee
which is static and has a conformally flat spatial part. In addition
one requires that the gauge field is static. In a bosonic background the
Killing spinor equation takes the form
\be
\left. 
\d \psi_{\m i} \right|_{\mscr{bosonic bg}}
= D_{\m} \e_i = \nabla_\m \e_i - \ft14 F^{-}_{\rho \sigma}
\g^{\rho} \g^{\sigma} \g_{\m} \ve_{ij} \e^j = 0 \;,
\ee
where $F^-_{\m\n}$ is the antiselfdual part of the field strength
and $\e_i$ and $\e^i$ are chiral projections of the two Majorana
spinors that parametrize the supersymmetry transformations.\footnote{
As already mentioned
we follow \cite{Gib:1981,GibHul:1982}, but use our own conventions, which
are explained in chapter \ref{ChapterN=2BlackHoles}.}
In order to get an overview over the possible solutions one can 
analyse the integrability conditon
\be
D_{[\m } D_{\n]} \e_i = 0 \;.
\ee
One class of solutions is obtained by imposing spherical
symmetry and invariance under all ${\cal N}=2$ supertransformations.
This corresponds to 
the maximal number of eight Killing 
spinors.
Then the geometry is the Bertotti-Robinson geometry $AdS^2 \times S^2$ 
with covariantly constant field strength. In the limit of infinite radius one
recovers flat Minkowksi space. These are the only static, spherically
symmetric and fully ${\cal N}=2$ supersymmetric field 
configurations.\footnote{This will be derived in 
chapter \ref{ChapterN=2BlackHoles} for a more general class of
${\cal N}=2$ supergravity Lagrangians. One can show an even
stronger result \cite{CardWKaeMoh}: the Bertotti-Robinson geometry
is the only stationary and $N=2$ supersymmetric field configuration
for the class of Lagrangians discussed in chapter
\ref{ChapterN=2BlackHoles}.} 
Since the field equations are automatically satisfied 
we will call them vacua. Observe that these are precisely 
the asymptotic geometries of the extremal Reissner-Nordstr{\o}m black hole
at the horizon and at infinity. Thus the black hole interpolates
between two vacua of ${\cal N}=2$ supergravity.

Another class of solutions is found when imposing that the 
background is conformastatic and that the Killing spinor
satisfies
\be
\e_i + i \g_0 \frac{\Sigma}{|\Sigma|} \ve_{ij} \e^j = 0 \;,
\label{KillingSpinorN=1}
\ee
where $\Sigma$ is allowed to depend on the 
fields.\footnote{See again chapter 
\ref{ChapterN=2BlackHoles} for the 
details.}
In this case half of the parameters are fixed in terms of the others
so that there are only four Killing spinors. Thus the solution
is invariant under half of the supersymmetry transformations, as expected
for a BPS soliton. It turns out that these solutions are precisely
the Majumdar-Papapetrou solutions discussed above. They are the most
general static supersymmetric solutions of the theory.
The explicit form of the Killing spinor is
\be
\e_i(\vec{x}) = H^{-1/2}( \vec{x} ) \e_i(\infty) \;,
\ee
where the asymptotic value of the spinor is constrained by
\be
\e_i(\infty) + i \g^0 \frac{Z}{|Z|} \ve_{ij} \e^j(\infty) =0 \,.
\ee
$Z$ is the central charge and $H$ is the harmonic function
of the Majumdar-Papapetrou solution. The relation between the
asymptotic transformation parameters can be directly understood
from the supersymmetry algebra. If the soliton is a supersymmetric state,
then the asymptotic transformation parameters must be 
null eigenvectors of the Bogomol'nyi matrix, which is the matrix
of all supersymmetry anticommutators, evaluated in the background.
This observation allows one to systematically construct supersymmetric
solitons, such as $p$-branes in more complicated supergravity theories,
see for example \cite{Tow:1997}.

Due to the invariance under half of the supertransformations, the
number of fermionic collective modes of the supersymmetric 
Reissner-Nordstr{\o}m black hole 
is reduced by a factor of $1/2$, so that there
are 4 instead of 8. By supertransformations one can generate the
corresponding 'half hypermultiplet'. Such a multiplet contains 
a fermion and two real scalars. Generically the multiplet 
carries some non-vanishing quantum number and therefore it 
is not CP-selfconjugate. Then one has to add the CP-conjugated
half hypermultiplet to restore CPT-invariance and this way one obtains a
hypermultiplet.
In our case the black hole solution carries electric or magnetic charge
and one gets a full hypermultiplet by adding 
the black hole of 
opposite charge and its supersymmetry partners. 
Thus the extremal Reissner-Nordstr{\o}m
black hole belongs to a hypermultiplet, like the monopoles and
dyons of ${\cal N}=2$ Super-Yang-Mills theories.

In conclusion we have seen that supersymmetry accounts for all the
particular features of the extremal Reissner-Nordstr{\o}m black hole.
This manifests itself nicely when expressing the metric,  
the mass and the entropy in terms of the central charge,
\be
ds^2 = - \left( 1 + \frac{|Z|}{r} \right)^{-2} dt^2 
+ \left(1 + \frac{|Z|}{r} \right)^2
( dr^2 + r^2 d\Omega^2 ) \;,
\ee
\be
M = |Z| \;, \;\;\;
{\cal S} = \pi |Z|^2 \;.
\ee
As a further application we would like to mention that the above methods
can also be used to prove statements about non-supersymmetric theories
of gravity. Using the embedding into a supersymmetric theory as 
a mere tool one can prove, under certain technical assumptions such as
the dominant energy condition, that the mass of a static, asymptotically
flat space-time in Einstein gravity coupled to matter 
is bounded by its total charge through
\be
M \geq |Q|
\ee
and that all static 
space-times saturating the bound posess four Killing spinors
and therefore belong to the Majumdar-Papapetrou class \cite{GibHul:1982}.

We already mentioned that the method that we explained here for the
extremal Reissner-Nordstr{\o}m black hole can also be used to 
systematically construct supersymmetric solitons of supergravity theories
in various dimensions. The knowledge of these supergravity solitons
was crucial for the study of string dualities and for establishing the
new M-theory paradigm that all five perturbatively consistent
string theories, together with eleven-dimensional supergravity,
are asymptotic expansions of one underlying 
theory \cite{HulTow:1994,Wit:1995}.

\subsection{Outlook}

In the following chapters we will extend and connect all the topics
introduced here. Instead of pure ${\cal N}=2$ supergravity we will
consider the generic low energy effective theory of a ${\cal N}=2$
string compactification, namely ${\cal N}=2$ supergravity coupled
to a model-dependent number of abelian vector multiplets and neutral
hypermultiplets. Since supersymmetry does not allow gauge-neutral
couplings of vector and hypermultiplets \cite{deWLauvP:1985}, 
one can find solutions
of the equations motion where the hypermultiplets are trivial.
Since this class of solutions is already very rich, and contains
a variety of charged black holes, we will neglect 
hypermultiplets in the following.\footnote{More recently we have 
shown that in a stationary space-time with residual
supersymmetry as specified in (\ref{KillingSpinorN=1}) the hypermultiplet scalars
are automatically constant \cite{CardWKaeMoh}.}  
In our discussion 
we will include a particular class of
higher curvature terms, which are quadratic in the Riemann tensor.
We will then study generalizations of the extremal Reissner-Nordstr{\o}m
black hole in these theories and in particular derive an entropy
formula for them according to the proposal of Wald. 

After embedding supergravity into string theory it is possible to 
lift some of our solutions 
to ten-dimensonial string theory
or eleven-dimensional M-theory. By counting
the collective modes of the higher dimensional solitons one gets
a microscopic entropy that can be compared to the macroscopic one.
Both expressions agree and we will see that this depends on 
various subtleties like the proper treatment of the higher curvature
terms.

\chapter{Four-Dimensional ${\cal N}=2$ Supergravity \label{ChapterFourDSuGra}}

In this chapter we review ${\cal N}=2$ supergravity coupled
to an arbitrary number of abelian vector multiplets. This is the part of
the low energy effective field theory of a ${\cal N}=2$ 
string compactification that we will need 
for the construction of black hole solutions.
The discussion is model independent,
concrete compactifications will be discussed later. 

The approach that we will describe is based on the
so-called superconformal multiplet calculus and was developed in
\cite{deWvHvP:1980,deWvHvP:1981,deWvP:1984,deWLauvP:1985}. It has
the advantage that it provides an off-shell formulation of the theory
and that one can include a particular class of curvature squared terms,
which are 
encoded in the Weyl multiplet. Both properties will be important when
we study black hole solutions later.
The whole vector multiplet Lagrangian is encoded in 
a single function of the vector multiplets, called the prepotential.
The scalar sigma-model must take values in a special K\"ahler manifold.
The structure of the Lagrangian and the role of the prepotential 
can be understood in terms of special K\"ahler geometry or
special geometry, for short \cite{deWvP:1984,CreEtAl:1985,deWLauvP:1985}.
We will also see that symplectic transformations, which generalize
electric - magnetic duality rotations, play an important role. They are
intimately related to the special K\"ahler geometry of the scalar
sector, because ${\cal N}=2$ vector multiplets contain both gauge fields
and scalars.

The two crucial ingredients of the approach are the construction
of extended conformal supergravity as a gauge theory of the
superconformal group \cite{FerKakTowVanN:1977}
and  the gauge equivalence of conformal and 
Poincar\'e supergravity.
It turns out that it is much easier to find off-shell
realizations of the ${\cal N}=2$ superconformal algebra. In particular
one can find smaller off-shell multiplets: Whereas the standard 
${\cal N}=2$
Poincar\'e supergravity multiplet \cite{dWvH:1979,FraVas:1979,deWvHvP:1980}
has $40+40$ off-shell degrees
of freedom,\footnote{It is known that smaller off-shell representations
of ${\cal N}=2$ Poincare\'e
supergravity exist: There are two $32+32$ representations 
\cite{Mue:1986,Mue:1987} and a $24+24$ representation 
\cite{AkeGriHasHer:1999}.
In the first two cases an action is also known. But for our 
purposes we need to know matter couplings and $R^2$-couplings.}
the minimal representation of ${\cal N}=2$ conformal supergravity
has only $32+32$. Moreover the minimal representation is reducible and
decomposes into the Weyl multiplet, with $24+24$ degrees of freedom
and a vector multiplet wit $8+8$.

The supersymmetry transformation rules have a simpler form
in the superconformal theory
and one can in a systematic way derive these rules and the 
Lagrangian starting from a gauge theory associated with the
${\cal N}=2$ superconformal algebra.
By adding multiplets which act as compensators one can construct
${\cal N}=2$ superconformal Lagrangians in such a way 
that they reduce upon partial gauge fixing 
to ${\cal N}=2$ super Poincar\'e Lagrangians.

The plan of this chapter is as follows.
We first
illustrate how gauge equivalence works using a non-supersymmetric
toy example. Then we review the construction of the relevant
superconformal off-shell multiplets, the superconformal action and
symplectic transformations. Finally we discuss the transition to
Poincar\'e supergravity and special geometry. The conventions are
those of \cite{Clausetal:1998,CardWMoh:1998/12}, which are somewhat
different from those in the original papers quoted above. Our presentation
has profited a lot from the nice review contained in \cite{Kleijn:1998}.
Reviews of ${\cal N}=2$ conformal supergravity are \cite{vP:1983,deW:1984}.

At the end of the chapter we make contact with the on-shell 
formulation of ${\cal N}=2$ supergravity developed in 
\cite{CasDauFer:1990,DauFerFre:1991}, which is also called the intrinsic
or geometric formulation. Reviews which also contain more references on this
approach to ${\cal N}=2$ supergravity are \cite{Fre:1995,Andetal:1996}. 
Our treatment of black holes in the next
chapter will be based on the off-shell formulation, but since most of the
literature on ${\cal N}=2$ black holes is based on the on-shell formulation
it is important for us to know how the two formulations are related.
The advantages and disadvantages of the two formalisms are
complementary. The superconformal off-shell formulation allows 
to include $R^2$-terms, but is tied to a parametrization of the scalar 
manifold in terms of so-called special coordinates. The geometric on-shell
formulation does not depend on special coordinates and one can 
describe the theory in terms of a section over the scalar manifold instead
of using a prepotential. This is relevant when constructing the most
general gauged ${\cal N}=2$ supergravity Lagrangian, but does not lead
to more general couplings of abelian vector multiplets. The reason
why we prefer the superconformal formulation is that
the geometric formulation only provides the on-shell supersymmetry 
transformation rules and that it has not been worked out how to include
$R^2$-terms.

\section{Gauge Equivalence}

Before entering into the details of the superconformal 
formalism we illustrate the concept of gauge equivalence
in a simple example. The Einstein-Hilbert action can
be obtained from the action of a scalar field coupled to
conformal gravity by partial gauge fixing of the conformal
symmetries. This is a standard example for gauge equivalence.
We follow \cite{Kleijn:1998,Moh:1986}. We try to focus
on the line of thought and omit several details, which can be
found either in the references or, for the more complicated ${\cal N}=2$
supersymmetric generalization, in later sections.

First we recall the 
Poincar\'e Lie algebra $\mbox{iso}(1,3)$:
\be
[P_{a},M_{bc}]=2 P_{[b} \eta_{c]a}, \;\;\;
[M_{ab},M_{bc}]=2 \eta_{[a[c} M_{b]d]},
\label{PoincAlg}
\ee
where $a,b,\ldots=0,\ldots,3$ are flat indices (tangent space indices).
$P_{a}$ generate the translations and 
$M_{bc}$ generate the Lorentz transformations.
This is extended to the Lie algebra of conformal transformations 
$\mbox{su}(2,2) \simeq \mbox{so}(2,4)$
by adding the generators $K_a$ of special conformal transformations
and $D$ of dilatations:
\be
\begin{array}{llll}
[K_{a},M_{bc}] = 2 K_{[b} \eta_{c]a}, &
[P_{a},K_{b}] = \eta_{ab} D - 2 M_{ab}, &
[D,P_{a}] = P_{a}, & [D,K_{a}] = - K_{a}. \\
\end{array}
\label{ConfAlg}
\ee
Finite special conformal transformations are singular
on hypersurfaces in Minkowski space, but we only consider
infinitesimal transformations here.

The first step is to 
construct a gauge theory of the conformal group. One introduces
connections $e_{\mu}^{\;\;a}, \omega_{\mu}^{ab}, f_{\mu}^{a},
b_{\mu}$ corresponding to $P_{a},M_{ab},K_{a},D$.
Here $\mu,\nu = t,x,y,z$ are space-time indices (curved indices,
world indices). 
The corresponding curvatures are denoted by
$R(P)_{\mu \nu}^{a}$, etc. At this point the superconformal
tranformations have been introduced as purely internal
transformations, i.e. they are space-time dependent but
do not act on space-time, but rather on a bundle over space-time.

The next step is to interpret the translational connection
$e_{\mu}^{\;\;a}$ as the vielbein\footnote{This means that
we require that $e_{\mu}^{\;\;a}$ is invertible.}, 
the local translations
as general coordinate transformations of space-time and
the local Lorentz transformations as acting on the 
tangent bundle of space-time, i.e. as local Lorentz 
transformations of the vielbein. As a consequence
the Lorentz connection $\omega_{\mu}^{ab}$ is identified
with the spin connection. 

In order to make these
identifications one has to impose so-called conventional
constraints on the conformal curvatures. Such constraints
have to be algebraic in order that they do not impose
dynamical equations on the field. Moreover they have to be
consistent with the underlying algebra. In the case of conventional
constraints this is guaranteed because 
one imposes relations that can be realized by redefinitions
of the connections. When imposing such conditions as constraints,
the Bianchi identities become non-trivial and it might be necessary
to modify the transformation rules of the remaining independent fields
in order to have consistent realization of the algebra.

In the case at hand the constraints have an obvious physical 
content.
In general relativity we know 
that the spin connection $\omega_{\mu}^{ab}$
can be expressed in terms of the vielbein $e_{\mu}^{\;\;a}$.
This has to be implemented by the constraints.
More generally 
the constraints have to be chosen 
such that all conformal transformations
become transformations acting in space-time. 
For example the local translations 
(P-transformations) have to become general coordinate 
transformations on {\em all} fields.

All this can be implemented by imposing the conventional constraints 
\be
R(P)_{\mu \nu}^{a} = 0, \;\;\;
e^{\nu}_{\;\;b} R(M)_{\mu \nu}^{ab} = 0 \;.
\ee
These relations can be solved for the dependent connections
$\omega_{\mu}^{ab}$ and $f_{\mu}^{\;\;a}$:
\be
\begin{array}{lcl}
\omega_{\mu}^{ab} &=& 
-2 e^{\nu [a} \der_{[\mu} e_{\nu]}^{\;\;b]}
- e^{\nu[a}e^{b]\sigma}e_{\mu c} \der_{\sigma} e_{\nu}^{\;\;c}
-2 e_{\mu}^{\;\;[a} e^{b]\nu} b_{\nu} \;,\\
f_{\mu}^{\;\;a} &=& \frac{1}{2}( R_{\mu \nu} - \frac{1}{6}
g_{\mu \nu}R) e^{\nu a} \;.\\
\end{array}
\ee
The spin-connection $\omega_\m^{ab}$ 
differs from the standard spin-connection
of general relativity by the additional term involving 
the dilatational connection $b_{\mu}$. This additional term
is needed to make the Riemann tensor 
covariant with respect to dilatations.

Consider now the case of a massless 
scalar field, which is taken to be $K$-invariant and to have 
Weyl weight (dilatational or $D$-weight)
$w=1$.  The minimal coupling to conformal gravity is given
by the Lagrangian
\be
{\cal L} = - e \phi D^{\mu}D_{\mu} \phi \;,
\label{ScalarConfGrav}
\ee
where $e=\det(e_{\mu}^{\;\;a})$ and $D_{\mu}$ is the
conformal covariant derivative. Under special conformal
transformations the dilatational 
gauge field $b_{\mu}$ transforms as
\be
\delta_{K} b_{\mu} = \Lambda_{K  \mu} \;,
\ee
where $\Lambda_{K \mu}$ are the transformation parameters
of the $K$-transformation. The first and second
covariant derivative of $\phi$ are:
\be
\begin{array}{lcl}
D_{\mu} \phi &=& \der_{\mu} \phi - b_{\mu} \phi\;, \\
D_{\mu} D^{a} \phi &=&
(\der_{\mu} - 2 b_{\mu}) D^{a} \phi - \omega_{\mu}^{ab} D_{b}
\phi + f_{\mu}^{\;\;a} \phi \;. \\
\end{array}
\ee
In the first line the term $-b_{\mu}$ provides the 
covariantization with respect to dilatations,
because $\phi$ has $w=1$. In the second line $D^{a} \phi$ is 
not invariant under $K$-transformations, because it contains
$b_{\mu}$. Therefore the  extra term involving
the $K$-connection $f_{\mu}^{\;\;a}$ is necessary
for making the second derivative covariant. 
This illustrates a general principle: In theories 
with several gauge symmetries and corresponding connections
the structure of covariant derivatives can be systematically
determined by considering the transformation properties
of the object one differentiates, see appendix \ref{AppSpecialGeometry}. 

The Lagrangian (\ref{ScalarConfGrav}) is conformally invariant and
in particular invariant under special conformal transformations.
Since $b_{\mu}$ is the only independent gauge field
that transforms under $K$, one knows a priori that the 
$b_{\mu}$-terms in (\ref{ScalarConfGrav}) must 
cancel out when expressing the dependent connections
in terms of the independent ones. Explicitly one finds:
\be
{\cal L} = -e \phi {\cal D}^{\mu} {\cal D}_{\mu} \phi -
e f_{\mu}^{\;\;\mu} \phi^{2} \;,
\ee
where ${\cal D}_{\mu} \phi = \der_{\mu} \phi$ is the
covariant derivative with respect to general coordinate
transformations. By partial integration and elimination of
the $K$-connection in terms of independent quantities one gets
\be
{\cal L} = e \der_{\mu} \phi \der^{\mu} \phi - \frac{1}{6} e
\phi^{2} R \;.
\ee
When gauge fixing the $D$-transformations by imposing
$\phi=\sqrt{6}/(\sqrt{2} \kappa)$ we get the Einstein-Hilbert
action
\be
{\cal L} = - e \frac{1}{2 \kappa^{2}} R \;.
\ee
We will call the special parametrization of a conformal theory where
only the Poincar\'e symmetry is manifest the {\em Poincar\'e frame}.

Conversely one can start with the Einstein-Hilbert action
and add a scalar with suitable transformation properties
to get the conformally invariant version of the 
Einstein-Hilbert action. The scalar acts as a {\em compensator}
in the sense that it is tailor-made to compensate
for the non-invariance of the Einstein-Hilbert action
under conformal transformations. Since one adds new symmetries
and new degrees of freedom in a balanced way, the 
total number of on-shell degrees of freedom remains 
unchanged, because the new degrees of freedom can
be gauged away. The two actions are said to be {\em gauge
equivalent}. 

Note that we got the Einstein-Hilbert action by starting from
a conformal matter action.
There is no explicit term involving the conformal curvatures but
the Einstein-Hilbert term is hidden in
the minimal coupling. Similarly the ${\cal N}=2$ supergravity 
action can be obtained from one minimally coupled conformal vector 
multiplet (together with a second compensating multiplet).

Explicit curvature terms can also be added to the
conformal action. They are not related to the Einstein-Hilbert term,
but play a different role. Gauge invariant curvature terms in
the conformal action are quadratic in the conformal curvatures 
and produce additional higher derivative terms in the Poincar\'e
action. In particular one gets the Weyl action (\ref{WeylAction}). 
In ${\cal N}=2$ supergravity
we will include terms quadratic in the Weyl multiplet in the superconformal
action and this way we will
obtain in the Poincar\'e frame additional curvature terms,
including terms quadratic in the Weyl tensor.

\section{Superconformal Multiplets}

\subsection{The Superconformal Algebra}

We now turn to the technical details of the superconformal
formalism. In this section we review the ${\cal N}=2$
superconformal algebra and introduce the so-called chiral notation.

We already displayed the relations of the Poincar\'e and of the
conformal Lie algebra in (\ref{PoincAlg}) and  (\ref{ConfAlg}).
The ${\cal N}$-fold extended superconformal algebra $\mbox{su}(2,2|{\cal N})$
is a supersymmetric extension of the conformal Lie algebra 
by adding $2 {\cal N}$ Majorana supercharges. 

Let us start with the Poincar\'e superalgebra, which contains
${\cal N}$ Majorana 
supercharges $q^i$, $i=1,\ldots,{\cal N}$, satisfying the
anticommutation relations \cite{HaaLopSoh:1975}
\be
\{q^i, q^j \} = 2 \gamma^a P_a \delta^{ij} + Z^{ij} \;,
\ee
where $Z^{ij}$ is a complex antisymmetric matrix of central charges.
In the absence of central charges the algebra is invariant
under the automorphism group (also called $R$-symmetry group)
$U({\cal N})_R \simeq SU({\cal N})_R \times U(1)_R$.\footnote{
There is one exception: For
${\cal N}=4$ the automorphism group is $SU(4)_R$ and not
$U(4)_R$ \cite{HaaLopSoh:1975}.} In the presence of central charges
the automorphism group is reduced to 
$USp({\cal N}) = U({\cal N}) \cap Sp({\cal N}, \mathbb{C} )$. Since the
supercharges are Majorana spinors, the action of 
$U({\cal N})_R$ is
chiral and the positive (left) and negative (right) chirality components 
transform in the fundamental and antifundamental representation,
respectively. 

It turns out to be  
useful to adopt the so-called {\em chiral notation},
which amounts to keeping track of spinor chiralities through writing
the $SU({\cal N})_R$ index as an upper or lower 
index \cite{deW:1979,deRvHdeWvP:1980}. Thus, upper and lower
$SU({\cal N})_R$ indices are correlated with a fixed spinor chirality
and with either the fundamental or antifundamental representation.
Note that depending on the spinor an upper index might be 
associated with left or with right chirality. The assignements are
listed in various tables in this chapter. As a consequence
$SU({\cal N})_R$ indices are raised and lowered by complex
conjugation.

The chiral 
projections of the supercharges $q^i$ are: 
\be
Q^i := \frac{1}{2} (\mathbb{I} + \gamma_5) q^i, \;\;\;
Q_i := \frac{1}{2} (\mathbb{I} - \gamma_5) q^i \;.
\ee
Remember that the chiral projections of a Majorana spinor are 
not independent of each other.
In the following we will omit several commutators which
do not contain independent information.
$\overline{\psi}$ denotes the Dirac conjugate
of the spinor $\psi$. For Majorana spinors, such as the $Q$- and 
$S$-supercharges, the Dirac conjugate equals the Majorana conjugate.
We refer to appendix \ref{AppSpaceTime} for an overview of our
spinor conventions.

The ${\cal N}$-fold extended superconformal algebra has
twice as many fermionic generators as the 
${\cal N}$-fold extended super Poincar\'e algebra \cite{HaaLopSoh:1975}.
The supercharges already present in the super Poincar\'e algebra
are called $Q$-supercharges $Q^i, Q_i$, whereas the additional supercharges
are called special supercharges or $S$-supercharges $S^i, S_i$. 
Both kinds of supercharges transform as Lorentz spinors:\footnote{
The Lorentz generators $\sigma_{ab}$ are defined in 
appendix \ref{AppSpaceTime}.}
\be
[M_{ab}, Q^i] = \frac{1}{2} \sigma_{ab} Q^i, \;\;\;
[M_{ab}, S^i] = \frac{1}{2} \sigma_{ab} S^i \;.
\ee

Closure of 
the algebra (see \ref{QSbar}) requires to include the 
$R$-symmetry algebra 
$U({\cal N})_R \simeq SU({\cal N})_R \times U(1)_R$ into
the superconformal algebra as well. The generators are denoted
by $V_{\Lambda}$ and $A$, respectively.
We now specialize to the case ${\cal N}=2$. 
In chiral notation the $SU(2)_R$ transformation rules 
read:
\be
\begin{array}{ll}
[V_{\Lambda}, Q]^i = i (\sigma_{\Lambda})^i_{\;\;j} Q^j, &
[V_{\Lambda}, Q]_i = i (\sigma_{\Lambda})_i^{\;\;j} Q_j, \\
{}[V_{\Lambda}, S]^i = i (\sigma_{\Lambda})^i_{\;\;j} S^j, &
[V_{\Lambda}, S]_i = i (\sigma_{\Lambda})_i^{\;\;j} S_j, \\
\end{array}
\ee
where the generators are normalized as 
$[V_{\Lambda}, V_{\Sigma}]=-2 \epsilon_{\Lambda \Sigma}^{\;\;\;\; \Xi}
V_{\Xi}$ ($\Lambda, \Sigma, \Xi = 1,2,3$). 

Under dilatations and $U(1)_R$ transformations the supercharges
transform as follows:
\be
\begin{array}{ll}
[D,Q^i] = \frac{1}{2} Q^i \;,& [A, Q^i] = - \frac{i}{2} Q^i\;, \\
{}[D,S^i] = \frac{1}{2} S^i\;, & [A, S^i] =  \frac{i}{2} S^i\;. \\
\end{array}
\ee
The $U(1)_R$ transformations are chiral and $S_i, Q_i$ have the
opposite charge (but the same dilatational weight).

The fermionic generators are related by $P_a$ and $K_a$:
\be
[K_a, Q^i] = \gamma_a S^i \;\;\;,
[P_a, S^i] = \frac{1}{2} \gamma_a Q^i \;.
\ee

The anticommutators of the $Q$- and $S$-supercharges 
close into translations and special conformal transformations,
respectively:
\be
\begin{array}{lcl}
\{ Q^i, \overline{Q}_j \} &=& - ( \mathbb{I}  - \gamma_5) \gamma^a P_a 
\delta^i_j \;, \\
\{ S^i, \overline{S}_j \} &=& - \frac{1}{2} 
(\mathbb{I} + \gamma_5) \gamma^a K_a 
\delta^i_j \;.\\
\end{array}
\ee
The mixed anticommutators yield other bosonic generators:
\be
\{Q^i, \overline{S}_j \} = \frac{1}{2} ( \mathbb{I}- \gamma_5 )
\left( 2 \sigma^{ab}M_{ab} + D - iA - 2 V^i_{\;\;j}
\right) \;.
\label{QSbar}
\ee
Finally we have a complex central charge $Z$, 
\be
\{Q^i, \overline{Q}^j \} = \frac{1}{2} (\mathbb{I} - \g_5)
\varepsilon^{ij} Z \;,
\label{QQbarZ}
\ee
where $\ve_{ij}$ is the antisymmetric tensor in $i,j=1,2$.

\subsection{The Weyl Multiplet}

In this section we review the construction of the minimal off-shell
representation of ${\cal N}=2$ conformal supergravity, the
so-called Weyl or superconformal gauge 
multiplet \cite{deWvHvP:1980,deWvHvP:1981,FraVas:1979},
\cite{Clausetal:1998,Kleijn:1998}.

The first step in constructing ${\cal N}=2$ conformal supergravity
is to build a superconformal gauge theory in which all generators
of the algebra act as {\em internal} symmetries. This is 
in principle a straightforward procedure: 
Starting from the algebra
one considers space-time dependent symmetry transformations and 
introduces gauge fields (connections). The gauge fields  
associated with translations ($P$), Lorentz transformations ($M$), 
dilatations ($D$),
special conformal transformations ($K$), $SU(2)$ and $U(1)$ 
transformations ($V$, $A$)
and $Q$- and $S$ supertransformations are denoted by
$e_{\mu}^{\;\;a}, \omega_{\mu}^{ab}, b_{\mu}, f_{\mu}^{\;\;a}, 
{\cal V}_{\mu\;\;j}^{\;\;i}, A_{\mu}, \psi_{\mu}^{\;\;i},
\phi_{\mu}^{\;\;i}$. 
The $SU(2)$ gauge field ${\cal V}_{\mu \;\;j}^{\;\;i}$ is
antihermitean and traceless:
\bea
{\cal V}_{\mu \;\;j}^{\;\;i}  + {\cal V}_{\mu j}^{\;\;\;\;i}
&=& 0 \;, \nonumber\\
{\cal V}_{\mu \;\;i}^{\;\;i} &=&0 \;. \\
\nonumber
\eea
The (space-time) dependent transformation parameters
corresponding to the transformations $T=P,M,D,K,Q,S,V,A$
are: $\xi^a, \ve^{ab}, \Lambda_D, \Lambda_K^a, \ve^i, \eta^i,
\Lambda_{V\;j}^{\;\;i}, \Lambda_A$. The chirality properties and
the Weyl and $U(1)$ weights of the gauge fields and transformation
parameters are listed in table \ref{Weights}.

The $Q\bar{Q}$ anticommutator (\ref{QQbarZ})
closes into the central charge and $Q$-transformations are 
realized as local transformations. Therefore
the central charge transformations are local as well and 
we have to introduce a gauge field for them.
This gauge field is not part of the Weyl multiplet but 
sits instead 
in a separate vector multiplet. Therefore the central charge gauge
transformations can be treated as additional abelian gauge symmetries.
This is consistent with the algebra because $Z$ commutes with all other
generators.

The covariant derivative with respect to
all superconformal transformations is
\be
D_{\mu} := \der_{\mu} - \sum_T \delta(h_{\mu}(T)) \;,
\ee
where the sum runs over all superconformal generators $T$.
The covariantization of a derivative
with respect to a transformation $T$
works by adding a 
gauge transformation with the gauge field $h_{\mu}(T)$ as parameter,
see appendix \ref{AppSpecialGeometry}. Some of the gauge fields
$h_{\mu}(T)$ appearing in the covariant derivative
differ from the fields $e_{\mu}^{\;\;a},\ldots$ listed above
by normalization factors, see table \ref{TableSCgaugefields}.

For later use we introduce another
covariant derivative ${\cal D}_{\mu}$, which is covariant with
respect to the bosonic $P,M,D,A,V$-transformations (and, if present,
gauge transformations), but which does not include covariantization
terms with repect to the fermionic $Q,S$- transformations and with
respect to the $K$-transformations.

Next one has to calculate the associated
field strengths (curvatures) $R_{\mu \nu}(T)$ and
to write down
the corresponding transformation rules. Note that the notation
$R_{\mu \nu}(T)$ is schematic, since most of the curvatures
carry additional indices. When refering to a specific field strength
we will always display all its bosonic indices (in practice
Lorentz and $SU(2)$ indices) while surpressing the fermionic
indices on the $Q$- and $S$-field strengths. 
The explicit expressions for the field strengths
are rather involved. Later we we will have to use modified field 
strengths,
which will be introduced below. Therefore we
will not need the explicit expressions for the $R_{\mu \nu}(T)$, which
can be found in reference \cite{deWvHvP:1980}.

The second step is to rebuild this superconformal gauge theory into 
a theory of
(conformal) supergravity. This means that the conformal symmetries 
and the supertransformations must
be realized as space-time symmetries rather than as internal ones.
In particular, local translations must be identified with local
coordinate transformations and local Lorentz transformations must
act as local rotations on the tangent bundle of space-time, whereas
$Q$-supertransformations must close into general coordinate transformations,
modulo other symmetries. As in the non-supersymmetric example discussed
earlier
the necessary identifications can be achieved
by imposing conventional constraints. 
Since the constraints are not invariant under supersymmetry
the dependent gauge fields that one obtains by solving
the constraints transform differently from the original
independent fields. In order to make the various
field strengths $R_{\mu \nu}(T)$
covariant with respect to the modified transformation
rules, additional terms have to be added. The new covariant
curvatures are denoted $\widehat{R}_{\mu \nu}(T)$. The Bianchi
identities are likewise modified. They are no longer identities
in the literal sense, but become non-trivial equations, whose
consistency has to be checked. All
these changes can be 
fine-tuned in such a way that the commutator of two supersymmetry
transformations closes into a general coordinate transformation.\footnote{
To be precise, the commutator of two $Q$-supertransformations
yields a so-called covariant general coordinate transformation plus
further symmetry transformations. This is discussed below
equation (\ref{qqcomb}).}
This requires the introduction of auxiliary fields, which together
with the independent gauge fields form a superconformal multiplet,
called the Weyl multiplet.
It is the basic multiplet of the theory, because it describes the
gravitational degrees of freedom.

One is interested in imposing as many 
constraints as possible in order to find a minimal representation
of ${\cal N}=2$ conformal supergravity, i.e. a representation
with the minimal number of degrees of freedom.
The constraints are not completely fixed by the above requirements. However a
suitable set of conventional constraints is known:
\bea
 & & R_{\mu \nu}(P) = 0 \;,\label{RP} \\
 & & \gamma^{\mu} \left(  \widehat{R}_{\mu \nu}(Q)^i + \sigma_{\mu \nu} \chi^i
\right) = 0 \;, \\
 & & e_{b}^{\;\;\nu} \widehat{R}_{\mu \nu}(M)_a^{\;\;b}  
- i \widetilde{\widehat{R}}_{\mu a}(A)
+ \ft18 T_{abij} T^{ij}_{\mu b} - \ft32 D e_{\mu a} = 0 \;.\\
\nonumber
\eea
We denote the dual tensor of $R$ by $\widetilde{R}$. Note that we use
a non-standard definition for the dual tensor, see 
appendix \ref{AppSpecialGeometry}.
The auxiliary fields consist of
an antiselfdual antisymmetric Lorentz tensor $T_{ab}^{ij}$, 
an $SU(2)$ doublet of Majorana spinors $\chi^i$ and a real scalar field $D$.
Note that the tensor field is antisymmetric in its $SU(2)$ indices
and, hence, is an $SU(2)$ singlet. 
In Minkowski space the selfdual and antiselfdual apart of an antisymmetric
tensor are related by complex conjugation:
\be
(T_{abij})^* = T_{ab}^{ij} \;.
\ee
For later use we introduce
\be
T^+_{ab} = T_{ab ij} \ve^{ij}\;, \;\;\;
T^-_{ab} = T_{ab}^{ij} \ve_{ij} \;.
\ee
Since the $SU(2)$ antisymmetric tensor is normalized by
$\ve_{ij} \ve^{ij}=2$, the inverse relations read
\be
T_{ab ij} = \ft12 T_{ab}^+ \ve_{ij} \;, \;\;\;
T_{ab}^{ij} = \ft12 T_{ab}^- \ve^{ij} \;.
\ee

We next list the explicit expressions for the following 
modified field strengths:
\bea \widehat{R}_{\mu\nu}(Q)^i &=&
     2{\cal D}_{[\mu}\psi_{\nu]}^i
     -\g_{[\mu}\phi_{\nu]}^i
     -\ft{1}{4}\s^{ab} T^{ij}_{ab} \g_{[\mu}\psi_{\nu]j}\,, \nonumber\\
     \widehat{R}_{\mu\nu}(A) &=&
     2\der_{[\mu}A_{\nu]}
     -i\Big(\ft{1}{2}\bar{\psi}_{[\mu}^i\phi_{\nu]i}
     +\ft{3}{4}\bar{\psi}_{[\mu}^i\g_{\nu]}\chi_i-{\rm 
h.c.}\Big)\,, \nonumber\\
     \widehat{R}_{\mu\nu}(V)^i{}_j &=& 2\der_{[\mu} {\cal V}_{\nu ]}^i{}_j 
     + {\cal V}_{[\mu }^i{}_k{\cal V}_{\nu ]}^k{}_j \nonumber\\
     && +\Big(2\bar{\psi}_{[\mu }^i\phi_{\nu ]j} 
     -3\bar{\psi}_{[\mu }^i\g_{\nu]}\chi_j
     -({\rm h.c.}\,;\,{\rm traceless}) \Big)\,. \nonumber\\
\widehat{R}_{\mu \nu}(M)^{ab} &=& 2 \der_{[\mu} \omega_{\nu]}^{ab} - 2 
\omega_{[\mu}^{ac} \omega_{\nu]}^{cb} - 4 f_{[\mu}^{\;\;[a} e_{\nu]}^{\;\;b]}
+ \left( \ov{\psi}_{[\mu}^{\;\;i} \sigma^{ab} \phi_{\nu]i} 
+ \mbox{h.c.} \right) \label{RhatM}\\
 & & \ft12 \ov{\psi}_{[\mu}^{\;\;i} T^{ab}_{ij} \psi_{\nu}^{\;\;j}
- \ft32 \ov{\psi}_{[\mu}^{\;\;i} \gamma_{\nu]} \sigma^{ab} \chi^i
- \ov{\psi}_{[\mu}^{\;\;i} \gamma_{\nu]} \widehat{R}^{ab}(Q)_i +
\mbox{h.c.} \;.\nonumber \\
\nonumber
\eea
By '$-({\rm h.c.}\,;\,{\rm traceless})$' we denote the projection
of a product of spinors in $SU(2)$ doublets onto the
antihermitean and traceless part. The explicit definition is
\be
\ov{\eta}^i \e_j - ( \mbox{h.c.; traceless}) =
\ov{\eta}^i \e_j - \ov{\eta}_j \e^i  - \ft12 \delta^i_j (
\ov{\eta}^k \e_k - \ov{\eta}_k \e^k) \;.
\ee 
The other modified field strengths can be found in 
\cite{deWvHvP:1980}.

The constraints can be used to express the gauge fields of
local Lorentz transformations $\omega_{\mu}^{ab}$, of special conformal
transformations $f_{\mu}^a$ and of $S$-supertransformations
$\phi_{\mu}^i$ in terms of the other gauge fields:
\bea \omega_\mu^{ab} &=& -2e^{\nu[a}\der_{[\mu}e_{\nu]}{}^{b]}
     -e^{\nu[a}e^{b]\s}e_{\mu c}\der_\s e_\nu{}^c
     -2e_\mu{}^{[a}e^{b]\nu}b_\nu   \label{omega} \\
     & & -\ft{1}{4}(2\bar{\psi}_\mu^i\g^{[a}\psi_i^{b]}
     +\bar{\psi}^{ai}\g_\mu\psi^b_i+{\rm h.c.}) \,,\nonumber\\
     \phi_\mu^i &=& (\s^{\rho\s}\g_\mu-\ft{1}{3}\g_\mu\s^{\rho\s})
     ({\cal D}_\rho\psi_\s^i-\ft{1}{8}\s^{ab} T^{ij}_{ab}\g_\rho\psi_{\s j}
     +\ft{1}{2}\s_{\rho\s}\chi^i) \,,\nonumber\\
     f_\mu{}^a &=& \ft12 \widehat{R}_{\mu}^{\;\;a} 
- \ft14 (D+ \ft13 \widehat{R}) \, e_\m^{\,a}  -
\ft12 i 
\widetilde R_{\m a}(A) + \ft1{16} T_{\m b}^{ij}\, T^{ab}_{ij} \,, 
\label{Kconnection}\\
\nonumber
\eea
where 
\be
\widehat{R}_{\mu}^{\;\;a} = \left. \widehat{R}(M)_{\mu \nu}^{\;\;\;\;ab} e_b^{\;\;\nu}
\right|_{f=0} 
\ee
is the Ricci tensor\footnote{This tensor is not symmetric, when
converting to pure world or pure tangent space indices.} 
constructed out
of the curvature tensor $\widehat{R}(M)_{\mu\nu}^{\;\;\;\;ab}$ by 
contraction, 
but with the terms involving the
conformal gauge field $f_{\mu}^{\;\;a}$ omitted. 
$\widehat{R} = \widehat{R}_{\mu}^{\;\;a} e_a^{\;\;\m}$ is the corresponding
Ricci scalar. In a bosonic 
background and after Poincar\'e gauge fixing these expressions reduce
to the standard Ricci tensor and Ricci scalar.

The independent
gauge fields are the ones of general coordinate transformations
$e_{\mu}^{\;\;a}$, dilatations $b_{\mu}$, chiral
$SU(2) \times U(1)$ transformations ${\cal V}_{\mu j}^i, A_{\mu}$ and
$Q$-supertransformations $\psi_{\mu}^i$. The full
content of the superconformal gauge multiplet is given by the
independent gauge fields together with the auxiliary fields
\be
\left( e_{\mu}^{\;\;a}, \psi_{\mu}^{\;i}, b_{\mu}, A_{\mu}, 
{\cal V}_{\mu \;\;j}^{\;\;i}, T_{ab}^{ij}, \chi^i, D \right) \;,
\ee
so that there are $24 + 24$ off-shell degrees of freedom. 
The Weyl and chiral weights, and the chirality
properties of the Weyl multiplet, of the dependent gauge fields and
of the supersymmetry transformation parameters are listed 
in table \ref{Weights}.

One of the constraints is to set the curvature associated with
local translations to zero, $R_{\mu \nu}(P)=0$ (\ref{RP}). This constraint
enables one to interpret the translational gauge field $e_{\mu}^{\;\;a}$
as a vielbein, because after imposing it this field 
transforms in the appropriate way under general coordinate transformations.
The internal indices $a,b,\ldots$ can now be seen as tangent space indices
and one can convert them into world indices using the vielbein. Note that
the vielbein carries a non-trivial Weyl weight, so that the Weyl weight
of a tensor changes when going from one type of indices to the other.

When solving the constraint one gets an expression for the Lorentz
connection $\omega_{\mu}^{ab}$
in terms of the vielbein, the dilational gauge field
$b_{\mu}$ and the $Q$-transformation gauge field 
$\psi_{\mu}^{\;\;i}$ (\ref{omega}). Thus, 
$\omega_{\mu}^{ab}(e,b,\psi)$ is a superconformal generalization of the
spin connection, in the sense that it does not only contain the standard
terms that express the spin connection in terms of the vierbein,
but also contains
further terms involving the dilatational and $Q$-supertransformation
gauge fields.
Likewise, the modfied
field strength $\widehat{R}_{\mu \nu}(M)^{ab}$ of Lorentz rotations
is a superconformal version of the Riemann tensor. The expression
(\ref{RhatM}) differs from the standard expression for the curvature
tensor in terms of the spin connection,
\be
R_{\mu \nu}^{\;\;\;\;ab} = 2 \der_{[\mu} \omega_{\nu]}^{ab} - 2 
\omega_{[\mu}^{ab} \omega_{\nu]}^{cb} \eta_{bc} \;,
\label{RiemannSpinConnection}
\ee
by additional terms involving the conformal gauge field
$f_{\mu}^{\;\;a}$ and the $Q$-gauge field $\psi_{\mu}^{\;\;i}$.
In the following we will refer to (\ref{RiemannSpinConnection})
as the Riemann tensor, because it reduces to the standard 
Riemann tensor when going to the super Poincar\'e theory by
appropriate gauge fixing.

As already mentioned the Bianchi identities have to be modified 
after imposing the constraints. The modified Bianchi identity 
for the field strength $R_{\mu \nu}(P)$ 
is an algebraic equation 
which can be used to express
the dilatational field strength in terms of the Lorentz field strength:
\be
\widehat{R}(M)_{[\m\n}^{ab} \,e_{\rho]\,b} = 
\widehat{R}_{[\m\n} (D) \,e_{\rho]}^{\,a} \,.
\label{BianchiRMRD}
\ee
For the standard Riemann tensor the left hand side would be zero.
Its non-vanishing reflects the presence of
the dilatational gauge field $b_{\mu}$. Formula (\ref{BianchiRMRD})
is the non-standard Bianchi identity\footnote{It is the integrability condition
for expressing the spin connection in terms of the vielbein.}
for the Riemann tensor in a theory with local scale invariance. 
Relation (\ref{BianchiRMRD}) leads to the pair-exchange property 
\be
\widehat{R}(M)_{ab}{}^{\!cd} - \widehat{R}(M)^{cd}_{\;\;ab} 
= 2 \Big( \d^{\;\;[c}_{[a} \widehat{R}(M)^{\;\;d]}_{b]} -  
\d_{\;\;[a}^{[c} \widehat{R}(M)_{\;\;b]}^{d]}\Big)  \,,
\ee
where 
$\widehat{R}(M)_{\mu}^{\;\;a} 
= \widehat{R}(M)_{\mu \nu}^{\;\;\;\;ab} e_b^{\;\;\nu}$.

We now list some properties of the field strength that we will
need later for constructing black hole solutions.
It is convenient to define the modified field strength
\be
{\cal R}(M)_{ab}{}^{\!cd} = \widehat{R}(M)_{ab}{}^{\!cd} 
+ \ft1{16}\Big( 
T^{ijcd}\,T_{ijab} + T^{ij}_{ab}\, T^{cd}_{ij}  \Big)\,,
\label{calRM}
\ee
where the $T^2$-modification cancels exactly the $T^2$-terms in the 
contribution to $\widehat{R}(M)$ from $f^{\,a}_\m$. 
${\cal R}(M)$ 
satisfies the following self conjugacy relations
\bea
\ft14 \varepsilon_{ab}{}^{ef} \,\varepsilon^{cd}{}_{gh} \,{\cal 
R}(M)_{ef}{}^{\!gh} &=& {\cal R}(M)_{ab}{}^{\!cd} \nonumber\\
\varepsilon_a^{\;\;ecd}\,{\cal R}(M)_{cdbe} &=& 
\varepsilon_a^{\;\;ecd} \,{\cal 
R}(M)_{becd}  = 2\widetilde{\widehat{R}}_{ab}(D) = 2i R_{ab}(A)\,.
\eea
In the second line the relations (\ref{BianchiRMRD}) and 
(\ref{Kconnection}) have been used.

It is also useful to introduce the following modified field
strength for $S$-supertransformations:
\be
{\cal R}(S)^i_{ab} =  R(S)^i_{ab} + \ft34 T^{ij}_{ab}\chi_j\,.  
\ee
The Bianchi identities and constraints imply that it satisfies 
\be
\g^a\widetilde {\cal R}(S)^i_{ab} = 2\, D^a\widetilde R(Q)_{ab}^i\,.
\ee
By contraction with $\g^b\g_{cd}$ one gets a relation between
${\cal R}(S)^i_{ab}$ and its dual:
\be
{\cal R}(S)_{ab}^i-\widetilde{\cal R}(S)_{ab}^i = 2\,\rlap/\!D(
R(Q)^i_{ab} + \ft34 \g_{ab} \chi^i)\,.
\ee
Like the corresponding gauge field $\phi^i_{\m}$, the field strengths
$R(S)^i_{ab}$ and ${\cal R}(S)^i_{ab}$ have negative chirality.

Finally we note that the $Q$-field strength is related 
to the auxiliary field $\chi^i$ by
\be
\g^a R(Q)_{ab}^i + \ft32 \g_b \chi^i = 0 \;, \mbox{   or   }
\chi^i =  \ft16 \g^{ab} R(Q)_{ab}^i \;.
\ee
Contraction with $\g^b \g_{ef}$ implies that
$R(Q)^i_{ab}$ is antiselfdual.  Like the corresponding gauge field
$\psi^i_{\m}$ the field strength $R(Q)^i_{ab}$ has positive chirality.

We now turn to the transformation properties of the fields under
supersymmetry.
As mentioned above $Q$-supertransformations have to close into
general coordinate transformations, modulo other symmetries. 
The precise form in which successive $Q$-supertransformations act is
the following:
\be [\d_Q(\e_1),\d_Q(\e_2)]=
    \d^{(cov)}(\xi)
    +\d_M(\ve)+\d_K(\La_K)+\d_S(\h)+\d_{{\rm gauge}} \,.
\label{qqcomb}
\ee
Three comments are in order. The first is that $Q$-transformations
do not close on a standard general coordinate transformation
$\delta_{\mscr{gct}}(\xi)$ but involve
a covariant general coordinate transformation, which
is defined by
\be \d^{(cov)}(\xi)=\d_{{\rm gct}}(\xi)
    +\sum_{T}\d_T (-\xi^\mu h_\mu(T)) \,.
\ee
The sum is over all superconformal transformations except
the general coordinate transformation. If the field to which the
above operator is applied transforms under additional
gauge symmetries, then these have to be included in the sum as well.
The second comment is that there are additional 
Lorentz, special conformal and $S$-supertransformations on the
right hand side of (\ref{qqcomb}). The transformation parameters
of these transformations and of the covariant general coordinate
transformation are given in terms of the parameters 
($\epsilon_1, \epsilon_2)$  of the two $Q$-supertransformations
by
\bea \xi^\mu &=& 2\,\bar{\e}_2^i\g^\mu\e_{1i}+{\rm h.c.}\,, \nonumber\\
     \ve^{ab} &=& \bar{\e}^i_1\e^j_2\,T^{ab}_{ij}+{\rm h.c.}\,, \nonumber\\
     \La_K^a &=& \bar{\e}^i_1\e^j_2\, D_bT^{ba}_{ij}
     -\ft32\bar{\e}_2^i\g^a\e_{1i}\,D+{\rm h.c.}\,, \nonumber\\
     \eta^i &=& 3\,\bar{\e}^i_{[1}\e^j_{2]}\,\chi_j  \,.
\label{qqparamsb}
\eea
The third remark is that whenever the field has additional gauge
symmetries, like central-charge gauge transformations or 
abelian or non-abelian gauge symmetries, then the gauge transformations
$\d_{{\rm gauge}}$ are in general present on the right hand side.
An example will be provided by the vector multiplets discussed in the
next section.

For completeness we also list the commutators between
$S$ and $Q$-supertransformations,
\bea 
[\d_S(\h), \d_Q(\e)] &=& \d_M \Big( 2 \bar{\h}^i\s^{ab}\e_i + {\rm h.c.} \Big)
+ \d_D \Big( \bar{\h}_i \e^i + {\rm h.c.} \Big)
+ \d_{A} \Big(  i\bar{\h}_i \e^i + {\rm h.c.} \Big) \nonumber\\
&& + \d_{V} \Big( -2  \bar{\h}^i \e_j -({\rm h.c.}\,;\,
{\rm traceless}) 
\Big) 
\label{SQcomm}
\eea
and between two $S$-supertransformations,
\be
[\d_S(\h_1),\d_S(\h_2)] = \d_K (\La^a_K) \,,\quad \mbox{ with } 
\L^a_K = \bar \eta_{2i}\g^a\eta^i_1 + {\rm h.c.}\,.
\ee

We conclude the section by presenting the
transformation rules of the Weyl multiplet and of the dependent
gauge fields under $Q$-, $S$- and $K$-transformations. These relations
are central for the construction of supersymmetric black hole solutions.

The
transformation rules for the components of the Weyl multiplet are:
\bea \d e_\mu{}^a &=&
     \bar{\e}^i\g^a\psi_{\mu i}+{\rm h.c.}\,, \nonumber\\
      \d\psi_\mu^i &=& 2{\cal D}_\mu\e^i
      -\ft18 \g_a\g_b  T^{ab\,ij}\g_\mu\e_j
      -\g_\mu\eta^i \,,\nonumber\\
      \d b_\mu &=&
      \ft12\bar{\e}^i\phi_{\mu i}
      -\ft34\bar{\e}^i\g_\mu\chi_i
      -\ft12\bar{\eta}^i\psi_{\mu i}+{\rm h.c.} +\L_K^a\,e_\m^a  
      \,,\nonumber\\   
      \d A_\mu &=& \ft{1}{2}i \bar{\e}^i\phi_{\mu i}
      +\ft{3}{4}i\bar{\e}^i\g_\mu\chi_i
      +\ft{1}{2}i \bar{\eta}^i\psi_{\mu i}+{\rm h.c.}\,, \nonumber\\
      \d {\cal V}_{\mu\,j}^i &=&
      2\bar{\e}_j\phi_\mu^i-3\bar{\e}_j\g_\mu\chi^i+2\bar{\eta}_j\psi_\mu^i
     -({\rm h.c.} \, ; \, {\rm traceless})\,,
      \nonumber\\
      \d T_{ab}^{ ij} &=&
      8 \bar{\e}^{[ i} \widehat{R}_{ab}(Q)^{j]}\,, \nonumber\\
      \d\chi^i &=& -\ft1{12}\g_a\g_b D\llap/ T^{ab\,ij}\e_j
      +\ft{1}{6}\widehat{R}(V)^i{}_{j\,\m\n}\g^\m\g^\n\e^j
      -\ft{1}{3}i \widehat{R}(A)_{\m\n} \g^\m\g^\n \e^i \nonumber\\
      && +D\,\e^i
      +\ft1{12} T^{ij}_{ab}\g^a\g^b\eta_j \,,\nonumber\\
      \d D &=& \bar\e^i D\llap/ \chi_i+{\rm h.c.}\,,
\label{transfo4}
\eea
and the transformation rules of the dependent gauge fields are:
\bea \d\omega_\mu^{ab} &=& -\bar{\e}^i\s^{ab}\phi_{\mu i} 
     -\ft12\bar{\e}^iT^{ab}_{ij}\psi_\mu^j
     +\ft32\bar{\e}^i\g_\mu\s^{ab}\chi_i \nonumber\\
     && +\bar{\e}^i\g_\mu\widehat{R}^{ab}(Q)_i 
    -\bar{\eta}^i\s^{ab}\psi_{\mu i} + {\rm h.c.}  + 
    2\L_K^{[a}\,e_\m^{b]} \,, \nonumber\\
     \d\phi_\mu^i &=& -2f_\mu^a\g_a\e^i
     -\ft14 D \llap/ T^{ij}_{cd} \s^{cd} \g_\mu\e_j
     +\ft32\big[(\bar{\chi}_j\g^a\e^j)\g_a\psi_\mu^i
     -(\bar{\chi}_j\g^a\psi_\mu^j)\g_a\e^i\big] \nonumber\\
     & & +\ft{1}{2}\widehat{R}(V)_{cd\;\;j}^{\;\;\;\;i} 
     \s^{cd} \g_\mu\e^j
     +i\widehat{R}(A)_{cd} \s^{cd} \g_\mu\e^i +2{\cal D}_\mu\eta^i 
     +\L^a_K \g_a\psi_\m^i \,,\nonumber\\
     \d f_\mu^a &=& -\ft12\bar{\e}^i\psi_\mu^j\, D_b T^{ba}_{ij}
     -\ft34e_\mu{}^a\bar{\e}^i D\llap/ \chi_i
     -\ft34\bar{\e}^i\g^a\psi_{\mu i}\,D 
       \nonumber\\
     & & +\bar{\e}^i\g_\mu D_b\widehat{R}^{ba}(Q)_i 
     +\ft12\bar{\eta}^i\g^a\phi_{\mu i}+ {\rm h.c.} +{\cal 
     D}_\mu\L^a_K \,.
\eea


\begin{table}

\begin{center}
\begin{tabular}{|c||cccccccc|ccc||cc|}
\hline
&
   \multicolumn{11}{c||}{Weyl multiplet} &
   \multicolumn{2}{c|}{parameters} \\
\hline
\hline
field          &
   $e_\mu{}^a$   &
   $\psi_\mu^i$  &
   $b_\mu$       &
   $A_\mu$       &
   ${{\cal V}_\mu}^i{}_j$ &
   $T_{ab}^{ij}$    &
   $\chi^i$      &
   $D$           &
   $\omega_\mu^{ab}$ &
   $f_\mu{}^a$   &
   $\phi_\mu^i$  &
   $\e^i$        &
   $\eta^i$ \\[.5mm]
\hline
\hline
$w$         & $-1$     & $-\ft12$  & $0$      & $0$      & $0$    
  & $1$      &  $\ft32$  & $2$      & $0$       & $1$      & 
$\ft12$  & $-\ft12$ & $\ft12$ \\[.5mm] 
 \hline
$c$         & $0$      & $-\ft12$  & $0$      & $0$      & $0$    
  & $-1$     &  $-\ft12$ & $0$      & $0$       & $0$      & 
$-\ft12$ & $-\ft12$ & $-\ft12$ \\[.5mm] 
\hline
$\gamma_5$&          & $+$       &          &          &          
&          & $+$       &          &           &          &    $-$ 
  &  $+$     &  $-$     \\[.5mm] 
\hline
\end{tabular}\\[.13in]
\caption{    Weyl and chiral weights 
             and fermion chirality 
             ($w$ and $c$ and $\gamma_5$, respectively)
             of the Weyl multiplet component fields
             and of the supersymmetry transformation 
             parameters.\label{Weights} }
\end{center}

\end{table}

\begin{table}

\begin{center}
\begin{tabular}{|l|l|l|l|l|l|l|l|l|} \hline 
Generator $T$ & $P^a$ & $M^{ab}$ & $D$ & $K^a$ & $Q^i$ & $S^i$ 
& $(V_{\Lambda})^i_{\;\;j}$ & $A$ \\ \hline
Connection $h_{\mu}(T)$ 
&$e_{\mu}^{\;\;a}$ & $\omega_{\mu}^{ab}$ & $b_{\mu}$ & $f_{\mu}^{\;\;a}$
& $\ft12 \psi_{\mu}^{\;\;i}$ & $\ft12 \phi_{\mu}^{\;\;i}$ 
& $- \ft12 {\cal V}_{\mu\;\;j}^{\;\;i}$ & $-i A_{\mu}$ \\  \hline
\mbox{Parameter} & $\xi^a$ & $\ve^{ab}$ & $\Lambda_D$ & $\Lambda_K^a$ &
$\ve^i$ & $\eta^i$ & $\Lambda_{V\;\;j}^{\;\;i}$ & $\Lambda_A$ \\ \hline
\end{tabular}
\end{center}
\caption{Table of superconformal gauge fields and transformation 
parameters \label{TableSCgaugefields}}
\end{table}

\subsection{Vector Multiplets}

We now turn to ${\cal N}=2$ vector multiplets
\cite{FirJen:1975,Fay:1976,GriSohWes:1978,BreSoh:1980,deWvHvP:1980,deWLauvP:1985,Clausetal:1998,Kleijn:1998},
restricting
ourselves to the case of abelian gauge symmetries. We consider
$N_V +1$ vector multiplets, labeled by an index $I=0,\ldots,N_V$.
One linear combination of the abelian gauge symmetries  
corresponds to the gauged central charge transformation, and the
corresponding field strength belongs to the graviphoton.
Note that we must have at least one vector multiplet in the theory
in order to make contact with ${\cal N}=2$ Poincar\'e supergravity,
because the Weyl multiplet does not account for the graviphoton. 

A conformal ${\cal N}=2$ vector multiplet 
\be
{\bf X}^I = \left( X^I,\Omega_i^I,W_{\mu}^I, Y_{ij}^I \right)
\ee
has $8+8$ off-shell degrees of freedom
and consists of a complex scalar $X^I$, an $SU(2)$ doublet of 
chiral fermions $\Omega_i^I$, called gaugini, a vector gauge field
$W_{\mu}^I$ and a real $SU(2)$ triplet of auxiliary scalars
$Y_{ij}^I$ (this means $Y_{ij}^I = Y_{ji}^I$ and 
$Y_{ij}^I = \ve_{ik} \ve_{jl} Y^{kl \;I}$). The Weyl and chiral weights
and the chirality properties of the component fields are listed in
table \ref{tableVM}. The superconformally 
covariant field strength is defined by
\be \F_{\mu\nu}^I=
    F_{\mu \nu}^I
    -\Big(\ve_{ij}\bar{\psi}_{[\mu}^i\g_{\nu]}\O^{j\,I}
    +\ve_{ij}\bar{X}^{I}\bar{\psi}_\mu^i\psi_\nu^j
    +\ft14\ve_{ij}\bar{X}^{I}T^{ij}_{\mu\nu}+{\rm h.c.}\Big) \,,
\label{calF}
\ee
where
\be
F_{\mu \nu}^I = 2\der_{[\mu}W_{\nu]}^I
\ee
is the standard abelian field strength. The covariant field strength
satisfies the Bianchi identity
\be
D^b\Big(\F^{+I}_{ab} -  \F^{-I}_{ab} +\ft14 X^I T_{ab\,ij} 
\varepsilon^{ij} -\ft14 \bar X^I T^{ij}_{ab} \varepsilon_{ij} \Big) = 
\ft34 \Big(\bar \chi^i\g_a \O^{Ij}\varepsilon_{ij} -\bar \chi_i\g_a 
\O_j^I\varepsilon^{ij}  \Big)\,. 
\label{BianchiVM}
\ee
The selfdual and antiselfdual part ${\cal F}^{\pm I}_{ab}$
of ${\cal F}^I_{ab}$ are defined according to the conventions explained
in appendix \ref{AppSpaceTime}.
The components of the multiplet transform as follows under supersymmetry:
\bea \d X^{I} &=& \bar{\e}^i\O_i^{\,I} \,,\nonumber\\
     \d\O_i^{\,I} &=& 2 D\llap/ X^{I}\e_i
     +\ft12 \ve_{ij} \F^{I\m\n-} \g_\m\g_\n \e^j
     +Y_{ij}^{\,I}\e^j
         +2X^{I}\eta_i\,,
     \nonumber\\
     \d W_\mu^I &=& \ve^{ij}\bar{\e}_i\g_\mu\O_j^{\,I}
     +2\ve_{ij}\bar{\e}^i\bar{X}^{I}\psi_\mu^j+ {\rm h.c.}\,, \nonumber\\
     \d Y_{ij}^{\,I} &=& 2\bar{\e}_{(i} D\llap/ \O_{j)}^I
     +2\ve_{ik}\ve_{jl}\bar{\e}^{(k} D\llap/ \O^{l)\,I} \;.
\label{vrules}\eea
These transformation rules satisfy relation (\ref{qqcomb}), 
including a field-dependent gauge transformation
on the right-hand side, which acts with the following parameter:
\be
\t^I=4\ve^{ij}\bar{\e}_{2i}\e_{1j}\,X^I+
 {\rm h.c.} \,.
\label{vgauge}
\ee
The covariant field strength transforms as follows under supersymmetry,
\be \d\F^{I}_{ab}=
    -2 \ve^{ij}\bar{\e}_i\g_{[a}D_{b]}\O_j^{\,I}
    -2\ve^{ij}\bar{\eta}_i\s_{ab}\O_j^{\,I} + {\rm h.c.}\,.
 \ee

\begin{table}
\begin{center}
\begin{tabular}{|c||cccc|}
\hline
&
   \multicolumn{4}{c|}{vector multiplet}  \\
\hline
\hline
field          &
   $X^I$         &
   $\O_i^{\,I}$      &
   $W_\mu^{\,I}$     &
   $Y_{ij}^{\,I}$    \\[.5mm]
\hline
\hline
w         & $1$ & $\ft32$  & $0$ & $2$ \\[.5mm]
\hline
c         & $-1$& $-\ft12$ & $0$ & $0$ \\[.5mm]
\hline
$\gamma_5$&     &  $+$     &     &     \\[.5mm]
\hline
\end{tabular}\\[.13in]
\caption{Weyl and chiral weights 
             and fermion chirality ($w$ and $c$ and $\gamma_5$, respectively)
             of the vector multiplet component fields \label{tableVM}}
\end{center}

\end{table}

\subsection{Chiral Multiplets}

The gauge invariant quantities of the Weyl multiplet
and of vector multiplets sit in chiral 
multiplets 
\cite{SalStr:1975,Fay:1976,deWvHvP:1980a,deWvHvP:1980a,deRvHdeWvP:1980,
BreSoh:1981,deWLauvP:1985,Clausetal:1998,Kleijn:1998}, which therefore
are the building blocks of the action. We will now 
discuss these multiplets.

A (left-handed) ${\cal N}=2$ chiral multiplet $\widehat{\bf A}$
is obtained from a general
scalar ${\cal N}=2$ superfield by imposing the chirality 
constraint\footnote{We put a hat on the chiral superfield
and on its components in order to have the same notation
as in the following sections and chapters, 
where a background chiral superfield
will play an important role.}
\be
{\bf D}^i \widehat{\bf A} =0 \;,
\label{ChiralConstraint}
\ee
where
\be
{\bf D}^i = \frac{\der}{ \der \bar{\theta}^i } + \g^\m \theta^i 
\frac{\der}{ \der x^\m }
\ee
is the right-handed superderivative in ${\cal N}=2$ superspace with
coordinates $(x^\m, \theta^i)$.
A chiral multiplet has $16 + 16$ off-shell components,
\be
\widehat{\bf A} =
(\widehat{A}, \widehat{\Psi}_i, \widehat{B}_{ij}, \widehat{F}^-_{ab}, \widehat{\Lambda}_i, 
\widehat{C}) \;,
\ee
namely two complex scalars $\widehat{A},\widehat{C}$, a complex $SU(2)$-triplet
of scalars $\widehat{B}_{ij}$, an antiselfdual Lorentz tensor $\widehat{F}^-_{ab}$
and two $SU(2)$ doublets of left-handed fermions 
$\widehat{\Psi}_i,\widehat{\Lambda}_i$.
The Lagrangian also contains the conjugated right-handed chiral multiplet.
When coupling to conformal supergravity one has to
assign Weyl and chiral weights $(w,c)$ to a chiral superfield.  The weights
of the component fields are fixed by the ones of the lowest component,
$(w,c)$ as indicated in table \ref{WeightsCM}. A consistent coupling
to conformal supergravity in addition requires $w=-c$.

In the special case $w=-c=1$ one can impose the further 
constraint \cite{deRvHdeWvP:1980}
\be
(\ve_{ij} \bar{\bf D}^i \sigma_{ab} {\bf D}^j)^2 ( \widehat{\bf A} )^{\star} =
\mp 96 \square  \widehat{\bf A} \;,
\label{Restriction}
\ee
which reduces the number of independent off-shell components to
$8+8$. The resulting multiplet is called the restricted chiral
multiplet. In the case of rigid ${\cal N}=2$ supersymmetry the constraint
allows one to express $\widehat{C}$ and $\widehat{\Lambda}_i$ in terms
$\widehat{A}$ and $\widehat{\Psi}_i$, respectively, and it imposes a
reality constraint on $\widehat{B}_{ij}$ and the standard Bianchi
identity of an abelian field strength on $\widehat{F}^-_{ab}$.
When coupling to ${\cal N}=2$ conformal supergravity the
equations are more complicated due to additional terms needed
for covariantization. The independent components 
$\widehat{A},\widehat{\Psi}_i, \widehat{F}^-_{ab}, \widehat{B}_{ij}$
of a reduced 
chiral multiplet can both in the rigid and in the local case
be identified with the convariant quantities 
$X^I, \Omega^I_i, {\cal F}^{-I}_{ab}, Y^I_{ij}$ 
associated with a vector multiplet. In particular $\widehat{F}^-_{ab}$
can be interpreted as a field strength, because it satisfies
the appropriate Bianchi identity (which, in the local case, is
(\ref{BianchiVM}).)

The covariant quantities of the Weyl multiplet
are associated with a reduced chiral multiplet 
${\bf W}_{ab}^{ij}$ \cite{BerdRdW:1981}.\footnote{${\bf W}_{ab}^{ij}$ 
denotes the full ${\cal N}=2$
superfield.}
This multiplet is obtained by imposing the constraints
(\ref{ChiralConstraint},\ref{Restriction}) on a ${\cal N}=2$
superfield which has an antiselfdual 
tensor field $\widehat{A}_{ab}^{ij}$ 
as its lowest component, with weights $w=-c=1$. The higher
components are a left-handed antisymmetric tensor spinor $\widehat{\Psi}_{ab}^i$,
a triplet of tensor fields $\widehat{B}_{ab\;\;j}^{\;\;\;\;i}$,
a tensor $\widehat{F}_{ab}^{\;\;\;\;cd}$, which is antisymmetric
and antiselfdual in both pairs of indices, a left-handed tensor-spinor
$\widehat{\chi}_{abi}$ and a tensor field $\widehat{C}_{abij}$.
The lowest component of ${\bf W}_{ab}^{ij}$ is the
auxiliary $T$-field,
$\widehat{A}_{ab}^{ij}= T_{ab}^{ij}$, whereas the higher components are
related to the $Q$-field strength $\widehat{R}(Q)_{ab}^i$, 
the $SU(2)$-field strength $\widehat{R}(V)_{ab\;\;j}^{\;\;\;\;i}$,
the modified Lorentz field strength
${\cal R}(M)_{ab}^{\;\;\;\;cd}$, 
the modified $S$-field strength 
${\cal R}(S)_{abi}$ and to
auxiliary fields.
We will not give the explicit relations here, because the superconformal
field strengths will finally enter the action in terms of yet another 
chiral multiplet, that we will discuss next. But note that 
all field strengths associated with independent 
gauge fields appear in ${\bf W}_{ab}^{ij}$.

Chiral multiplets can be multiplied, and the product is another
chiral multiplet. Weyl and chiral weights behave additive in products.
Therefore the product of two reduced chiral multiplets is a non-reduced
chiral multiplet \cite{deRvHdeWvP:1980}. 
The multiplet which contains the superconformal
field strength in the action is a non-reduced chiral multiplet
${\bf W}^2=(\widehat{A},\ldots )$ 
of weights $w=-c=2$, which is obtained by contracting the
superfield ${\bf W}_{ab}^{ij}$ with itself in the following way:
\be
{\bf W}^2 = \ve_{ik} \ve_{jl} {\bf W}_{ab}^{ij} {\bf W}^{abkl} \;.
\ee
By a long and tedious calculation the components of this
multiplet are found to be the following \cite{BerdRdW:1981,CardWMoh:1998/12}:
\bea
\widehat A   &=& (\varepsilon_{ij}\,T^{ij}_{ab})^2\,,\nonumber \\
\widehat \Psi_i &=& 16\, \varepsilon_{ij}R(Q)^j_{ab} \,T^{klab} \,
\varepsilon_{kl} \,,\nonumber\\  
\widehat B_{ij}  &=& -16 \,\varepsilon_{k(i}R(V)^k{}_{j)ab} \,
T^{lmab}\,\varepsilon_{lm} -64 \,\varepsilon_{ik}\varepsilon_{jl} 
\bar R(Q)^k_{ab} R(Q)^{l\,ab}   \,,\nonumber\\
\widehat F^{-ab}  &=& -16 \,{\cal R}(M)_{cd}{}^{\!ab} \,
T^{klcd}\,\varepsilon_{kl}  -16 \,\varepsilon_{ij}\, \bar R(Q)^i_{cd} 
\gamma^{ab} R(Q)^{j\,cd}  \,,\nonumber\\
\widehat \Lambda_i &=&32\, \varepsilon_{ij} \,\g_{ab} R(Q)_{cd}^j\, 
{\cal R}(M)_{cd}{}^{\!ab} 
+16\,({\cal R}(S)_{ab\,i} +3 \g_{[a} {\cal D}_{b]}\chi_i) \, 
T^{klab}\, \varepsilon_{kl} \nonumber\\
&& -64\, R(V)_{ab}{}^{\!k}_i \,\varepsilon_{kl}\,R(Q)^l_{ab}   \,,\nonumber\\
\widehat C &=&  64\, {\cal R}(M)^-_{cd}{}^{\!ab}\, {\cal 
R}(M)^-_{cd}{}^{\!ab}  + 32\, R(V)^-_{ab}{}^{\!k}{}_l^{~} \, 
R(V)^-_{ab}{}^{\!l}{}_k^{~} \nonumber \\
&& - 32\, T^{ij\,ab} \, D_a \,D^cT_{cb\,ij} +128 \, \bar {\cal 
R}(S)^{ab}_i \,R(Q)_{ab}^i  +384 \,\bar R(Q)^{ab\,i} 
\gamma_aD_b\chi_i   \,.   
\label{WeylMultComp}
\eea
The highest component $\widehat{C}$ contains terms quadratic and linear in the
curvature and plays a central role in the ${\cal N}=2$ Lagrangian and
in the computation of black hole entropy. We will see that the
matching between macroscopic and microscopic black hole entropy 
depends on the precise value of the coefficients in $\widehat{C}$.

\begin{table}
\begin{center}
\begin{tabular}{|c||c|c|c|c|c|c|} \hline
Field & $\widehat{A}$ & $\widehat{\Psi}_i$ & $\widehat{B}_{ij}$ & $\widehat{F}^-_{ab}$ & 
$\widehat{\Lambda}_i$ & $\widehat{C}$ \\  \hline \hline
Weyl weight & $w$ & $w+\ft12$ & $w+1$ & $w+1$ & $w+\ft32$ 
&  $w+2$ \\  \hline
Chiral weight & $c$ & $c+\ft12$ & $c+1$ & $c+1$ & $c+\ft32$ 
&  $c+2$ \\ \hline
Chirality& & + & &  & + & \\ \hline
\end{tabular}
\end{center}
\caption{Weyl and chiral weights and chiralities
of the components of a chiral multiplet \label{WeightsCM}}

\end{table}

\subsection{The Non-linear Multiplet}

We have to introduce one further multiplet, which will 
be used later to consistently gauge-fix the superconformal
theory to a super Poincar\'e theory. This multiplet was introduced
for precisely this purpose in \cite{deWvHvP:1981}.
It is called
the non-linear multiplet, because some of its components transform
into products of other components. For all other multiplets 
considered before the non-linear terms in the transformation rules
are entirely due to superconformal covariantizations.

A non-linear multiplet has the following components:
\be
( \Phi^i_{\;\;\alpha}, \lambda^i, M^{ij}, V_a)
\ee
The field $\Phi=\Phi^i_{\;\;\alpha}$ is an $SU(2)$ matrix of scalar fields.
Note that it has, besides the $SU(2)$ index $i$ a second
index $\alpha = 1,2$. This index is associated with an additional
rigid $SU(2)$ symmetry, which acts from the right, whereas the local
$SU(2)$, which is part of the superconformal group, acts from the left.
The hermitean conjugate is denoted $\Phi^+=\Phi^{\alpha}_{\;\;i}$. Since 
the matrix is an element of $SU(2)$, we have the constraints
\be
\Phi \Phi^+ = \mathbb{I} = \Phi^+ \Phi 
\mbox{   and   }
\det \Phi = 1 \;.
\ee
Therefore $\Phi$ describes 3 real scalars. The scalars have weigth zero.
The other components are a spinor doublet $\lambda^i$, 
a complex antisymmetric matrix $M^{ij}$ of Lorentz scalars and a real 
Lorentz vector $V_a$. The weights and chiral properties are listed in
table \ref{NonLin}.

\begin{table}
\begin{center}
\begin{tabular}{|c|| c | c | c | c |} \hline
Field &
$\Phi^i_{\;\;\alpha}$ &
$\lambda^i$ &
$M^{ij}$ & 
$V_a$ \\ \hline \hline
Weyl weight & 0 & $\ft12$ & 1 & 1 \\ \hline
Chiral weight & 0 & $- \ft12$ & $-1$ & 0 \\ \hline
Chirality &  & $-$ & & \\ \hline
\end{tabular}
\end{center}

\caption{Weyl weights, chiral weights and chirality of the
components of the non-linear multiplet \label{NonLin}}
\end{table}

Naive counting yields $9+8$ degrees of freedom. This indicates the
presence of a constraint. When constructing the transformation
rules 
\bea
\d \Phi^i_{\;\;\alpha} &=& \left( 2 \ov{\e}^i \lambda_j  - \d^i_j
\ov{\e}^k \lambda_k - \mbox{h.c.} \right) \Phi^j_{\;\;\alpha}\; \nonumber\\
\d \lambda^i &=& - \ft12 \g^a V_a \e^i - \ft12 M^{ij} \e_j 
- 2 \lambda^i ( \ov{\lambda}^j \e_j + \ov{\lambda}_j \e^j )
+ \g^a \e^i ( \ov{\lambda}^j \g_a \lambda_j ) \; ,\nonumber \\
 & & - 2 \sigma_{ab} \e_j \ov{\lambda}^j \sigma^{ab} \lambda^i
+ \Phi^i_{\;\;\alpha} \g^a D_a \Phi^{\alpha}_{\;\;j} \e^j + \eta^i \;,
\nonumber \\
\d M^{ij} &=& 6 \ov{\e}^{[i} \chi^{j]}  + \ov{\e}^k \sigma^{ab}
T^{-ij}_{ab} \lambda_k - 2 \ov{\e}^{[i} \g^a V_a \lambda^{j]}
-2 \ov{\e}^k \lambda_k M^{ij} \;,\nonumber \\
 & & + 4 \ov{\e}^{[i} ( \g^a D_a \lambda^{j]} + \Phi^{j]}_{\;\;\alpha}
\g^a D_a \Phi^{\alpha}_{\;\;k} \lambda^k ) \; \nonumber \\
\d V_a &=& \left[ \ft32 \ov{\e}^i \g_a \chi_i - \ft14 \ov{\e}^i
\g_a \sigma_{bc} T^{+ bc}_{ij} \lambda^j - \ov{\e}^i \g_a \g^b V_b
\lambda_i + \ov{\e}^i \g_a \lambda^j M_{ij} \right. \;,\nonumber \\
 & & \left. + 4 \ov{\e}^i \sigma_{ab} D^b \lambda_i 
+ 2 \ov{\e}_i \g_a \Phi^i_{\;\;\alpha} \g^b D_b \Phi^{\alpha}_{\;\;j}
\lambda^j - \ov{\lambda}_i \g_a \eta^i + \mbox{h.c.} \right] + 
2 \Lambda_{K a}\;, \\
\nonumber
\eea
one finds that one needs to impose the
supersymmetric constraint
\be
D^a V_a - 3 D - \ft12 V^a V_a - \ft14 |M_{ij}|^2 + D^a \Phi^i_{\;\;\alpha}
D_a \Phi^{\alpha}_{\;\;i} + \mbox{fermions} = 0
\ee
in order to close the algebra. This is interpreted as a constraint
on the vector $V_a$ which reduces the degrees of freedom to
$8+8$. When considering the coupling to the minimal ${\cal N}=2$
representation, we will give a somewhat different interpretation.

A particular property of the vector $V_a$ is that it transforms
under special conformal transformations, 
\be
\delta_K V_a = 2 \Lambda_{K a} \;.
\ee
The other components
transform trivially, like almost all other independent fields in 
the other superconformal multiplets.
As a consequence the covariant derivative of $V_a$ contains the
K-transformation gauge field:
\be
D^a V_a = {\cal D}^a V_a - 2 f^a_a + \mbox{fermionic terms} \;.
\ee
This will be used later.

\section{Superconformal Actions}

In the preceeding sections we outlined the construction of 
various off-shell representations of the ${\cal N}=2$ superconformal
algebra. The next step is to find the action. We will first
explain the basic ideas then describe how to
find the action for $N_V+1$ abelian vector multiplets coupled
to conformal supergravity.

An elementary method for constructing invariant actions 
is the
well known {\em Noether method}. One first writes down all the terms
that one wants to have in the Lagrangian. Then one iteratively adds
terms to compensate for the non-invariance of the terms already
present until one has found an invariant.

Since the method is tedious, especially for theories with a large
number of degrees of freedom, additional methods are helpfull.
In the context of rigid supersymmetry, for instance, one is used
to the fact that the highest component of a chiral superfield transforms
into a total derivative, $\delta_Q(\epsilon)C = \der_{\mu}(\cdots)$.
Therefore $\int d^4 x C$ is an invariant and a candidate for an
invariant action, though it is of course not guaranteed to be a 
physically sensitive choice. When coupling to ${\cal N}=2$
conformal supergravity things are more complicated: There are
additional terms present in the transformation rules and one wants
to have an invariant with respect to all superconformal transformations.
Using the Noether method one can find the necessary covariantizations.
The result is a so called {\em density formula}, which specifies
a quantity that transforms into a total deriviative under all
superconformal transformations. In the case of a chiral multiplet
the density formula takes the form\footnote{In the construction
of the action we need to distinguish two chiral multiplets by notation.
The chiral multiplet appearing in the density formula and a
background chiral multiplet which encodes the higher curvature 
terms. They are denoted by $(A,\ldots,C)$ and $(\widehat{A}, \ldots, \widehat{C})$,
respectively.} \cite{deRvHdeWvP:1980}, \cite{Kleijn:1998}:
\bea
e^{-1} {\cal L} &=& C - \ve^{ij} \overline{\psi}_i^\m \g_\m \Lambda_j
- \ft14 \overline{\psi}_{\m i} \sigma_{ab} T^{ab}_{jk} \g^\m
\psi_l \ve^{ij} \ve^{kl} \nonumber \\
 & & - \ft1{16} A ( T_{abij} \ve^{ij})^2  - \overline{\psi}_{\m i}
\sigma^{\m\n} \psi_{\n i} B_{kl} \ve^{ik} \ve^{jl} \nonumber \\
 & & \overline{\psi}_{\m i} \psi_{\n j} \ve^{ij}( F^{- \m\n}
- \ft12 A T^{\m \n}_{kl} \ve^{kl}) \nonumber \\
 & &- \ft12 \ve^{ij} \ve^{kl} e^{-1} \ve^{\m\n\rho \sigma}
\overline{\psi}_{\m i} \psi_{\n j} ( \overline{\psi}_{\rho k} 
\g_{\sigma} \psi_l + \overline{\psi}_{\rho k} \psi_{\sigma l} A) 
+ \mbox{h.c.}\;.
\label{ChiralDensity}\\ 
\nonumber  
\eea
Note that the weight of the chiral multiplet is not arbitrary. 
The action $S=\int d^4x  {\cal L}$ has to be Weyl and chirally invariant.
Since the integration measure has weights $w=-4,c=0$ it follows that
the $C$  must have $w_C = 4, c_C = 0$. This means that the 
lowest component has weights $w_A = - c_A =2$. As a consequence 
one needs a weight 2 chiral superfield to define the Lagrangian.

Similar density formulae can be derived for other multiplets.
In order to put them to use one needs another ingredient.
In practice one knows a density formula, like the one 
for a chiral
multiplet, and wants to use it to construct an action for one
or several other multiplets, like a collection
of vector multiplets. 
Then one needs to know how to construct a (weight 2)
chiral multiplet (or another multiplet for which a density formula is known)
out of the vector multiplets (or out of other whatever multiplets 
are to appear in the action). The techniques that enable one to get
one multiplet out of another are known as the {\em multiplet calculus},
and in our particular context as the ${\cal N}=2$ superconformal
multiplet calculus \cite{deWLauvP:1985}.

We will now outline the construction of an action for
several abelian vector multiplets using the chiral density formula.
The most general choice for the 
lowest component of the chiral multiplet in the density formula 
is a function 
of the vector multiplet scalars $X^I$, $I=0,\ldots,N_V$ 
\cite{deRvHdeWvP:1980,deWLauvP:1985}:
\be
A \sim F(X^I) \;.
\ee
The function $F(X^{I})$,
which is called the {\em prepotential}, is subject to two restrictions:
First, it needs to be {\em holomorphic} in the sense that
it does not depend on the complex conjugated scalars $\ov{X}^I$.
The second restriction, which does not apply to the
case of rigid supersymmetry, follows from the fact that $A$ must
have weight 2 ($w=-c=2$) in the presence of conformal supergravity.
Since a vector multiplet has weight 1 ($X$ has $w=-c=1$) this implies
that the function $F(X^I)$ must be homogenous of degree 2:
\be
F(\lambda X^I) = \lambda^2 F(X^I) \;,
\ee
for all $\lambda \in \mathbb{C} - \{0 \}$.

The chiral density and 
the action depend on the Weyl multiplet only through the 
superconformal covariantizations. When gauge-fixing the theory
to obtain ${\cal N}=2$ Poincar\'e supergravity coupled to 
$N_V$ vector multiplets\footnote{Actually this requires to add 
one further compensating multiplet. We will come back to this
later.} one obtains the Einstein-Hilbert term basically in the
same way as in the toy example discussed earlier and there are
no higher powers of the Riemann tensor in the action. 
The theory also contains minimal terms for the other fields, only, where
minimal
means terms with up to two derivatives and up to four fermions.

In order to have higher curvature terms in the Lagrangian 
one has to include
explicit couplings to the Weyl multiplet. Since the covariant
quantities of the Weyl multiplet sit in the chiral multiplet
${\bf W}^2$, only this multiplet can appear in the Lagrangian.
For simplicity we will call the ${\bf W}^2$-multiplet the Weyl multiplet
in the following. It will be clear from the context whether we
refer to the multiplet of superconformal gauge fields or to the
corresponding chiral multiplet. The problem of 
finding the coupling of vector multiplets to the Weyl multiplet
is the same as finding the coupling to a background chiral
multiplet, because the only relevant fact for constructing the
coupling is the type of multiplet we want to couple to. Since the
$R^2$-Lagrangian is very complicated and non-linear it is 
useful to work with a chiral background field as long as possible
and to plug in the explicit expressions for the Weyl multiplet
only at the end. The coupling to the background field is described by 
a function $F$, which now also depends on the lowest component of
the chiral multiplet 
$\widehat{\bf A} = (\widehat{A}, \ldots)$ \cite{deW:1996/02}.
The function must be holomorphic and homogenous
of second degree in both $X^I$ and $\widehat{A}$:
\be
F(\lambda X^I, \lambda^w \widehat{A}) = \lambda^2 F(X^I, \widehat{A})\;,
\label{HomTwo}
\ee
where $w$ is the weight of the chiral background field.\footnote{ 
For $w\not=1$ this is a slightly generalized definition of
'homogenous', which allows the variable $\widehat{A}$ to be 'weighted'.}

The homogenity of $F(X^I, \widehat{A})$ implies several useful
identities between the function and its derivatives. We use
the following notation for derivatives:
\be
F_I := \frac{\der}{\der X^I} F(X^I,\widehat{A}), \;\;\;
F_{\widehat{A}} := \frac{\der}{\der \widehat{A} } F(X^I,\widehat{A}),
\ee
and consequently for higher deriviatives: 
\be
F_{I_1 \cdots I_k \widehat{A} \cdots \widehat{A}} =
\frac{\der}{\der X^{I_1}} \cdots
\frac{\der}{\der X^{I_k}}
\frac{\der}{\der \widehat{A} } \cdots
\frac{\der}{\der \widehat{A} }  F(X^I,\widehat{A}) \,.
\ee
Then, by differentiating the defining relation (\ref{HomTwo})
with respect to $\lambda$, and setting $\lambda = 1$, we get
\be
X^I F_I + w \widehat{A} F_{\widehat{A}} =2 F(X^I,\widehat{A}) \;.
\ee
This is an alternative definition of a homogenous function
(of degree 2), the so-called Euler relation.
Further identities are found by taking derivatives with respect
to $X^I$ or $\widehat{A}$, for example
\be
X^I F_{IJ} + w \widehat{A} F_{J\widehat{A}} =  F_{J} \;.
\ee
When assuming that the function $F(X^I,\widehat{A})$ has a power
expansion around $\widehat{A}=0$ one defines a family of 
functions $F^{(g)}(X^I)$ by
\be
F(X^I,\widehat{A}) = \sum_{g=0}^{\infty} F^{(g)}(X^I) \widehat{A}^{g} \;.
\ee
Note that the $F^{(g)}$ are homogenous of degree $2-gw$ in $X^I$.
When the background field is taken to be the Weyl multiplet 
(${\bf W}^2$ multiplet), then $w=2$. The first function
$F^{(0)}(X^I)$ in the expansion is the prepotential, 
and controls the minimal terms in the action.

The full ${\cal N}=2$ superconformally invariant action for
$N_V+1$ vector multiplets coupled to conformal supergravity
is obtained by using the chiral density formula 
(\ref{ChiralDensity})
for a chiral
multiplet with lowest component $A=-\ft{i}2 F(X^I,\widehat{A})$.
Since we will be interested in black holes, which are 
purely bosonic solutions to the field equations, we will
only need to display the bosonic part, which reads \cite{deW:1996/02}:
\bea
e^{-1} {\cal L} &\sim& \Big[
i  \ov{F}_I X^I ( \ft16 R - D )
+ i  {\cal D}_{\mu} F_I {\cal D}^{\mu} \ov{X}^I 
 \nonumber \\
 & & + \ft14 i F_{IJ} ( F^{-I}_{ab} - \ft14 \ov{X}^I T^{ij}_{ab} \ve_{ij})
 ( F^{-J}_{ab} - \ft14 \ov{X}^J T^{ij}_{ab} \ve_{ij})
- \ft18 i F_I ( F^{+I}_{ab} - \ft{1}{4}X^I T_{abij} \ve^{ij}) T^{ab}_{ij}
 \ve^{ij} \nonumber \\
 & & - \ft18 i F_{IJ} Y^I_{ij} Y^{J ij}  
  -\ft{i}{32}F(T_{abij}\ve^{ij})^2 \nonumber \\
 & & + \ft12 i F_{\widehat{A}} \widehat{C} 
- \ft18 i F_{\widehat{A} \widehat{A}}
(\ve^{ik} \ve^{jl} \widehat{B}_{ij} \widehat{B}_{kl} - 2 \widehat{F}^-_{ab}
\widehat{F}^-_{ab}) 
+ \ft12 i \widehat{F}^-_{ab} F_{\widehat{A}I} ( F^{-I}_{ab} - \ft14 \ov{X}^I
T^{ij}_{ab} \ve_{ij})  \nonumber\\
 & &  - \ft14 i \widehat{B}_{ij} F_{\widehat{A}I} Y^{Iij} \Big]
+ \mbox{h.c.}  \\
\nonumber
\eea
Let us comment on the various terms of this lenghty expression.
The two terms in the first line result from decomposing 
the term $i \ov{F}_I \square_{\mscr{sc}} X^I + \mbox{h.c.}$, where
$\square_{\mscr{sc}}$ is the superconformal d'Alembertian,
$\square_{\mscr{sc}}=D^aD_a$, in terms of expressions 
which are natural from the super Poincar\'e perspective. 
One piece can be written in terms
of the bosonic covariant derivative ${\cal D}_{\mu}$ and after
a partial integration one gets the second term in the first line. 
The rest contains
the auxiliary $D$ field plus a curvature piece which modulo
fermionic pieces is proportional to the Ricci scalar associated with
the spin connection $\omega_{\mu}^{ab}$. As in the toy example 
the special conformal gauge field has dropped out. Note however that
we are still in a conformally invariant theory and therefore the
spin connection still contains the dilatational gauge field.
These considerations explain the first term in the first line, 
which has a piece that should become the Einstein-Hilbert term when 
going to the 
super Poincar\'e theory. It is accompanied by an awkward looking term
linear in the auxiliary field $D$. We will have to deal with this term
later on.

In the following lines we get terms involving the gauge fields.
Note that we have  rewritten the
superconformally 
covariant field strengths ${\cal F}_{ab}^I$ 
in terms of the standard field strength $F^I_{ab}$ 
in order to make
explicit the dependence on the auxiliary field $T_{ab}^{ij}$. 
The terms in the last two lines explicitly involve the chiral background
field. The most important term is 
$\ft12 i F_{\widehat{A}} \widehat{C} + \mbox{h.c.}$ which,
according to (\ref{WeylMultComp}), 
contains curvature squared terms. 
We will have a closer look at these terms later. 
The minimally coupled case is obtained by setting the chiral background
to zero. Then, the last two lines are absent and the function
$F(X^I, \widehat{A})$ reduces to the prepotential $F(X^I)$.

\section{Symplectic Reparametrizations \label{SectionSymplectic}}

As is well known, in extended supergravity models the full set of 
field equations (including the Bianchi identities) is invariant
under continuous transformations which 
nowadays are called duality transformations \cite{GaiZum:1981}.
These transformations generalize the electric-magnetic
duality transformations of Maxwell electrodynamics and like
them they are - in general - not symmetries, because they are
not invariances of the action and act non-trivially
on the couplings. In supergravity theories couplings are
scalar field dependent, and therefore supersymmetry
implies that the scalars must transform as well. A particular
structure occurs in the case of ${\cal N}=2$ vector multiplets,
where gauge fields and scalars sit in the same supermultiplet.
In this case the duality transformations manifest
themselves as {\em symplectic reparametrizations} as we will
review in this section \cite{deWvP:1984}. We will follow the
work \cite{deW:1996/02} of de Wit, who analysed symplectic reparametrizations
in a chiral background.

Since duality transformations generalize electric-magnetic 
duality, they can only exist if the action 
depends on the field strength, but not explicitly
depends on the gauge potential.
Therefore they are restricted to the case of abelian
vector multiplets.
This is precisely the case we are interested in.

The action that we found in the last section contains terms
quadratic and linear in the field strength $F^{I-}_{\mu \nu}$.
It is convenient to introduce a so-called dual field 
strength\footnote{Note that this is in general not the
Hodge dual.
See appendix \ref{AppAbelianGaugeFields} for more details.}
by
\be
G^{-\mu \nu}_{I} = 
\frac{2i}{e} \frac{\der {\cal L}}{\der F^{I-}_{\mu \nu}} \;,
\ee
so that the Lagrangian takes the form
\be
e^{-1} {\cal L} \sim - \ft{i}{2} (F^{-I}_{\mu \nu} G^{-\mu \nu}_{I}
- \mbox{h.c.} ) + \cdots \;.
\ee
For the superconformal action whose bosonic part was
displayed in the last section the dual gauge field
takes the form
\be
G^{-}_{\mu \nu I} = F_{IJ} F^{-J}_{\mu \nu} + {\cal O}^-_{\mu \nu I} \;,
\label{DefG}
\ee
where ${\cal O}^-_{\mu \nu I}$ are all the terms that couple linearly
to the field strength. When restricting ourselves to the bosonic
fields this term takes the form
\be
{\cal O}^{-}_{\mu \nu I} = \ft14 
( \ov{F}_{I} - F_{IJ} \ov{X}^{J}) T^{ij}_{\mu \nu} \ve_{ij} 
+  \widehat{F}^{-}_{\mu \nu} F_{I \widehat{A}}\;.
\label{Ominus}
\ee
In the full theory this term also contains fermions.
Clearly fermions have to transform under duality too, as they
sit in the same supermultiplet as the gauge fields. But since we are
interested in bosonic field configurations, only, we will restrict
our discussion of duality to the bosonic fields.
We also note that
the second derivatives $F_{IJ}$ of the function
$F(X^I,\widehat{A})$ obviously 
encode the field dependent couplings and 
$\theta$ angles.\footnote{This term gets modified
when eliminating the auxiliary fields. Then $F_{IJ}$ is replaced
by another quantity usually denoted $\ov{\cal N}_{IJ}$. 
We will come back to this later.}
  
The 
field equations and Bianchi identities take the form\footnote{
See appendix \ref{AppAbelianGaugeFields} for more details.}
\bea
{\cal D}_{\mu}( G^{- \mu \nu}_I - G^{+ \mu \nu}_I)&=&
0 \;, \nonumber \\
{\cal D}_{\mu}( F^{-I \mu \nu} - F^{+I \mu \nu})&=&
0 \;.  \\
\nonumber
\eea
The combined set of equations is manifestly invariant under
the duality rotation
\be
\left( \begin{array}{c}
F^{\pm I}_{\mu \nu} \\ G^{\pm}_{J \mu \nu} 
\end{array} \right)
\rightarrow
\left( \begin{array}{cc}
U^I_{\;\;K} &  Z^{IL} \\ W_{JK} & V_J^{\;\;L} \\
\end{array} \right)
\left( \begin{array}{c}
F^{\pm K}_{\mu \nu} \\ G^{\pm}_{L \mu \nu} \\
\end{array} \right) = 
\left( \begin{array}{c}
\breve{F}^{\pm I}_{\mu \nu} \\ \breve{G}^{\pm}_{J \mu \nu}\\
\end{array} \right) \;,
\ee
if the matrix
\be
{\cal O} = \left( \begin{array}{cc}
U^I_{\;\;K} &  Z^{IL} \\ W_{JK} & V_J^{\;\;L} \\
\end{array} \right) 
\ee
is real and 
invertible, ${\cal O} \in \mbox{GL}(2N_{V}+2, \mathbb{R})$.
In fact the choice of ${\cal O}$ is even more restricted,
because we want to relate 
the transformed set of equations to a dual Lagrangian
\be
e^{-1} \breve{\cal L} 
\sim - \frac{i}{2} (\breve{F}^{-I}_{\mu \nu} \breve{G}^{-\mu \nu}_{I}
- \mbox{h.c.} ) \;.
\ee
This Lagrangian must again have the structure that we 
found in the last section and therefore
\be
\breve{G}^{-}_{\mu \nu I} = 
\breve{F}_{IJ} \breve{F}^{-J}_{\mu \nu} + \breve{\cal O}^-_{\mu \nu I} \;.
\ee
This implies that $F_{IJ}$ has to transform as
\be
\breve{F}_{IJ} = [ V F + W ]_{IL}
[(U + Z F)^{-1}]^{L}_{\;\;J} \;,
\label{Fmatrixtilde}
\ee
i.e. by a projective linear transformation and 
${\cal O}^{-I}_{\mu \nu}$ must transform as
\be
\breve{\cal O}^-_{\mu \nu I} =
{\cal O}^-_{\mu \nu J} [( U + Z F)^{-1}]^J_{\;\;I} \;.
\label{Otilde}
\ee
Inside such matrix equations we use the notation
$F=(F_{IJ})$. The symmetric matrix
$F_{IJ}$ has to be mapped 
to a symmetric matrix $\breve{F}_{IJ}$.
This implies that (up to a uniform scale transformation, that we
neglect) the matrix ${\cal O}$ has to be symplectic,
${\cal O} \in \mbox{Sp}(2N_V+2,\mathbb{R})$:
\be
{\cal O}^T \Omega {\cal O} = \Omega,\;\;\;
\mbox{where} \;\;\;
\Omega = \left( \begin{array}{cc}
0 & \mathbb{I} \\ -\mathbb{I} & 0 \\
\end{array} \right) \;.
\ee
In terms of the block matrices this means
\be
U^T W - W^T U = 0 = Z^T V - V^T Z \mbox{   and   }
U^T V - W^T Z = \mathbb{I} \;.
\ee
We have now to remind ourselves that $F_{IJ}$ are the second
derivatives of the function $F(X^I,\widehat{A})$. Since $F_{IJ}$
transforms, the function itself has to transform in such a way
that the correct transformation of $F_{IJ}$ follows.
As a consequence various 
terms in the action besides 
the ones discussed so far transform under
duality as well. This is no surprise, since the
field strengths sit in the same ${\cal N}=2$ multiplet as the
scalars. What we have to find is a transformation rule for
the $X^I$ that precisely induces the correct transformation
rule of $F_{IJ}$. The additional chiral background field
$\widehat{A}$ sits in a different supermultiplet 
and therefore is inert
under duality rotations. Note however that it will enter into
the transformation rules because of its appearence in 
$F(X^I, \widehat{A})$.

In order to find the transformation rule of the scalars 
it is convenient
first to focus on the expression $(X^I, F_J)^T$ and to postpone
making the connection to the function $F(X^I,\widehat{A})$. The reason
is that the correct transformation law for $F_{IJ}$ is induced
by simply requiring that $(X^I,F_J)$ transforms linearly under
symplectic transformations:
\be
\left( \begin{array}{l}
X^I \\ F_J \\ \end{array} \right) 
\longrightarrow
\left( \begin{array}{cc}
U^I_{\;\;K} &  Z^{IL} \\ W_{JK} & V_J^{\;\;L} \\
\end{array} \right)
\left( \begin{array}{c}
X^{K} \\ F_{L} \\
\end{array} \right) = 
\left( \begin{array}{l}
\breve{X}^I \\ \breve{F}_J \\ \end{array} \right)  \;.
\label{TransfSect}
\ee
Computing $\breve{F}_{IJ} = \ft{\der}{\der \breve{X}^J} \breve{F}_I$
one indeed finds (\ref{Fmatrixtilde}). 
Quantities transforming linearly under the symplectic group are
called {\em symplectic vectors}. Given two symplectic vectors
$v,w$ one can form a symplectic scalar by taking the symplectic
scalar product, $-i\ov{v}^T \Omega w$. The symplectic scalar
product of $(X^I, F_J)$ with itself, 
$-i(F_I \ov{X}^I - X^I \ov{F}_I)$ enters the action as the coefficient
of the Ricci scalar. Thus the invariance of this term is manifest.
Moreover this is another way of seeing why one cannot allow
more general duality rotations than symplectic ones.

The next step is to take into account that $X^I$ and $F_J$ are
not independent quantities. $F_J$ is a function of the $X^I$ (and
of the chiral background) and 
therefore all transformations
have to be induced by transformations of the scalars $X^I$.
From (\ref{TransfSect}) we can read off that the new scalars
are given by the field-dependent transformation
\be
\breve{X}^I = U^I_{\;\;J}X^J + Z^{IK} F_{K} =
(U^I_{\;\;J} + Z^{IK}F_{KJ})X^J + 
w \widehat{A} Z^{IK} F_{K\widehat{A}} \;.
\ee
This transformation has to be invertible,\footnote{This changes when
we go to the super Poincar\'e theory, because the gauge fixing imposes one
relation among the scalars. We will come back to this later.}
because the number
of independent scalar fields should not change under a duality
transformation. 
Finally one can find the new function $\breve{F}(\breve{X}^I, \widehat{A})$
by integrating (\ref{TransfSect}):
\bea
\breve{F}(\breve{X}, \widehat{A}) &=& F(X, \widehat{A}) - \ft12 X^I F_I
+ \ft12 (U^T W)_{IJ} X^I X^J
+ \ft12 (U^T V + W^T Z)_{I}^{\;\;K} X^I F_K \nonumber \\
& & + \ft12 (Z^T V)^{IJ} F_I F_J \;.\\ 
\nonumber
\eea
The integration constant is fixed by the homogenity 
properties of
the function. Note that $F(X,\widehat{A})$ does not transform as 
a function under symplectic transformations
$F(X,\widehat{A})\not=\breve{F}(\breve{X},\widehat{A})$. An example
of an expression transforming as a function is
\be
{F}({X},\widehat{A}) - \ft12 {X}^I {F}_I =
\breve{F}(\breve{X},\widehat{A}) - \ft12 \breve{X}^I \breve{F}_I \;.
\ee
Such objects, which turn out to be rare, are called
{\em symplectic functions}. Note that being a symplectic function
refers to a simple, tensoriel transformation behaviour, which
is not to be confused  with an invariance property. 
An expression is called 
{\em invariant} (with respect to one, several or all
symplectic transformations)
if the functional dependence on the variables is
not changed under the transformation, 
\be
\breve{f}(\breve{X},\widehat{A}) = f(\breve{X},\widehat{A}) \;.
\ee
If the function $F(X,\widehat{A})$ is invariant under a 
duality transformation, then this transformation is a
symmetry of the theory. Note that this is possible even
though $F(X,\widehat{A})$ is not a symplectic function.
The $T$-duality symmetries of string theory,
which in ${\cal N}=2$ compactifications of heterotic string theories
are realized as a subset of the symplectic reparametrizations,
are examples of such duality symmetries.

In order to write down the transformation rules
for some of the derivatives of $F(X,\widehat{A})$ it
is convenient to introduce the following 
definitions:
\bea
{\cal S}(X,\widehat{A})^{I}_{\;\;J} &:=& 
\ft{ \der \breve{X}^{I} }{\der X^{J}}
= U^{I}_{\;\;J} + Z^{IK} F_{KJ} \;, \nonumber \\
{\cal Z}^{IJ} &:= & [{\cal S}^{-1}]^{I}_{\;\;K} Z^{KJ} \;,
\nonumber \\
N_{IJ} &:= & 2 \mbox{Im} F_{IJ} \;,\nonumber \\
N^{IJ} & := & [N^{-1}]^{IJ} \;. \nonumber \\
\eea
For the lowest derivatives of $F(X,\widehat{A})$ one finds
the following rules:
\bea
\breve{N}_{IJ} &=& N_{KL} [ \ov{\cal S}^{-1}]^{K}_{\;\;I}
[{\cal S}^{-1}]^{L}_{\;\;J}\;, \nonumber \\
\breve{N}^{IJ} &=& N^{KL} \ov{\cal S}^{I}_{\;\;K}
{\cal S}^{J}_{\;\;L} \;, \nonumber \\
\breve{F}_{IJK} &=& F_{MNP} [ {\cal S}^{-1}]^{M}_{\;\;I}
[ {\cal S}^{-1}]^{N}_{\;\;J}
[ {\cal S}^{-1}]^{P}_{\;\;K} \;, \nonumber \\
\eea
and
\bea
\breve{F}_{\widehat{A}} &=& F_{\widehat{A}} \;,\nonumber \\
\breve{F}_{\widehat{A}I} &=& F_{\widehat{A}J} 
[ {\cal S}^{-1}]^{J}_{\;\;I}\;, \nonumber \\
\breve{F}_{I} - \breve{F}_{IJ} \breve{X}^{J} &=&
[F_{J} - F_{JK} X^{K}] [ {\cal S}^{-1}]^{J}_{\;\;I}\;, \nonumber \\
\breve{F}_{I} - \breve{\ov{F} }_{IJ} \breve{X}^{J} &=&
[F_{J} - \ov{F}_{JK} X^{K}] 
[ \ov{\cal S}^{-1}]^{J}_{\;\;I}\;, \nonumber \\
\breve{F}_{\widehat{A} \widehat{A}} &=& F_{\widehat{A} \widehat{A}}
- F_{\widehat{A} I} F_{\widehat{A}J} {\cal Z}^{IJ} \nonumber \\
\label{SymplTransfDerivF}
\eea
and so on.
Note that $F_{\widehat{A}}$ is a symplectic function, whereas
all other derivatives of $F(X,\widehat{A})$ are not.
The above formulae are sufficient to verify that
the terms ${\cal O}^{-}_{I \mu \nu}$ (\ref{Ominus})
transform as required by (\ref{Otilde}).

Though it can be shown that the field equations 
are invariant under continuous $Sp(2N_V+2,\mathbb{R})$ transformations,
one expects that this group is broken to a discrete subgroup,
denoted by $Sp(2 N_V +2, \mathbb{Z})$ at the non-perturbative level.
One way of seeing this is to make the connection with
electric-magnetic duality transformations more explicit. To do
so one writes the vector kinetic term as
\be
e^{-1} {\cal L} = \ft18 N_{IJ} F^I_{\mu \nu} F^J_{\mu \nu}
+ \ft{i}4 \Theta_{IJ} F^I_{\mu \nu} \widetilde{F}^{\mu \nu J} \;,
\ee
where $N_{IJ}$ was defined above and $\Theta_{IJ} = \mbox{Re} F_{IJ}$.
Obviously, $N_{IJ}$ encodes the gauge couplings, whereas 
$\Theta_{IJ}$ are the $\Theta$-angles. The symplectic transformations
contain a subgroup of the form $U=V=\mathbb{I},Z=\mathbb{O}$ and
$W_{IJ} = \Delta \Theta_{IJ}$, which acts by constant shifts of the
$\Theta$-angles. Perturbatively such shifts can be ignored but 
non-perturbatively this is not guaranteed. If, for instance, 
the $U(1)$ effective field theory considered here comes from
a spontanously broken $SU(2)$ Yang-Mills theory, then there will
be instantons, and this restricts the $\Theta$-shifts to a discrete
subgroup.

Another way of seeing the reduction from a continuous to a discrete
group is to take into account the presence of electric and magnetic 
charges. If the effective $U(1)$ theory is obtained by integrating
out the massive degrees of freedom of a string theory, then 
the theory contains electrically and magnetically charged massive
states. Part of them are elementary string states, wheras
the rest is realized as solitons. String dualities require that 
at the non-perturbative level both electric and magnetic charges 
with respect
to all $U(1)$ factors exist. Thus, at the non-perturbative level
one has to take into account that the theory contains electric
and magnetic charges. According to the generalized Dirac
quantization rule \cite{Dir:1931,Sch:1966,Zwa:1968,Wit:1979} 
the allowed values of electric and magnetic
charges are discrete and form a lattice.

The magnetic and electric
charges $(p^I,q_J)$ 
are the sources for the gauge fields $(F^I_{\mu \nu}, G_{\mu \nu J})$.
They are defined by
\bea
p^I &=& \ft1{4\pi} \oint F^I = \ft14 \int_0^{2\pi}  \int_0 ^{\pi}
F_{23}^I r^2 \sin \theta  d \theta d \phi \nonumber \\
q_J &=&  \ft1{4\pi} \oint G_I = \ft14 \int_0^{2\pi}  \int_0 ^{\pi}
G_{J 23} r^2 \sin \theta  d \theta d \phi \nonumber \\
\label{DefCharges}
\eea 
where $F_{ab}^I, G_{J ab}$, $a,b =0,1,2,3$ are the tangent space components
of the field strength. 
The definition of charges is 
such that gauge fields with the asymptotic behaviour
\be
F_{23}^I \simeq_{r \rightarrow \infty} \frac{p^I}{r^2}, \;\;\;
G_{I 23} \simeq_{r \rightarrow \infty} \frac{q_I}{r^2}, \;\;\;
\ee
carry charges $(p^I,q_I)$. 
Note that $(p^I,q_J)$ transforms as 
 a vector under symplectic $Sp(2N_V+2,\mathbb{R})$
transformations. A general continous symplectic transformation
will not map the charge lattice onto itself but will deform it into 
a different
lattice. This means that the spectrum of admissible charged states
is not invariant. If such states are actually present, as it happens
in string theory, then one has to restrict the symplectic transformations
to a discrete subgoup $Sp(2N_V+2,\mathbb{Z})$, which by definition
maps the charge lattice to itself.
When normalizing the gauge fields appropriately,
this is just the subgroup of matrices with integer entries.
Again we have reached the conclusion that the symplectic
group is reduced to a discrete subgroup. Since duality rotations are
not automatically symmetries of one given Lagrangian, this discrete
group is the maximal possible group of duality symmetries. In
string compactifications one can have, depending on the amount
of unbroken supersymmetry various discrete duality symmetries, which
are called T-, S- and U-duality. These are always proper subgroups
of the discrete symplectic group.

\section{Poincar\'e Supergravity}

\subsection{Poincar\'e Gauge Fixing \label{SectionPoincareGaugeFixing}}

Our motivation for constructing ${\cal N}=2$ conformal supergravity
is not to study it as a theory in its own right but to use it as a tool
for dealing with ${\cal N}=2$ Poincar\'e supergravity. We now will
discuss how to go from the superconformal to the super Poincar\'e
theory by imposing gauge conditions, following \cite{deWvHvP:1981}, 
\cite{Kleijn:1998}. More generally, we would like
to know how to go back and forth between the two formulations, because
this gives us the option of analysing questions about the Poincar\'e
theory in the conformal set-up. Since the Poincar\'e theory is a
gauge fixed version of the conformal theory, what we have to do
is to find appropriate gauge conditions and to identify gauge invariant
quantities. The usage of the gauge invariant, conformal formulation
is sometimes advantageous because here more symmetries are realized
in a simple linear way, quantities transform in a simple and 
systematic way and the off-shell multiplets are smaller. In the
gauge-fixed Poincar\'e theory symmetries and in particular supersymmetry
are realized in a more complicated non-linear way. Moreover the multiplets
are larger and since the graviphoton now belongs to the gravity
supermultiplet instead of sitting in a separate vector multiplet,
symplectic reparametrizations are more complicated.

We first sketch the gauge fixing procedure using the most simple example, 
the construction
of pure ${\cal N}=2$ supergravity (without higher derivative terms).
Then we give a more detailed account for the case where
an arbitrary number of abelian vector multiplets is present
together with a chiral background describing higher derivative terms.

The standard 
${\cal N}=2$ Poincar\'e supergravity multiplet 
has $40 + 40$ off-shell
degrees of freedom 
\cite{dWvH:1979,FraVas:1979,deWvHvP:1980}.\footnote{As mentioned
at the beginning of the chapter there exist smaller off-shell
representations.
Within the superconformal approach we only know how to make contact
with the $40+40$ multiplet.} 
The physical degrees of freedom
are the graviton $e_{\mu}^{\;\;a}$, two gravitini $\psi^i_\m$ and
one gauge field, the graviphoton.
On the superconformal side we have to use the Weyl multiplet,
which has $24 + 24$ off-shell degrees of freedom,
to describe the graviton and the gravitini. Since this does not
account for the graviphoton, the natural thing is to add
one vector multiplet $(X,F_{\mu \nu},\ldots)$. The resulting
reducible representation with $32+32$ degrees of freedom is
called the ${\cal N}=2$ minimal field representation, because
it is mandatory if one wants to describe ${\cal N }=2$ 
Poincar\'e supergravity. 
The coupling
to the Weyl multiplet is described by a prepotential, which has
to be holomorphic and homogenous of degree 2. 
If only one vector multiplet is present, then
up to normalization
the unique choice therefore is $F(X)=X^2$. 
The vector multiplet
contains a scalar $X$, which cannot be a physcial degree of freedom
in the Poincar\'e theory. But from the toy example discussed 
earlier we expect that dilatational gauge fixing will be done by
setting this scalar to a constant. This is indeed the case, but
we will not enter into the details here, because we will discuss
this in a more general situation later. At this point we have 
accounted for all physical degrees of freedom of the Poincar\'e
theory. But nevertheless it is clear that this cannot be the full
story. 

The most obvious indication is that the minimal field 
representation has $8+8$ degrees of freedom less then the
$40+40$ of the off-shell super Poincar\'e multiplet.
A closer inspection shows that these
additional degrees of freedom are indeed needed to accomplish
the Poincar\'e gauge fixing, because the fields in the
minimal field representation cannot be used as compensators
for the chiral $SU(2)$ symmetry. A related point, that we already
mentioned, is that the action for the minimal field representation
(plus $N_V \geq 0$ additional vector multiplets) has an 
awkward term linear in the auxiliary $D$-field, that leads to
inconsistent equations of motions unless further terms are 
added.

Thus one always has to add further superconformal matter multiplets
to the minimal field representation in order to describe
Poincar\'e supergravity. 
There are three known choices for a second compensator multiplet
with $8+8$ degrees of freedom: One can use
a so-called non-linear
multiplet, a hypermultiplet or a tensor multiplet.

We will now go through the gauge fixing procedure in some more
detail and consider the more general case, where one starts
with the Weyl multiplet and  $(N_V+1)$ abelian vector multiplets. 
In this case one linear combination
of the field strengths provides the graviphoton whereas
the corresponding scalar becomes dependent on the others through
a gauge condition. This way one arrives at ${\cal N}=2$
supergravity coupled to $N_V$ vector multiplets. The couplings
are encoded in the function $F(X^I, \widehat{A})$, where the chiral
background field $\widehat{\bf A}$ is identified with the Weyl multiplet
${\bf W}^2$.

The first step is to break
special conformal invariance by imposing the $K$-gauge $b_{\mu}=0$. 
Since this constraint is not invariant under the remaining 
transformations, one has to modify the rules for
$Q$- and $S$-supertransformations and for dilatations by
a compensating, field dependent $K$-transformation.
Note, however, that this does not modify the transformation
properties of the other components of the Weyl multiplet and 
of the components of the 
vector multiplet, because these are $K$-independent. Only $b_{\mu}$
and the composite gauge fields transform non-trivially under
special conformal transformations.

The second step is to gauge-fix dilatations.
From the toy example we expect that fixing the dilatations will 
eliminate one scalar field and will lead to a standard Einstein-Hilbert
term with a constant coefficient instead of a field dependent one.
Therefore the natural choice is the $D$-gauge
\be
-i (X^I \ov{F}_I - F_I \ov{X}^I) = 1 ( = m_{\mscr{Planck}}^2) \;.
\ee
This condition is manifestly symplectic. Moreover it is dimensionful
as expected for a constraint that breaks scale invariance. 
The constant on the right hand side should therefore be related
to the natural scale of Poincar\'e gravity, the Planck mass.
We will postpone fixing the overall normalization of our Lagrangian,
because there will be another contribution to the Einstein-Hilbert
term. Except where dimensional analysis is required we will use
Planckian units, $m_{\mscr{Planck}}=1$.
One can now proceed by
fixing the chiral $U(1)$ symmetry, for example by imposing
the $A$-gauge
\be
X^0 = \ov{X}^0 \;.
\ee
Both the $D$- and the $A$-gauge are constraints on the scalars 
$X^I$. As a consequence the Poincar\'e theory has only 
$N_V$ independent scalars. Further analysis leads to the notion
of special K\"ahler geometry, that we will discuss
in \ref{SectionSpecialGeneral}. At this point two comments are in order:
First, for many purposes it is useful not to fix
the gauge but to 
stay in the superconformal setup and to work with 
appropriate gauge invariant variables under dilatations and $U(1)$ 
transformations. 
Second, the above gauge choices are not
unique and depending on the problem under considerations different
choices might be useful.

In order to break $S$-supersymmetry one imposes 
another constraint, called the $S$-gauge.
This constraint can be solved by eliminating one 
of the vector multiplet fermions. 
As a result the physical fields now precisely correspond
to the ${\cal N}=2$ Poincar\'e gravity multiplet plus 
$N_V$ vector multiplets. To be precise the $S$-gauge also breaks
$Q$-supersymmetry, but a combination of a $Q$-transformation and
a compensating $S$-transformation is preserved. Furthermore one
has to combine this with a compensating $K$-transformation which
restores the $K$-gauge. Therefore the Poincar\'e supertransformations
take the form
\be
\delta_Q^{\mscr{Poincar\'e}}(\epsilon) = \delta_Q (\epsilon)
+ \delta_S(\eta) + \delta_K(\Lambda_K) \;,
\ee
with suitable field dependent choinces of $\eta$ and $\Lambda_K$.
The above formula is an example of a {\em decomposition rule}, that
is a rule which displays a combination of symmetries that is
left unbroken by a gauge choice. Obviously supersymmetry is
realized in a much more complicated way in the Poincar\'e theory
than in the conformal theory.

Finally we have to discuss the gauge-fixing of chiral 
$SU(2)$ transformations. We already mentioned that one
has to add another multiplet in order to provide the
necessary compensators. For definiteness we will use
the non-linear multiplet. The correct coupling can
be found by the following reasoning: As already mentioned
the Lagrangian of the minimal field representation 
(with $N_V$ vector multiplets
added) has a term linear in the auxiliary $D$ field,
\be
e^{-1} {\cal L} \sim -i (X^I \ov{F}_I - F_I \ov{X}^I) ( D- \ft16 R) 
+ \cdots
\label{MinRandD}
\ee
which leads
to inconsistent equations of motion. We have displayed the Einstein-Hilbert
term as well for reasons that will become obvious in a second.
The idea is to couple
the non-linear multiplet such that all linear terms involving
$D$ are precisely canceled. We now remind ourselves that the 
non-linear multiplet is subject to the constraint
\be
D^a V_a - 3 D - \ft12 V^a V_a - \ft14 | M_{ij} |^2 + D^a \Phi^i_{\;\;\alpha}
D_a \Phi^{\alpha}_{\;\;i} + \cdots = 0 \;,
\ee
where we neglected the fermionic terms, because we are only inerested in the
bosonic Lagrangian. The constraint can be interpreted as a constraint on
the vector $V_a$, but now we take it as the defining equation
of $D$ in terms of the non-linear multiplet and take
$V_a$ to be unconstrained. When coupling
the non-linear multiplet to the ${\cal N}=2$ minimal field representation
this has the net effect of adding $8+8$ degrees of freedom, so that
one has a total of $40 + 40$ (not counting the additional $N_V$ 
vector multiplets).
Since the constraint contains $D$ linearly one takes the Lagrangian
of the non-linear multiplet to be proportional to the constraint. 
The normalization of this term 
is chosen such that the terms involving $D$ cancel in the full Lagrangian. 
At this point one has to keep in
mind that $V_a$ transforms under $K$-transformations. As a consequence
its covariant derivative contains a term linear in $D$:
\be
D^a V_a = {\cal D}^a V_a  - 2 f^a_a + \cdots =
{\cal D}^a  V_a + 2 D - \ft13 R + \cdots
\ee
and the term to be added to the Lagrangian is proportional to
\be
{\cal D}^a V_a - D - \ft13 R  
- \ft12 V^a V_a - \ft14 | M_{ij} |^2 + D^a \Phi^i_{\;\;\alpha}
D_a \Phi^{\alpha}_{\;\;i} + \mbox{fermions} \;.
\ee
Comparing this to (\ref{MinRandD}) we see that we have to
add the term
\be
e^{-1} \Delta {\cal L} = 
- i(X^I \ov{F}_I - F_I \ov{X}^I) ({\cal D}^a V_a  -\ft13 R - D + \cdots)
\ee
on the right hand side. The resulting bosonic Lagrangian
is
\bea
8 \pi e^{-1} {\cal L} &=&
( - i (  X^I\ov{F}_I - F_I \ov{X}^I )) \cdot  ( - \ft12 R ) \nonumber \\
 & & + \big[ i  {\cal D}_{\mu} F_I {\cal D}^{\mu} \ov{X}^I 
 \nonumber \\
 & & + \ft14 i F_{IJ} ( F^{-I}_{ab} - \ft14 \ov{X}^I T^{ij}_{ab} \ve_{ij})
 ( F^{-J}_{ab} - \ft14 \ov{X}^J T^{ij}_{ab} \ve_{ij})
- \ft18 i F_I ( F^{+I}_{ab} - \ft{1}{4}X^I T_{abij} \ve^{ij}) T^{ab}_{ij}
 \ve^{ij} \nonumber \\
 & & - \ft18 i F_{IJ} Y^I_{ij} Y^{J ij}  
-\ft{i}{32}F(T_{abij}\ve^{ij})^2 \nonumber \\
 & & + \ft12 i F_{\widehat{A}} \widehat{C} - \ft18 i F_{\widehat{A} \widehat{A}}
(\ve^{ik} \ve^{jl} \widehat{B}_{ij} \widehat{B}_{kl} - 2 \widehat{F}^-_{ab}
\widehat{F}^-_{ab}) 
+ \ft12 i \widehat{F}^-_{ab} F_{\widehat{A}I} ( F^{-I}_{ab} - \ft14 \ov{X}^I
T^{ij}_{ab} \ve_{ij})  \nonumber\\
 & &  - \ft14 i \widehat{B}_{ij} F_{\widehat{A}I} Y^{Iij}
+ \mbox{h.c.} \big]  \nonumber\\
 & & -i (  X^I\ov{F}_I - F_I \ov{X}^I ) \cdot ({\cal D}^a V_a 
- \ft12 V^a V_a - \ft14 | M_{ij} |^2 + D^a \Phi^i_{\;\;\alpha}
D_a \Phi^{\alpha}_{\;\;i}) \;. \label{ActionWVNL}\\ 
\nonumber
\eea
Note that we now have fixed the overall normalization of our Lagrangian.
We have chosen it such that when imposing the $D$-gauge
$-i(X^I\ov{F}_I - F_I \ov{X}^I ) = m_{\mscr{Planck}}^2$ the 
Einstein-Hilbert term takes the form
\be
e^{-1} {\cal L}_{EH} = -\frac{1}{2 \kappa^2} R \;,
\ee
where $\kappa$ is the gravitational coupling, which is related
to Newtons constant by $\kappa^2 = 8 \pi G_N$. In natural units
($\hbar = c = 1$) we have 
$G_N = l_{\mscr{Planck}}^2 = m_{\mscr{Planck}}^{-2}$ so that
\be
e^{-1} {\cal L}_{EH} = -\frac{m_{\mscr{Planck}}^2}{16 \pi} R \;,
\ee
which explains the factor $8 \pi$ on the left hand side of
(\ref{ActionWVNL}).

After discussing the Einstein-Hilbert term we next
display the curvature squared terms in
the Lagrangian. To do so we need to work out the term
\be
8 \pi e^{-1} {\cal L} =  \ft{i}2 F_{\widehat{A}} \widehat{C} + \mbox{h.c.}
+ \cdots \;.
\ee
First recall that the highest component $\widehat{C}$
of the ${\bf W}^2$ superfield is given by
\be
\widehat{C} = 64 {\cal R}(M)^{-\;\;ab}_{cd} {\cal R}(M)^{-\;\;cd}_{ab}
+ \cdots \;,
\ee
where the neglected terms are at most linear in the curvature. Next
recall that
\be
{\cal R}(M)_{ab}^{\;\;\;\;cd} = R_{ab}^{\;\;\;\;cd} - 4 
\d_{[a}^{\;\;[c} f_{b]}^{\;\;d]} + \cdots \;,
\ee
where the omitted terms do not depend on the curvature.
But the $K$-gauge field $f_{a}^{\;\;b}$ is a composite field which
depends on the curvature. Substituting the explicit form we find
\be
{\cal R}(M)_{ab}^{\;\;\;\;cd} = C_{ab}^{\;\;\;\;cd} + \cdots \;,
\ee
where $C_{ab}^{\;\;\;\;cd}$ reduces to the Weyl tensor in the
Poincar\'e frame. As a consequence we 
have\footnote{Recall that $C^-_{abcd}$ is antiselfdual in both pairs of
indices.} 
\be
\widehat{C} = 64 C^{-\;\;ab}_{cd} C^{-\;\;cd}_{ab} + \cdots 
=: 64 (C^-_{abcd})^2 + \cdots \;.
\ee
This motivates the name Weyl multiplet.

Next the function $F_{\widehat{A}}$ is the derivative of 
$F(X,\widehat{A})$ with respect to 
$\widehat{A} = T^-_{ab} T^{-ab}=: (T^-_{ab})^2$,
where $T^-_{ab}$ is the auxiliary $T$-field. Thus
\be
F_{\widehat{A}} = F^{(1)}(X) + 2 F^{(2)}(X) (T^-_{ab})^2 +  \cdots \;.
\ee
Therefore we get a series of interaction terms which are quadratic in the
Weyl tensor and even powers in the $T$-field, with field dependent
couplings $F^{(g\geq 1)}$:
\be
8 \pi e^{-1} {\cal L} = 32 i \left( F^{(1)}(X) (C^-)^2
+ 2 F^{(2)}(X) (C^-)^2 (T^-)^2 + 3 F^{(3)}(X) (C^-)^2 (T^-)^4 + \cdots
\right) + \mbox{h.c.}
\label{WeylPowersInAction}
\ee

For $g=1$ we get purely gravitational couplings, which can be 
rewritten as
\be
8 \pi e^{-1} {\cal L} = - 32 \,\mbox{Im} \; F^{(1)}(X) (C_{abdc})^2 - 32 i
\, \mbox{Re} \; F^{(1)}(X) C_{abcd} \widetilde{ C}^{abcd} \;.
\ee
The first term is the Weyl action with a field dependent 
coupling. 
The second term is real (in Minkowski signature) in view
of our definition of the dual tensor, which includes a factor $i$.
It is a  
'gravitational $\theta$-term', because 
$C_{abcd} \widetilde{C}^{abcd}= R_{abcd}\widetilde{R}^{abcd}$ and
$R_{abcd}\widetilde{R}^{abcd}$ is proportional to
the Hirzebruch signature density. We already remarked 
in chapter \ref{ChapterGravity} that this action resembles
a Yang-Mills action in many respects.

We also note that our Lagrangian contains various other higher
derivative terms. For example when expanding the 'gauge kinetic term'
we find a series of higher powers of the auxiliary $T$-field:
\be
8 \pi e^{-1} {\cal L}= \ft{i}4 F^{I-}_{ab} F^{J-ab} \left(
F_{IJ}^{(0)}(X) + F^{(1)}_{IJ}(X) (T^-_{ab})^2 + F^{(2)}_{IJ}(X)
(T^-_{ab})^4 + \cdots \right) + \mbox{h.c.}  
\ee

Finally we comment on the terms in the last line of (\ref{ActionWVNL}), 
which come
from the second compensating multiplet.
We note that when imposing the $D$-gauge the term
${\cal D}^a V_a$ becomes a total derivative and can be dropped.
The local chiral $SU(2)$ invariance can be 
gauge-fixed by imposing the $V$-gauge
\be
\Phi^i_{\;\;\alpha} = \delta^i_{\alpha} \;.
\ee
Once this is imposed the theory is only invariant under
rigid $SU(2)_R$ rotations, which are automorphisms of the
super Poincar\'e algebra.

\subsection{Special Geometry in Special 
Coordinates \label{SectionSpecialSpecial}}

The next step is to eliminate the auxiliary fields by their equations of
motion and to obtain an on-shell formulation of the theory in terms
of unconstrained physical fields.
This can be done in closed form for the minimal
terms, i.e. as long as the chiral background $\widehat{\bf A} = {\bf W}^2$
is absent. Therefore we first restrict ourselves to a discussion
of the minimal terms related to the prepotential
$F(X^I, \widehat{A}=0) = F^{(0)}(X^I)$ and comment on the higher 
derivative terms later on. We will see that the target space of the
scalar sigma model is restricted to be a {\em special K\"ahler manifold},
and that the structure of the whole theory is governed by {\em special
geometry} \cite{deWvP:1984,CreEtAl:1985,deWLauvP:1985}.

The physical fields of the theory are the graviton $e_{\mu}^{\;\;a}$,
two gravitini $\psi^i_\m$, $N_V +1$ gauge fields $W_{\mu}^I$,
and $N_V$ gaugini-doublets and scalars described by $\Omega^I_i$
and $X^I$, subject to one complex constraint. Finding an on-shell
formulation in terms of unconstrained fields includes the 
following: The auxiliary fields $T^{ij}_{ab}, \chi^i, D$ and
$Y^I_{ij}$ have to be eliminated, as well as the gauge fields 
$b_{\mu},A_{\mu},{\cal V}_{\mu\;\;j}^{\;\;i}$ of the local 
dilatational and $U(1) \otimes SU(2)$ gauge symmetry 
and the scalars and fermions have to be expressed in
terms of unconstrained fields. We will only discuss the purely
bosonic part of the Lagrangian. The fields $b_{\mu},D$ have already
been dealt with by imposing the $K$-gauge and by coupling to the
non-linear multiplet, respectively. Moreover it turns out that
${\cal V}_{\mu \;\;j}^{\;\;i}$ and $Y^I_{ij}$ are only relevant
for fermionic terms.\footnote{The equations of motion of
${\cal V}_{\mu \;\;j}^{\;\;i}$ also involve hypermultiplet scalars,
but we are concerned with the vector multiplets, only.}
It remains to deal with the $U(1)$ connection
$A_{\mu}$ and with the auxiliary field $T^{ij}_{ab}$.

Solving the equation of motion for the $U(1)$ gauge field
gives\footnote{In chapter \ref{ChapterGravity} the graviphoton
was denoted $A_\m$. In this chapter $A_\m$ denotes the composite
$U(1)$ connection, which is a completely different object. 
The graviphoton will only enter via its field strength in the
following.} 
\be
A_{\mu} = \ft{1}{2} \ft{ \ov{F}_I 
\stackrel{\leftrightarrow}{\der}_\m X^I -
\ov{X}^I \stackrel{\leftrightarrow}{\der}_\m F_I}
{-i (X^I \ov{F}_I - F_I \ov{X}^I)} \;.
\label{AasComposite}
\ee
In this formula neither the $D$- nor the $A$-gauge has been imposed.
Note that the field has Weyl and chiral weight 0 and therefore
is gauge invariant. The above formula is used when working
in the superconformal setup. A gauge fixed version is obtained 
by imposing the $D$-gauge, which sets the denominator to unity.
In absence of a chiral background we can use the homogenity
properties $F_I = F_{IJ}X^J$ and $F_{IJK}X^K=0$
of the prepotential to rewrite the numerator and we get:
\be
A_{\mu} = \ft{i}2 N_{IJ}  (X^I \der_{\mu} \ov{X}^J - 
\ov{X}^I \der_{\mu} X^J ) \;.
\ee
As a result the scalar kinetic term 
\be
8 \pi e^{-1} {\cal L}_{\mscr{scalar}} =
i {\cal D}_\m F_I {\cal D}^\m \ov{X}^I + \mbox{h.c.}
\ee
takes the form
\be
8 \pi e^{-1} {\cal L}_{\mscr{scalar}} =
- N_{IJ} \der_\m X^I \der^\m \ov{X}^J + \ft14
( N_{IJ} (X^I \der_\m \ov{X}^J - \ov{X}^I \der^\m X^J ))^2\;,
\ee
where we again used homogenity properties (in particular that the
$D$-gauge takes the form $N_{IJ}X^I \ov{X}^J = -1$).

One could procede fixing the $U(1)$ gauge invariance
by a suitable gauge condition and then solve the two real gauge
conditions in terms of $N_V$ unconstrained scalars. In practice
it is more convenient to leave the local $U(1)$ invariance intact.
This is related to the fact that scalar fields in a locally
supersymmetric theory are described by non-linear sigma models
with a K\"ahler manifold as target space.\footnote{In the case
of abelian vector multiplets in ${\cal N}=2$ supergravity no
scalar potential is possible.} Such models have an
invariance under K\"ahler transformations, which are closely related to
the local $U(1)$ transformations. Therefore it is natural
to leave this symmetry intact. The corresponding gauge field
is a composite field given by (\ref{AasComposite}).

There is an elegant way of reformulating the theory in terms
of $N_V$ unconstrained scalars (modulo local $U(1)$ transformations).
It is based on the observation that the $D$-gauge condition has
a geometrical interpretation, because it is a normalization
condition for the scalars $X^I$. 
One interprets
the $X^I$ as homogenous coordinates of the scalar manifold and 
introduces inhomogenous coordinates by
\be
Z^0 = 1 = \frac{X^0}{X^0}\;, \;\;\;
Z^A = \frac{X^A}{X^0} \;,
\ee
where $A=1,\ldots, N_V$.
Since these equations are $U(1)$ invariant one can use them to
reconstruct
the $X^I$ from a given set of inhomogenous coordinates
$Z^A$ up to an irrelevant phase. The $Z^A$ are called
{\em special coordinates} of the scalar manifold, because they
provide a coordinatization that is distinguished by its relation
to the ${\cal N}=2$ multiplet structure. General coordinatizations
will be discussed in the next section.

We can now rewrite the scalar kinetic term using the unconstrained
fields $Z^A$. First note that one can use the homogenity 
of $F(X^I)$ to rewrite it as a 
function of the $Z^I$:
\be
F(X^0,\ldots,X^{N_V}) = (X^0)^2 F(1,Z^1,\ldots,Z^{N_V})=:
(X^0)^2 \widetilde{F}(Z^1,\ldots,Z^{N_V}) \;.
\ee
For simplicity we will not introduce a new symbol for the 
prepotential as a function of $Z^I$ and simply write $F(Z)$
instead of $\widetilde{F}(Z)$. Since the $r$-th derivative is homogenous
of degree $2-r$ we can likewise rewrite them:
\be 
F(X) = (X^0)^{2} F(Z), \;\;\;
F_I(X) = X^0 F_I(Z), \;\;\;
F_{IJ}(X) = F_{IJ}(Z)\;,
\ee
etc.

It is useful to introduce the matrix
\be
{\cal M}_{I \ov{J}} := N_{IJ} + N_{IK}\ov{X}^K N_{JL} X^L \;,
\ee
which has two null directions that are identified by the transversality
equations
\be
X^I {\cal M}_{I \ov{J}} = 0 = {\cal M}_{I \ov{J}} \ov{X}^J \;.
\ee
Using this matrix the scalar kinetic term can be rewritten
\be
8 \pi e^{-1} {\cal L}_{\mscr{scalar}} = - {\cal M}_{I \ov{J}}
\der_\m X^I \der^\m \ov{X}^J \;,
\ee
as can be verified using homogenity and the $D$-gauge
(note in particular $N_{IJ}X^I \ov{X}^J = -1$ combined with
$F_{IJK}X^K = 0$ implies
$N_{IL}X^L \der_{\m} \ov{X}^I = - N_{IL} \ov{X}^L \der_\m X^I$).
Using the transversality equations this implies
\be
8 \pi e^{-1} {\cal L}_{\mscr{scalar}} = (Z^K N_{KL} \ov{Z}^L)^{-1}
{\cal M}_{I \ov{J}}
\der_\m Z^I \der^\m \ov{Z}^J \;,
\ee
where we used the $D$-gauge
\be
N_{IJ} Z^I \ov{Z}^J = -|X^0|^{-2} \;.
\ee
${\cal M}_{I \ov{J}}$ can be expressed in terms of
$Z^I$ as
\be
{\cal M}_{I \ov{J}} = N_{IJ} - 
\frac{N_{IK} \ov{Z}^K N_{JL} Z^L}{
Z^M N_{MN} \ov{Z}^N} \;.
\ee
Then it is easy to show that
\be
K(Z,\ov{Z}) = - \log( - N_{IJ} Z^I \ov{Z}^J )
= - \log( -i (Z^I \overline{F}_I - \overline{Z}^I F_I ))
\label{KaehlerPotentialSpecialCoordinates}
\ee
is a K\"ahler potential for the scalar metric,
because
\be
\der_I \der_{\ov{J}} K = - (Z^K N_{KL} \ov{Z}^L)^{-1} 
{\cal M}_{I \ov{J}} \;.
\ee
Therefore the scalar kinetic term is a K\"ahler sigma model:
\be
8 \pi e^{-1} {\cal L}_{\mscr{scalar}} =
- g_{I \ov{J}} \der_\m Z^I \der^\m \ov{Z}^J, \;\;\;
g_{I\ov{J}} = \der_I \der_{\ov{J}} K \;.
\ee

The target space of a scalar sigma model coupled to 
supergravity is not only restricted to be K\"ahler, but
has to be K\"ahler-Hodge. This means that the K\"ahler metric
comes from the fibre metric of a $U(1)$ bundle over the
scalar manifold. The sigma model found above is of that type
and the scalars $X^I$ are, speaking geometrically, sections
of this $U(1)$ bundle. We will discuss this in the
next section, when we investigate the intrinsic structure
of the scalar manifold.

The Riemann curvature tensor of the scalar metric takes a
remarkably simple form:
\be
R^{A\;\;\;\;D}_{\;\;BC} = - 2 \delta^A_{\;\;(B}
\delta^D_{\;\;C)} - e^{2K} Q_{BCE} \ov{Q}^{EAD} \;,
\label{special1}
\ee
where
\be
Q_{ABC} := i F_{IJK}(X(Z)) \frac{\der X^I(Z)}{\der Z^A}
\frac{\der X^J(Z)}{\der Z^B}\frac{\der X^K(Z)}{\der Z^C} \;.
\ee
A K\"ahler-Hodge manifold where the curvature tensor takes the
special form (\ref{special1}) is called a {\em special K\"ahler 
manifold}.
The above construction shows that in ${\cal N}=2$ Poincar\'e 
supergravity the geometry of the scalar manifold is even more
restricted then in the case ${\cal N}=1$. This is so because 
vector multiplets contain both scalars and gauge fields. We will
see in the next section how the additional geometric structure is
related to symplectic invariance.

We now consider the vector kinetic term and the auxiliary 
$T$-field. The corresponding terms in the Lagrangian can be
reorganized in the following way:
\be
8 \pi e^{-1} {\cal L}_{\mscr{vect}} =
\ft{i}4 F_{IJ} F^{-I}_{ab} F^{-J ab} + \ft{i}8(
\ov{F}_I - F_{IJ} \ov{X}^J) F_{ab}^{-I} T^{- ab}
- \ft1{64} N_{IJ} \ov{X}^I \ov{X}^J T^-_{ab} T^{-ab}
+ \mbox{h.c.}\;.
\ee
We can now solve for the auxiliary $T$-field:
\be
T^-_{ab} = 4 (\ov{X}^K N_{KL} \ov{X}^L )^{-1} N_{IJ} \ov{X}^J F_{ab}^{-I}\;.
\ee
Plugging this back into the Lagrangian we get
\be
8 \pi e^{-1} {\cal L}_{\mscr{vect}} =
\ft{i}4 \ov{\cal N}_{IJ} F^{-I}_{ab} F^{-J ab} - \ft{i}4 {\cal N}_{IJ} 
F^{+I}_{ab} F^{+Jab} \;,
\ee
where
\be
{\cal N}_{IJ} = \ov{F}_{IJ} + 
i \frac{N_{IK}Z^K N_{JL} Z^L}{Z^M N_{MN} Z^N} \;.
\ee
After the elimination of the auxiliary $T$-field the dual gauge field
takes the form
\be
G_{I ab}^- = \overline{{\cal N}}_{IJ} F^{-J}_{ab} \;.
\ee
Comparing to the bosonic off-shell Lagrangian we see that
all terms linear in the field strengths
become quadratic and the field dependent coupling and $\Theta$-angles
take a more complicated form than in the off-shell formulation. 
But the theory still has symplectic reparametrization invariance.
In particular the 
matrix ${\cal N}_{IJ}$ transforms by projective transformations, so
that formulae of section \ref{SectionSymplectic} 
carry over. Note that the matrix
${\cal N}_{IJ}$ is invertible, because all the $N_V +1$ field strengths
are independent.

The gauge field
sector of the Poincar\'e theory is more complicated because
one linear combination of the gauge fields belongs to the
gravity multiplet. In the superconformal setup the $N_V+1$ field strengths
together with their duals form a symplectic vector, whereas the 
Weyl multiplet is invariant. In the Poincar\'e theory the 
gravity supermultiplet has to be invariant and therefore the
graviphoton is given by the symplectically invariant combination
\be
T^{\mscr{GP}-}_{ab} = F_I F^-_{I ab} - X^I G_{I ab}^- \;.
\ee
In the off-shell formulation the auxiliary 
field $T^-_{ab}$ satisfies formally the same relation and is therefore often
simply called the graviphoton. Note however that
in the off-shell formalism the definition of $G_{I ab}$ depends
itself on $T^-_{ab}$ and therefore the relation is implicit.
Moreover in the presence of $R^2$-terms the auxiliary $T$-field 
cannot be solved for explicitly but only iteratively. Therefore 
the relation between $T$-field and graviphoton
is complicated.

The graviphoton field of ${\cal N}=2$ supergravity gauges the central
charge transformations. The associated conserved charge 
\be
Z = \ft1{4\pi} \oint T^- = \lim_{r \rightarrow \infty}
\ft1{4 \pi} \int_0^{2\pi} \int_0^{\pi}
T^-_{23} r^2 \sin \theta d\theta d\phi
\label{centralcharge}
\ee
is the central charge of the ${\cal N}=2$ supersymmetry 
algebra \cite{Tei:1977}.

\subsection{Special Geometry in General Coordinates \label{SectionSpecialGeneral}}

The special coordinates $Z^{A}$ are non-generic 
holomorphic coordinates, singled out by their relation to
the ${\cal N}=2$ vector supermultiplets. In order to 
analyse intrinsic geometric properties of the scalar
manifold we now introduce a set $z^{A}$ of generic
local holomorphic coordinates. The special coordinates, being
a specific system of holomorphic coordinates, are
holomorphically related to them, $Z^{I}=Z^{I}(z)$.

Let us first clarify the relation between local $U(1)$ transformations
and K\"ahler transformations, following \cite{CraRooTrovP:1997}. 
The scalars
$X^{I}$ transform as follows under local
scale ($D$) and $U(1)$ ($A$) transformations:
\be
X^I  \rightarrow e^{\Lambda_D(z) - i \Lambda_A(z)} X^I \;.
\ee
One can split the $X^I$ as
\be
X^I = a X^I(z) \;,
\ee
which introduces a holomorphic ambiguity
\be 
X^{I}(z) \rightarrow e^{\Lambda(z)} X^{I}(z) \;,\;\;\;
a \rightarrow e^{-\Lambda(z)} a \;.
\label{HolSym}
\ee
The motivation behind the notation $X^I(z)$ will become obvious later.
The extra symmetry (\ref{HolSym}) can be used to take $X^I(z)$ to be
invariant under scale and $U(1)$ transformations. 
The $D$-gauge implies $|a|^{2} = e^{K}$, where
\be
K(z,\ov{z}) = - \log \left( -i[X^{I}(z) \ov{F}_{I}(\ov{X}(\ov{z})) -
F_{I}(X(z)) \ov{X}^{I}(\ov{z})]\right) \;.
\label{KaehlerPotentialHolomorphicSections}
\ee
The $U(1)$ invariance can be fixed by choosing $a$ to be
real and positive, $a=e^{K/2}$:
\be
X^{I} = e^{\ft12 K(z,\ov{z})} X^{I}(z) \;.
\ee
After that, there is still a residual invariance
under combined $A$ and holomorphic transformations
\be
-i \Lambda_A(z) = \ft12 \left( f(z) - \overline{f}(\overline{z})
\right), \;\;\;
\Lambda(z) = f(z) \;,
\ee
which act by
\be
X^{I} \rightarrow e^{i \mscr{Im} f(z) } X^{I}\;,
\;\;\;
a \rightarrow e^{- \mscr{Re} f(z) } a \;, \;\;\;
\;\;\;X^{I}(z) \rightarrow e^{f(z)} X^{I}(z) \;.
\label{CombinedTransformation}
\ee
This invariance can be fixed by imposing a constraint 
on the $X^{I}(z)$. One possible choice is to take
them to be special coordinates,
\be
X^{0}(z) = Z^{0}=1 \mbox{   and   }X^{A}(z) = Z^{A} \;.
\ee
Comparing to (\ref{KaehlerPotentialSpecialCoordinates})
we see that the function (\ref{KaehlerPotentialHolomorphicSections})
is the K\"ahler potential of the scalar sigma model.

We could instead impose another condition, but since 
we know that the scalar manifold is a holomorphic
hypersurface the constraint must take the form
$g(X^{I}(z)) = \mbox{const}$, with holomorphic $g$.
Changing the condition amounts to a holomorphic reparametrization
of the hypersurface, $X^I(z) \rightarrow e^{f(z)} X^I(z)$, which
acts on the $X^I$ as a local $U(1)$ transformation, according
to (\ref{CombinedTransformation}). Note that the $X^I(z)$ are
holomorphically related to the generic holomorphic coordinates $z^A$.
This motivates the notation we have chosen.

The transformation (\ref{CombinedTransformation})
acts on the K\"ahler potential as a K\"ahler transformation
\be
K(z,\ov{z}) \rightarrow K - f - \ov{f} \;,
\ee
which leaves the metric of the scalar sigma model invariant.
This is the link between local $U(1)$ transformations and
K\"ahler transformations that we promised to explain.

The next step is to find out how the
theory behaves under a general holomorphic reparametrization
$z \rightarrow z'(z)$. This way 
one can arrive at an intrinsic characterization
of special K\"ahler geometry and
see that the 
fields $X^I$ and $X^I(z)$ can be geometrically characterized as
sections of certain bundles over the scalar manifold.

Note that in the minimal coupling case the knowledge of the 
symplectic vector $(X^{I},F_{J})$ is sufficient to write down 
the on-shell action. In the absence of the chiral background $2 F= F_{I} X^I$
and therefore we can take $(X^I, F_J)$ as the defining data.\footnote{As
we will see later, there exist vectors of the form $(X^I,F_J)$, which are
not related to a prepotential $F(X)$. But it turns out that such choices are
related by symplectic transformations to symplectic vectors which come
from a prepotential. Also note that the existence of a prepotential is
indispensable for the off-shell formulation. Only after elimination
of the auxiliary fields the theory can be formulated purely in terms
of $(X^I,F_J)$.}
Using the newly defined $X^{I}(z)$ and
the homogenity of the prepotential we can likewise
describe it by $(X^{I}(z),F_{J}(X(z)))$. Since the
form of the action cannot depend on how
we parametrize the scalar manifold, a reparametrization 
$z \rightarrow z'$ must yield a new vector 
$(X^{I}(z') , F_{J}(X(z')))$. The most general way
in which the two vectors can be related is a combination
of a K\"ahler transformation with a symplectic transformation:
\be
\left( \begin{array}{c} 
X^{I}(z') \\
F_{J}(X(z')) \\
\end{array} \right) =
e^{f(z)} {\cal O}^{IL}_{JK}
\left(  \begin{array}{c}
X^{K}(z) \\
F_{L}(X(z)) \end{array} \right) \;.
\ee
For $(X^{I},F_{J})$ the K\"ahler transformation acts
by the induced $U(1)$ transformation:
\be
\left( \begin{array}{c} 
(X^{I})' \\
(F_{J})' \\
\end{array} \right) =
e^{i \mscr{Im} f(z)} {\cal O}^{IL}_{JK}
\left(  \begin{array}{c}
X^{K} \\
F_{L} \\ \end{array} \right) \;.
\ee
This provides a geometric characterization of the
symplectic vectors: $(X^{I}(z),F_{J}(X(z)))$ is a
section of a bundle ${\cal L} \otimes {\cal H}$ over
the scalar manifold, where ${\cal L}$ is a holomorphic
line bundle and ${\cal H}$ is a flat symplectic vector
bundle. $(X^{I},F_{J})$ is a section of a related bundle
${\cal P} \otimes {\cal H}$, where ${\cal  P}$ is the
principal $U(1)$ bundle associated to ${\cal L}$.

We now collect how various terms in the Lagrangian look,
when expressed in terms of general coordinates. 
The 
scalar kinetic term is a K\"ahler sigma-model,
\be
e^{-1} {\cal L}_{\mscr{scalar}} \sim - g_{A \ov{B}} \der_\m z^A
\der^\m \ov{z}^{\ov{B}}, \;\;\;
g_{A \ov{B}} = \der_A \der_{\ov{B}} K(z,\ov{z})\;,
\ee
with K\"ahler potential
\be
K = - \log \left( -i [X^I(z) \ov{F}_I(\ov{X}( \ov{z})) 
- F_I(X(z)) \ov{X}^I (\ov{z}) ] \right)
\ee
and the vector kinetic term is
\be
e^{-1} {\cal L}_{\mscr{vect}} \sim
\ft{i}4 \ov{\cal N}_{IJ} F^{-I}_{ab} F^{-Iab} - \ft{i}4
{\cal N}_{IJ} F^{I+}_{ab} F^{I+ ab} \;,
\ee
with
\be
{\cal N}_{IJ} = \ov{F}_{IJ} + i 
\frac{N_{IK} X^K(z) N_{JL} X^L(z)}{X^M(z) N_{MN} X^N(z)} \;.
\ee
In order to define a physical model
the section  $(X^I(z),F_J(z))$ has to be chosen such that the
various kinetic terms are positive definite. From the
spin-two, spin-one and spin-zero kinetic terms one gets three conditions:
\[ -i (X^I(z) \ov{F}_I (\ov{X}(\ov{z}) 
- F_I(X(z)) \ov{X}^I(\ov{z})) > 0 \;,
\] 
and the matrices $\mbox{Im} \; {\cal N}_{IJ}$ and
$g_{A \ov{B}}(z, \ov{z})$ must be negatvie definite and 
positive definite, respectively.
It can be shown that the second condition is implied by the
first and third \cite{CreEtAl:1985}.

Let us investigate the bundles ${\cal L}$ and ${\cal P}$
a little closer \cite{Fre:1995}. A section
$\phi(z,\ov{z})$ of ${\cal L}^p$, the $p$-th power of ${\cal L}$ is
an object that transforms according to
\be
\phi(z,\ov{z}) \rightarrow e^{pf(z)} \phi(z,\ov{z}) \;.
\ee
A natural connection on ${\cal L}^p$ is  found by covariantizing
the complex partial derivatives with respect to the K\"ahler transformations:
\be
D_A \phi = (\der_A + p (\der_A K)) \phi, \;\;\;
D_{\ov{A}} \phi = \der_{\ov{A}} \phi \;.
\ee
Note that for ${\cal L}^p$ covariantly holomorphic is the same
as holomorphic, $D_{\ov{A}} \phi = \der_{\ov{A}} \phi = 0$.
The sections $X^I(z)$ of ${\cal L}$ are holomorphic, 
$\der_{\ov{A}} X^I(z) = 0$.

Sections $\varphi(z,\ov{z})$ of ${\cal P}^p$, the 
$p$-th power of the associated principal $U(1)$ bundle 
transform as
\be
\varphi(z,\ov{z}) \rightarrow e^{i \,p\, \mscr{Im}\, f(z)} 
\varphi(z,\ov{z}) \;.
\ee
Such sections can be obtained from sections of ${\cal L}^p$ by
$\varphi = e^{p/2 K} \phi$. The $U(1)$ covariant derivatives
are 
\be
D'_A \varphi = (\der_A + \ft{p}2 (\der_A K)) \varphi, \;\;\;
D'_{\ov{A}} \varphi = (\der_{\ov{A}} - 
\ft{p}2 (\der_{\ov{A}} K)) \varphi \,.
\ee
A section of ${\cal P}^p$ is covariantly holomorphic,
$D'_{\ov{A}} \varphi = 0$ if and only if the corresponding
section of ${\cal L}^p$ is holomorphic. In particular the
sections $X^I$ of ${\cal P}$ are covariantly holomorphic.

The bundles ${\cal L}$ and ${\cal P}$ are nontrivial and 
the connections introduced above can be used to compute their
first Chern classes, or, in physical terms, the integrated 
field strength associated with the K\"ahler gauge field.

The connection one-form of the holomorphic connection on ${\cal L}$ is
$\Theta = \der K = \der_A K dz^A$, and the associated connection one-form
$Q$ on ${\cal P}$ is the imaginary part $Q = \mbox{Im} \Theta$.
By definition the first Chern classes are
the $H^2(M,{\bf R})_{\mscr{de Rham}}$
cohomology classes of the appropriately normalized 
curvatures ($M$ denotes the scalar manifold):
\be 
c_1 ({\cal L}) = 2 \pi i \left[ \ov{\der} \Theta
\right] \;.
\ee
The K\"ahler two-form ${\bf K}$ of the scalar manifold $M$ 
is
\be
{\bf K} = \ft{i}{2 \pi} g_{A \ov{B}} dz^A \wedge d \ov{z}^{\ov{B}}
= \ft{i}{2 \pi} \der \ov{\der} K \;.
\ee
The curvature of ${\cal L}$ and the K\"ahler two-form are
related by
\be
2 \pi i \ov{\der} \Theta = {\bf K}
\ee
and therefore the K\"ahler class is integral:
\be
[ {\bf K} ] = c_1({\cal L}) \;.
\ee
K\"ahler manifolds with this property are called 
K\"ahler-Hodge manifolds in the mathematical literature.
They have the particular property that the K\"ahler metric
comes from the fibre metric of a holomorphic line bundle
${\cal L}$ over the manifold. The fibre metric on ${\cal L}_z$
is just the
exponential $e^K$ of the K\"ahler potential at fixed $z$.
Alternatively one can express this in terms of the 
associated $U(1)$ principal bundle which has the curvature
\be
dQ = 2 \pi {\bf K} \;.
\ee
In the physics literature a manifold is called K\"ahler-Hodge if
it is an admissible scalar manifold in ${\cal N}=1$ 
supergravity.\footnote{We refer to \cite{BinGirGri:2000} for
a review of ${\cal N}=1$ supergravity, its matter couplings
and its geometric structure.}
This is more restrictive than the mathematical definition,
because of the presence of fermions which are sections of
${\cal L}^{1/2}$. K\"ahler invariance in the presence of fermions
requires a compensating chiral rotation, which is only well defined
if the curvature on ${\cal L}^{1/2}$ is normalized as
$\int F = 2 \pi i n$, with $n \in {\bf Z}$. (In mathematical terms
this means that one can take the square root of ${\cal L}$ such
that ${\cal L}^{1/2}$ is a well defined holomorphic line bundle.)
As a consequence the
K\"ahler form must have even integer cohomology and not just
integer cohomology, because $[{\bf K}] = 2 c_1 ({\cal L}^{1/2})$
\cite{CraRooTrovP:1997}.

We finally note that the K\"ahler connection one-form $Q$
is related to the $U(1)$ connection constructed earlier.
In (\ref{AasComposite})
the $U(1)$ connection was expressed in terms of $(X^I, F_J)$ by 
its equation of motion. Rewriting this in terms of 
$\left(X^I(z), F_J(X(z)) \right)$
we find
\be
A_{\mu} = \ft12 e^K \left( \ov{F}_I ( \ov{X}(\ov{z}) ) 
\stackrel{\leftrightarrow}{\der}_\m X^I(z) -
\ov{X}^I(\ov{z}) \stackrel{\leftrightarrow}{\der}_\m F^I(X(z)) 
\right) \;.
\label{KaehlerConnection}
\ee
The connection one-form $Q$ is related to a composite 
space-time gauge field
by
\be
Q_\m = - \ft{i}2 ( \der_A K \der_\m z^A - \der_{\ov{A}} K \der_\m
\ov{z}^{\ov{A}} ) \;.
\label{KaehlerConnectionSpecial}
\ee
Using
\be
\der_A K \der_\m z^A = i e^K \left( \ov{F}_I(\ov{X}(\ov{z})) 
\der_\m X^I - \ov{X}^I \der_\m F_I (X(z)) \right)
\ee
we find $A_\m = Q_\m$.

So far we have focussed on the bundles ${\cal L}, {\cal P}$ related
to the K\"ahler-Hodge structure required by a consistent coupling 
to ${\cal N}=1$ supergravity. We next turn to the symplectic
bundle ${\cal H}$ which is the additional structure required
by coupling vector multiplets to ${\cal N}=2$ supergravity. 
Here scalars and vectors sit in the same multiplet and the
consistent action of symplectic reparametrizations puts
additional restrictions on the scalar manifold. The admissible
manifolds of vector multiplet scalars in ${\cal N}=2$
supergravity are called special K\"ahler manifolds.
We already gave a definition in terms of the
Riemann curvature tensor in special coordinates. 
The first intrinsic definition
was given by Strominger \cite{Str:1990}: A K\"ahler-Hodge manifold is
special K\"ahler if it allows a flat symplectic vector
bundle ${\cal H}$ with a holomorphic section $v$, $\ov{\der}v=0$,
such that the K\"ahler form ${\bf K}$ can be expressed 
in terms of the section as
\be
{\bf K} = -i \der \ov{\der} \log ( -i \la \ov{v}, v \ra ) \,,
\ee
where $\la u, w \ra = u^T \Omega w$ is the symplectic bilinear 
form.
As pointed out in \cite{CraRooTrovP:1997} one has to 
include the additional condition
\be
\la v, \der_A v \ra = 0
\ee
in order to guarantee that the matrix ${\cal N}_{IJ}$
is symmetric: for models constructed using Kaluza-Klein
compactification on Calabi-Yau-threefolds this holds 
automatically, but it is not guaranteed in general.

Various other definitions can be given. For example  one can
work with the principal bundle ${\cal P}$ instead of the
line bundle ${\cal L}$. Then holomorphicity conditions
are replaced by covariant holomorphicity properties.
Other definitions use the characterization through the
curvature tensor or the existence of special coordinates.
We will not enter the details here but refer to 
\cite{CraRooTrovP:1997,Fre:1995}. More about the geometric
formulation of ${\cal N}=2$ supergravity can be found in
\cite{CasDauFer:1990,DauFerFre:1991,Andetal:1996}.

Finally we have to discuss one further point.
The intrinsic definition reviewed above
does not directly refer to a prepotential, but only
required the existence of a secion $v$ of 
${\cal L} \otimes {\cal H}$ with certain properties.
It is not true that all such sections take the
form $(X^I(z), F_J(X(z)))$, where $X^I(z)$ are coordinates
and $F_J(X(z))$ is the gradient of a prepotential \cite{CerDauFervP:1995}. 
For instance, if one starts with a section with prepotential and
performs a symplectic transformation where 
the new $\breve{X}^I(z)$ are not invertibly related
with the old $X^I(z)$,
\be
\det \frac{\der \breve{X}^I}{ \der X^J} = 0 \;,
\ee
then one finds that the new would-be prepotential
vanishes identically, $\breve{F}=0$. Note that in
the superconformal case the $X^I$ were treated as $N_V+1$ independent
scalars, so that it was natural to require that symplectic transformations
are related to invertible transformations of the scalars.
In the super Poincar\'e situation, where 
only $N_V$ scalars are independent, there is no
reason for imposing this condition.
As a consequence models exist which cannot
be described by a prepotential. Nevertheless it can be
shown that every such model can be reparametrized by
a symplectic transformation such that in the new basis
a prepotential exists. In fact, the existence of a description
through special coordinates and a prepotential provides one
of the possible definitions of special K\"ahler 
geometry \cite{CraRooTrovP:1997}.
We will see later that for some models obtained from string
theory the description in a basis without prepotential is
the natural one. Finally we would like to point out that
the equivalence of models without prepotential to models
with prepotential requires the possibility of 
symplectic reparametrizations.
If these are no longer possible, for example
when considering gauged supergravity, then the equivalence
cannot be expected to hold.

\subsection{Consequences of the Presence of Higher Derivative Terms}

In the last two sections we discussed the minimal terms
of the action, which are controlled by the prepotential
$F^{(0)}$. Non-minimal higher derivative terms are encoded
in the higher terms $F^{(g>0)}(X)$ of the function
$F(X,\widehat{A}=W^2)$. Once these terms are taken into account
several things change. The first thing is that the Lagrangian
becomes much more complicated. In particular one cannot
eliminate the auxiliary fields in closed form. A closer look
shows that the Lagrangian now contains derivatives of 
the auxiliary fields, which asks for an interpretation.
Auxiliary fields are expected to have algebraic equations of
motion, so that they do not introduce new degrees of freedom
into the theory. We already discussed in chaper \ref{ChapterGravity}
that the $R^2$-terms themselves also introduce new degrees of freedom
which are in conflict with perturbative unitarity.
The resolution of the problem is the same in both cases:
The Lagrangian is not to be interpreted as a fundamental 
Lagrangian but as an effective Lagrangian of the
underlying fundamental theory that we believe to be string theory.
The expansion in terms of derivatives is a low energy expansion,
because every derivative is suppressed by a factor of
$m_{\mscr{Planck}}$. The effective action is used to compute
corrections iteratively, oder by order in $m_{\mscr{Planck}}$.
The fundamental theory has to provide the explicit form of the 
functions $F^{(g)}(X)$.
As we will see later these quantities 
can be computed in string perturbation theory.

Since the auxiliary fields can only be solved for iteratively
we have to insist on the existence of a prepotential. As we
explained  in the last section, this is no loss in generality
at least as long as we consider ungauged supergravity.
In view of the complicated dependence of the Lagrangian
on the auxiliary fields, we will try to avoid using its
explicit form as much as we can. In particular we will 
not eliminate the auxiliary fields, and we will
work in the superconformal setup where symmetries are realized
in a more simple way. The computation of black hole entropy 
will show that this is an effective way for solving 
problems explicitly.

\chapter{Four-Dimensional 
${\cal N}=2$ Black Holes \label{ChapterN=2BlackHoles}}

In this chapter we discuss extremal black holes in ${\cal N}=2$ supergravity
with $N_V+1$ vector multiplets. The discussion is
model-independent, because we work with a general function 
$F(X,\widehat{A})$. We derive one of our main results,
the model-independent entropy formula (\ref{ModIndEntropyFormula}). 
This result and its derivation
were briefly described in the letter
\cite{CardWMoh:1998/12}. Here we will give a detailed presentation.
The main line of thought is the following: As we discussed in
chapter \ref{ChapterGravity} the extremal Reissner-Nordstr{\o}m
black hole is a BPS-soliton which interpolates between
two ${\cal N}=2$ supersymmetric vacua, flat space at infinity
and $AdS^2 \times S^2$ at the horizon \cite{Gib:1981}.
The same is true when considering ${\cal N}=2$ supergravity coupled
to vector multiplets with a general prepotential $F^{(0)}(X)$:
The event horizon of a static and spherically symmetric
BPS black hole is fully
${\cal N}=2$ supersymmetric and the solution therefore 
interpolates between two ${\cal N}=2$ 
vacua \cite{FerKalStr:1995,Str:1996,FerKal:1996/02,FerKal:1996/03}.
Naturally one expects that this property will persist when
$R^2$-terms are switched on. One can give a general argument
for this: In the sections 
\ref{SectionStabilizationEquations} and \ref{SectionBlackHoleSolutions}
we will review how the asymptotic behaviour of BPS black holes
is determined by the so-called {\em stabilization equations}. 
Symplectic covariance requires that 
these equations
have to be modified in presence of $R^2$-terms 
\cite{BehCardWLueMohSab:1998,CardWMoh:1998/12}.
The modified stabilization equations then imply universal behaviour and 
full ${\cal N}=2$ supersymmetry on the horizon.

In absence of
$R^2$-terms one can derive the stabilization equations from the
BPS condition. Moreover one can find an expression for the full
interpolating black hole solution, 
which is determined by the {\em generalized stabilization equations} 
\cite{Sab:1997/03,Sab:1997/04,BehLueSab:1997}. 
The same is expected when $R^2$-terms
are present and we will argue in section 
\ref{SectionBlackHoleSolutions} that the result is predicted by
symplectic invariance. A derivation of the stabilization equations
and of the full black hole solution in presence of $R^2$-terms is
currently under investigation and the 
result will be the subject of a future publication \cite{CardWKaeMoh}.
Here we will determine the near horizon geometry by imposing
full ${\cal N}=2$ supersymmetry on a static and spherically
symmetric field configuration.
This is already a complicated problem and we will devote
section \ref{SectionNearHorizon} to a detailed discussion. 
The result is that the unique static and spherically symmetric
${\cal N}=2$ vacuum is the Bertotti-Robinson geometry
$AdS^2 \times S^2$. The geometry and the corresponding gauge and
scalar fields can be expressed
in terms of a single field $Z$, which is related to the central charge.

In section \ref{SectionEntropyFormula} we use Wald's entropy
formular to compute the black hole entropy corresponding to our near
horizon solution. We call the resulting formula model-independent,
because it is valid for arbitrary $F(X^I,\widehat{A})$.
The formula is manifestly
covariant under symplectic reparametrizations and
the entropy is uniquely determined by the value of the
field $Z$ at the horizon. This field
is a function of the electric and magnetic charges and of the
scalar fields.

In section \ref{SectionStabilizationEquations} we review
the {\em supersymmetric attractor mechanism} and how it
leads to the {\em stabilization equations} which determine
the near horizon behaviour of the scalar fields in terms
of the charges 
\cite{FerKalStr:1995,Str:1996,FerKal:1996/02,FerKal:1996/03}. 
Based on symplectic invariance we propose
a generalization of the stabilization equations
to the case with $R^2$-terms 
\cite{BehCardWLueMohSab:1998,CardWMoh:1998/12}.  
Combining the stabilization equations
with our model-independent entropy formula we see that
the black hole entropy is determined by the charges.
In order to arrive at explicit
expressions one has to specify concrete models by an explicit choice of the
function $F(X,\widehat{A})$ and then to solve the stabilization equations. 
This will be done in chapter \ref{ChapterStringBHs}.

In section \ref{SectionBlackHoleSolutions} we review for completeness
the structure of the full explicit black hole solutions in 
${\cal N}=2$ supergravity without $R^2$-terms. We explain how the
the solution is expressed in terms of
harmonic functions using the so-called 
{\em generalized stabilization equations} 
\cite{Sab:1997/03,Sab:1997/04,BehLueSab:1997}.
As already mentioned the generalization of these results to the case 
with $R^2$-terms 
is currently under investigation \cite{CardWKaeMoh}.
Here we will restrict ourselves to a few remarks which are 
based on symplectic invariance.

\section{The Near Horizon Geometry \label{SectionNearHorizon}}

We now turn to the classification of all fully supersymmetric,
static and spherically symmetric field configurations of
${\cal N}=2$ supergravity with $N_V+1$ vector multiplets, based
on a general function $F(X,\widehat{A})$, where $\widehat{A}$ will
eventually be identified with lowest component of
the Weyl multiplet ${\bf W}^2$.

By reparametrizations a static and spherically symmetric metric can
be brought to the form
\be
ds^2 = - e^{2g(r)} dt^2 + e^{2 f(r)} \left(dr^2 + r^2 ( \sin^2 \theta
d \phi^2 + d \theta^2)  \right) \;.
\label{metricansatz}
\ee
In absence of higher curvature terms it was shown by Tod \cite{Tod:1983} 
that
for a superymmetric static metric the two functions $f$ and $g$ are
not independent, but related through $f=-g$. We will not make
the assumption that this generalizes to our case but rather 
prove it. Therefore in our ansatz (\ref{metricansatz})
$f$ and $g$ are independent functions.

Since we are looking for bosonic field configurations which are
invariant under all ${\cal N}=2$ supertransformations, nontrivial
constraints arise from the condition
that variations of fermions with arbitrary transformation parameter
$\e^i$ must vanish when evaluated in the background. 
The bosonic fields themselves vary into fermions,
which by assumption vanish in the background. The vanishing of the
fermionic variations imposes conditions on the bosonic background 
and we have to find the most 
general static and spherically symmetric 
bosonic background which satisfies them.
Terms 
which are at least quadratic in the fermions are
irrelevant for our problem because they vanish and 
transform under supersymmetry
into objects which vanish in a bosonic background. Therefore we will
save work in this chapter by systematically ignoring all higher
order fermionic terms. This will be applied from now on 
without further notice.

We will work in the superconformal off-shell formulation and in a general 
chiral background field. This way we can avoid to deal with
the complications of the action and the equations of motion 
that we described earlier. We require that the variations of
all fermionic quantities vanish in the bosonic background for
arbitrary choice of the supersymmetry parameter. As we will see
the vanishing of the variation of a fermion does not necessarily 
imply that the variation of the covariant derivative of the fermion
vanishes. For example the vanishing of the variation of the 
covariant derivatives
of the gaugini gives new conditions on the background, which are equivalent
to the field equations and Bianchi identities of the gauge fields.
Therefore we have to continue analysing fermionic variations until
no new conditions can arise, because the background is completely determined. 
Since we are imposing all conditions on the bosonic background which
are compatible with ${\cal N}=2$ supersymmetry, we know that the
background must satisfy the equations of motion.
Thus by imposing full supersymmetry we 
do not need to solve the equations of motion directly.
This is similar to the analysis of fully supersymmetric
compactifications of eleven-dimensional supergravity 
in \cite{BirEngdWNic:1983}, where
unbroken supersymmetry requires that the variation of the supercovariant
gravitino field strength vanishes. In that case it follows 
from this single condition that all equations of motion are satisfied.

Since we keep superconformal invariance intact while solving the Killing
spinor equations we profit from the simpler structure
of the supersymmetry variations and from the fact that the superconformal
off-shell multiplets are smaller then their Poincar\'e counterparts.
In order to apply our results to black holes in ${\cal N}=2$
Poincar\'e supergravity we finally have to fix the extra conformal
symmetries or to consider suitable gauge invariant
quantities. As we saw earlier dilation invariance is fixed by setting
\be
e^{- {\cal K}} := i \left( F_I(X, \widehat{A}) \ov{X}^I -
X^I \ov{F}_I ( \ov{X}, \ov{\widehat{A}}) \right)
\ee
to a constant. In the superconformal situation $e^{-{\cal K}}$
is a symplectic function
of Weyl weight $w=2$ and chiral weight $c=0$. It appears
in various places, for example in the Einstein-Hilbert term,
in order to preserve dilatational invariance. We will use
it to construct dilatation-invariant quantities. 
The notation ${\cal K}$ is chosen because this quantity
resembles the K\"ahler potential. Note however that it does
not reduce to the K\"ahler potential when setting $\widehat{A}=0$
because it involves the covariantly holomorphic section $(X^I,F_J)$
and not the holomorphic one, $(X^I(z), F_J(X(z))$.

In the superconformal setup the special $S$-supertransformations
are still present. Therefore we cannot require strict invariance under
$Q$-supertransformations, but only $Q$-invariance up to an uniform
$S$-supertransformation.\footnote{This corresponds to the fact that
Poincar\'e $Q$-supertransformations are combinations of 
$Q$- and $S$-supertransformations and $K$-transformations.
For the problem studied in this chapter the $K$-transformations
are not relevant, since all the quantities whose supersymmetry
variation we require to vanish are $K$-invariant.} An elegant
way to deal with this is to find $S$-invariant spinors. This
can be done using a spinor $\zeta_i$ which under $S$-supersymmetry
transforms into the $S$-variation parameter $\eta_i$. Given any 
spinor that transforms under $S$, one can form a suitable combination
with $\zeta_i$ such that the $S$-variations mutually cancel.
A spinor with the required inhomogenous transformation under
$S$ can be found and is related by supersymmetry to the function
${\cal K}$:
\be
\zeta_i = - \left( \Omega^I_i \ft{\der}{\der X^I} +
\widehat{\psi}_i \ft{\der}{\der \widehat{A}} \right) {\cal K}
= - i e^{\cal K} \left( (\ov{F}_I - \ov{X}^J F_{IJ}) \Omega^I_i
-\ov{X}^I F_{I\widehat{A}} \widehat{\psi}_i \right) \;.
\ee
The derivative operator acts on the symplectically invariant
${\cal K}$ in such a way that $\zeta_i$ is symplectically 
invariant. The behaviour of $\zeta_i$ under $Q$- and $S$-transformations
is
\bea
\delta \zeta_i  &=& -2 i e^{\cal K} ( \ov{F}_I \gamma^a D_a
X^I - \ov{X}^I \gamma^a D_a  F_I ) \epsilon_i - i e^{\cal K}
\left( (\ov{F}_I - \ov{X}^J F_{JI} ) Y^I_{ij} - \ov{X}^I F_{I\widehat{A}}
\widehat{B}_{ij} \right) \epsilon^j \nonumber \\
 & & - \ft12 i \ve_{ij} {\cal F}^-_{ab} 
\gamma^{ab} \epsilon^j + 2 \eta_i \:, \\
\nonumber 
\eea
where we defined
\be
{\cal F}^-_{ab} := e^{\cal K} \left( \ov{F}_I F^{-I}_{ab}
- \ov{X}^I G^-_{ab I} \right) \;.
\label{calFminus}
\ee
Note that $\zeta_i$ has the required behaviour under 
$S$-transformations. Now we can form $S$-invariant spinors and
require that their $Q$-variations vanish exactly.

\subsection{The Gaugino Variations}

We start our analysis with the gaugini. The $S$-invariant 
combination is given by $\Omega_i^I - X^I \zeta_i$. The 
$Q$ variation has to vanish:
\bea
\delta( \Omega^I_i - X^I \zeta_i ) &=&
 2 D^a X^I \g_a \e_i  + \ft12 \ve_{ij} {\cal F}^{-Iab} \gamma_{ab} \e^j
+ Y^I_{ij} \e^j 
\nonumber \\
 & & + i e^{\cal K} X^I \big[ 2 (\ov{F}_J D^a X^J - \ov{X}^J D^a F_J) \gamma_a 
\e_i + \ft12 ( \ov{F}_J F^{-J}_{ab} - \ov{X}^J G^{-}_{Jab}) \g^{ab}
\ve_{ij} \e^j \big] 
\nonumber \\
 & & + X^I 
( (\ov{F}_J - \ov{X}^K F_{KJ} ) Y^J_{ij} - \ov{X}^J F_{J\widehat{A}}
\widehat{B}_{ij}) \e^j \nonumber \\
 & \stackrel{!}{=} & 0 \;.\\
\nonumber
\eea
Since this must hold for all choices of $\e^i$ the
coefficients of terms with a different structure concerning spinor
indices (which we did not write out explicitly)
and $SU(2)$ indices ($i,j$)
have to vanish separately. There are three independent
types of terms: those proportional to $\g^a \e_i $, to  $\e^j$ and to
$\gamma_{ab} \ve_{ij} \e^j$. Thus we get three equations:
\bea
D_a X^I + i e^{\cal K} X^I (\ov{F}_J D_a X^J - \ov{X}^J D_a F_J)
 &=&0 \;, \label{gaugino1} \\
{\cal F}^{I-}_{ab} + i e^{\cal K} X^I ( \ov{F}_J F^{-J}_{ab} -
\ov{X}^I G^-_{Iab}) &=& 0  \;,\label{gaugino2} \\
Y^I_{ij} + i e^{\cal K} X^I \left( 
( \ov{F}_J - \ov{X}^J F_{JK} ) Y^K_{ij} - \ov{X}^J F_{J \widehat{A}}
\widehat{B}_{ij} \right) &=& 0 \;.\label{gaugino3} \\
\nonumber
\eea

The first equation can be brought to a simpler form by writing
out the covariant derivatives and reorganizing terms using 
the composite connection
\be
{\cal A}_a = \ft12 e^{\cal K} \left( \ov{X}^J 
\stackrel{\leftrightarrow}{\der}_a F_J - \ov{F}_J 
\stackrel{\leftrightarrow}{\der}_a X^J \right) \;.
\label{CompositeConnection}
\ee
Using this equation (\ref{gaugino1}) takes the form 
\be
( \der_a -i {\cal A}_a ) \left( e^{ {\cal K}/2 }  X^I \right) = 0\;.
\label{gaugino1a}
\ee
Note that this equation is Weyl and $U(1)$ invariant. Moreover 
it is obvious that there is an integrability condition
\be
\der_{[a} {\cal A}_{b]} = 0\;,
\label{ConnectionFlat}
\ee
which tells us that the connection ${\cal A}_a$ is flat.
Using the quantity ${\cal F}^-_{ab}$ introduced in (\ref{calFminus})
we can also write equation (\ref{gaugino2}) in more suggestive form:
\be
{\cal F}^{I-}_{ab} = - i X^I {\cal F}^-_{ab} \;.
\label{gaugino2a}
\ee

\subsection{The Background Spinor Variation}

Next we look at the variation of the background
spinor $\widehat{\psi}_i$. The structure is the same as for 
the gaugino variation.
In a completely analogous way we get three equations. The
first can be brought to the form
\be
(\der_a -w i {\cal A}_\m ) \left( e^{w {\cal K}/2} \widehat{A} 
\right) = 0 \;.
\label{background1}
\ee
Combining this with (\ref{gaugino1a}) 
and using homogenity of $F(X,\widehat{A})$ we
find
\be
(\der_a - i {\cal A}_a)  \left( e^{ {\cal K}/2} F_I \right) = 0 \;.
\label{background1a}
\ee
The second equation can be brought to the form
\be
\widehat{F}^-_{ab} = - i w \widehat{A} {\cal F}^-_{ab} \;.
\label{background2a}
\ee
The third equation can be combined with (\ref{gaugino2a})
resulting in
\be
\widehat{B}_{ij} X^I = w \widehat{A} Y^I_{ij} \;.
\label{gaugino3a}
\ee

\subsection{The Gravitini Variations}

We now turn to the variation of the gravitini $\psi_a^i$.
Since the gravitini are gauge fields 
it is to restrictive to set them to zero, instead of requiring
that they are pure gauge. Therefore we will only require that
the corresponding gauge-invariant quantity,
the gravitino field strength $R^i_{ab}(Q)$ has a vanishing
$Q$-variation modulo an $S$-variation. As before we first 
have to find the appropriate $S$-invariant object, which is
$R^i_{ab}(Q) - \ft1{16} T^{cd ij} \gamma_{cd} \gamma_{ab} \zeta_j$.
The resulting expression is somewhat more complicated than the ones
we encountered before. One can use the selfduality and chirality
properties of the various quantities and the $\g$-matrix identities
listed in appendix \ref{AppSpaceTime}
to simplify it. After collecting terms with
the same spinor and $SU(2)$ index structure the result takes the form
\bea
& &\delta R_{ab}^i(Q) - \ft1{16} T^{cd ij} \g_{cd} 
\g_{ab} \delta \zeta_j \nonumber \\
 &=& A_{ab}^e \gamma_e \ve^{ij} \e_j
 + B_{ab}^{cd} \sigma_{cd} \e^i + \ve^{ij} C_{ab jk} \e^k  
+D_{ab\:\:j}^{cdi} \sigma_{cd} \e^j + E_{ab} \e^i \nonumber \\
& \stackrel{!}{=} & 0 \;.\\
\nonumber
\eea
Explicit expressions will be given soon. All tensors are antisymmetric
and antiselfdual in the index pairs $a,b$ and $c,d$. The tensors
$C_{abjk}$ and $D_{ab \;\;j}^{cdi}$ are symmetric and 
antihermitean-traceless in the $SU(2)$ indices, respectively.
All five terms have to vanish independently. 

The first condition $A_{ab}^c=0$ gives an equation for the
covariant derivative of the auxiliary tensor $T^-_{ab}$:
\be
{\cal D}_c T_{ab}^{-}= i e^{\cal K} 
( \ov{X}^J {\cal D}_d F_J - \ov{F}_J {\cal D}_d X^J )
\left( \delta_c^d T^{-}_{ab}- 2 \delta_{[a}^d T^{-}_{b]c}
+ 2 \eta_{c [a} T_{b]}^{- d} 
\right) \;. 
\label{DTequation}
\ee
From $E_{ab}=0$ we get:
\be
T_a^{-d} {\cal F}_{db}^- - T_{b}^{-d} {\cal F}^-_{da} = 0 \;.
\label{Eequation}
\ee

Next we have $D_{ab\;\;j}^{cdi}=0$ which after a few manipulations
gives
\be
R(V)_{ab\;\;j}^{\;\;\;\;i} = 0 \;.
\ee

Then we have $B_{ab}^{cd}=0$. With some effort and using (\ref{Eequation})
this can be
brought to the form
\be
{\cal R}(M)_{ab}^{-\;cd} = \ft{-i}{16} T^{-ef} {\cal F}^-_{ef}
( \delta_{[a}^c \delta_{b]}^d - \ft12 \ve_{ab}^{\;\;\;\;cd})
+ \ft{i}8 [ T^-_{ab} {\cal F}^{-cd} + T^{-cd} {\cal F}^{-}_{ab} ] \;.
\label{RMequation}
\ee
The expression ${\cal R}(M)_{ab}^{-\;cd}$ is related to the
modified Lorentz field strength ${\cal R}(M)_{ab}^{\;\;\;\;cd}$
defined in (\ref{calRM}) by antiselfdual projection in both pairs
of indices. The terms on the right hand side are manifestly
antiselfdual in both $a,b$ and $c,d$.

We can get more explicit information out of this tensor equation
by making contractions. First we substitute the explicit 
form (\ref{Kconnection}) 
of the $K$-connection into the definition (\ref{calRM}) of
${\cal R}(M)_{ab}^{\;\;\;\;cd}$ and we define the tensor
\be
C_{ab}^{\;\;\;\;cd} = R_{ab}^{\;\;\;\;cd}
- 2 \d_{[a}^{[c} R_{b]}^{\;\;d]} + \ft13 R \d_{[a}^{[c}
\d_{b]}^{d]} \;,
\ee
which in the Poincar\'e frame, i.e. after gauge fixing the
conformal symmetries, becomes the Weyl tensor.
Plugging this into (\ref{calRM}) we get:
\be
{\cal R}(M)_{ab}^{\;\;cd} = C_{ab}^{\;\;\;\;cd}
+ D \d_{[a}^{[c} \d_{b]}^{d]} + 2 i \d_{[a}^{[c} \widetilde{R}(A)_{b]}^{\;\;d]} \;.
\ee
Next we perform contractions that project onto the 
terms containing $R(A)_{ab}$ and $D$, respectively.
First note that
\be
\ve^{gb}_{\;\;\;\;cd} {\cal R}(M)_{ab}^{\;\;\;\;cd}
= 2i R(A)^g_{\;\;a} = 2 \widetilde{R}(D)^g_{\;\;a}\;.
\ee
The first equations follows by computing the contraction, the
second one is a consequence of the Bianchi identity (\ref{BianchiRMRD}).
Performing the same contraction on the right hand side of
(\ref{RMequation}) one gets zero and therefore we have
\be
R(A)_{ab} = 0 = R(D)_{ab} \;.
\ee
Next we consider the trace part ${\cal R}(M)_{ab}^{\;\;\;\;ab}$
to get information about $D$, using that the Weyltensor
is traceless. This gives
\be
D=-\ft{i}{24} T^{-ab} {\cal F}^-_{ab}
\label{Dequation}
\ee
and plugging the result back we find
\be
{\cal R}(M)_{ab}^{-\;\;cd} = C_{ab}^{-\;\;cd}
- \ft{i}{48} T^{-ef} {\cal F}^-_{ef} ( \d_{[a}^{c}
\d_{b]}^d - \ft12 \ve_{ab}^{\;\;\;\;cd})
\ee
(note that going from ${\cal R}(M)_{ab}^{\;\;\;\;cd}$
to ${\cal R}(M)_{ab}^{-\;\;cd}$ involves the antiselfdual
projection) and by (\ref{RMequation}) we find that the Weyl tensor
is
\be
C_{ab}^{-\;cd} = \ft{-i}{24} T^{-ef} {\cal F}^-_{ef}
( \delta_{[a}^c \delta_{b]}^d - \ft12 \ve_{ab}^{\;\;\;\;cd})
+ \ft{i}8 [ T^-_{ab} {\cal F}^{-cd} + T^{-cd} {\cal F}^{-}_{ab} ] \;. 
\label{Cequation}
\ee
For generic antiselfdual tensors the right hand side has all
algebraic symmetries of the Weyl tensor. 
But if we impose in addition spherical symmetry on this equation and
look at the equation component by component, we find that it can
only be satisfied trivially. To see this we first note that the Weyl 
tensor of the metric (\ref{metricansatz}) has only one independent
non-vanishing component:
\be
C_{01}^{01} = C_{23}^{23} = -2C_{02}^{02} =-2C_{03}^{03} = 
-2C_{12}^{12} = -2C_{13}^{13}  \;,
\ee
whereas all other independent components vanish. In order to use
this in equation (\ref{Cequation}) one has to project onto the 
antiselfdual part in both pairs of indices. This gives
\be
C_{01}^{-\;\;01}= \ft12 C_{01}^{\;\;\;\;01}\;,
C_{02}^{-\;\;02} = - \ft14 C_{01}^{\;\;\;\;01} \;, \ldots
\ee
On the right hand side one uses that an antiselfdual antisymmetric
tensor only has one independent component $T^-_{01} = i T^-_{23}$.
Evaluation (\ref{Cequation}) for the two components of the antiselfdual
Weyl tensor given above implies that either $T^-_{01} =0$ or
${\cal F}^-_{01}=0$ and therefore
\be
C_{ab}^{\;\;\;\;cd} = 0
\mbox{  and   }
T^-_{ab} {\cal F}^{-cd} = 0 \;.
\ee
By (\ref{Dequation}) this also implies
\be
D=0\;.
\ee

Finally $C_{abjk}=0$ implies
\be
T_{ab}^- e^{\cal K} [ (\ov{F}_I - \ov{X}^J F_{JI}) Y^I_{jk}
- \ov{X}^I F_{I\widehat{A}} \widehat{B}_{jk} ] = 0 \:.
\ee
We will see in the following sections 
that $T_{ab}^-=0$ leads to flat space. Now we take it to
be non-vanishing and use that
$F(X,\widehat{A})$ is an arbitrary function. Combining this with
(\ref{gaugino3a}) implies
\be
Y^I_{ij} = 0 = \widehat{B}_{ij} \;.
\ee

\subsection{Variation of the Spinor of the Second Compensating
Multiplet}

The inclusion of the second compensating multiplet is not only
needed for the consistency of the whole construction, but it also
provides additional information for our problem. We take the 
non-linear multiplet as the second compensator. The S-invariant
variation of its spinor yields
\be
{\cal F}^-_{ab} =  0
\ee
and
\be
A_a + {\cal A}_a = 0 \;.
\label{AandCalA}
\ee
The first equation implies 
\be
{\cal F}^{-I}_{ab} = 
\widehat{F}^{-I}_{ab} =  {\cal R}(M)^{-\;\;cd}_{ab}=0 \;,
\label{compensator1a}
\ee
using (\ref{gaugino2a}), (\ref{background2a}) and (\ref{RMequation}).
Since ${\cal A}_a$ as defined in (\ref{CompositeConnection})
equals the K\"ahler connection (\ref{KaehlerConnection})
up to sign, we recognize that (\ref{AandCalA}) is the equation of 
motion for the $U(1)$ gauge field.

Using the relation between $A_a$ and ${\cal A}_a$ we can
rewrite the equations (\ref{gaugino1a}), (\ref{background1}) and
(\ref{background1a}) as
\be
{\cal D}_a (e^{{\cal K}/2} X^I) = {\cal D}_a ( e^{\cal K}/2 F_I )
= {\cal D}_a ( e^{w {\cal K}/2}  \widehat{A})  = 0 \;.
\ee
As a consequence of (\ref{ConnectionFlat})
the $U(1)$ connection is flat. Now we pick
a gauge where $A_a=0$ and go to the Poincar\'e frame by imposing
the $K$- and $D$-gauge,
\be
b_{\m}= 0 \mbox{   and   } e^{-\cal K} = m_{\mscr{Planck}}^2 \;.
\ee
Since the fields $X^I, F_J$ and $\widehat{A}$ are Lorentz scalars
we are left with
\be
\der_a X^I = \der_a F_I = \der_a \widehat{A} \;,
\ee
implying that the scalars $X^I$ and the background field $\widehat{A}$ 
are constant. 
Constancy of $X^I$ and $F_J$ implies ${\cal D}_c T_{ab}^i= 0$ 
by (\ref{DTequation}).
When we go to the Poincar\'e frame and impose the gauge $A_a=0$ as above 
then the covariant derivative still contains the spin connection because
$T^-_{ab}$ is a Lorentz tensor.
But using the explicit form of the spin connection for a static 
spherically symmetric metric (\ref{SphericConnection})
one can verify
that $T_{ab}^-$ is actually constant, $\der_c T_{ab}^-=0$.

\subsection{Variation of the Derivative of the Spinor
$\zeta_i$}

Since all fermionic quantities have to vanish in our ${\cal N}=2$
invariant background, derivatives of spinors must have 
a vanishing variation. Usually this does not lead to new conditions,
but in our case it does. When imposing that the $Q$-variation
of $D_a \zeta_i$ vanishes up to a uniform S-transformation one
gets an equation for the $K$-connection:
\be
f_{ab} - \ft12 {\cal D}_a {\cal D}_b {\cal K} - \ft14 {\cal D}_a {\cal K}
{\cal D}_b {\cal K} +  \ft18 \d_{ab} {\cal D}_c {\cal K} {\cal D}^c {\cal K}
= 0 \;.
\ee
In the Poincar\'e frame, when imposing the $K$- and $D$-gauge
conditions, we have
\be
{\cal D}_a {\cal K} = \der_a {\cal K} = 0
\ee
and therefore the $K$-gauge field vanishes
\be
f_{ab}= 0 \;.
\ee
This yields an important information about the geometry of the
background. Using the explicit form (\ref{Kconnection}) of the 
composite $K$-gauge field we find
\be
R_{ab} = \ft18 T_{a}^{ijc} T_{cbij} = \ft1{16} T_a^{-c} T_{cb}^+
\ee
for the Ricci tensor 
and the corresponding Ricci scalar vanishes
\be
R = \ft1{16} T_a^{-c} T_{ca}^+ =0
\ee
by an identity which is generally valid for (anti-)selfdual tensors (see
\ref{IdASDT}).
The Riemann tensor takes the same form as in the
case of the near horizon geometry of an extremal 
Reissner-Nordstr{\o}m black hole:
The Weyl tensor and Ricci scalar vanish while the traceless part of 
the Ricci tensor is non-trivial. Since the Ricci tensor 
is given in terms of the
auxiliary tensor $T^-_{ab}$ we have to 
find the explicit form of this field.

\subsection{The Relation between the Auxiliary $T$-Field and the 
Gauge Fields}

The auxiliary $T$-field is related to the gauge fields. First recall
that the dual gauge field $G^-_{I ab}$ is given by (\ref{DefG},
\ref{Ominus})
\be
G^-_{I ab} = F_{IJ} F^{-J}_{ab} + \ft14 ( \ov{F}_I - F_{IJ} \ov{X}^J )
T^-_{ab} + \widehat{F}^-_{ab} F_{I \widehat{A}} \;.
\ee
Now we contract this with $X^I$, use the homogenity of $F(X,\widehat{A})$
amd formula (\ref{compensator1a})
to derive
\be
T^-_{ab} = 4i e^{\cal K} \left( F_I F^{-I}_{ab} - X^I G^-_{Iab}
\right)  \;.
\ee
Since the definition of 
$G^-_{Iab}$ involves $T^-_{ab}$,
this is not an explicit expression for the $T$-field.

\subsection{The Gauge Field Equations of Motion}

The gauge field equations of motion and the Bianchi identities
have the following form:
\bea
{\cal D}^{\m} ( G^-_{I \m \n} - G^+_{I\m\n} ) &=& 0 \;,\nonumber \\
{\cal D}^{\m} ( F^{I-}_{\m\n} - F^{I+}_{\m\n} ) &=& 
0 \;. \label{GFeom+BI}\\
\nonumber
\eea
They can be derived from the Lagrangian, as discussed in
chapter \ref{ChapterFourDSuGra} and appendix \ref{AppAbelianGaugeFields}.
Alternatively these equations can be derived by setting the 
$Q$-variation of the
covariant derivatives of the gaugini to zero (modulo the usual 
uniform $S$-transformation). This illustrates that
full supersymmetry implies the equations of motion.

We can solve the equations of motion in a spherically symmetric and
static background.  Note that the field strengthes 
$F^I_{ab}$ and $G_{Iab}$ are not independent. When solving the
equations we have to
pick a set of independent components. Taking $F^I_{ab}$ (or
$G^{Iab}$) as independent 
is inconvenient because then one of the two equations
is very complicated. Instead on takes $F^I_{23}$ and $G_{I23}$.
Then all the equations take the form of Bianchi identities
and can be solved as discussed 
in appendix \ref{AppAbelianGaugeFields}. For the spherically
symmetric case one finds:
\bea
G^-_{I 01} - G^+_{I 01} = i G_{I 23} &=& i \frac{e^{-2f(r)}}{r^2} 
q_I \;,\nonumber \\
F^{I-}_{01} - F^{I+}_{ 01} = i F^I_{23} &=& i \frac{e^{-2f(r)}}{r^2} 
p^I \;, \\
\nonumber
\eea
(using flat indices)
with constants $p^I,q_I$, which, according to our definition 
(\ref{DefCharges}) are the magnetic and electric charges.

Now we use that ${\cal F}^+_{ab}=0$ implies
\be
F_I F^{I+}_{ab} - X^I G^+_{Iab} = 0 \;,
\ee
to rewrite the expression for $T$
\be
T^-_{ab} = 4i e^{\cal K} (F_I F^{I-}_{ab} - X^I G^-_{Iab}) 
= 4i e^{\cal K} (F_I F^{I}_{ab} - X^I G_{Iab}) \;.
\ee
Next we can use our solution for the gauge fields 
to express the $T$-field as
\be
T^-_{01}  = i T^-_{23} = - 4 e^{ {\cal K}/2 } Z 
\frac{ e^{-2f(r)}}{r^2} \;,
\ee
where we defined
\be
Z = e^{ {\cal K}/2} (p^I F_I - q_I X^I)  \;.
\label{Zfunction}
\ee
The symbol $Z$ was chosen because this quantity resembles the central
charge. Note however that $Z$ as defined here is not a number but a
field. We will see later that in an asymptotically
flat geometry the value of $Z$ at infinity is the central charge 
as defined in (\ref{centralcharge}). 

We already argued that in a supersymmetric, static and 
spherically symmetric 
background the quantities $X^I, F_J, Z, {\cal K}, T_{ab}^-$ are 
constant in the Poincar\'e frame. This implies 
that $r^2 e^{2f(r)}$ must be constant and is given by
\be
r^2 e^{2 f(r)}= - 4 e^{{\cal K}/2} \frac{Z}{T^-_{01}} \;,
\label{r22f}
\ee 
so that we have fixed one
of the unknown functions in (\ref{metricansatz}).
Obviously the constants on the right hand side must be related
such that a real positive number results. In order to find this relation
we now turn to a detailed investigation of the  metric.

\subsection{The Metric}

In order to proceed systematically we start by computing the 
curvature components of a spherically symmetric, 
static metric (\ref{metricansatz}).
The non--vanishing components of the Ricci tensor, with tangent
space indices $a,b=0,\ldots,3$ and the Ricci scalar are
\begin{eqnarray}
R_0^0 &=& \Big[ g'' + g'(g'+f') + {2\over r} g'\Big] {\rm 
e}^{-2f}\,,\nonumber\\
R_1^1 &=& \Big[2 f'' +  g'' + g'(g'-f') + {2\over r} f'\Big] {\rm 
e}^{-2f}\,,\nonumber\\
R_2^2 &=& R_3^3= \Big[ f'' + f'(g'+f') + {1\over r} (3f' +g') 
\Big] {\rm e}^{-2f}\,,\nonumber\\
R&=& \Big[ 2(2 f''+g'')  + 2(f'^2+ g'^2 +f'\,g')+ {4\over r} (2f' 
+ g') \Big] e^{-2f}  
\label{RicciTensor}
\end{eqnarray}
and the components of the Weyl tensor (again with tangent space indices)
are equal to
\begin{eqnarray}
C_{01}^{01} = C_{23}^{23} = -2C_{02}^{02} =-2C_{03}^{03} = 
-2C_{12}^{12} = -2C_{13}^{13}= && \nonumber\\
 \frac{1}{3} \Big[-f'' + g'' + (g'-f')^2 +{1\over 
r}(f'-g') \Big] e^{-2f}\,. &&
\end{eqnarray}
We saw that full ${\cal N}=2$ supersymmetry implies
\be
R^a_b = - \frac{1}{16} T^{-ac} T_{bc}^+\;\;\;\mbox{and}\;\;\;
C^{ab}_{cd} = 0 \;.
\ee
Moreover 
$T_{ab}^-$ is constant and has only one independent non-vanishing
component. Therefore all nonvanishing components of the Ricci
tensor are equal up to sign
\be
R^0_0 = R^1_1 = - R^2_2 = - R^3_3 = \frac{1}{16} | T_{01}^-  |^2
= \mbox{constant} \;.
\ee
Using (\ref{RicciTensor}) and the equations $R_0^0 = R_1^1$ and
$R_0^0 = - R_2^2$ we get
\begin{eqnarray}
f'' - f' g' + \frac{1}{r} ( f' - g' ) &=&0 \; ,\\
f'' + g'' + (f' + g')^2 + \frac{3}{r} ( f' + g') &=&0 \;.
\end{eqnarray}
In addition Weyl flatness implies
\be
f'' - g'' - ( f' - g')^2 - \frac{1}{r} ( f' - g') = 0 \;.
\label{Weyl}
\ee
Finally we have one inhomogenous equation, $R_0^0 =$ constant.
Using the other equations, it can be brought to the form
\be
( f' (f' -g') + \frac{1}{r} ( 2 f' - g')) e^{-2f} = 
\mbox{constant} \;.
\ee
There is one linear combination of the three homogenous equations
such that the second derivatives drop out:
\be
f' g' + \frac{1}{r} g' = 0 \;.
\ee
Thus either $g' = 0$ or $f' = \frac{-1}{r}$. The first case yields
flat space. 
In the second case we have
\be
e^f = \frac{c}{r} \;.
\ee
This is the same result (\ref{r22f})
that we got from the gauge field equations,
where the constant took the value 
$c= \sqrt{-4 e^{{\cal K}/2} \frac{Z}{T^-_{01}}}$. We will check
that this value of $c$ is consistent with what we get by solving
the conditions on the Ricci and Weyl tensor.

We proceed by using 
that the Weyl flatness condition
implies 
\be
e^{g-f} = Ar^2 + B \;.
\ee
Defining $a=cA$ and $b=cB$ we find the metric
\be
ds^2 = - (ar + \frac{b}{r})^2 dt^2 + \frac{c^2}{r^2} (d r^2 + r^2
d\Omega^2)  \;.
\label{GeneralMetric}
\ee
Space--time factorizes into a sphere parametrized by $(\phi,\theta)$
times another two-surface parametrized by $(t,r)$.
One can check that the Ricci scalar 
vanishes and that the Ricci
tensor (with flat indices) is constant and only depends on $c$ whereas the 
coefficients $a,b$ drop out. 
This suggests that the constants $a,b$ do not have an invariant
meaning, 
but that they can be changed by coordinate transformations, whereas
$c$ has an invariant meaning and parametrizes a family of 
inequivalent (non-isometric) metrics.
In order to prove this we have computed
the Killing vectors associated with the $(t,r)$ surface.
The result is that for all admissible choices (we have to request $a\not=0$ or 
$b\not=0$ to exclude singular cases 
and we have $c\not=0$ ) three Killing vectors exist, and that
the isometries satisfy the Lie algebra 
$sl(2,\mathbb{R}) \simeq so(2,1)$. In view of the signature of
the $(t,r)$ surface this leaves us with the unique possiblity
of two-dimensional anti de Sitter space 
$AdS^2 \simeq SO(2,1) / SO(1,1) \simeq Sl(2,\mathbb{R}) / U(1)$.
The full metric is the Bertotti-Robinson geometry $AdS^2 \times S^2$.
The standard parametrization is obtained by setting $b=0$ and
rescaling $t$ such that $a=c^{-1}$:
\be
ds^{2} = - \frac{r^{2}}{c^2} dt^{2} + c^2 \frac{dr^{2}}{r^{2}} + 
c^2 ( \sin^2 \theta d\phi^2 + d \theta^2 ) \;.
\label{standard}
\ee
It is possible to find the  explicit coordinate transformation that 
eliminates $b$, but the expression is rather complicated and so we
do not write it down. The constant $c$ specifies the
radius of the geometry. We have found two expressions for the constant,
one from the Ricci tensor and one from the gauge field equations.
The formula for the Ricci tensor implies
\be
e^{2g(r)} = e^{-2f(r)} = \frac{r^2}{c^2} = \ft1{16} | T^-_{01} |^2 r^2 \;,
\ee
whereas the gauge field equations gave us
\be
e^{-2 f(r)}  = - \ft14 e^{-{\cal K}/2} T^-_{01} Z^{-1} r^2 \;.
\label{fTZ}
\ee
This implies the relation
\be
T^-_{01} = -4 e^{-{\cal K}/2} \ov{Z}^{-1}  \,.
\label{valueTminus}
\ee
Thus the two constants $T^-_{01}, Z$ are related\footnote{
We note in passing that the phase relation between 
$T^-_{01}$ and $Z$ is such that the right hand side of
(\ref{fTZ}) is positive, and that the constant $c$, when
computed from the gauge field equation as 
$c= \sqrt{-4 e^{{\cal K}/2} \frac{Z}{T^-_{01}}}$ is real, $c=|Z|$.
}
and one can express the 
radius either in terms of $T^-_{ab}$ or in terms of
$|Z|$:
\be
e^{2g(r)} = e^{-2f(r)} = \frac{r^2}{c^2} =  
e^{-{\cal K}} \frac{r^2}{|Z|^2} \;.
\label{NearHorizon}
\ee
The later form is familiar from the case without chiral 
background. $Z$ is related to the ${\cal N}=2$ central charge.
Therefore it is convenient to express all the other constants
in terms of $Z$.

Our result describes the near horizon geometry of a 
static extremal black hole in Poincar\'e supergravity. 
The global form of the solution is (\ref{metricansatz})
with two functions $f_{\mscr{BH}}(r), g_{\mscr{BH}}(r)$. 
In the near horizon region $r \rightarrow 0$ the
black hole geometry approaches the 
$AdS^2\times S^2$ solution (\ref{NearHorizon}) 
\be
(r^2 e^{2f(r)_{\mscr{BH}}}) \rightarrow_{r \rightarrow 0}
r^2 e^{2f(r)} = e^{{\cal K}} |Z|^2 \mbox{   and   }
f_{\mscr{BH}}(r) \simeq - g_{\mscr{BH}}(r) \;.
\ee
Therefore the size of the event horizon is $A = 4 \pi |Z|^2$.
In order to describe $R^2$-corrections we identify the chiral 
background multiplet with the Weyl multiplet. The background scalar
is
\be
\widehat{A} = T^{ij}_{ab} T^{kl ab}\ve_{ik} \ve_{jl}
= - 64 e^{-{\cal K}} \ov{Z}^{-2} 
\label{valueAhat}
\ee
In conclusion we have now fully specified the near horizon solution
in terms of the field $Z$, which depends on the charges 
and on the scalar fields.
In the Bertotti-Robinson geometry $Z$ is constant, but in the full
black hole solution, which only approaches the Bertotti-Robinson geometry
asymptotically, it is in general an $r$-dependent quantity through the
$r$-dependence of the scalar fields. 

We conclude this lengthy section by collecting the formulae
which describe a static and spherically symmetric ${\cal N}=2$
vacuum and, simultanously, the near horizon geometry of a BPS black 
hole. The metric, gauge fields and scalars are
\be
\begin{array}{l}
ds^2 = - e^{-2f(r)} dt^2 + e^{2f(r)} \left( dr^2 + r^2 d \Omega^2 \right) \\
G_{I23} = \frac{e^{-2 f(r)}}{r^2} q_I \;,\;\;\;
F^I_{23} = \frac{e^{-2 f(r)}}{r^2} p^I \;,\;\;\;
X^I = \mbox{const.}\;, \\
\end{array}
\label{BRvacuum}
\ee
where the function $f(r)$ is related to the field
$Z=p^I F_I(X,\widehat{A}) - q_I X^I$ by
\be
e^{2f(r)} r^2 = e^{{\cal K}} |Z|^2 \;.
\label{BRvacuum1}
\ee
The field $e^{{\cal K}}$ is the compensator for dilation invariance and
becomes a constant in the Poincar\'e frame. The values of the
auxiliary $T$-field and of the background scalar $\widehat{A}$ are
given in (\ref{valueTminus}) and (\ref{valueAhat}) in terms
of $Z$. When we substitute the expression for $\widehat{A}$ into the
definition of $Z$ and use the homogenity of $F_I(X,\widehat{A})$ we get
an equation for $|Z|^2$ in terms of the $X^I$, whereas the phase
of $Z$ remains arbitrary (see also the beginning of chapter
\ref{ChapterStringBHs}). 
We will see in the next section that the $X^I$ and therefore
$|Z|^2$ and the complete solution 
can be expressed in terms of the electric and magnetic charges.

\section{The Entropy}

\subsection{The Entropy Formula \label{SectionEntropyFormula}}

With the Poincar\'e frame action ((\ref{ActionWVNL}), subject to
gauge conditons)
and the near horizon solution 
at our disposal we can now use Wald's formula to compute the entropy.
First remember that the entropy is given by (\ref{WaldEntropy})
\bea
{\cal S} = 2 \pi \oint d^2x \sqrt{h} \varepsilon_{ab}  \varepsilon_{cd}
\frac{\der {\cal L}_{\rm Poinc}}{\der R_{abcd}} \;\;\;,
\eea
where the integral is over the event horizon, $h$ is the
absolute value of the 
determinant of the pulled-back metric and $\ve_{ab}$ is the
binormal. We have already normalized the Poincar\'e action
such that we can use the conventions of chapter \ref{ChapterGravity}.
First we have to compute the partial derivative with respect to 
the Riemann tensor. To do so we have to remind ourselves that the
components of the chiral background field are related to the Weyl 
multiplet and therefore a lot of the terms depend on the Riemann
tensor. However there are no derivatives of the Riemann tensor present
so that the version (\ref{WaldEntropy}) of the entropy formula
applies. The result is:
\bea
\frac{\der {\cal L}_{Poinc}}{\der R_{abcd}} 
= - \frac{1}{2 \kappa^2} \eta^{ac} \eta^{bd} + \frac{i}{16 \pi} \left(
F_{\widehat{A}I} {\cal F}_{ef}^{I-} \frac{\der {\widehat F}^-_{ef}}{\der
R_{abcd}} + F_{\widehat{A} \widehat{A}}  {\widehat F}^-_{ef}
 \frac{\der {\widehat F}^-_{ef}}{\der
R_{abcd}} +  F_{\widehat{A}} \frac{\der {\widehat C}}{\der
R_{abcd}}
- h.c. \right) \;\;\;.
\label{variat}
\eea
The first term comes from the Einstein-Hilbert term and gives the
standard Bekenstein-Hawking entropy. The other terms are deviations
from the area law that result from the additional curvature terms.
Using that for our
solution ${\cal F}^{-I}_{ab} = \widehat{F}_{ab} = 0$ the formula
simplifies:
\bea
\frac{\der {\cal L}_{Poinc}}{\der R_{abcd}} 
= - \frac{1}{2 \kappa^2} \eta^{ac} \eta^{bd} 
+ \frac{i}{16 \pi}
(F_A  \frac{\der {\widehat C}}{\der
R_{abcd}} - h.c.)    \;\;\;
\eea
Now remind the transformation rules (\ref{SymplTransfDerivF})
of the function $F(X,\widehat{A})$ and of its derivatives. Most of
these quantities transform in a complicated, non-covariant way
under symplectic transformation. This is true in particular 
for $F_{\widehat{A}I}$ and $F_{\widehat{A} \widehat{A}}$, but luckily
these quantities do not contribute because there coefficients vanish.
The remaining correction term involves $F_{\widehat{A}}$ which
is a symplectic function.

It remains to compute $\ft{\der \widehat{C}}{ \der R_{abcd}}$.
We now use that in a bosonic background
\bea
D_a(D^c T_{cbij}) = {\cal D}_a({\cal D}^c T_{cbij}) - f_a\;^c T_{cbij} \;,
\eea
to rewrite (the bosonic terms in) ${\widehat C}$ as\footnote{
The commutator part of the double derivative vanishes,
$T^{ab-} [{\cal D}_a, {\cal D}^c] T^+_{cb} = 0$, due to
(anti-)selfduality of $T^{\pm}_{ab}$.}
\bea
{\widehat C} &=& - 8 T^{ab-} \{{\cal D}_a , {\cal D}^c \} T_{cb}^+ 
+ 16 T^{ab-}
f_a\;^c T_{cb}\;^+ + 64 {\cal R}(M)_{cd}^{-ab}  {\cal R}(M)_{cd}^{-ab} 
\nonumber\\
&+& 32 R(V)_{ab\;k}^{-l} R(V)_{ab\;l}^{-k} \;\;\;.
\eea
We compute
\bea
\frac{\der {\widehat C}}{\der R_{ab}\,^{cd}} 
&=& 
8 T^{af-} T_{cf}\,^{+} \delta^b_d + 128
 {\cal R}(M) ^{-mn}\,_{lq} 
\frac{\der {\cal R}(M)_{mn}^{-\;lq}}{\der  R_{ab}\,^{cd}} \;.\;\;
\eea
Using that $ {\cal R}(M)_{ab}^{-cd} =0$ on the horizon we find
\bea
\epsilon_{ab} \epsilon^{cd} 
\frac{\der {\widehat C}}{\der R_{ab}\,^{cd}} = 
16 \; T^{01+} T^{01-}  \;.
\;\;\;
\eea 
Performing the integral over the horizon
and expressing everything in terms of $Z$ we finally get
our model-independent entropy formula
\be
{\cal S} = \pi (G_N^{-1} |Z|^2 - 256 \;\mbox{Im}\; F_{\widehat{A}}(X,\widehat{A}))
\;,\;\;\;
\mbox{where}\;\;\; \widehat{A} = - 64 \overline{Z}^{-2} e^{-{\cal K}} \;.
\label{ModIndEntropyFormula}
\ee
This is one of our central results \cite{CardWMoh:1998/12}.
Note that we expressed the gravitational coupling $\kappa$ 
by Newtons constant. The dilatational gauge fixing 
relates $e^{-{\cal K}}$ to the Planck scale and therefore
to $G_N$: $e^{-{\cal K}} = m_{\mscr{Planck}}^2 = G_N^{-1} = 8 \pi
\kappa^{-2}$. The formula is manifestly convariant with respect
to symplectic transformations.

In view of the complications we had to go through this is a
strikingly simple result. All modifications through
additional curvature terms are captured by the function
$F_{\widehat{A}}$, and we did not need to specify this function so that
the formula is completely model-independent. 
Now recall that $Z$ is a function of the $X^I$ and of the charges
$p^I,q_I$
and that the fields $X^I$, which themselves are
functions of the unconstrained physical scalars.
Therefore the 
entropy is a complicated function of the charges and 
of the scalars on the horizon. Normally one would expect
that the scalars on the horizon can take
arbitrary or at least a continuous range of values so that
the entropy can vary continuously. Such a behaviour would
clash with the intended interpretation of the entropy as the 
degeneracy of the macroscopic state of the black hole.
The next step is to argue that the scalars cannot take
arbitrary values on the horizon, but are themselves determined
by the charges.

\subsection{The Stabilization 
Equations \label{SectionStabilizationEquations}}

In absence of higher derivative terms a set of beautiful relations
have been derived, which express the scalars on the
event horizon, $z^A_{\mscr{hor}} = (X^A/X^0)_{\mscr{hor}}$ 
in terms of the charges 
$(p^I,q_J)$ \cite{FerKalStr:1995,Str:1996,FerKal:1996/02,
FerKal:1996/03}.\footnote{We set 
$e^{\cal K}=1$ for the rest of this chapter.}
These relations are called the {\em stabilization equations} and
they take the form
\be
\left[
\ov{Z} \left( \begin{array}{c}
X^I \\ F_J \\
\end{array} \right)  - 
Z \left( \begin{array} {c}
\ov{X}^I \\ \ov{F}_J \\
\end{array} \right) \right]_{\mscr{hor}}
= i \left( \begin{array}{c}
p^I \\ q_J \\
\end{array} \right) \;,
\label{Stab}
\ee
where $F(X)$ is the prepotential, $Z$ is the function defined in 
(\ref{Zfunction}) and the expression on the left side is evaluated on 
the event horizon. The name stabilization equations
is motiviated by the fact that in the presence of non-trivial 
scalar fields one needs to impose conditions on their near horizon 
behaviour in order to avoid that they take singular values on the horizon.
In the context of string compactifications, where the scalar fields
are moduli, such singularities can be understood as compactification
artefacts. The values of the moduli specify the four-dimensional coupling
and the geometry of the internal compact space. Singularities in these
quantities do not signal a breakdown of the underlying microscopic
higher-dimensional theory, but a breakdown of the effective
four-dimensional description due to decompactification or strong
four-dimensional coupling. Therefore the singularity can be removed
by lifting the solution to a higher-dimensional solution of the
fundamental theory. This can be explicitly studied in the case
of toroidal compactifications. Conversely one also sees that 
the compactification of a regular higher dimensional solution
often leads to a singular lower dimensional solution, unless 
suitable conditions are imposed which stabilize the moduli.\footnote{
From now on we will call scalars with flat potentials moduli, even 
when not explicitly referring to a string compactification.}

BPS black holes in ${\cal N}=2$ theories do not result from simple
toroidal compactifications. But one can analyse the condition for
having regular moduli using four-dimensional supergravity. The result
is that on the horizon full ${\cal N}=2$ supersymmetry must be
restored, implying that the solution interpolates between
two vacua, Minkowski space and the Bertotti-Robinson geometry. Moreover
this implies the stabilization equations (\ref{Stab}).

The equations impose $2N_V+2$ real conditions
on $2N_V+2$ complex quantities $X^I,F_J$. Since the $F_J$ are
functions of the $X^I$ only half of these quantities are independent,
and generically these equations will
fix the moduli
$z^A_{\mscr{hor}}$ at the horizon
in terms of charges $(p^I,q_J)$. If the prepotential is sufficiently
simple one can solve the equations explicitly. We will see examples
of this when discussing concrete models in chapter 
\ref{ChapterStringBHs}. 
The behaviour of the moduli in the asymptotically flat region
at infinity, $z^A_{\infty}=(X^A/X^0)_{\infty}$, 
is completely different in that the values of the moduli
are not fixed, but can be chosen arbitrarily. The scalar equations
of motion in a static spherically symmetric BPS background can be interpreted
as a dynamical system with the radius $r$ as 'time'. They describe
the evolution of the moduli as functions of $r$ from their
arbitrary values at infinity to their fixed values at the horizon.
Thus the dynamical system exhibits {\em fixed point behaviour}.
Since the fixed points are attractive, they form 
an attractor and the mechanism has been called 
the {\em supersymmetric attractor mechanism}.

The asymptotic form of the metric at the horizon is
\be
ds^2 = - \frac{r^2}{|Z|^2} dt^2 + \frac{|Z|^2}{r^2} d\vec{x}^2 \;,
\ee
and therefore the Bekenstein-Hawking entropy is given by
\be
{\cal S} = \frac{A}{4} = \pi | Z|^2_{\mscr{hor}} \;,
\ee
where $|Z|_{\mscr{hor}}$ is the value of $|Z|$ on the horizon.
Since on the horizon the scalar fields are fixed in terms of the
charges, it follows that 
$|Z|_{\mscr{hor}}$ and the entropy
are exclusively determined by the charges.

The stabilization equations can be reformulated as an
extremalization condition on $|Z|$, evaluated
on the horizon. In fact it was first directly observed that for
static supersymmetric backgrounds the dynamical system 
governing the $r$ evolution of the function $f(r)$ shows attractive
fixed point behaviour \cite{FerKalStr:1995}. 
Later it was shown that the fixed points
are determined by the stationary points of $|Z|$ as a function of the moduli,
keeping the charges fixed,
by 
\be
\der_A | Z(z,\ov{z},p^I,q_J) | = 0 \;,
\label{extremal}
\ee
where $\der_A$ is the partial
derivative with respect to the scalars $z^A$ 
\cite{FerKal:1996/02,FerKal:1996/03}. It was also shown
that this is equivalent to the stabilization equations (\ref{Stab}).
In practice it is much easier to solve the stabilization equations
than to perform the explicit extremization \cite{BehCardWKalLueMoh:1996}. 
This is the reason why
we started our discussion with them. Note also that they can
be derived directly from the Killing spinor equations, without
first deriving the equivalent statement (\ref{extremal}) 
\cite{Sab:1997/03,Sab:1997/04,BehLueSab:1997}.

It turns out that $|Z|$ is not just stationary at the fixed point 
but in fact takes 
its minimum, which
motivates the term {\em minimal area principle} \cite{FerGibKal:1997}. 
It is known
from explicit examples that several minima with the same 
charges may exist \cite{Moo:1998,KalLinShm:1999,Kal:1999}. 
Thus in general every fixed point has a finite
{\em bassin of attraction} and is characterized by an 
{\em area code}. It might also happen that the fixed point sits
on the boundary of the moduli space. The global aspects of the attractor
mechanism are not yet fully 
understood. We refer to \cite{Moo:1998} for a more complete discussion, 
which also describes the deep relations of the attractor mechanism 
to number theory and to the geometry of Calabi-Yau manifolds.
We also note that the attractor mechanism applies generally to
BPS solitons in extended supergravity. Besides 
four-dimensional and five-dimensional BPS black holes in
${\cal N}=2,4,8$ supergravity 
\cite{FerKal:1996/02,FerKal:1996/03,ChaFerGibKal:1996}
this applies more generally to BPS $p$-branes \cite{AndDauFer:1996}.
We refer to \cite{DauFre:1998} for a review.

One can give yet another derivation of the stabilization equations,
which is based on electric-magnetic duality, or symplectic 
covariance \cite{BehCardWKalLueMoh:1996}.
Given a black hole solution that in the near horizon area
exhibits full ${\cal N}=2$ supersymmetry we know that the solution
is described by two symplectic vectors, 
$(p^I,q_J)$ and $(X^I, F_J)$. Assuming that the solution is 
uniquely determined by the charges the two vectors must
be related. The only possible relation admitted by symplectic
invariance is that they are proportional.
Since $(p^I,q_J)$ is real whereas $(X^I,F_J)$
is complex the constant of proportionality must be complex.
Using the $D$-gauge condition this constant is fixed to be $Z$
and this way one obtains (\ref{Stab}). This argument has the
advantage that it can be directly applied to the case
with $R^2$-terms without having to find the full interpolating
solution. The only possibility which respects 
symplectic covariance is \cite{BehCardWLueMohSab:1998}:
\be
\left[
\overline{Z} \left( \begin{array}{c}
X^I \\ F_{J}(X, \widehat{A}) \\ \end{array} \right) -
Z \left( \begin{array}{c}
\ov{X}^I \\ \ov{F}_{J}(\ov{X}, \ov{\widehat{A}}) \\ \end{array} \right)
\right]_{\mscr{hor}}
= i \left( \begin{array}{c}
p^I \\ q_J \\
\end{array} \right) \;.
\label{StabEquatWithR2}
\ee
This form of the stabilization equation will be used later
to determine the black hole entropy in terms of the charges.

\section{Black hole Solutions \label{SectionBlackHoleSolutions}}

So far we have restricted our attention to the near horizon 
geometry and to the black hole entropy. In this section we will
give a short review of the results obtained in 
\cite{Sab:1997/03,Sab:1997/04,BehLueSab:1997}
on the full BPS black hole solution. 
These results were obtained using the on-shell formulation and 
refer to the minimal Lagrangian without $R^2$-terms. 
At the end we will briefly comment on the case with $R^2$-terms.

BPS solutions can be found by looking for a static (or stationary)
and asymptotically flat geometry that is invariant under 
half of the ${\cal N}=2$ supersymmetries. This means that
the supersymmetry variations of the fermions vanish in the
bosonic background for specific choices of the $\e_i$, where
four of the eight real components are fixed in terms of the
others. The resulting geometry has four Killing spinors rather
then eight and is invariant under four of the eight supersymmetry
transformations.

The analysis can be carried out without imposing spherical 
symmetry.\footnote{In our analysis of ${\cal N}=2$ supersymmetry
we imposed spherical symmetry to simplify the problem. The generalization
to multi-centered BPS solutions is currently under investigation 
\cite{CardWKaeMoh}.}
One starts with a metric that is static and has a conformally
flat space part ('conformastatic')
\be
ds^2 = - e^{2g(\vec{x})} dt^2 + e^{2f(\vec{x})} d\vec{x}^2
\label{staticmetric}
\ee
with two unknown functions. 
It was shown by Tod \cite{Tod:1983} that the most general static supersymmetric
metric is of that type and that the  
two functions in the supersymmetric case must be related by $f=-g$.

The projection which identifies the four unbroken supersymmetries
can be found in various ways. One can start from the supersymmetry
algebra and ask for a massive state at rest which saturates the
BPS bound. This implies that the Bogomolnyi matrix, which
is the matrix of all $Q$-anticommutators, evaluated in the background,
must have a zero mode. This implies a condition on the asymptotic
form of the Killing spinor, which expresses half of the components
in terms of the others. Alternatively one can look at the
$Q$-variations of the gravitini field strengthes. If one does not
impose full ${\cal N}=2$ supersymmetry, then one obains
a relation between the components of the
Killing spinor, which fixes half of them. This is an
integrability condition for the vanishing of the gravitini
variations. One can also directly solve the gravitini variations 
without looking at the integrability condition, 
as for example done in \cite{Sab:1997/03,Sab:1997/04,BehLueSab:1997}.

In the case of static (and stationary) 
BPS backgrounds the projection takes the form
\be
\e_i = i \g_0 \ve_{ij} \e^j
\label{projbreak}
\ee
(up to a phase factor that we will discuss later).
Moreover the vanishing of the gaugino variations implies the
so-called {\em generalized stabilization equations} \cite{Sab:1997/03}
\be
i(X^I(z) - \ov{X}^I(\ov{z})) = H^I, \;\;\;
i(F_J(z) - \ov{F}_J(\ov{z})) = H_J
\label{GenStabHol}
\ee
where $(X^I(z), F_J(z))$ is the holomorphic section.
The functions $H^I(\vec{x}),H_J(\vec{x})$ 
are related to the magnetic parts of the
gauge fields by
\be
F^I_{mn} = \ft12 \e_{mnp} \der_p H^I\;, \;\;\;
G_{J mn} = \ft12 \e_{mnp} \der_p H_I\;, \;\;\;
\ee
and therefore form a symplectic vector. Note that
$m,n=x,y,z$ are spatial world indices and that
the $\e$-symbol is the standard real one, $\e_{xyz}=1$.
When imposing the gauge field equations of motion the
functions $H^I,H_J$ must be harmonic. The generalized
stabilization equations are the rationale behind the observation
that the full black hole solution is obtained from the
near horizon solution by replacing electric and magnetic charges
with harmonic functions \cite{Beh:1996}.

The static supersymmetric
metric takes the form (\ref{staticmetric})
with
\be
e^{-2g} = e^{2f} = i \left( \ov{X}^I(\ov{z}) F_I(z)
- X^I(z) \ov{F}_I(\ov{z}) \right) = e^{-K}\;,
\ee
where $K$ is the K\"ahler potential. As a further condition
of unbroken supersymmetry in a static background one finds
that the K\"ahler connection 
(\ref{KaehlerConnection},\ref{KaehlerConnectionSpecial})
has to vanish, 
\be
A_\m = - \ft{i}2 ( \der_A K \der_\m z^A - \der_{\ov{A}} K \der_\m
\ov{z}^{\ov{A}} ) =0\;.
\ee
Moreover the $Z$-field has to be real.

Depending on the choice of harmonic functions one can describe
a single BPS black hole or a static ensemble of BPS black holes.
The above analysis has even been generalized to stationary
BPS backgrounds and to gravitational instantons. 
In these cases the K\"ahler connection is non-vanishing
and determines the off-diagonal components $g_{tm}(\vec{x})$
of the metric.  The resulting solutions
describe rotating geometries, 
Eguchi-Hanson and Taub-NUT spaces \cite{BehLueSab:1997}.

We will consider only static solutions in the following. 
Let us first reformulate the above solution in 
order to bring it closer to the formalism we use in this paper.
In particular we use the covariantly holomorphic
section $(X^I,F_J)$ instead of the holomorphic one. We would also like
to formulate the solution in a manifestly $U(1)$ and K\"ahler
invariant way. 
In the above solution
the metric is given in terms of the K\"ahler potential. As we 
discussed earlier, a general holomorphic reparametrization of 
the scalar manifold acts on the holomorphic section as a 
symplectic transformation accompanied by a K\"ahler transformation.
The K\"ahler potential is symplectically invariant but is changed
by the K\"ahler transformation. Since the function $e^{2f}$ is
invariant under reparametrizations of the scalar manifold it seems
that the solution imposes a 'gauge fixing condition' that picks
a particular class of parametrizations of the scalar manifold.
A related point is that the 
projection (\ref{projbreak}) explicitly
breaks chiral $U(1)$ invariance and thus fixes the phase of 
$Z$, such that $Z$ is real.
One can, however, replace (\ref{projbreak}) by the $U(1)$ covariant version
\be
\e_i =  \g_0 \frac{\overline{\Sigma}}{|\Sigma|} \ve_{ij} \e^j
\label{projcov}
\ee
where $\Sigma$ is a field of $U(1)$ weight $-1$, which is later
on determined by the Killing spinor equations.

The calculation of \cite{BehLueSab:1997} can now be recast in a 
$U(1)$ invariant way. Alternatively one can  
do the calculation in the off-shell formulation, which allows
to include the Weyl background \cite{CardWKaeMoh}.

In both cases the Killing equations imply 
generalized stabilization equations of the form
\be
\ov{\Sigma} \left( \begin{array}{c}
X^I \\ F_J \\ \end{array} \right) -
\Sigma \left( \begin{array}{c}
\ov{X}^I \\ \ov{F}_J \\ \end{array} \right)
= i \left( \begin{array}{c}
H^I \\ H_J \\ \end{array} \right) \;,
\label{GeneralizedStabEqs}
\ee
with 
\be
\Sigma = H^I  F_I - H_I X^I \;.
\ee
The metric is given by
\be
e^{-2g}= e^{2f} = | \Sigma |^2 \;.
\label{MetricSigma}
\ee
In addition one finds that the K\"ahler connection takes the
form 
\be
A_r = \ft{i}2 \der_r \log \frac{\Sigma}{\ov{\Sigma}}
\label{KaehlerConSusyBG}
\ee
and that $\ov{\Sigma}Z$ has to be real.

Note that (\ref{MetricSigma}) is invariant under reparametrizations
of the scalar manifold because the covariantly holomorphic section
transforms by a symplectic transformation combined with a 
local $U(1)$ rotation, and $\Sigma$ is symplectically invariant
while the phase is irrelevant for $|\Sigma|$. The original solution
is recovered in a parametrization where $\Sigma = \ov{\Sigma} =
-e^{-K/2}$: With this condition the generalized stabilization 
equations take the form (\ref{GenStabHol}), the metric is 
given in terms of the K\"ahler potential, the K\"ahler connection
vanishes  and the $Z$-field is real.

We now take a closer look onto static single-centered solutions, which
are obtained by choosing the harmonic functions
\be
H^I = h^I + \frac{p^I}{r}, \;\;\; H_I = h_I +  \frac{q_I}{r} \;.
\ee
First we analyse the behaviour near the horizon to make contact
with the discussion of the last section. We find that
\be
(r \Sigma)_{\mscr{hor}} = Z_{\mscr{hor}}\;
\ee
and  therefore the generalized stabilization equations evaluated at the
horizon are the stabilization equation (\ref{Stab}).

The Bekenstein-Hawking entropy 
${\cal S}=\ft{A}{4}$ can be extracted from the behaviour
of the metric at the horizon,
\be
A= 4 \pi (e^{2f} r^2)_{r=0} \;.
\ee
Using that
\be
(e^{2f} r^2)_{r=0} = ( |\Sigma|^2 r^2 )_{r=0} =
| Z |^2_{r=0} = | Z |^2_{\mscr{hor}}
\ee
we obtain
\be
{\cal S} = \pi | Z |^2_{\mscr{hor}} \;,
\ee
where $|Z|_{\mscr{hor}}$ only depends on the charges $p^I,q_J$ as
a consequence of the stabilization equations.

Let us now study the asymptotic behaviour at infinity
and compute the ADM-mass. The asymptotic value of $\Sigma$ is
\be
\Sigma_{\infty}  = h^I F_I(\infty)  - h_I X^I(\infty) \;.
\ee
In order to have the correct normalization of the metric at infinity
we need to impose
\be
e^{2g(\infty)} = e^{2f(\infty)} = | \Sigma |^2_{\infty} = 1 \;.
\ee
This yields one real condition on the parameters $h^I, h_J$, which
characterize the behaviour of the harmonic functions at infinity.
There is a second condition which results from the fact that
the K\"ahler connection $A_{\m}$ has to take the particular form 
(\ref{KaehlerConSusyBG}) in
a supersymmetric static
background. This implies
\be
h^I q_I - h_I p^I =0 \;.
\ee
Thus the family of solutions is parametrized by 
$h^I, h_J, p^I, q_J$ subject to two real constraints. 
The parameters $p^I,q_J$ are discrete by Dirac quantization,
and independend because they are the electric and magnetic charges
with respect to the $N_V+1$ gauge fields in the theory. The parameters
$h^I,h_J$ are continuous and subject to 2 real constraints. This
corresponds to the fact that the theory contains $N_V+1$ independent
gauge fields, but only $N_V$ independent 
complex scalar fields. The scalars can take
arbitrary (or at least a continuous range\footnote{
Remember that the scalar manifold can be an arbitrary
special K\"ahler manifold.
The precise range of value of the scalar fields depends on how one
parametrizes this manifold.} of) values at infinity.
In other words the solution is parametrized by the charges $p^I,q_J$
and the
moduli at infinity, $z^A_{\infty}$. The ADM mass is obtained by expanding
the metric component $g_{tt}$ to lowest order in $\ft1r$:
\be
- e^{2g} = - \left( 1 - 2 \frac{M_{ADM}}{r} + {\cal O}(\ft1{r^2})
\right) \;.
\ee
Using the explicit form
of $\Sigma$ one can show that
\be
M_{ADM} = \ov{\Sigma}_{\infty} Z_{\infty} \;.
\ee

Since $|\Sigma|_{\infty}=1$ we can rewrite the ADM mass:
\be
M_{ADM} = |Z|_{\infty} = | p^I F_I(\infty) - q_I X^I (\infty)| \;.
\ee
This is in fact the BPS mass formula
\be
M_{BPS}= |Z|\;,
\ee
where $Z$ is the central charge defined in (\ref{centralcharge}).
To make this explicit one uses that the relation
$F_I F^{+I}_{ab} - X^I G^+_{Iab}=0$ that we derived earlier for 
${\cal N}=2$ vacua also holds for BPS configurations with four
Killing spinors \cite{BehLueSab:1997}. 
Thus one has
\be
T^-_{ab}= F_I F^{I}_{ab} - X^I G_{Iab} \;.
\ee
The asymptotic form of the gauge field is
\be
T^-_{23}  \simeq_{r \rightarrow \infty} 
\frac{p^I F_I(\infty)  - q_I X^I(\infty) }{r^2} 
= \frac{Z_{\infty}}{r^2} \;,
\ee
where we used the definition of $p^I,q_J$ in terms of $F^I_{ab}, G_{Iab}$.
Then the  central charge is
\be
Z = \ft1{4\pi} \oint T^- = \ft1{4\pi} \int_{S^2_{\infty}} 
T^-_{23} r^2 d\Omega = Z_{\infty} \;.
\ee
Therefore the asymptotic value
$Z_{\infty}$ of the $Z$-field (\ref{Zfunction})
is the central charge (\ref{centralcharge}) carried by the BPS
solution and we obtain the standard BPS mass formula, as expected.
Obviously the mass depends both on the charges $p^I,q_J$ and
on the moduli $z^A_{\infty}$.

For any given set of charges $p^I, q_J$ one can obtain a
particularly simple solution by setting the scalar fields
to constant values. Consistency then requires that these
values are precisely the fixed point values that 
supersymmetry dictates, 
$z^A(r) = z^A_{\mscr{hor}}=z^A_{\mscr{FP}}(p^I,q_J)$. This implies that
$Z_{\infty} = Z_{\mscr{hor}}$ and since $|Z|_{\mscr{hor}}$ 
takes the minimal 
value (as a function of the moduli $z^A$), 
it follows that the ADM mass is minimized as a function
of the moduli $z^A_{\infty}$. It is of course plausible that
for black hole solutions with scalar fields the minimum of 
the mass is obtained when the scalars are constant.
Such black hole solutions have been called
{\em double extremal} \cite{KalShmWon:1996}, motivated by the following set of
inequalities:
\be
M_{BH}( z_{\infty}, p^I, q_J ) \geq
M_{BPS} (z_{\infty}, p^I, q_J) \geq
M_{FP} (z_{\mscr{FP}}(p^I, q_J), p^I, q_J) \;.
\ee
This means that given a general (non-extremal) static black hole
with charges $p^I,q_J$ and generic values of the moduli $z^A_{\infty}$
one can minimize the mass, while keeping charges and moduli fix,
by going to the extremal limit, where the black hole becomes 
supersymmetric. Then one can further minimize the
BPS mass over the moduli space, which yields the black hole with
the minimal possible mass for the given set of charges. For constant
scalars the relation between $\Sigma$ and $Z$ simplifies with the
result
\be 
e^{-2g}= e^{2f} = | \Sigma |^2 = \left( 1 + \frac{|Z|}{r} \right)^2 \;.
\ee
Therefore the metric of a double extremal black hole is the 
extremal Reissner-Nordstr{\o}m metric.

In the above discussion we did not consider $R^2$-terms
but worked with the mininal Lagrangian. Given the form 
of the solution we expect that in presence of
$R^2$-terms the solution is modified by replacing the prepotential
$F(X)$ by the full function $F(X,\widehat{A})$. As is obvious from our
discussion of ${\cal N}=2$ solutions in presence of $R^2$-terms
it is, however, difficult and involved to actually prove this
statement. We will not enter the investigation here, 
but refer to \cite{CardWKaeMoh}.

Finally we would like to point out that there are 
further interesting topics in relation to 
$D=4,{\cal N}=2$ black holes, which we did not touch in the 
short review given above. Extremal black holes in five-dimensional
${\cal N}=2$ supergravity (8 real supercharges) are very similar
to their four-dimensional counterparts, because of the so-called
very special geometry \cite{deWvP:1992}, which controls the vector 
multiplet couplings. This has been used to construct and analyse
extremal black hole solutions \cite{Sab:1997/08,ChaSab:1998,ChaSab:1999}.
In the context of string or M-theory compactifications on Calabi-Yau
threefolds black holes can be used as probes to investigate topological
phase transitions 
\cite{ChoEtAl:1997,BehLueSab:1997,GaiMahMohSab:1998,Moo:1998}.
One can use T-duality to map extremal black holes onto
Euclidean wormholes, which are generalizations of the ten-dimensional
type IIB D-instanton \cite{BehGaiLueMahMoh:1997}.
We refer to \cite{DauFre:1998} for 
a nice review of BPS black holes in extended supergravity 
and to \cite{BerTri:1999,BerTri:2000} for a recent account on the
role of so-called generating solutions for both macroscopic and
microscopic aspects of BPS black holes in $N=8$ supergravity. 

During the last year the metric on the moduli space of 
multiblack hole solutions has been studied for both five-
and four-dimensional ${\cal N}=2$ supergravity coupled to
vector multiplets \cite{GutPap:1999,MalSprStr:1999,GutPap:2000}.
The extension of these results to the case with $R^2$ terms is
currently under investigation \cite{CardWKaeMoh2}. The dynamics
of near coincident black hole is described by superconformal
quantum mechanics on the moduli space, see \cite{StrEtAl:1999} for a 
review.

\section{More recent results on stationary BPS solutions in presence
of $R^2$-terms \label{AddedSection} }

More recently we have succeeded in finding all stationary 
space-times with residual supersymmetry embedded according to
(\ref{projcov}), for an arbitrary function
$F(X,\wh{A})$. We refer the reader to \cite{CardWKaeMoh}
for the details and only summarize the most important results here.
The most general stationary solution can be fully specified 
in terms of $2N_V + 2$ harmonic functions, subject to the 
generalized stabilization equations (\ref{GeneralizedStabEqs})
with the prepotential $F(X)$ replaced by the full function
$F(X^I, \wh{A})$, as conjectured above. We emphasize that the
generalized stabilization equations are not only sufficient,
but also necessary for having partial supersymmetry. 
A particular subclass of solutions describes
single-  and multi-centered
configurations of extremal black holes and these solution approach
the Bertotti-Robinson solution on their horizons.
In \cite{CardWKaeMoh} we included an arbitrary number of 
neutral hypermultiplets into the analysis and found 
that the hypermultiplet scalars have to be constant
in stationary space-times with residual supersymmetry 
embedded according to (\ref{projcov}). Thus including 
hypermultiplets does not lead to more general solutions.

In \cite{CardWKaeMoh} we also proved two further results
about solutions with full $N=2$ supersymmetry (again for
arbitrary $F(X,\wh{A})$). The first is that
the Bertotti-Robinson solution is the unique stationary $N=2$
solution. The properties of spherical symmetry and staticity,
which we imposed separately in section
\ref{SectionNearHorizon}
follow in fact automatically from full supersymmetry. 
The second new result is that in any stationary
$N=2$ solution, except the limiting case of Minkowski space,
the scalar fields are determined in terms of the electric and magnetic
charges by the stabilization equations (\ref{StabEquatWithR2}).
In other words one does not need top invoke arguments based on the flow
from a BPS solution to a fully supersymmetric configuration
in order to derive (\ref{StabEquatWithR2}). The stabilization equations
are a necessary condition for having full ${\cal N}=2$
supersymmetry, and the scalars must take their fixed-point values,
if non-trivial gauge fields are present. If there are no charges, one
gets flat Minkowski space and the scalars can take arbitrary constant
values.

What is the origin of this additional condition for $N=2$ solutions?
As explained above, new conditions may arise when considering the
variations of covariant derivatives of fermions. In \cite{CardWKaeMoh}
we succeeded in finding all such conditions, and it turned out
that there is one additional condition, which finally implies the
stabilization equations. In order to find all conditions coming
from derivatives of fermions in a transparent and systematic way,
it is advantagous to use a hypermultiplet rather than a 
non-linear multiplet as the second compensating multiplet. 
Therefore we do not enter into the details here, but refer the
reader once again to \cite{CardWKaeMoh}.

\chapter{Fourdimensional String and M-theory Compactifications
  \label{ChapterFourDimensionalStringsAndM}}

In the last two chapters we discussed ${\cal N}=2$ supergravity and
its black hole solutions. Since supergravity is not consistent as a
quantum theory, it must be embedded into a larger consistent theory.
The most promissing candidate is string theory, and in this chapter 
we review how four-dimensional ${\cal N}=2$ supergravity arises as the 
low energy effective theory of compactifications of ten-dimensional
string theories and of eleven-dimensional M-theory.

\section{Type II String Theory on a Calabi-Yau Threefold}

One way to obtain four-dimensional ${\cal N}=2$ supergravity
from string theory is by compactifying type II string theory on
a complex-three-dimensional Calabi-Yau manifold. Such spaces
are K\"ahler manifolds with vanishing first Chern 
class.\footnote{We also include the condition $h_{1,0}=0$  in order
to exclude from the definition 
the six-torus $T^6$ and $K3 \times T^2$, where $K3$ 
denotes the $K3$ surface.}
Since they have holonomy group $SU(3)$ they posess Killing spinors
and the number of supercharges of the
four-dimensional theory is $1/4$ of the number
of supercharges of the ten-dimensional theory. 
In the following we review some facts about Calabi-Yau compactifications
of string theory. For a more complete account we refer to
the reviews \cite{Can:1987,Hue:1992,Gre:1997,Asp:2000}.

The four-dimensional theory has a massless spectrum that consists
of the ${\cal N}=2$ supergravity multiplet plus a model-dependent
number of abelian vector multiplets and neutral hypermultiplets.
A hypermultiplet contains two Weyl spinors and four real scalars
as its on-shell degrees of freedom.
The kinetic term of the hypermultiplet scalars 
is a non-linear sigma-model with a target space that
is restricted to be quaternionic by local ${\cal N}=2$ 
supersymmetry \cite{BagWit:1983}.
Local ${\cal N}=2$ supersymmetry forbids neutral couplings between
vector and hypermultiplets \cite{deWLauvP:1985}. 
Since the string compactification for generic moduli only contains
gauge-neutral fields, the total moduli space factorizes into
a product of the special K\"ahler manifold of vector multiplet moduli and
the quaternionic manifold of hypermultiplet moduli. 
This factorization breaks down at special points in the moduli space, 
where the Calabi-Yau manifold becomes singular in such a way 
that the moduli space metric and the low energy effective action
are singular, too.\footnote{There exist milder singularities, such as
orbifold singularities and flop transitions, which do not give rise
to singularities of string theory.}
The most popular case of such a singularity 
is the conifold singularity \cite{CanOssGrePar:1991}.
The conifold point and other more complicated singular points
in the moduli space
are at finite distance from regular points. The fact that string backgrounds
can become singular under finite changes of the parameters was long
considered as a severe problem.
Then it was observed by Strominger that the
conifold singularity could be physically explained by the presence of a 
charged hypermultiplet that becomes massless at the conifold point
\cite{Str:1995}. The additional massless state corresponds to a 
type IIB threebrane, which is wrapped on the three-cycle which 
degenerates at the conifold point. When the additional state is taken
into account, all physical quantities behave smooth at the conifold point.
Subsequent work
generalized this to other types of singularities. In particular it
was shown that more complicated singularities correspond to 
multicritical points in the scalar potential of ${\cal N}=2$
supergravity coupled to charged matter \cite{GreMorStr:1995}. 
At these points one has the
option to go to different branches of the theory, such as the Coulomb
and Higgs branch, which we discussed in chapter \ref{ChapterGravity}.
The corresponding geometric mechanism is that a singularity can
be resolved in several non-equivalent ways. The transition between
branches of the scalar potential corresponds to a topological
phase transition, i.e. a change in the topology of the Calabi-Yau
manifold. In such transitions physics, or more precisely the
low energy effective action is smooth. We refer to \cite{Gre:1997}
for a review and references. In the following 
we will only consider generic compactifications where all fields
in the Lagrangian are gauge-neutral and the moduli space factorizes
into a vector multiplet and a hypermultiplet part.

In type IIA compactifications the numbers $N_V,N_H$ of vector and 
hypermultiplets are given by
\be
N_V = h_{1,1} \mbox{   and   }
N_H = h_{2,1} + 1 \;,
\ee
where $h_{1,1},h_{2,1}$ are Hodge numbers of the Calabi-Yau manifold.
Part of the scalar fields are geometric moduli of the internal space.
The $h_{1,1}$ vector multiplet moduli describe deformations of the
K\"ahler structure and of the internal part of the stringy
$B$-field. Both data can be conveniently combined into the so-called
complexified K\"ahler structure. Among the $4 h_{2,1}$ real hypermultiplet
moduli $2h_{2,1}$ describe deformations of the complex structure whereas
the other $2h_{2,1}$ moduli come from the fields of the Ramond-Ramond
sector. There is one additional hypermultiplet which is 
always present, even when the Calabi-Yau manifold has a unique
complex structure, $h_{2,1}=0$. It is called the universal hypermultiplet
because of its independence from Calabi-Yau data.
One of its scalars is the dilaton $\phi$, whose
vacuum expectation value is related to the four-dimensional 
IIA string coupling by $g_{IIA} = e^{\langle \phi \rangle}$.
In addition it contains the stringy
axion, which is obtained from the space-time part of the 
$B$-field by Hodge duality, and two scalars from the
Ramond-Ramond sector.

In type IIB compactifications the role of vector and hypermultiplets
is reversed,
\be
N_V^{(B)} = h_{2,1} \mbox{   and   }
N_H^{(B)} = h_{1,1} + 1 \;.
\ee
By mirror symmetry type IIB string theory on a Calabi-Yau manifold is
equivalent to type IIA string theory on a so-called mirror 
manifold \cite{LerVafWar:1989,GrePle:1990}, see \cite{HosKleThe:1994,Gre:1997}
for a review.
A Calabi-Yau manifold and its mirror are related by exchanging
the roles of the complex structure moduli and 
(complexified) K\"ahler moduli 
spaces.\footnote{The precise statement of mirror symmetry is that 
IIA and IIB string theory in the corresponding
string backgrounds are equivalent. String backgrounds are defined in terms of
conformal field theories. It is important to take into accoung
stringy $\alpha'$ corrections of the classical geometry. Moreover
one has to include regions of the moduli space, so-called non-geometric
phases, which do not have a geometrical interpretation in terms of a
Calabi-Yau sigma model.}
In particular the Hodge numbers are related by
$\widetilde{h}_{1,1}=h_{2,1}$ and $\widetilde{h}_{2,1}= h_{1,1}$. 
Due to mirror symmetry we can restrict our attention to one
of the two type II theories. For our purposes it is convenient
to consider the type IIA theory.

The part of the effective Lagrangian that we are interested in 
is the vector multiplet sector which  
is encoded in the function $F(X,\widehat{A})$. By expansion
of the function,
\be
F(X,\widehat{A}) = \sum_{g=0}^{\infty} F^{(g)}(X) \widehat{A}^g \;,
\ee
one finds the prepotential $F^{(0)}(X)$ which describes the minimal
part of the Lagrangian and the coefficients $F^{(g \geq 1)}(X)$ of
higher derivative couplings of the form $C^2 T^{2g-2}$.
In string perturbation theory
one can compute on-shell scattering amplitudes and obtain 
an effective action. This action has ambiguities because
string perturbation theory has only access to on-shell quantities.
Therefore terms in the effective action are only known up to terms
which vanish on-shell. For example the difference between the
square of the Riemann tensor and the square of the Weyl tensor
vanishes in a Ricci-flat background and therefore string
perturbation cannot decide 
whether the curvature squared term with coupling $F^{(1)}(X)$
involves the Riemann or the Weyl tensor.
We can invoke the off-shell formalism of chapter \ref{ChapterFourDSuGra}
to identify this term as the square of the Weyl tensor. 
Conversely, the effective supergravity Lagrangian of chapter
\ref{ChapterFourDSuGra}
needs the function $F^{(1)}(X)$ as input from string theory, because
supergravity is not a consistent quantum theory. As discussed in chapter
\ref{ChapterFourDSuGra} the relation between the auxiliary $T$-field
and the graviphoton is complicated in the off-shell formulation.
In \cite{AntGavNarTay:1993} a relation is found by requiring that
the effective action reproduces the on-shell scattering amplitudes.
In this context the graviphoton is defined through its vertex operator.

The string computation involves the physical scalars $z^A$ 
rather then the sections $X^I$. In order to rewrite the couplings
$F^{(g)}(X)$ we introduce the holomorphic sections $X^I(z)$ by
\be
X^I = m_{\mscr{Planck}} e^{\ft12 K(z,\ov{z}) } X^I(z) \;,
\ee
where $K(z,\ov{z})$ is the K\"ahler potential. Note that we
have restored the Planck mass in order to be able to do 
dimensional analysis later. We go to special coordinates $z^A$ by
imposing $X^0(z)=1, X^A(z) = z^A$.
Next we use that $F^{(g)}(X)$ is homogenous of degree $2-2g$ and define
\be
{\cal F}^{(g)}(z) = i [m_{\mscr{Planck}}]^{g-1} e^{-(1-g)K} F^{(g)}(X) \;.
\ee
In the language of chapter \ref{ChapterFourDSuGra} the quantity
${\cal F}^{(g)}(z)$ is a holomorphic section
of ${\cal L}^{2(1-g)}$, whereas $F^{(g)}$ is a covariantly holomorphic
section of ${\cal P}^{2(1-g)}$.

In order to display the dependence of $F^{(g)}(X)$ on the string coupling
$g_S$, we have to replace  the Planck mass by the string mass 
$m_{\mscr{String}} = m_{\mscr{Planck}} g_S$. This yields
\be
F^{(g)}(X) = -i m_{\mscr{String}}^{2-2g} g_S^{-2+2g} e^{(1-g)K} 
{\cal F}^{(g)}(z) \;.
\ee 
Next we use that the string coupling is given by the vacuum expectation
value of the dilaton and that in type II compactifications 
the dilaton sits 
in a hypermultiplet. The K\"ahler potential $K$ and the function
${\cal F}^{(g)}(z)$ depend only on the K\"ahler moduli and not
on hypermultiplet moduli like the dilaton. Therefore the factorization of the
moduli space into vector and hypermultiplet moduli has the far reaching
consequence that the dependence of the terms in the vector multiplet
part of the Lagrangian is very simple:
$F^{(g)}(X)$
depends on the string coupling only through the factor
$g_S^{2-2g}$ required by dimensional analysis. This term 
is precisely generated at the
$g$-loop level of type II perturbation theory. Supersymmetry leaves no
room for perturbative or non-perturbative corrections. In particular
the prepotential ${\cal F}^{(0)}(z)$
can be computed exactly at string tree level.
But quantities related to the hypermultiplet sector
can depend on the dilaton in a non-trivial way and are therefore 
subject to complicated perturbative and non-perturbative
quantum corrections. But this will not concern us here, since black 
hole solutions exclusively depend on quantities in the vector multiplet
sector.

In string theory we have a second type of corrections in addition
to loop corrections, namely $\alpha'$-corrections. At given loop order
these are additional stringy corrections to point particle behaviour,
which give rise to higher derivative terms in the effective action.
These terms are suppressed by additional powers of the string scale,
which can be either parametrized by the string mass $m_{\mscr{String}}$
or by the parameter $\alpha'$ which has dimension length squared,
$m_{\mscr{String}}^{-1} \sim \sqrt{\alpha'}$. One way to understand
$\alpha'$-corrections is to look at strings moving in non-trivial
background fields, in particular in a curved background geometry. 
In this case the worldsheet action of the string is a non-linear 
sigma-model, and therefore the two-dimensional worldsheet theory
has non-trivial quantum corrections. The role of the dimensionless
loop counting parameter is played by the curvature of the manifold measured 
in units of $\alpha'$. In the case of strings propagating in a
Calabi-Yau manifold the information about the size and curvature of the
manifold is encoded in the K\"ahler moduli. Since the K\"ahler moduli
sit in vector multiplets, all quantities in this sector have a non-trivial
dependence on the K\"ahler structure and are subject to perturbative
and non-perturbative $\alpha'$-corrections. The perturbative corrections are
given by world-sheet loops, whereas the non-perturbative corrections
are due to world-sheet instantons.

World-sheet instantons arise from  non-trivial embeddings of the string 
world-sheet into the Calabi-Yau manifold and produce
saddle points in the string path integral. More specifically the 
string world sheet is a genus $g$ Riemann surface and can be
holomorphically mapped onto two-cycles inside the Calabi-Yau space.
The mappings are classified by their degrees $d_1,\ldots, d_{h_{1,1}}$,
where the degree $d_i$ specifies how many times the world sheet is
wrapped around the $i$-th generator of the homology group 
$H_2(X,\mathbb{Z})$. These generators provide a basis in which every
two-cycle can be expanded (modulo homology). The number of 
genus $g$ instantons, i.e. the number of distinct holomorphic 
mappings of the genus $g$ world-sheet onto holomorphic
two-cycles with degrees $d_1,\ldots, d_{h_{1,1}}$ is denoted
by $n^{(g)}_{d_1, \ldots d_{h_{1,1}} }$. These numbers 'count' holomorphic 
genus $g$ curves of given homology and degrees in the Calabi-Yau 
manifold.\footnote{This is at least the intuitive interpretation of
these numbers. The mathematics behind it is more complicated.  
For example, holomorphic curves are not necessarily isolated
but can form continuous families, so that one needs to generalize
the notion of counting. We will not need to describe this here and
refer the interested reader to the literature, see for example 
\cite{BerCecOogVaf:1992}.}

Obviously it is a hopeless task to compute all the world sheet intstanton
corrections explicitly term by term. But here mirror symmetry 
enables one to find the full result. When switching from
type IIA theory to type IIB theory on the mirror manifold the vector
multiplet moduli are now complex structure moduli. But since the
world-sheet corrections are controlled by the K\"ahler moduli, all
quantities in the complex structure moduli space do not receive 
$\alpha'$-corrections and can be calculated exactly
at tree level in $\alpha'$.
In geometric terms this means that while the K\"ahler structure
of the Calabi-Yau manifold is modified when probed by a string
instead of a point particle, the complex structure is not.
One can now try to compute the functions 
${\cal F}^{(g)}(z)$ in the type IIB theory and then switch to
IIA variables by the mirror map that connects the two theories.
This is still a complicated problem, because one has 
to compute a string $g$-loop diagram in a Calabi-Yau background
geometry. In the case of the prepotential ${\cal F}^{(0)}(z)$ special 
geometry can be exploited. The holomorphic section $(X^I(z), F_J(X(z)))$ 
is related to the periods of the holomorphic $(3,0)$-form of
the Calabi-Yau manifold and the so-called Picard-Fuchs equations
can be used to compute it \cite{CanOssGrePar:1991,HosKleTheYau:1993}. 
The computation of the function ${\cal F}^{(1)}(z)$ 
\cite{BerCecOogVaf:1993/02} is related to
the supersymmetric index of \cite{CecFenIntVaf:1992}. The higher
functions ${\cal F}^{(g)}(z)$, $g>1$ satisfy a holomorphic anomaly 
equation and information about them can be obtained by
using various techniques in particular the topological twisting
of the string sigma model and the computation of special
(so-called topological) scattering 
amplitudes \cite{BerCecOogVaf:1992,AntGavNarTay:1993}.

Let us next have a look at concrete formulae for the functions
${\cal F}^{(g)}(z)$. We will use the 
type IIA description. Since the Calabi-Yau metric has to be positive
definite, the space of K\"ahler deformations has the structure of a
cone. We use the convention that the K\"ahler moduli sit in the
imaginary part of the complex field $z^A$. The variables are chosen
such that the K\"ahler cone is given by
\be
\mbox{Im} \; z^A > 0 \;.
\ee
The real parts of the $z^A$ contain the zero modes of the internal
part of the stringy $B$-field. The gauge symmetry associated with
this field translates into a Peccei-Quinn symmetry of
$\mbox{Re}\; z^A$.

We are now prepared to present the general form of the IIA 
prepotential \cite{CanOssGrePar:1991,HosKleTheYau:1993}:
\be
{\cal F}^{(0)}_{IIA} = - i\ft16 C_{ABC} z^A z^B z^C
- i\frac{\chi \zeta(3)}{2 (2 \pi)^3}
+ i\frac{1}{ (2  \pi)^3} \sum_{ \{d_i \} } n^{(0)}_{ \{d_i \}  }
\mbox{Li}_3 \left( \exp \left[ i \sum_A d_A z^A \right] \right) \;,
\ee
where $\{ d_i \} := \{ d_1, \ldots, d_{h_{1,1}} \}$.
In this formula $\mbox{Li}_3$ denotes the third polylogarithmic
function, see appendix \ref{AppPolylog}. 
The coefficients $C_{ABC}, \chi, n^{(0)}_{ \{ d_i \} }$
are topological data of the Calabi-Yau space, namely its triple
intersection numbers, the Euler number and the rational or 
genus zero world-sheet instanton numbers, respectively.
The first term arises at tree level in $\alpha'$, whereas the
second term is a loop correction. The third term encodes
the world-sheet instanton corrections that we discussed above.
The classical or large volume limit corresponds to taking
all K\"ahler moduli to be large, $\mbox{Im} \;z^A \rightarrow \infty$.
This corresponds to a region deep inside the K\"ahler cone, as opposed
to the boundaries $\mbox{Im} \; z^A =0$. Geometrically,  large K\"ahler moduli
mean that the sizes of the manifold and of all its two- and four cycles
are large (in units of $\alpha'$) and therefore curvature is small.
In the limit $\mbox{Im} \; z^A \rightarrow \infty$ the world-sheet instanton
corrections are exponentially small, whereas the classical cubic
term is very large. Since the leading classical piece contains the
triple intersection form, one can interpret the $\alpha'$-corrections
as a 'quantum deformation' of the standard triple intersection form. Thus the
formula for the prepotential summarizes in which way the geometry
seen by strings differs from classical geometry.\footnote{Though the
name quantum geometry is common for this deformation, one should keep
in mind that 'quantum' refers to world-sheet and not to space-time
properties. One might prefer to call this 'stringy geometry' as 
opposed to 'point particle geometry', i.e. the geometry seen by
point particles.}

Let us next display the structure of the $C^2$-coupling 
\cite{BerCecOogVaf:1993/02}:
\bea
{\cal F}^{(1)(\mscr{hol})}_{IIA}(z) &=& -i \sum_A z^A c_{2A} 
- \frac{1}{\pi} \sum_{ \{ d_i \} } \left\{
12 n^{(1)}_{d_1,\ldots} \log \left[ \widetilde{\eta} \left(
\exp \left[ i \sum_A d_A z^A \right] \right) \right] \nonumber \right.\\
 & & \left. + n^{(0)}_{ \{ d_i \} } \log \left[ 1 - \exp \left( i
\sum_A d_A z^A
\right) \right] \right\} \\
\nonumber
\eea
where $\widetilde{\eta}(q) = \prod_{m=1}^{\infty} (1 -q^m)$. 
Among the topological data that enter this time are the second Chern class
numbers $c_{2A}$, which are the expansion coefficients of the second
Chern class in a basis of $H^4(X,\mathbb{Z})$ that is dual to
the chosen basis of $H_2(X,\mathbb{Z})$. The corresponding term arises
at tree level in $\alpha'$ and is dominating in the large volume limit.
There are no $\alpha'$-loop corrections but world sheet instantons
of genus 0 and 1, called rational and elliptic instantons, respectively. 
Naively one would only expect elliptic instantons according to the
discussion given above. The appearence of rational instantons is
related to one of the many subtleties that we do not discuss 
explicitly here, namely
to the proper treatment of degenerate curves, where the 
so-called 'bubbling phenomenon' has to be taken into 
account \cite{BerCecOogVaf:1993/02}. 

Also note that we
put an additional label 'hol' on the function in order to notify that
we have only displayed the holomorphic part of ${\cal F}^{(1)}(z)$.
The function also has a non-holomorphic part
which is determined by the so-called holomorphic
anomaly equation \cite{BerCecOogVaf:1993/02}. 
From the point of view of the string world sheet 
the non-holomorphic contribution comes from a contact term and the
holomorphic anomaly equation is a modified superconformal Ward identity.
From the space-time point of view
the fact that the full coupling ${\cal F}^{(1)}(z)$ is 
non-holomorphic is due to non-trivial infrared 
physics \cite{AntGavNar:1992}. As we
saw in chapter \ref{ChapterFourDSuGra} the couplings occuring in the 
most general
local supersymmetric Lagrangian are necessarily holomorphic 
functions of the moduli. In quantum field theory the term
effective action refers to the generating functional of 1PI
Greens functions. 
If massless particles are present this effective
action is in general non-local, due to the fact that one integrates
out all modes, including the massless ones. 
This has the consequence
that the physical couplings of supersymmetric field theories depend
in a non-holomorphic way on the moduli, as discussed for 
gauge couplings in  \cite{ShiVai:1986,KapLou:1994} and
for gravitational couplings in \cite{AntGavNar:1992,AntGavNarTay:1993}.
In string perturbation theory one computes on-shell amplitudes, which
automatically incorporate the loops of massless particles. Therefore
the couplings extracted from such a computation are the physical,
non-holomorphic couplings. A local supersymmetric 
Lagrangian cannot properly account for the non-holomorphic part of
the coupling. The corresponding effective action is therefore
interpreted as a Wilsonian effective action, i.e. an effective 
action where only the massive modes above a certain infrared
cut-off have been integrated out. The corresponding holomorphic
couplings are called Wilsonian couplings. Note that
these remarks do not only apply to ${\cal N}=2$ compactifications but
to supersymmetric string effective actions in general,
see \cite{KapLou:1995} for an overview.

Finally we have to discuss the higher couplings ${\cal F}^{(g>1)}(z)$
\cite{BerCecOogVaf:1992,AntGavNarTay:1993,GopVaf:1998/09,GopVaf:1998/12}.
These functions do not have a contribution
at tree level in $\alpha'$ and the leading term is an $\alpha'$-loop 
correction which yields a term proportional to $\chi \zeta(3)$.
The non-perturbative terms involve genus $g$ world sheet instantons,
but there are additional subtleties such
as 'bubblings' which lead to further contributions.
Like the function ${\cal F}^{(1)}(z)$ all the higher functions 
receive non-holomorphic contributions, which are subject to 
a hierarchy of holomorphic anomaly equations.
In chapter \ref{ChapterStringBHs}
we will discuss the contribution of non-holomorphic
terms to the black hole entropy.

The derivation of both the macroscopic and microscopic black hole entropy
that we will discuss in the next chapter is performed in the large
volume limit. The most basic approximation consists of taking only the
terms which arise at tree level in $\alpha'$. In this case only the 
leading parts of ${\cal F}^{(0)}(z)$ and ${\cal F}^{(1)}(z)$ contribute,
\be
{\cal F}^{(0)}(z) = - i \ft16 C_{ABC} z^A z^B z^C \mbox{   and   }
{\cal F}^{(1)}(z) = - i c_{2A} z^A \;.
\ee

\section{M-Theory on a Calabi-Yau Threefold times $S^1$}

The most striking development in the discovery of string dualities
was that the strong coupling limit of ten-dimensional IIA string theory
is an eleven-dimensional theory, which at low energies is effectively
described by eleven-dimensional supergravity \cite{Wit:1995}. The full
theory behind this effective theory cannot be a perturbative supersymmetric
string
theory, because one is beyond the critical dimension $D=10$. The new
theory has been called eleven-dimensional M-theory, and subsequent 
work has led to the conclusion that all five consistent perturbative
string theories describe asymptotic expansions around special points
in the full moduli space of M-theory \cite{Pol:1998}.

In the duality between ten-dimensional IIA string theory and 
eleven-dimensional M-theory the ten-dimensional IIA string coupling
is related to the (geodesic) radius of the eleventh dimension by
\be
g_{IIA}^2 = \left( \frac{R_{11}}{L_{11}} \right)^3 \;,
\label{IIAcouplingMradius}
\ee
where $L_{11}$ is the eleven-dimensional Planck length, which is related to 
the Regge parameter $\alpha'$ of IIA string theory by
\be
L_{11}^3 = \alpha' R_{11} 
\ee
and to the eleven-dimensional gravitational coupling by
$\kappa_{11}^2 = L_{11}^9$.

As a consequence of equation (\ref{IIAcouplingMradius}) weak IIA coupling
corresponds to a small eleventh dimension, which is invisible in
IIA perturbation theory. Conversely large coupling corresponds to
the decompactification of the extra dimension and at sufficiently small
energies eleven-dimensional supergravity describes the strong coupling
behaviour of the IIA string. 

Upon compactification on a manifold $X$, IIA string theory on $X$ and
M-theory on $X\times S^1$ are 'on the same moduli'. This means that
they are part of one single theory and describe two different regimes,
namely weak ten-dimensional IIA coupling and strong ten-dimensional IIA
coupling, respectively.

We consider now the case that $X$ is a Calabi-Yau threefold. 
The compactification of eleven-dimensional supergravity on a
Calabi-Yau threefold \cite{CadCerDauFer:1995}
yields minimal five-dimensional supergravity
coupled to $N_V^{(5)}$ abelian vector and $N_H^{(5)}$ 
neutral hypermultiplets, where
\be
N_V^{(5)} = h_{1,1}-1 \;, \;\;\;
N_H^{(5)} = h_{2,1}+1 \;
\ee
and $h_{1,1}, h_{2,1}$ are Hodge numbers of the Calabi-Yau space.
By further compactification on a circle one gets four-dimensional
${\cal N}=2$ supergravity coupled to $N_V = h_{1,1}$ vector and
$N_H = h_{2,1} +1$ hypermultiplets, which is indeed the same
spectrum that one gets by compactification of type IIA string theory
on the same threefold.

The vector multiplet couplings are fully specified by the
triple intersection numbers $C_{IJK}$ \cite{GueSieTow:1984,GueSieTow:1985}.
The geometric structure behind the vector multiplet couplings
of five-dimensional supergravity is very special geometry \cite{deWvP:1992}.
In this language the particular feature compared to four-dimensional
${\cal N}=2$ supergravity is that the prepotential is purely cubic.
This has a nice interpretation in terms of the M-theory limit of
the IIA string \cite{Wit:1996}: In this case the limit of
strong ten-dimensional IIA coupling can be reinterpreted as the large
volume limit of the Calabi-Yau space. Here 'large' refers to measuring
the metric in stringy $\alpha'$-units and therefore is equivalent
to taking the limit $\alpha' \rightarrow 0$ in which all perturbative
and non-perturbative $\alpha'$-corrections vanish. Thus one is left with
the classical part of the prepotential.

When studying the theory at finite $S^1$ radius, then 
$\alpha'$-corrections
are present and correspond in the M-theory language to perturbative
and non-perturbative corrections involving the Kaluza-Klein modes
and solitons of M-theory. This has recently been used to obtain 
information about the four-dimensional vector multiplet couplings
${\cal F}^{(g)}(z)$ from the M-theory perspective 
\cite{GopVaf:1998/09,GopVaf:1998/12}.
In particular one-loop calculations involving Kaluza-Klein
modes of the five-dimensional theory and wrapped M2-branes
have been used to compute the leading parts of ${\cal F}^{(g)}(z)$
for all $g$. These computations confirm and extend known results and are 
sensitive to effects like the 'bubblings' we mentioned in the
last section.

Later on we will consider a contribution to the function
$F(X,\widehat{A})$ which takes the form $G(X^0, \widehat{A})$ and 
encodes subleading contributions from all genera $g$ in the 
large volume limit. For this function an integral representation
was derived in \cite{GopVaf:1998/09}
by a computation which is similar to the Schwinger calculation of
charged particle creation in an external field.
We will discuss how such a contribution modifies the black hole entropy.

\section{Heterotic String Theory on $K3 \times T^2$}

So far we constructed four-dimensional ${\cal N}=2$ string vacua
using theories with 32 supercharges. Alternatively one can start with
the heterotic or with the type I string, which only posess 16 supercharges.
Then one needs to compactify on a manifold which preserves $1/2$
of the supersymmetries of 
the ten-dimensional theory. The most simple choice, besides singular
spaces such as orbifolds, is $K3 \times T^2$, where $K3$ is a 
$K3$-surface and $T^2$ is the two-dimensional torus. A 
$K3$-surface is a complex K\"ahler surface with vanishing
first Chern class or, in other words, a Calabi-Yau twofold.
The holonomy group is $SU(2)$.
In contrast to Calabi-Yau threefolds the topology of such
surfaces are unique, but deformations of the complex structure
and of the (complexified) K\"ahler structure exist. 
We refer to \cite{Asp:1996,Asp:2000} for a review of string theory in
$K3$ backgrounds.

Heterotic and type I string theories have a gauge group,
$E_8 \times E_8$ or $SO(32)$, in ten dimensions. In order to obtain
consistent compactifications one has to switch on a non-trivial
gauge field configuration if the space-time geometry is curved.
In our case one has to choose an instanton configuration inside
the $K3$. The low-energy effective theory (at generic points in the
moduli space) only depends on the topological class of the gauge field,
i.e. on the instanton number (the second Chern class of the gauge bundle).

All three options, the heterotic string theories with gauge groups
$E_8 \times E_8$ or $SO(32)$ and the type I string theory
with gauge group $SO(32)$ are related by dualities. 
Whereas the two heterotic theories are related by T-duality after
compactification on a circle, the type I theory is $S$-dual
to the heterotic $SO(32)$ theory. When compactifyed on $K3 \times T^2$
all three theories are 'on the same moduli'. We will choose the
perspective of the heterotic $E_8 \times E_8$ theory. Moreover we
will restrict ourselves to so-called perturbative vacua, i.e.
to compactifications without additional $p$-branes in the
compact part of space-time. 

By compactification of the $E_8 \times E_8$ theory on $K3$
one obtains minimal, chiral six-dimensional ${\cal N}=(0,1)$
supergravity coupled to $N_T^{(6)}=1$ tensor, $N_V^{(6)}$
vector and $N_H^{(6)}$ hypermultiplets. The resulting models
have been studied in the context 
of F-theory 
\cite{MorVaf:1996/02,MorVaf:1996/03,LouSonTheYan:1996}.

The tensor multiplet
contains the six-dimensional dilaton. 
In order to obtain 
a consistent, in particular anomaly-free theory one has
to switch on $E_8 \times E_8$
instantons with instanton numbers $N_I^{(1)} + N_I^{(2)}=24$
in the $K3$. The gauge group depends on the 
instanton numbers $(N_I^{(1)}, N_I^{(2)})$ and on the position
in moduli space. In six dimensions a vector multiplet does not
contain scalars, so that the moduli sit in tensor and hypermultiplets.
For the perturbative vacua that we consider here the dilaton
is the only tensor multiplet modulus.
The hypermultiplet 
moduli space contains the moduli of the $K3$ and of the instantons.
At a generic positon in moduli space the gauge group and
massless spectrum are minimal. 
We assume to be at a generic point 
in the hypermultiplet moduli space. Then, the three models with
instanton numbers $(12,12)$, $(13,11)$ and $(14,10)$ have the same
massless spectrum: The gauge group is broken completely, 
$N_V^{(6)}=0$ and there are $N_H^{(6)}=244$ neutral 
hypermultiplets. If the instanton numbers are distributed more
asymmetrically, then the gauge group is not broken completely, so that
one has $N_V^{(6)} \not=0$.

By compactification on $K3 \times S^1$ one obtains
minimal five-dimensional supergravity coupled to 
$N_V^{(5)}=N_V^{(6)} + 2$ vector multiplets and 
$N_H^{(5)} = N_H^{(6)}$ hypermultiplets. One of the additional
vector multiplets is related to the usual Kaluza-Klein vector 
whereas the other is obtained by Hodge dualization of the
six-dimensional tensor multiplet. A five-dimensional vector multiplet
contains one real scalar. At generic points in the moduli space 
the gauge group is abelian and the hypermultiplets are neutral.
As in the case of five-dimensional M-theory compactifications the
vector multiplet couplings are described by very special geometry.

By compactification on $K3 \times T^2$ one obtains 
four-dimensional ${\cal N}=2$ supergravity coupled to
$N_V^{(4)}=N_V^{(6)}+3$ vector and $N_H^{(4)}=N_H^{(5)}$ hypermultiplets. 
Again we have gained one vector multiplet by Kaluza-Klein reduction.
The scalars in the vector multiplets are now complex.

This counting applies to the three models mentioned above,
where the six-dimensional gauge group is completely broken
for generic moduli. 
For the other models  with instanton numbers
$(15,9), (16,8), \ldots$ one has an unbroken non-abelian 
group $SU(3), SO(8), \ldots$
in six dimensions. After compactification one can use
the two versions of the Higgs mechanim available 
in four-dimensional ${\cal N}=2$ theories ('going to the
Coulomb- or Higgsbranch') that we briefly mentioned in chapter
\ref{ChapterGravity} to break the gauge group.
For generic moduli one is left with abelian vector and neutral
hypermultiplets. Thus the effective action is of the type we have studied.

In the case of the models with instanton numbers $(12,12)$,
$(13,11)$ and $(14,10)$ we obtain precisely three vector multiplets,
and therefore gauge group $U(1)^4$, where the extra $U(1)$ is
due to the graviphoton. These models are called the three
parameter models, because they have three complex vector
multiplet moduli.\footnote{Since all models are equivalent at
the classical level, one sometimes refers to them simply
as 'the' three parameter model. It turns out that at the 
non-perturbative level there are actually two inequivalent
models, see below and \cite{LouSonTheYan:1996}.} 
The number of neutral hypermultiplets is
$244$. Fortunately this sector will not concern us here.
The other models have larger abelian gauge groups, which
are relics of the six-dimensional gauge group. All models form
branches of one single moduli space.

By a slightly modified 
scheme of compactification one can construct another class of
models where an 
even smaller vector multiplet sector is possible \cite{KacVaf:1995}:
One first compactifies on $T^2$ to $D=8$
and freezes one modulus by imposing that the K\"ahler modulus
$T$ and the complex structure modulus $U$ are equal, $T=U$, 
so that the model has an
enhanced gauge group $SU(2)$ related to the $T^2$ in 
$D=8$.\footnote{We are using the standard parametrization
of the moduli of a torus. See for example \cite{CarLueMoh:1995}.}
Now one compactifies on $K3$ and puts instantons both in
the $E_8 \times E_8$ and in the $SU(2)$. When distributing
the instanton numbers sufficiently symmetric and going to
generic moduli one arrives at the
so-called two-paramter model with generic gauge group $U(1)^3$.

In contradistinction to type II compactifications the dilaton
sits in one of the vector multiplets. Therefore the prepotential
and all the higher couplings get perturbative and non-perturbative
corrections. We will proceed step by step and start with the
tree level prepotential.

\subsection{The Tree Level Prepotential}

The tree level prepotential takes a universal form that
only depends on the number of vector moduli \cite{FervP:1989}.
In order to describe the models in terms of the physical 
scalars we introduce
\be
{\cal F}^{(0)}(z) = i m_{\mscr{Planck}}^{-1} e^{-K} F^{(0)}( X(z)) \,,
\ee
where $K(z,\ov{z})$ is the K\"ahler potential and 
$X^I(z)$ are the holomorphic sections. In the following we set
$m_{\mscr{Planck}}=1$ and drop the label $(0)$ on the
prepotential. We introduce special coordinates 
$X^0(z)=1$ and $X^A(z)=z^A$. The heterotic tree level
potential is purely cubic, ${\cal F}(z) = i D_{ABC} z^A z^B z^C$,
with a special form of $D_{ABC}$ that we will discuss in a
minute. First we remark that the standard convention in heterotic
models is to parametrize the moduli by $S^A = -i z^A$. Thus it
is the imaginary part rather than the real part that has
a Peccei-Quinn symmetry.\footnote{In the most simple cases
the $z^A$ parametrize the upper complex half plane, whereas the
$S^A$ paramatrize the right half plane, $\mbox{Re}\;S^A > 0$.}
The prepotential takes the form ${\cal F} = D_{ABC} S^A S^B S^C$.
In order to understand the special form that $D_{ABC}$ takes
in heterotic tree level models, let us look at the 
physical and geometrical interpretation of the three moduli
of the three parameter model.

One of the moduli contains the 
four-dimensional dilaton
$\phi^{(4)}$ and the universal axion $a$, which is obtained
by dualizing the universal $B_{\m \n}$-field. At the supermultiplet
level
the dilaton originally resides in the tensor multiplet, which,
below six dimensions, can be dualized into a vector multiplet.
One combines the dilaton and axion 
into a four-dimensional complex dilaton
\be
S^1= S = 4 \pi e^{-2 \phi^{(4)}} + i a = \frac{4\pi}{g_S^2} - i 
\frac{\theta}{2 \pi}\;,
\ee
where $g_S$ is the four-dimensional heterotic tree level coupling and
$\theta$ is the $\theta$-angle associated with the universal axion $a$. 
The other two universal moduli are related to 
the two-torus. They are the complexified
K\"ahler modulus $S^2 = T$ and the complex structure modulus
$S^3=U$ of the torus. The standard notation $S,T,U$
motivates the synonym $STU$ model for the three parameter model. 
Models with more vector multiplets have additional moduli
$S^{3+i}=V^i$, which parametrize Wilson lines on the torus.

We now come back to the prepotential. The tree level
part must be linear in $S$ in order to have the right dependence
on the string coupling. The quadratic
polynomial depending on the geometric moduli
$T^a=T,U,V^i$ is unique:
\be
{\cal F}(S,T^a)  = - S T^a \eta_{ab} T^b = -S \left(TU - \sum_i V^i V^i\right)\;.
\ee
Note that 
\be
(\eta_{ab}) = \left( \begin{array}{ccc}
0 & \frac{1}{2} & 0 \\
\frac{1}{2} & 0 & 0 \\
0 & 0 & -\delta_{ij} \\
\end{array} \right) \;.
\label{EtaMatrix}
\ee

One can now compute the K\"ahler potential, with the 
result that the moduli space locally takes the form
\be
{\cal M } \sim \left.\frac{SU(1,1)}{U(1)} \right|_S \times 
\left. \frac{SO(2, N_V -1)}{SO(2) \times SO(N_V-1)} \right|_{T,U,V_i} \;.
\ee
The first factor is related to the dilaton, which at tree level
does not mix with the moduli. For the $STU$-model one gets
\be
{\cal M } \sim \left. \frac{SU(1,1)}{U(1)} \right|^3_{S,T,U}
\ee
by using the local isomorphism $SO(2,2) \sim SU(1,1) \times SU(1,1)$.

The global structure of the moduli space is determined by T-duality.
We refer to \cite{GivPorRab:1994} for a review. 
In compactifications on $S^1$ T-duality relates the radius
to its inverse in string units. For torus compactifications
this generalizes to a more complicated
discrete group, which in particular contains large diffeomorphisms
of the tours, the inversion of radii and discrete axion-like shift 
symmetries of the imaginary parts of the moduli. In heterotic
compactifications the group also contains transformations acting on
the Wilson lines. In our case 
the discrete group is $SO(2,N_V-1,\mathbb{Z})$. In ${\cal N}=2$ 
compactifications T-dualities are realized as
a subset of the symplectic transformations
\cite{CerDauFervP:1995,dWKapLouLue:1995}. In contradistinction
to generic symplectic transformations, T-dualities are true
symmetries of the theory, at least at the tree and perturbative 
level. Therefore physical observables have to be 
T-duality invariant. In order to display the symplectic action
of T-duality, we have to describe the theory in terms of 
symplectic sections. For simplicity we will consider
the $STU$ model. In this case the prepotential is
\be
F(X(z)) = -\frac{X^1(z) X^2(z) X^3(z)}{X^0(z)} \leftrightarrow 
{\cal F}(S,T,U) = -STU
\label{PrepHetTree}
\ee
and therefore the holomorphic section, written in special
coordinates is
\be
(X^I,F_J) = (1,iS,iT,iU,-iSTU,TU,SU,ST) \;.
\ee
It turns out that this parametrization is not the
natural one from the point of view of the heterotic string.
When computing the associated gauge couplings in the perturbative
limit $S \rightarrow \infty$,  the coupling of the gauge field
in the dilaton vector multiplet blows up, whereas all other 
gauge couplings go to zero.
The natural behaviour, where all gauge couplings
are small for $S\rightarrow \infty$, is obtained when going to another
section by a symplectic reparametrization,
\be
(P^I, Q_J) = (1, -TU, iT, iU, -iSTU, iS, SU, ST) \;.
\ee
Note that $Q_J$ is not the gradient of a prepotential $F(P)$. This
is the standard example for a section without 
prepotential \cite{CerDauFervP:1995}.
Since the two parametrizations are related to one another
in a simple way 
we find it convenient to refer to (\ref{PrepHetTree}) as 
'the prepotential'. The above symplectic transformation amounts
to an electric - magnetic duality in the $U(1)$ associated with
$X^1,F_1$. The new symplectic vector of 
electric and magnetic charges is denoted by 
\be
(M^I, N_J) = (q_0, -p^1, q_2, q_3, p^0, q_1, p^2, p^3) \;,
\ee
where $q^I, p_J$ are the electric and magnetic charges 
of the old section.

We now review how T-duality is realized in the $STU$ model. 
The T-duality group is
\be
O(2,2,\mathbb{Z}) = (SL(2,\mathbb{Z})_T \otimes SL(2,\mathbb{Z})_U)
\times \mathbb{Z}_2^{T-U} \;.
\ee
The two $SL(2,\mathbb{Z})$ groups act by fractional linear transformations
\be
\begin{array}{ll}
T \rightarrow \frac{a T - ib}{icT +d}\;, \;\;\; & U \rightarrow U \;, \\
T \rightarrow T  \;,\;\; & U \rightarrow \frac{a' U - ib'}{ic'U +d'} \;.\\
\end{array}
\ee
where
\be
\left( \begin{array}{cc}
a&b\\c&d\\
\end{array} \right) \;,\;\;\;
\left( \begin{array}{cc}
a'&b'\\c'&d'\\
\end{array} \right)
\in SL(2,\mathbb{Z})    \;.
\ee
The $\mathbb{Z}_2$ group exchanges the K\"ahler and complex structure
modulus
\be
T \leftrightarrow U \;.
\ee
This is mirror symmetry for $T^2$.
All these transformations leave the dilaton invariant, $S \rightarrow S$.

The above transformations act as symplectic transformations
on the sections. In the heterotic basis $(P^I, Q_J)$ 
all T-duality transformations take the particular form
\be
\Gamma_{\mscr{class}} = \left( \begin{array}{cc}
{\bf U} & \mathbb{O} \\
\mathbb{O} & {\bf U}^{T,-1} \\
\end{array} \right) \;,
\ee
where ${\bf U} \in O(2,2,\mathbb{Z})$.
Transformations of this special structure
leave the Lagrangian
invariant.\footnote{This does not only hold for T-duality
transformations but for all symplectic
transformations with ${\bf U} \in GL(4,\mathbb{Z})$.} 
This is related to the fact that the prepotential does
not have non-trivial monodromies at the classical level.

The prepotential and section of the classical $STU$-model
obviously have a higher symmetry. The corresponding 
triality group 
\be
(SL(2,\mathbb{Z})_T \otimes SL(2,\mathbb{Z})_U \otimes SL(2,\mathbb{Z})_S)
\times \mathbb{Z}_2^{T-U} \times \mathbb{Z}_2^{S-T} \times 
\mathbb{Z}_2^{S-U} 
\ee
contains in addition to the full T-duality group also 
S-duality
\be
S \rightarrow \frac{a S - ib}{icS +d} \;,\;\;
T \rightarrow T \;,\;\; U \rightarrow U
\ee
and the exchange symmetries
\be
S \leftrightarrow T, \;\;\; S \leftrightarrow U \;.
\ee
S-duality acts on the heterotic section by 
\be
\Gamma = \left( \begin{array}{cc}
a \mathbb{I} & b \mathbb{H} \\
c \mathbb{H} & d \mathbb{I} \\
\end{array} \right) \;,
\ee
where $\mathbb{H} = \eta \oplus \eta$ and
\be
\eta = \left( \begin{array}{cc}
0&1 \\ 1&0 \\
\end{array} \right) \;.
\ee
Such symplectic transformations are not invariances of the 
Lagrangian. In particular they map small to large string coupling.
The symplectic matrices for the two exchange symmetries involving
$S$ are of the same type. Note that it is not guaranteed or even likely
that such transformations are symmetries of the full non-perturbative
theory. For specific models it is known that 
discrete non-perturbative $S-T$ exchange symmetries exist
\cite{KleLerMay:1995,AntPar:1995,CarCurLueMoh:1996}. But such
symmetries have to be established case by case using heterotic
- type II duality (to be reviewed below). In generic 
${\cal N}=2$ models one does not expect S-duality
at the non-perturbative level. Nevertheless there are several situations
where the tree-level formulae displayed above are useful.
First we can describe a
subsector of a heterotic ${\cal N}=4$ compactification.
In this case one has evidence for 
unbroken S-duality at the non-perturbative
level \cite{HulTow:1994}. 
Moreover the full triality group of the $STU$-model is related
to heterotic - IIA - IIB triality present in a subsector of 
four-dimensional ${\cal N}=4$ compactifications \cite{DufLiuRah:1995}.
Second there are so called finite or ${\cal N}=4$ like
${\cal N}=2$ models, like the FHSV model
\cite{FerHarStrVaf:1995}, which exhibit S-duality. 
Finally the triality group simply is a symmetry of the tree level 
$STU$-model and therefore physical observables like the black hole entropy
should be invariant when computed
in the classical approximation. We will see that this
is indeed the case.

At special
points in the vector multiplet moduli space one obtains
non-abelian gauge groups. In the $STU$-model a $U(1)^2$
part of the abelian gauge group is enhanced to $SU(2) \times U(1)$
on the subspace $T=U$ whereas it is enhanced to $SU(2)^2$ and
$SU(3)$ at $T=U=1$ and $T=U=e^{i \pi/6}$, respectively. Similarly
one obtains larger non-abelian gauge groups in the other models
by switching off the Wilson lines.
The points of
enhanced, non-abelian gauge symmetry are fixed points under some
of the T-duality transformations. The fixed point transformations
can be identified with the Weyl group of the non-abelian group.
This indicates that T-duality is a discrete remnant of a local
gauge symmetry of string theory.

\subsection{The Perturbative Heterotic Prepotential}

Let us next review the perturbative corrections to the
prepotential 
\cite{CerDauFervP:1995,dWKapLouLue:1995,AntFerGavNarTay:1995,HarMoo:1995}.
The one loop contribution $h(T^a)$
to the prepotential is independent
of $g_S$ and therefore it can only depend on the (geometric)
moduli $T^a$ but not on the dilaton $S$:
\be
{\cal F}_{\mscr{1 loop}}(S, T^a) = - S T^a T^b \eta_{ab} + h(T^a) \;.
\ee
The function $h(T^a)$ can be computed in string
perturbation theory and takes the following form \cite{HarMoo:1995}:
\be
h(T^a) = p(T^a) - c - d \sum_{n_a \in \Gamma} 
c(n_a) \mbox{Li}_3 \left(   e^{-n_a T^a} \right) \;.
\ee
The first term $p(T^a)$ is a cubic polynomial in $T^a$ which
arises at $\alpha'$-tree level. In the case of the $STU$-model 
the combined tree and one loop cubic part of the prepotential
is
\be
{\cal F}(S,T,U) = -STU - \ft13 U^3 \;.
\ee
To be precise this is the expression in the region of moduli
space where $\mbox{Re} \; T > \mbox{Re}\; U$. This specification
is necessary since it turns out that the full one-loop function
has branch cuts. In the  other 'Weyl chamber',
where $\mbox{Re}\; U > \mbox{Re} \; T$, the cubic part
is ${\cal F}(S,T,U) = -STU - \ft13 T^3$.

The second term $c$ is an $\alpha'$-loop correction. 
It is a constant and in the $STU$-model it is 
proportional to $c(0) \zeta(3)$, 
where $c(0)$ is defined below in (\ref{Defqcoeff}).
The third term represents non-perturbative contributions in $\alpha'$.
At this point we have to explain the origin of the $\alpha'$-corrections.
Above we argued that $\alpha'$-corrections are related to
K\"ahler moduli and we also emphasized that the vector and
hypermultiplet moduli spaces factorize. Non-trivial $\alpha'$
corrections are to be expected from the K\"ahler moduli of the
$K3$, because this space has non-trivial curvature. But these
moduli sit in hypermultiplets, so how can they contribute to the
vector multiplet sector? The reason is as follows \cite{HarMoo:1995}: 
The loop corrections
to the prepotential depend on BPS states, only. But in ${\cal N}=2$
theories the number of BPS states is not conserved in moduli space.
As we explained in chapter \ref{ChapterGravity}
short vector multiplets and hypermultiplets,
which both are BPS multiplets, can combine into long vector multiplets,
which are not BPS mulitplets. When moving through moduli space the number
of BPS states changes in a very complicated, 'chaotic' manner.
But the difference of the number of short vector and hypermultiplets
is conserved and it turns out that $h(T^a)$ only depends
on this quantity. Moreover the correction does not explicitly depend
on the $K3$-moduli and can be computed in an orbifold limit of the $K3$.
Let us now investigate the third
term of the prepotential more closely: $d$ is an overall constant. The 
sum runs over a model dependent set of integers denoted by $\Gamma$ and the
coefficients $c(n_a)$ encode the corrections. The structure
behind these coefficients is an algebraic one: One can associate
an infinite dimensional Lie algebra, called the BPS algebra
with the BPS states. In particular cases this algebra is
a generalized Kac Moody algebra in the sense of Borcherds 
\cite{HarMoo:1995}.
The set $\Gamma$ parametrizes the positive roots of the BPS algebra.
As we already remarked above the explicit expressions for the
prepotential refer to subregions of the moduli space such
as $\mbox{Re} \; T > \mbox{Re}\; U$. These regions define
the Weyl chambers of the BPS algebra.
Furthermore, there is a deep 
relation between the BPS algebra and automorphic forms
of the T-duality group. Since T-duality is a symmetry of string
theory to all orders in perturbation theory, the perturbative
prepotential must transform in an appropriate way. This is
encoded in the above formula by the fact that the coefficients
$c(n_a)$, where $n_a$ runs over the positive roots of the
BPS algebra are expansion coefficients of automorphic forms.
To give an explicit example we consider the $STU$-model and
display the non-cubic part (thus including the constant part) 
\cite{HarMoo:1995}:
\be
{\cal F} = \cdots - \frac{1}{(2 \pi)^4} \left(
\sum_{k,l \geq 0} c(kl) \mbox{Li}_3 \left( e^{-2 \pi (kT + lU)}
\right) + \mbox{Li}_3 \left( e^{- 2 \pi(T-U)} \right) \right) \;.
\ee
In this case the coefficients $c(n)$ are defined by
\be
\frac{E_4 E_6}{\eta^{24}} = \sum_{n \geq -1} c(n) q^n 
= \frac{1}{q} - 240 - 141444 q - 8529280 q^2 - \cdots \;,
\label{Defqcoeff}
\ee
where $E_4,E_6$ are normalized Eisenstein series 
and $\eta$ is the Dedekind $\eta$-function, see appendix 
\ref{AppModularGeometry}.
In the above expression for the prepotential we separated
the part that describes the behaviour at $T=U$, which is a branch
locus of the function. The branch cut is related to the presence of
two extra charged massless vector multiplets which enhance one of
the $U(1)$ gauge groups to $SU(2)$. The gauge couplings calculated
from the prepotential contain the corresponding threshold 
correction. Near the branch cut the threshold correction
diverges logarithmically  like $\log(T-U)$, because $T-U$ is
the associated Higgs field. This form is correct up to finite
terms, which are needed to make the threshold correction compatible
with T-duality. The covariant form of the threshold correction is
$\log( j(iT) - j(iU))$, where $j(z)$ is the modular invariant
$j$-function. For generic $T \simeq U$ the expression
$j(iT) - j(iU)$ behaves like $T-U$. There are, however, the
special points $T=U=1$ and $T=U=e^{i\pi/6}$ in the moduli space, where
one finds the higher rank non-abelian gauge groups $SU(2)^2$
and $SU(3)$. Instead of two extra vector multiplets one
gets four and six extra multiplets, respectively. The behaviour
of the $j$ function and of its derivatives at $z=1,e^{i\pi /6}$ 
precisely captures this. Namely $j(iT) - j(iU) \simeq (T-U)^2$
for $T\simeq U \simeq 1$, whereas $j(iT) - j(iU) \simeq (T-U)^3$
for $T\simeq U \simeq e^{i \pi /6}$. Therefore the coefficients
of the threshold corrections $\log( j(iT) - j(iU) )$ 
have relative weights $1:2:3$, reflecting the numbers of extra
states at the threshold \cite{CarLueMoh:1995}.

In the above example we saw how a particular T-duality
transformation acts on the prepotential as a monodromy
transformation. More generally the whole T-duality group
now acts as a group of non-trivial monodromies.
It is convenient to consider the 
action of the monodromy transformations 
on the symplectic section instead of looking
at the action on the prepotential. In the 
heterotic basis for the section this takes the form
\be
\Gamma_{\mscr{pert}} = \left( \begin{array}{cc}
{\bf U} & \mathbb{O} \\
{\bf U}^{T,-1} \Lambda & {\bf U}^{T,-1} \\
\end{array} \right) \;,
\label{PertMonodr}
\ee
where ${\bf U}$ is the $O(2,N_V-1,\mathbb{Z})$ matrix of classical
T-duality wheras the symmetric matrix $\Lambda$ encodes the
perturbative modification. Symplectic transformations 
of the type (\ref{PertMonodr}) are called 'perturbative
transformations'. They have the special property that the
Lagrangian is invariant up to a total derivative.\footnote{
This is still true when ${\bf U}$ is not a T-duality transformation,
but a general matrix ${\bf U} \in GL(N_V+1,\mathbb{Z})$.}
This shows explicitly that T-duality transformations are
true symmetries at the perturbative level.  

Perturbative symplectic transformations such as T-dualities
have the property that the upper components $P^I$ of the
symplectic section still transform like under the corresponding
classical transformation, whereas the lower components 
$Q_J$ have a modified transformation law. This implies
that the geometric moduli $T^a=T,U,V^i$ transform
according to the classical T-duality rules, whereas the
conjugated variables $F_I$, $I\not=1$ transform in a modified
way. This precisely reflects that the prepotential now has
non-trivial monodromy properties. The group structure
of the perturbative monodromy group was analysed in
\cite{AntFerGavNarTay:1995}, with the result that the 
reflections with respect to the special loci are promoted to 
braidings. This precisely accounts for the non-trivial monodromies
around the branch loci.

The dilaton $S$, which was
invariant under classical T-duality sits in a lower component
of the heterotic section. Therefore its transformation behaviour
is modified, and  $S$ transforms non-trivially under T-duality
at the quantum level. Moreover the
metric of the moduli space does not factorize any more
into a direct product. 
The K\"ahler potential now takes the form
\be
K = - \log[ (S + \ov{S} ) + V_{GS} (T^i, \ov{T}^i) ] 
- \log[ (T^i + \ov{T}^i )(T^j + \ov{T}^j) \eta_{ij} ] \;,
\ee
where $V_{GS}(T^i, \ov{T}^i)$ is the so-called Green-Schwarz term
\be
V_{GS}(T^i, \ov{T}^i) = \frac{  2 (h + \ov{h}) - (T^k + \ov{T}^k)
( \der_{T^k} h + \der_{\ov{T}^k} \ov{h})}
{(T^i + \ov{T}^i )(T^j + \ov{T}^j) \eta_{ij}} \;,
\ee
which encodes the dilaton - moduli mixing \cite{dWKapLouLue:1995}.
At the tree level the dilaton was connected to the string coupling
in a simple way. At the loop level this relation has to be modified,
because the string coupling has to be T-duality invariant, whereas
the dilaton is not. The correct, T-duality invariant, perturbative
string coupling can be defined, by using the fact that the Green-Schwarz
term precisely compensates the non-invariance of $S$:
\be
\frac{4 \pi}{g^2_{\mscr{pert}}} = \ft12 ( S + \ov{S} + V_{GS}(T^i,
\ov{T}^i ) ) \;.
\ee
Since the new perturbative coupling is related to the special coordinates
$T^i$ in a non-linear way,  the simple
relation to the ${\cal N}=2$ supermultiplet structure is lost.
This is the price to be payed for having an invariant
coupling.\footnote{In the literature one often defines
an invariant dilaton, which contains both the perturbative 
string coupling and the $\theta$ angle. Note that there are
several variants, called the 'invariant dilaton' or the 
'quasi-invariant dilaton', which differ by conventional choices.}

In the above paragraphs we already used the expression 'perturbative'
synonymous to 'one-loop'. This is correct because there
are no higher loop corrections to the prepotential:
\be
{\cal F}_{\mscr{pert}}(S,T^a) = {\cal F}_{\mscr{1 loop}}(S,T^a) \;.
\ee
The underlying non-renormalization theorem is easily understood
in terms of string theory and of the holomorphicity properties
of the ${\cal N}=2$ action \cite{dWKapLouLue:1995}. 
The string coupling is inversely
related to the real part of the dilaton, up to a moduli dependent
correction, the Green-Schwarz term. The $g$-loop contribution
to the Lagrangian is proportional to
$g_S^{-2 + 2g}\sim (\mbox{Re}\;S)^{1-g}$. 
In string theory the dilaton $S$ has a continuous Peccei-Quinn 
symmetry, which is not broken in perturbation theory. 
But ${\cal N}=2$ supersymmetry implies that 
the action can be derived from a prepotential, which depends
holomorphically on special coordinates such as $S$. 
Higher loop corrections to the prepotential would either violate
the Peccei-Quinn symmetry, through negative powers of $S$ in the
prepotential, or violate holomorphicity, by a prepotential that
depends on $\mbox{Re}\;S$ rather than on $S$. Therefore there cannot
be higher loop contributions to the prepotential, but only 
tree and one loop contributions, which are compatible with both
Peccei-Quinn symmetry and holomorphicity.

\subsection{The Non-Perturbative Heterotic Prepotential}

At the non-perturbative level the prepotential receives further 
corrections by space-time instantons.
Such contributions depend on
the dilaton in the form $e^{-S}$:
\be
{\cal F}(S,T^a) = {\cal F}_{\mscr{pert}}(S,T^a) + f^{NP}(e^{-S},T^a) \;.
\ee
In string theory space-time instantons result from 
Euclidean $p$-branes, which are wrapped on $p+1$ cycles
of the compactification manifold. In the heterotic
string theory the instantons result from wrapping the heterotic
fivebrane around $K3 \times T^2$. Obviously such instanton corrections
cannot exist in $D>4$ dimensions. This fits with the observation
that the prepotential of minimal five-dimensional supergravity
is restricted to be purely cubic. Supersymmetry does not leave
room for non-perturbative corrections and higher order
$\alpha'$ corrections. The only non-perturbative
effects in five dimensions 
are discontinuous changes of the cubic coefficients
$C_{IJK}$ at the boundaries of the Weyl chambers. 
This has been shown explicitly by considering the decompactification
limit of the heterotic string \cite{AntFerTay:1995}. 
In the M-theory language the discontinuous changes of the triple
intersection numbers $C_{IJK}$ arise from flop transitions at 
boundaries of the K\"ahler cone \cite{Wit:1996}.

There is at time no direct way to compute
the instanton corrections to the four-dimensional heterotic prepotential.
One can, however, compute them indirectly
by using the duality between heterotic string theory on 
$K3 \times T^2$ and type IIA string theory on a Calabi-Yau 
threefold \cite{KacVaf:1995,FerHarStrVaf:1995}.
This duality can be motivated taking the six-dimensional duality
between the heterotic string on $T^4$ and the IIA string on $K3$
and expanding it adiabatically over
a holomorphic two-sphere $P^1$ \cite{VafWit:1995}. 
Globally the resulting geometries are taken to be fibrations rather
than direct products. This way the
compactification manifolds become $K3 \times T^2$ on the heterotic
side and a Calabi-Yau threefold on the type IIA side. The 
Calabi-Yau space has the special structure of a $K3$-fibration,
which means that it looks, locally and around a generic point, like
a product of $P^1$ and $K3$.

One can also directly show in four dimensions that the
Calabi-Yau space of a IIA compactification
must be a $K3$ fibration in order to have a 
perturbative heterotic dual theory \cite{AspLou:1995}. 
The basic observation is that the cubic part of the 
heterotic prepotential takes the special form
\be
{\cal F}^{(\mscr{het})}(S,T^a) 
= - S T^a T^b \eta_{ab} - D_{abc} T^a T^b T^c \;.
\ee
Comparing this to the cubic part of the IIA 
prepotential\footnote{The relative factor $i$ is related to
the different conventions for heterotic and type IIA moduli,
that we explained above.}
\be
{\cal F}^{(\mscr{IIA})} = i \ft16 C_{ABC} z^A z^B z^C \;,
\ee
where $C_{ABC}$ are the triple intersection numbers it is obvious
that the intersection form must take a special form and that
the dilaton plays a distinguished role. It turns out that
the heterotic dilaton corresponds to the IIA K\"ahler modulus that measures the
size of the $P^1$ basis of the $K3$ fibration.
Since in the IIA theory all vector multiplet moduli are 
K\"ahler moduli, whereas the IIA dilaton sits in a hypermultiplet
one can now compute the exact prepotential at tree level in
IIA perturbation theory. The heterotic - IIA duality was
therefore called  a 'second quantized mirror map' in \cite{FerHarStrVaf:1995}.

The weak coupling limit of the heterotic string, where 
heterotic perturbation theory is valid, corresponds to a
particular large volume limit in the IIA moduli space, namely the limit
of a large $P^1$. In order to ensure that the expansion
for the IIA prepotential converges the other moduli have to be
in the interior of the K\"ahler cone. Thus the perturbative 
heterotic limit corresponds to the IIA large volume limit.
In this
limit the prepotentials (and the higher derivative couplings)
can be compared for concrete dual pairs. The particular
family of heterotic theories with instanton numbers
$(12 +n, 12 -n)$ is dual to a particular family of Calabi-Yau
threefolds which can be constructed using methods of toric
geometry \cite{MorVaf:1996/02,MorVaf:1996/03,LouSonTheYan:1996}. 
These Calabi-Yau spaces are simultanously 
$K3$-fibrations over $P^1$ and elliptic fibrations over the 
Hirzebruch surfaces $\mathbb{F}_n$. We remarked earlier that
the three heterotic models with $n=0,1,2$ had the same
cubic part of the prepotential and therefore were
equivalent in the perturbative limit. At the non-perturbative level
the models with $n=0,2$ turn out to live in the same moduli space,
whereas the model with $n=1$ is a separate model \cite{LouSonTheYan:1996}.

One of the best and most studied 
examples for testing the duality is provided
by the $(14,10)$ model 
\cite{KacVaf:1995,KleLerMay:1995,KacKleLerMayVaf:1995}.
The dual Calabi-Yau
space to the $(14,10)$ model is a $K3$-fibration which 
is also an elliptic fibration over the Hirzebruch surface
$\mathbb{F}_2$. The model can be realized as a degree 24
hypersurface in weighted projective space with weights
$(1,1,2,8,12)$. As the most simple test of the conjectured duality
one might note that this Calabi-Yau space has
$h_{1,1}=3$ and $h_{2,1}=243$, which, taking into account
the universal hypermultiplet yields $N_V=3$ vector and
$N_H=244$ hypermultiplets. The triple intersection form,
in a suitable parametrization, yields the same cubic term
in the prepotential as the heterotic one. The constant term
in the prepotential is, on the IIA side proportional to 
$\zeta(3)$ times the Euler number, which is $2(h_{1,1}-h_{2,1})=480$.
This term also agrees with its heterotic counterpart.
Moreover one can compare the next expansion
coefficients of the heterotic and the IIA prepotential and finds that they 
agree.

Taking the duality for granted, the IIA tree level prepotential 
encodes the full non-perturbative physics of the heterotic theory.
The resulting physics makes perfect sense and is a generalization
of the Seiberg-Witten description \cite{SeiWit:1994}
of ${\cal N}=2$ Super-Yang-Mills
theory. Conversely one can recover the Seiberg-Witten prepotential
of super Yang-Mills theory by decoupling gravity and the massive
string modes \cite{KacKleLerMayVaf:1995}. Let us briefly indicate
the picture that heterotic - type II duality provides for the
dynamics of the heterotic string.
In the full non-perturbative theory the structure and physical
interpretation of the special loci in moduli space are different from
those of 
the perturbative theory. In particular the monodromy matrices 
corresponding to the singular loci now take the form
\be
\Gamma_{\mscr{nonpert}} = \left( \begin{array}{cc}
{\bf U} & {\bf Z} \\ {\bf W} & {\bf V} \\
\end{array} \right) \;,
\ee
which does not correspond to a symmetry of the perturbative
Lagrangian.  The physics associated to special loci in moduli
space can be understood in terms
of a dual local Lagrangian. To find the dual description one has to 
perform  a symplectic transformation
such that the particular monodromy one is interested in takes
the form of a perturbative monodromy. This amounts to an electric - magnetic
duality transformation, such that the new Lagrangian is weakly coupled
in the vicinity of the special locus.
One example is provided by the locus $T=U$, which in heterotic
perturbation theory corresponds to non-abelian gauge symmetry enhancement.
In the full theory the special locus is modified and 
the associated monodromies are of the non-perturbative type
from the point of view of the heterotic Lagrangian. 
The physics can be best understood by going to the type IIB
description by mirror symmetry \cite{Str:1995}. Then the special locus 
is part of the conifold locus, where the complex structure of
the mirror Calabi-Yau degenerates. This degeneration corresponds
to a vanishing three-cycle. The IIB theory has solitonic 
D3-branes, which can be wrapped holomorphically on the
three-cycle to give pointlike BPS states from the four-dimensional
point of view. As long as the three-cycle is large the BPS state
is very heavy and corresponds to an extreme black hole of 
precisely the type we discuss in this paper. If, however,
the three-cycle shrinks,
the BPS state becomes very light and should be treated like an
elementary particle. As usual for BPS black holes the correponding
multiplet is a hypermultiplet and the charge carried by the
state is magnetic or dyonic in terms of the heterotic basis for
the charges. Thus, instead of massless charged vector multiplets
one gets massless magnetic and dyonic hypermultiplets. This is
obviously the generalization of the Seiberg-Witten solution of
${\cal N}=2$ $SU(2)$ super Yang-Mills theory to the case of
local supersymmetry. There are various extensions of this 
to more complicated situations, which all have relations to field theory. 
We refer to \cite{Gre:1997,Kle:1997,May:1998} for review and references.

We already mentioned that 
the non-perturbative corrections change the topology of the moduli
space. This results in what is called non-perturbative breaking of
T-duality in \cite{AspPle:1999,Asp:2000}. 
To understand what this means we recall that the
original moduli space was the quotient space of a 
covering 'Teichm\"uller' space by the T-duality group. Our intuitive
geometric notions are tied to the Teichm\"uller space:
We imagine the radius of a dimension to be a number between
0 and $\infty$. Then the statement of T-duality is that small
and big radii are 'actually the same' or 'physically 
indistinguishable'. But to describe inequivalent physics one can
directly work with the moduli space, without reference to the
Teichm\"uller space. If one decides to describe the physics
of a compactification
using the domain which contains the point $r=\infty$ then there
is in principle no need to talk about radii smaller then the
minimal one. The statement of non-perturbative T-duality breaking
points out that it might happen, and in theories with
8 supercharges will generically happen, that the non-perturbative
moduli space is not the quotient of a modified Teichm\"uller space
by a modified T-duality group. Since the moduli space encodes
all physics there is no actual problem with this. But we loose
the intuitive interpretation based on the Teichm\"uller space.

We also note that at the non-perturbative level 
we still have discrete invariances.
As we discussed above, T-duality transformations act on the 
perturbative prepotential as monodromies.
At the non-perturbative level the monodromy group is modified but still
exists, and one might
consider the full mondromy group as 
the non-perturbative generalization of the 
T-duality group. In particular the monodromy group
encodes exact discrete symmetries of
the theory, like the exact $S-T$ exchange symmetries mentioned earlier.
If the Calabi-Yau space
is realized as an algebraic variety one can define a duality group
$\Gamma$, which gets contributions from the monodromy group
$\Gamma_M$ of the prepotential and from the group of symmetries of
the defining the equations, $\Gamma_W$. In \cite{LerSmiWar:1992} 
it was argued that
the duality group is given by the semidirect product 
$\Gamma_M \times \Gamma_W$. For one-moduli examples this is 
known to be true, see \cite{GivPorRab:1994} for discussion and references.
In comparison to perturbative heterotic T-duality groups it is 
interesting that in both cases braid groups seem to play a role.
The relations between the heterotic tree level and perturbative
T-duality group is given by replacing certain Weyl reflections by
braidings \cite{AntFerGavNarTay:1995}, wheras
in \cite{CerDauReg:1994}
it was shown that the duality group of a specific two-parameter Calabi-Yau
moduli space is the central extension of a braid group.

\subsection{Higher Derivative Couplings in Heterotic 
Compactifications}

So far we only discussed the heterotic prepotential. In order 
to analyse the perturbative structure of the higher order 
couplings, we must look at their dependence on the 
string coupling $g_{S}$.
We saw above that the perturbative K\"ahler 
potential is
\be
K(S, \ov{S}, T^a, \ov{T}^a) = \log g_S^2 + \widehat{K}(T^a, \ov{T}^a) \;.
\ee
Therefore the  couplings take the form
\be
F^{(g)}(X) = -i m^{2-2g}_{\mscr{String}} e^{(1-g) \widehat{K}} 
{\cal F}^{(g)} (S,T^a) \;.
\ee
This time the dependence on the string coupling or dilaton is through
the function ${\cal F}^{(g)}(S,T^a)$. Fortunately the dependence
on $S$ is restricted by holomorphicity considerations as we discussed
for the particular case of the prepotential. It can be
linear, constant or of the form $e^{-S}$, corresponding to tree level,
one loop and non-perturbative contributions. By explicit calculation
one finds that the higher couplings ${\cal F}^{(g>1)}$ do not have
a tree level term, whereas the tree level part of ${\cal F}^{(1)}$
is universal \cite{KapLouThe:1995,MarMoo:1998}. 
In summary one finds the following structure:
\bea
{\cal F}^{(0)}(S, T^a) &=& - S T^a T^b \eta_{ab} + h^{(0)} ( T^a)
+ f^{(0)}( e^{-S}, T^a ) \nonumber \\
{\cal F}^{(1)}(S, T^a) &=& 24 S  + h^{(1)} ( T^a)
+ f^{(1)}( e^{-S}, T^a ) \nonumber \\
{\cal F}^{(g>1)} (S,T^a) &=&  h^{(g>1)} ( T^a)
+ f^{(g>1)}( e^{-S}, T^a ) \;.\\
\nonumber
\eea
The prepotential was discussed at length in the preceeding
paragraphs. Let us now look at the $C^2$-coupling ${\cal F}^{(1)}(S,T^a)$.
For concreteness we consider the $(14,10)$ $STU$ model.
Then \cite{CarCurLueMohRey:1995,CarCurLueMoh:1996}
\bea
{\cal F}^{(1)}(S,T,U)= 24 S_{\mscr{inv}} + \frac{b_{\mscr{grav}}}{8 \pi^2}
\log \eta^{-2}(iT) \eta^{-2}(iU) + \frac{1}{2 \pi^2} 
\log( j(iT) - j(iU) ) \;,
\eea
where the invariant dilaton is defined by
\bea
S_{\mscr{inv}} &=& S - \frac{1}{2} \frac{\der h^{(1)}(T,U) }{\der T \der U}
- \frac{1}{8 \pi^2} \log ( j(iT) - j(iU) )  \nonumber \\
 &= & S + p(T,U) + \frac{1}{8 \pi^2} \sum_{k,l \geq 0}
kl c(kl) \mbox{Li}_1 \left( e^{- 2 \pi(kT +lU)} \right) \nonumber \\
 & & - \frac{1}{8 \pi^2} \mbox{Li}_1 \left( e^{- 2 \pi (T-U)} \right)
- \frac{1}{8 \pi^2} \log( j(iT) - j(iU))\;, \\
\nonumber
\eea
where $p(T,U)$ is a linear polynomial, $c(n)$ are the 
modular coefficients defined in (\ref{Defqcoeff})  and $\mbox{Li}_1$ is
the first polylogarithmic function, see appendix \ref{AppPolylog}.
The number $b_{\mscr{grav}}$ is the gravitational $\beta$-function
coefficient, associated with the one-loop running of the $C^2$ 
term. 
Note that there is no constant term in ${\cal F}^{(1)}(S,T,U)$
and that we find again threshold corrections
proportional to $\log( j(iT) - j(iU))$  which correspond to 
perturbative gauge symmetry enhancement. 

The comparison of heterotic and type II ${\cal F}^{(1)}$-functions
provides further evidence for heterotic - type II duality
\cite{KapLouThe:1995,Cur:1995,Cur:1995a,CarCurLueMohRey:1995,CarCurLueMoh:1996}.
One can in particular compare the expansion of 
the perturbative heterotic ${\cal F}^{(1)}(S,T^a)$ 
to that of its IIA counterpart in the
large volume limit. The IIA function counts rational and 
elliptic curves inside the Calabi-Yau, whereas the heterotic function
is related to expansion coefficients of modular forms.
For the $(14,10)$ model a detailed comparison of the heterotic and IIA
${\cal F}^{(1)}$-functions has been performed in \cite{CarCurLueMoh:1996}, 
with the result that both functions agree to the order one can compute.

The perturbative part of the higher couplings ${\cal F}^{(g>1)}(S,T^a)$
has also been computed in the $(14,10)$ model \cite{MarMoo:1998}. 
The resulting expressions are 
very complicated. They have no polynomial piece in the moduli,
but start with a constant piece 
and include in their holomorphic
part a sum over polylogs $\mbox{Li}_{3-2g}$, with coefficients
related to modular forms. As far as a comparison is possible
everything is consistent with heterotic - IIA duality.
In particular the constant term is always related to the
Euler number of the dual Calabi-Yau space with the correct
$g$-dependent prefactor. The structure of the polylog terms
fits with the counting of higher genus holomorphic curves in the
dual Calabi-Yau.

Another way of testing heterotic - type II duality using 
the higher couplings ${\cal F}^{(g)}(z)$ is provided by the holomorphic
anomaly \cite{AntGavNarTay:1995,dWCarLueMohRey:1996}.
As in the case of type II theory, one has to distinguish between
Wilsonian and physical couplings in the heterotic theory. 
The full physical couplings
have a non-holomorphic part. In the perturbative heterotic theory the
non-holomorphic pieces are necessary to make the couplings
covariant with respect to symplectic transformations \cite{deW:1996/02}. As we
discussed in chapter \ref{ChapterFourDSuGra} the holomorphic couplings are 
in general not symplectic functions. The physical couplings, however,
have to transform as symplectic functions, and they have
to be automorphic functions of the T-duality group.
The lack of symplectic covariance of the holomorphic couplings 
is encoded in a symplectic anomaly equation \cite{deW:1996/02}, 
which coincides with the large radius limit of the type II holomorphic
anomaly equation.
For lower $g$ the anomaly equation together with other physical input
fixes the non-holomorphic piece uniquely.
This has been used to predict the non-holomorphic parts
of the higher IIA couplings on the basis of heterotic T-duality in
\cite{dWCarLueMohRey:1996}.

Finally we would like to stress that 
from the mathematical point of view heterotic - IIA 
duality predicts a deep relation between automorphic forms and
Calabi-Yau geometry. As we saw the expansion coefficients
of the couplings count holomorphic curves inside the 
Calabi-Yau manifold. The heterotic couplings
are constrained by T-duality to be automorphic forms
of the T-duality group and are related to infinite dimensional
Lie algebras. Duality predicts that
'holomorphic curves are counted by (expansion coefficients of)
automorphic forms'. This is a highly non-trivial statement, but 
confirmed by explicit computations.

\subsection{Compactifications with ${\cal N}=4$ Supersymmetry}

The main focus of this paper is on four-dimensional string
vacua with ${\cal N}=2$ supersymmetry, but it is instructive to 
consider certain aspects of ${\cal N}=4$ compactifications in 
parallel. The most simple realization 
is provided by compactifying
the heterotic string on a six-torus $T^6$. One can take any of the
ten-dimensional gauge groups because the two theories are on
the same moduli after compactification by virtue of T-duality.

For generic moduli the massless spectrum consists of 
${\cal N}=4$ supergravity coupled to 22 abelian ${\cal N}=4$
vector multiplets. 
The gravity multiplet contains 6 graviphotons,
resulting in a gauge group $U(1)^{28}$. 
The minimal terms in the Lagrangian are
uniquely fixed by local ${\cal N}=4$ supersymmetry \cite{deR:1985}.
The moduli
space is locally
\be
{\cal M} = \left. \frac{SU(1,1)}{U(1)} \right|_S \times
\left. \frac{SO(6,22)}{SO(6) \times SO(22)} \right|_{T^a} \;,
\ee
where $S$ is the dilaton and $T^a$ are $132$ geometric
moduli which encode deformations of the metric of the torus, of
the background $B$-field and of the Wilson lines. The global
structure is determined by the combined  S- and T-duality
groups
\be
SL(2,\mathbb{Z})_S \times SO(6,22,\mathbb{Z})_{T} \;.
\ee
In contradistinction to the ${\cal N}=2$ case the two-derivative
part of the Lagrangian is not corrected and 
the S- and T-duality group are symmetries of the full 
theory \cite{Sen:1994}. As in the ${\cal N}=2$ case the 
theory contains a higher 
derivative gravitational $C^2$-term with a field
dependent coupling ${\cal F}^{(1)}(S)$. At tree level
this function is \cite{KapLouThe:1995}
\be
{\cal F}^{(1)}(S) = 24 S \;.
\ee
There are no perturbative corrections to this result and
in particular the gravitational $\beta$-function vanishes
because the coefficient $b_{\mscr{grav}}$ vanishes, as a result
of the matter content of the theory \cite{CarCurLueMohRey:1995}. 
But there has to
be a non-perturbative correction, because the above
coupling is not S-duality invariant \cite{HarMoo:1996}. 
This is different for 
the two-derivative terms which are S-duality invariant at
tree level and do not get any perturbative or
non-perturbative correction. By looking for an S-duality invariant
completion one finds 
\be
{\cal F}^{(1)(\mscr{hol})}(S) = \frac{24}{2 \pi i} \log \eta( 4 \pi i S)\;,
\ee
which has the correct behaviour under S-duality and 
reduces to the tree level result in the large S 
limit.\footnote{This form of $S$-dependence is also predicted
by considering the so-called topological free energy 
\cite{CarCurLueMohRey:1995}. The fact that such non-perturbative
terms have to be present was pointed out in \cite{HarMoo:1996}.}
More precisely this is the holomorphic, Wilsonian piece of
the coupling. We will discuss the full coupling in ${\cal N}=2$
language later.

The above reasoning was based on symmetry arguments. Using the
${\cal N}=4$ version of heterotic - IIA duality one can
verify the result by explicit computation. 
The heterotic string
on $T^4$ is equivalent to the IIA string on $K3$
\cite{HulTow:1994,Wit:1995,Sen:1995,HarStr:1995} and after
compactification to four dimensions the heterotic string on
$T^6$ and the type II string on $K3 \times T^2$ are on the same moduli.
Since under the heterotic - type IIA map the heterotic dilaton is identified
with the K\"ahler modulus of the $T^2$ one can compute the
correction to ${\cal F}^{(1)}(S)$ in IIA perturbation theory
by a one-loop calculation  \cite{HarMoo:1996}, which is similar to the 
heterotic ${\cal N}=2$ one-loop calculation that we discussed before.
The corrections are elliptic ($g=1$) world sheet instantons
from the IIA perspective and they are non-perturbative 
instanton corrections for the heterotic string, which come
from Euclidean fivebranes wrapped around the torus.

Let us now describe the ${\cal N}=4$ theory in the ${\cal N}=2$
language. The ${\cal N}=4$ gravity multiplet splits into
various ${\cal N}=2$ multiplets: one gets the ${\cal N}=2$
gravity multiplet which contains one of the six ${\cal N}=4$
graviphotons. Then one gets two ${\cal N}=2$ gravitini 
multiplets, which absorb four of the graviphotons.
The remaining ${\cal N}=4$ graviphoton sits in a vector multiplet,
together with the dilaton. The 22 ${\cal N}=4$ vector multiplets
decompose into 22 ${\cal N}=2$ vector and 22 ${\cal N}=2$
hypermultiplets. A certain subsector of this, containing
${\cal N}=2$ supergravity coupled to $N_V=23$ vector multiplets,
with moduli space
\be
{\cal M} = \frac{SU(1,1)}{U(1)} \times 
\frac{SO(2,N_V-1)}{SO(2) \times SO(N_V-1)} \;,
\ee
can be described using our ${\cal N}=2$ formalism.
One should note that the prepotential and the higher couplings of the
${\cal N}=4$ and ${\cal N}=2$ theories
differ beyond the tree level.
For example the heterotic ${\cal N}=2$ 
function ${\cal F}^{(1)}(S,T^a)$ has corrections that
depend on all the moduli and not only on $S$ and one does not
expect S-duality. Nevertheless we can use our ${\cal N}=2$ 
formalism to get information about the ${\cal N}=4$
theory, and in particular about its black hole solutions, as we will
see in the next chapter.

Finally there are special
heterotic ${\cal N}=2$ compactifications which have
non-renormalization properties similar to ${\cal N}=4$
models. 
These are stringy analogues of finite ${\cal N}=2$
gauge theories, where the matter content is chosen such that the
$\beta$-function vanishes. The prime example of such a special
heterotic ${\cal N}=2$ compactification is the so-called
FHSV model \cite{FerHarStrVaf:1995}, 
which has 12 vector and 11 hypermultiplets.
The dual Calabi-Yau space has $h_{1,1}=11=h_{2,1}$ and therefore
its Euler number vanishes, $\chi=0$. In this model the tree level
moduli space is exact and S- and T-duality are non-perturbative
symmetries. Like in the ${\cal N}=4$ case the $C^2$-coupling
gets corrections beyond the tree level, and the corrections
are precisely such that the function ${\cal F}^{(1)}(S,T^a)$ becomes
invariant under the dualities \cite{HarMoo:1996a}. 
The coupling does not only depend on the 
dilaton but also on the other 10 moduli through an 
automorphic form of the T-duality group $O(2,10,\mathbb{Z})$. 
The algebraic
structure behind the formula is controlled by the
so-called 'faked Monster Lie superalgebra'.

\chapter{Four-Dimensional ${\cal N}=2$ Black Holes in String and
M-Theory \label{ChapterStringBHs}}

In this chapter we describe the computation of the black hole entropy
for concrete models both from the macroscopic and the microscopic point
of view. In the first section the macroscopic entropy is computed by
solving the stabilization equations and plugging the result
into the entropy formula (\ref{ModIndEntropyFormula}). We start with
black holes in type II string theory and M-theory. After discussing
general properties of the entropy formula we consider the large
radius limit of a general Calabi-Yau compactification. The corresponding
$F$-function is leading order in $\alpha'$ and receives contributions
at tree and one-loop level in $g_{IIA}$. The one-loop term is
linear in $C^2$. It turns out that we can also include a class of
higher order terms, which can be computed in the M-theory picture
\cite{GopVaf:1998/09,GopVaf:1998/12}. For the most general
black hole, with no restriction on the charges, we can bring
the entropy to form (\ref{EntropyMacroIIGeneral}). Fully explicit
solutions are obtained after imposing the vanishing of some of the
charges or reality constraints on the solution. We discuss 
three specific cases. The most important result is formula
(\ref{EntropyCYmacro}) for the macroscopic black hole entropy 
of the large volume limit of a Calabi-Yau compactification, with
charge $p^0=0$. This is the most general case where the corresponding
microscopic entropy is known from \cite{MalStrWit:1997,Vaf:1997}.

Next we discuss black holes in heterotic ${\cal N}=2$ compactifications.
In view of heterotic - IIA duality this is a special case of the
discussion of type II black holes, but it deserves an explicit study
for the following reason: In perturbative
heterotic models T-duality is an exact symmetry and therefore one
expects that the entropy formula is T-duality invariant. Moreover we
can profit from the specific form of the heterotic prepotential and
find solutions for the entropy without imposing constraints on the 
charges.
We show in detail how one can make
T-duality invariance manifest. 
In particular we show that the contribution of the perturbative
prepotential to the entropy, including all $\alpha'$-corrections,
can be expressed in terms of the perturbative heterotic string coupling 
by formula (\ref{EntropyCoupling}). Next we include the higher
couplings, but now restrict ourselves to the leading order in
$\alpha'$. In this case we find a manifestly 
T-dual formula by combining all terms into invariants of the 
T-duality group in formula (\ref{entrofin}).
These two results  extend the known results on the
U-duality invariance of the entropy of ${\cal N}=8$ black holes 
\cite{KalKol:1996} and on the 
T- and S-duality invariance of the entropy of ${\cal N}=4$ black 
holes \cite{CveYou:1995,CveTse:1995}. 
We also discuss certain non-perturbative and non-holomorphic
contributions to the entropy. This is illustrated by finding the
manifestly T- and S-duality invariant formula (\ref{entropiaynonholo})
for the entropy of black holes in ${\cal N}=4$ compactifications
in presence of higher curvature terms. 

In the second part of the chapter we review the computation of
the microscopic black hole entropy \cite{MalStrWit:1997,Vaf:1997}. 
Following \cite{CardWMoh:1999/06}
we discuss ${\cal N}=8$ and ${\cal N}=4$ compactifications on
$T^6$ and $K3 \times T^2$ in parallel with generic 
${\cal N}=2$ compactifications on Calabi-Yau threefolds.
The result is 
the microscopic entropy formula 
(\ref{MicroEntropy}), which is valid 
for black holes with charge $p^0=0$
in the large radius limit of Calabi-Yau compactifications. 
This result agrees with the corresponding
macroscopic entropy (\ref{EntropyCYmacro}) and we discuss in
detail why this is a highly non-trivial result, which on the macroscopic
side depends crucially on the correct treatment of the higher curvature
terms and on the use of the modified entropy formula 
(\ref{WaldEntropy}) instead of the
Bekenstein-Hawking area law.

\section{The Macroscopic Black Hole Entropy}

\subsection{Black Holes in Type II String Theory and in M-Theory}

We have already seen in chapter \ref{ChapterN=2BlackHoles} 
that the stabilization equations (\ref{Stab})
together with the
model-independent entropy formula (\ref{ModIndEntropyFormula})
fix the macroscopic entropy
in terms of the electric and magnetic charges carried by the black hole.
In order to give an explicit expression for the entropy
one has to solve the stabilization equations for a 
specific choice of the 
function $F(X,\widehat{A})$. Due to the complicated nature
of this function, explicit solutions can only be obtained for sufficiently
simple choices of $F(X,\widehat{A})$ and for black holes which only
carry a subset of the possible charges. Our presentation follows
\cite{CardWMoh:1999/06}, where more details can be found. 
Results on the entropy
of Calabi-Yau black holes in absence of higher curvature 
terms were obtained in \cite{BehCardWKalLueMoh:1996,Shm:1996}.

As a first step it is convenient to introduce rescaled $U(1)$-invariant
variables by
\be
Y^I = e^{{\cal K}/2} \ov{Z} X^I \mbox{   and   }
\Upsilon = e^{{\cal K}} \ov{Z}^2 \widehat{A} \;.
\ee
Then the stabilization equations take the form
\bea
Y^I - \ov{Y}^I &=& ip^I \;,\\ \nonumber
F_J (Y,\Upsilon) - \ov{F}_J ( \ov{Y}, \ov{\Upsilon}) &=& iq_J \;,\\
\nonumber 
\eea
where we used the homogenity properties of $F(X,\widehat{A})$. We also note that
\be
|Z|^2 = p^I F_I (Y,\Upsilon) - q_I Y^I \;.
\ee
The model-independent entropy formula takes the form
\be
{\cal S} = \pi \left[ |Z|^2 +4 \; \mbox{Im} \left( \Upsilon 
F_{\Upsilon}(Y,\Upsilon) \right)
\right] \;,
\ee
where $\Upsilon=-64$. Before turning to explicit models
we note that the charge dependence of the 
entropy formula follows a general pattern: 
Let $Q$ be a generic charge
and recall that the function $F(Y,\Upsilon)$ has an expansion
$F(Y,\Upsilon) = \sum_{g\geq 0} F^{(g)}(Y) \Upsilon^g$. Now suppose
that we solve the stabilization equations iteratively in $\Upsilon$.
To leading order, $Y^I$ and $F_J$ are proportional to $Q$ and
the resulting entropy will therefore be proportional to $Q^2$, i.e.
it is quadratic in the charges. By iteration we then find
\be
{\cal S} = \pi \sum_{g=0}^{\infty} a_g Q^{2-2g} \;,
\label{EntropyEvenPowers}
\ee 
where the constant coefficient $a_g$ is related to the contribution
of the function $F^{(g)}(Y)$.

We now turn to specific models.
The most important model that we have to discuss is based on the function
\be
F(Y,\Upsilon) = D_{ABC} \frac{Y^A Y^B Y^C}{Y^0} + D_A \frac{Y^A}{Y^0}
\Upsilon \;.
\ee
This function describes the large volume limit of Calabi-Yau compactifications
of IIA string theory or eleven-dimensional M-theory. It contains all terms
which are (in type IIA language) tree-level in $\alpha'$. The first and
second term are tree level and one-loop in the type IIA coupling,
respectively. The second term is related
to a $C^2$-term of the form $D_A z^A C^2$
in the effective Lagrangian, where $z^A = Y^A / Y^0$. 
The coefficients $D_{ABC},
D_A$ are related to the triple intersection numbers and second
Chern class number of the Calabi-Yau space by
\be
D_{ABC} = - \ft16 C_{ABC} \mbox{   and   }
D_A = - \ft1{24} \ft1{64} c_{2A} \;.
\ee

In our discussion of the stabilization equation we will also include
a function $G(Y^0,\Upsilon)$, which contains terms of higher order
in $\Upsilon$, 
\be
F(Y,\Upsilon) = D_{ABC} \frac{Y^A Y^B Y^C}{Y^0} + D_A \frac{Y^A}{Y^0}
\Upsilon + G(Y^0,\Upsilon) \;,
\ee
where
\be
Y^0 G_0 + 2 \Upsilon G_{\Upsilon} = 2 G \;.
\label{Ghomogen}
\ee
The contributions encoded in $G(Y^0,\Upsilon)$ are in the M-theory
picture due to loop and non-perturbative corrections of 
Kaluza-Klein modes of the graviton (D0-branes) around the 
M-theory circle.

One now computes the explicit expression for $F_0$ and $F_A$ and gets
formulae for $q_0$ and $q_A$ from the stabilization equations.
Note that the function $G(Y^0,\Upsilon)$ only enters into the
expression for $q_0$, because $G(Y^0, \Upsilon)$ does not depend
on $Y^A$. The entropy takes the form\footnote{We do not make use
of the fact that $\Upsilon$ is real on the horizon, because the
formula looks more symmetric in terms of $\Upsilon$ and 
$\overline{\Upsilon}$.}
\be
{\cal S}= \pi \left[ |Z|^2 -2i D_A \left( \frac{Y^A}{\ov{Y^0}} \Upsilon -
\frac{\ov{Y}^A}{Y^0} \ov{\Upsilon} \right) -2i ( \Upsilon G_{\Upsilon}
- \ov{\Upsilon} G_{\ov{\Upsilon}} ) \right]   \;,
\label{EntropyMacroIIGeneral}\ee
where $Z$ is given through the formula $|Z|^2= p^I F_I - q_I Y^I$ by
\bea
|Z|^2 &=&  i \,D_{ABC} \Big[ {3\,Y^AY^B\bar Y^C\over Y^0} -{3\,\bar
Y^A\bar Y^B Y^C\over \bar Y^0} - 
{Y^AY^BY^C \bar Y^0\over (Y^0)^2} + {\bar Y^A\bar Y^B\bar Y^C
Y^0\over (\bar Y^0)^2} \Big] \nonumber \\
&& + i D_A\Big [  {\bar Y^A \,\Upsilon \over Y^0} 
- {Y^A \,\bar \Upsilon \over \bar Y^0} 
- {Y^A \bar Y^0 \,\Upsilon \over (Y^0)^2} 
+ {\bar Y^A Y^0\,\bar \Upsilon \over (\bar Y^0)^2}\Big]
\nonumber \\[2mm]
&& + \ft12 i (Y^0+ \bar Y^0)(G_0-\bar G_0) +  \ft12 p^0(G_0+\bar
G_0)   \,. \label{ZZZ}
\eea

Explicit solutions can be found only in cases where one does not
take the most general configuration of charges. The three cases where
solutions are available are (i) $p^0=0$, (ii) so-called axion-free black holes
where the moduli $z^A=Y^A/Y^0$ are purely imaginary and (iii) black holes
with $\mbox{Re} \; Y^0 = 0$.

\subsubsection{Black Holes with $p^0=0$}

When setting $p^0=0$ it imediately follows that $Y^0$ is real,
$Y^0=\ov{Y}^0$. Then the stabilization equation involving $q_A$
simplifies and one can derive
\be
Y^A = \ft16 Y^0 D^{AB} q_B + \ft12 i p^A \;,
\ee
where we defined $D^{AB}$ by
\be
D_{AB} := D_{ABC} p^C \mbox{  and   } D_{AB}D^{BC} = \d_A^C \;.
\ee
Using the $q_0$ stabilization equation one finds
\be
4\, (Y^0)^2 = {D_{ABC}\,p^Ap^Bp^C - 4 D_A p^A\,\Upsilon  \over \widehat q_0 +i
(G_0-\bar G_0) }\,,
\label{stabyg}
\ee
where $\widehat q_0 \equiv q_0 + \ft1{12} D^{AB}\,q_Aq_B$.
Furthermore one determines
\be
|Z|^2 = - {D_{ABC}\,p^Ap^Bp^C - 2D_Ap^A\,\Upsilon \over Y^0}  +
iY^0\,(G_0-\bar G_0)\,,  
\ee
and the entropy is given by
\be
{\cal S} = - 4\pi \, Y^0\,\widehat q_0 - i \pi \,( 3Y^0 \,G_0 + 2
\Upsilon\,G_\Upsilon - \mbox{ h.c.} )\,.
\label{ent2g}
\ee
One can now solve (\ref{stabyg}) iteratively for $Y^0$ in order to
express everything in terms of the charges. First note that one 
already has an explicit expression if $G(Y^0,\Upsilon)=0$. We denote
this value by $y^0$: 
\be
y^0 = \ft12 \sqrt{ \frac{ D_{ABC} p^A p^B p^C - 4 D_A p^A \Upsilon}{
\widehat{q}_0 } } \;.
\ee
The corresponding zeroth order value for $Y^A$ is 
$y^A = \ft16 D^{AB} q_B + \ft{i}2 p^A$ so that the
moduli $z^A = y^A / y^0$ are
\be
z^A= i p^A \sqrt{ \frac{ \widehat{q}_0 }{
D_{ABC} p^A p^B p^C - 4 D_A p^A \Upsilon}  }
+ \ft16 D^{AB} q_B \;.
\label{modzero}
\ee

Next observe that we have to restrict the charges to be either positive
or negative in order to make sure that the right hand side 
of (\ref{stabyg}) is positive.  We take the magnetic charges 
$p^A$ to be positive.
Then $D_{ABC}p^A p^B p^C = - \ft16 C_{ABC}p^A p^B p^C$  and
$-4 D_A p^A \Upsilon = - \ft16 c_{2A}p^A$ are negative and we have 
to take $\widehat{q}_0$ to be negative, $\widehat{q}_0 <0$. 
The fact that we have to specify the signs of the charges 
reflects that the K\"ahler moduli
space is a cone. With the above choice the square root
in (\ref{modzero}) is real and $z^A$ has a positive
imaginary part. This means that the physical range of parameters
is $\mbox{Im}\; z^A > 0$ and that $\mbox{Im}\; z^A = 0$ is
the boundary of the K\"ahler cone.

We have to make sure that the large volume
approximation is valid near the horizon. In the type IIA picture where
the vector multiplet moduli are K\"ahler moduli this 
is the case when the moduli near the horizon take values 
deep inside the K\"ahler cone, $\mbox{ Im} \; z^A >> 0$.
This is achieved by taking $|\widehat{q}_0|$ to be much larger than the
other charges:  $|\widehat{q}_0 | >> | p^A |$.
Next we have to make sure that the curvature at the horizon is
small so that the higher curvature terms can be treated as small
corrections. This can be done by making all charges large, so that we
have to impose
\be
|\widehat{q}_0| >> |p^A| >> 0
\ee
on the charges.

Finally we have to impose that the function $G(Y^0,\Upsilon)$
can be treated as a small perturbation, 
so that an iterative solution of (\ref{stabyg}) makes sense.
This imposes the constraint
$| \mbox{Im} \; (G_0(y^0, \Upsilon) ) | << |\widehat{q}_0|$.

The first iterative step then yields
\be
Y^0 = y^0 \Big( 1 + \ft{1}{2} \, \frac{i (G_0 (y^0,\Upsilon) - {\bar G}_0
({\bar y}^0, {\bar \Upsilon}))}{|{\widehat q}_0|} + \cdots  \Big) \;.
\label{yyfirst}
\ee
Inserting this into (\ref{ent2g}) gives the entropy formula
\be
{\cal S} =
2 \pi \sqrt{|{\widehat q}_0| ( - \ft16 D_{ABC} \,p^A p^B p^C -256 D_A\, p^A)} 
- 2 \pi i ( G(y^0, \Upsilon) - {\bar G}
({\bar y}^0, {\bar \Upsilon}))  + \cdots \;\;\;.
\label{EntropyPNull=Null}
\ee
The zeroth order approximation, which corresponds to the tree order
in $\alpha'$ is given by setting $G(Y^0,\Upsilon)=0$.
When expressing the coefficients $D_{ABC}, D_A$ in terms of the
topological quantities $C_{ABC}, c_{2A}$ the entropy takes the form
\bea
{\cal S} =
2 \pi \sqrt{\ft{1}{6}|{\widehat q}_0| (C_{ABC} \,p^A p^B p^C + c_{2A}\, p^A)} \;.
\label{EntropyCYmacro}
\eea
This formula agrees with a microscopic entropy formula derived
from state counting, as we will see in section \ref{Comparison}.

\subsubsection{Axion-Free Black Holes}

We now turn to 
axion-free black holes, which are characterized by
the condition that the moduli $z^A= Y^A/Y^0$ are imaginary. Using $\bar Y^0
Y^A +Y^0 \bar Y^A=0$, it follows that 
\be
Y^A = i p^A\,{Y^0\over \lambda}\,,\qquad \mbox{and}\quad Y^0 = \ft12
( \lambda + ip^0)\,,
\ee
where $\lambda$ is real constant.
The charges $q_A$ are given by
\be
q_A = -3 \,D_{ABC}\,p^Bp^C\,{p^0\over \lambda^2} - 4 D_A\,
{p^0\,\Upsilon\over  \lambda^2 + (p^0)^2}\,.
\ee
Therefore the charges are tightly constrained. We have found a
quadratic equation for $\lambda$, which 
fixes all the moduli in terms of the charges. The stabilization
equation for $q_0$ shows that $q_0$ is not independent,
but is given by 
\be
q_Ap^A + 3 q_0p^0 = - D_Ap^A\,\Upsilon\,{16\,p^0\over \lambda^2 + (p^0)^2} 
- 3 i p^0(G_0-\bar G_0)\,. 
\ee
In this case we get $|Z|$ from (\ref{ZZZ}) 
\bea
|Z|^2 &=& -2 \,D_{ABC}\,p^Ap^Bp^C {\lambda^2 + (p^0)^2\over \lambda^3} +
4\, D_Ap^A \,\Upsilon {\lambda^2 - (p^0)^2\over \lambda( \lambda^2 +
(p^0)^2) } \nonumber \\
&& + \ft12 i\lambda (G_0-\bar G_0) +  \ft12 p^0 (G_0+\bar G_0) 
\,,
\eea
and the entropy is
\bea
{\cal S} &=& \pi\,\Big[ -2 \,D_{ABC}\,p^Ap^Bp^C {\lambda^2 + (p^0)^2\over
\lambda^3} + 8\, D_A p^A \,\Upsilon {\lambda\over  \lambda^2 +
(p^0)^2 } \nonumber \\
&&\hspace{5mm}  + \ft12 i\lambda (G_0-\bar G_0) +  \ft12 p^0 (G_0+\bar
G_0)   -2i \Upsilon (
\,G_\Upsilon - \,\bar G_{\Upsilon})  \Big] \,. 
\label{EntropyAxionFree}
\eea
For large $Y^0$ the leading part of $G$ is
$G = i c (Y^0)^2$, with $c$ real \cite{GopVaf:1998/09}.
Taking $G = i c (Y^0)^2$
and $D_A=0$ the result (\ref{EntropyAxionFree}) of
\cite{CardWMoh:1999/06}
reduces to the one
obtained in \cite{BehCardWKalLueMoh:1996}.

\subsubsection{Black Holes with ${\rm Re} \,Y^0 =0$}

The third case where we can have an explicit solution is
${\rm Re}\, Y^0 =0$ which implies
$Y^0 = \ft{1}{2}i p^0$. The stabilization equations 
imply
\bea
p^0 \,q_A &=&  - 6 \,D_{ABC}(Y^B Y^C+\bar Y^B\bar
Y^C) - 4D_A\,\Upsilon \,,\nonumber \\
(p^0)^2\,q_0 &=& 4\, D_{ABC}\,p^A(Y^B Y^C+Y^B \bar Y^C+ \bar Y^B\bar
Y^C) + 4D_Ap^A \,\Upsilon -i (p^0)^2 \,(G_0-\bar G_0)\,. \quad~
\label{stabrmy}
\eea
Again the charge $q_0$ is not independent. 
The constraint is
\be
p^0\, p^I\,q_I =  2 \,D_{ABC}\,p^Ap^Bp^C -i (p^0)^2 \,(G_0-\bar
G_0)\,. 
\label{charconstr}
\ee

{From} (\ref{ZZZ}) and (\ref{EntropyMacroIIGeneral}) we obtain
\bea
|Z|^2 &=& {2\over p^0} D_{ABC}(Y+\bar Y)^A (Y+\bar Y)^B(Y+\bar Y)^C +
{4\over p^0} D_A(Y+\bar Y)^A \,\Upsilon
 + \ft12 p^0 (G_0+\bar G_0)\,, \nonumber\\
{\cal S}&=& {2\pi \over p^0}\, D_{ABC}(Y+\bar Y)^A (Y+\bar Y)^B(Y+\bar
Y)^C  -2i\pi (G-\bar G) \,,
\label{EntropyRe=0}
\eea
where we made use of the homogenity property (\ref{Ghomogen}) for $G$. 

The equations (\ref{stabrmy}) are quadratic equations for the scalars $Y^A$
and can be used to express them in terms of the charges. However, we
do not wish to pursue this in full generality.  Below we will determine the
value of the $Y^A$ for type-II models with a dual heterotic description.

We conclude our discussion of IIA black holes with a remark on the
dependence of the entropy on the topology of the Calabi-Yau space.
When considering all contributions to the couplings that arise
at tree level and loop level in $\alpha'$ we find that the topological
quantities involved are the intersection numbers, the second Chern class
and the Euler number. The entropy depends on these model dependent
data and on the charges
\be
{\cal S} = {\cal S}(p^I, q_J | C_{ABC}, c_{2A}, \chi) \;.
\ee 
It is amusing to note that the topological quantities 
$C_{ABC}, c_{2A}, \chi$ are necessary and sufficient, according
to Wall's theorem \cite{Wal:1966}, to determine the underlying 
Calabi-Yau space up to homotopy. When world-sheet instantons are taken 
into account, then the entropy also depends on finer data,
namely the world-sheet instanton number.

\subsection{Black Holes in Heterotic String Theory}

We now turn to the discussion of black holes in ${\cal N}=2$
heterotic string compactifications. As a consequence of heterotic
- type IIA duality this is, technically, a special case 
of the IIA string compactifications, where the Calabi-Yau space
is restricted to be a $K3$ fibration. But concerning the
physics involved, heterotic black holes are worth to be studied
in their own right.

\subsubsection{Tree Level Black Holes}

We start our investigation by considering black hole solutions
based on the tree-level prepotential
\be
F = - \frac{Y^1 Y^a \eta_{ab} Y^b}{Y^0} \;,
\label{prepwithWL}
\ee
where
\be
Y^a \eta_{ab} Y^b = Y^2 Y^3 - \sum_i (Y^{3+i})^2 \;.
\ee
We use the conventions of \cite{CardWMoh:1999/06}. The results
of this subsection were obtained in 
\cite{BehKalRahShmWon:1996,CarLueMoh:1996}.

The conventions for heterotic models are explained in
chapter \ref{ChapterFourDimensionalStringsAndM}. 
The physical moduli are $S=-i Y^1/Y^0$
and $T^a = -i Y^{1+a}/Y^0$, where $T^a = T,U,V^i$.
A special case of this class is the $STU$-model, where
\be
F = - \frac{Y^1 Y^2 Y^3}{Y^0} \;.
\ee
Heterotic tree level black holes have the special property that
one can solve the stabilization equations explicitly without
restricting the choice of the non-vanishing charges. In order to show
that the resulting entropy is manifestly T-duality invariant, and
in the case of the $STU$-model even is triality invariant, we
introduce certain invariants of the T-duality group
$O(2,N_V-1, \mathbb{Z})$. First recall that the heterotic
charges $(N^I,M_J)$ (the charges associated with the symplectic
basis $P^I,Q_J$ adapted to heterotic perturbation theory) differ from
the IIA charges $p^I,q_J$ (the charges associated with the 
symplectic section $Y^I,F_J$ determined by the prepotential) by a
symplectic transformation:
\bea
(N^I) &=& (p^0, q_1, p^2, \ldots) \;,\nonumber \\
(M_J) &=& (q_0, -p^1, q_2, \ldots) \;. \\
\nonumber 
\eea
T-duality acts on the section $(P^I,Q_J)$ by a specific subset of
the 'classical' symplectic transformations introduced in chapter
\ref{ChapterFourDimensionalStringsAndM}.
The action on the symplectic vector of charges is read off from
the expression for the BPS-mass, which has to be symplectically
invariant:
\be
M^2_{BPS} =  N_I Q^I - M^I P_I \;, 
\ee
where the ($U(1)$ invariant) 
section $(Q^I, P_J)$ is evaluated at spatial infinity.
This implies that under the transformation
\be
\left( \begin{array}{c}
Q^I \\ P_J \\
\end{array} \right) \rightarrow \Gamma
\left( \begin{array}{c}
Q^I \\ P_J \\
\end{array} \right)
\ee
the charges transform as
\be
(N^I, M_J) \rightarrow (N^I, M_J) \Gamma^T \;.
\ee
For a classical T-duality transformation this imples
\be
(N^I) \rightarrow (N^I) {\bf U}^T \mbox{   and   }
(M_J) \rightarrow (M_J) {\bf U}^{-1} \;,
\ee
where ${\bf U} \in O(2,N_V-1,\mathbb{Z})$. This gives rise to
three obvious invariants;
\bea
\langle M,M \rangle 
&=& 2 \Big(M_0 M_1 + \ft{1}{4} M_a \eta^{ab} M_b\Big) = 
2\Big( - q_0 p^1 + \ft{1}{4}
 q_a \eta^{ab} q_b \Big) \;\;,
\nonumber\\
\langle N,N \rangle &=& 2\Big( N^0 N^1 + N^a \eta_{ab} N^b \Big) = 
2 \Big(  p^0 q_1 +  p^a \eta_{ab} p^b \Big) \;\;,
\nonumber\\
M \cdot N &=& M_I N^I = q_0 p^0 - q_1 p^1 + q_2 p^2 + \cdots + q_{n}
p^{n} \;\;.
\label{invcomb}
\eea
Here $\eta^{ab}$ is the inverse matrix of $\eta_{ab}$ (\ref{EtaMatrix}):
\be
(\eta^{ab}) = \left( \begin{array}{ccc}
0&2&0\\
2&0&0\\
0&0&-\d^{ij}\\
\end{array} \right) \;.
\ee
The first two invariants use the pseudo-orthogonality of
${\bf U}$, whereas the third invariant is based on the fact
that $M$ and $N$ transform with contragradient matrices.

One can now use the stabilization equations to establish
\bea
Y^a &=& \frac{1}{S + \ov{S}} \left[  - \ft12 \eta^{ab} q_b - i
\ov{S} p^a \right] \;,\nonumber \\
|Z|^2 &=& (S + \ov{S}) \left( \ov{Y}^a \eta_{ab} Y^b -
\frac{ \ov{Y}^0}{Y^0}  Y^a \eta_{ab} Y^b + \mbox{h.c.} \right) \;,
\nonumber \\
q_1p^0 &=& -\left( \frac{ \ov{Y}^0}{Y^0} - 1 \right) 
Y^a \eta_{ab} Y^b + \mbox{h.c.} \;.\\
\nonumber
\eea
These equations can be combined into
\be
|Z|^2 = \frac{S +\ov{S}}{2} \la N, N \ra \;,
\ee
which implies (since no higher derivative terms are considered)
\be
{\cal S} = \frac{\pi}{2} ( S + \ov{S}) \la N, N \ra \;.
\ee
This relation is very interesting, because it relates the entropy
to the tree level string coupling evaluated on the event horizon
\be
{\cal S } = \la N, N \ra 
\left. \frac{4 \pi^2}{g_S^2} \right|_{\mscr{Horizon}} \;.
\ee
We will see that this relation generalizes to the full
perturbative level, where one cannot, in general, compute
the moduli, the dilaton and the entropy as functions of the charges.

In the tree level case one can use the remaining stabilization equations to
express the dilaton explicitly in terms of charges
\be
S = i \frac{M \cdot N}{\la N, N \ra}
+ \sqrt{ \frac{ \la M, M \ra}{\la N, N \ra} - \frac{(M\cdot N)^2}{\la N,
N \ra^2 } }
\ee
and therefore the entropy is
\be
{\cal S} = \pi \sqrt{ \la M, M \ra \la N, N \ra - 
( M \cdot N)^2 } \;.
\ee
This formula is manifestly T-duality invariant.
For the special case of the $STU$-model one can verify
that the entropy is invariant under the full triality group.
If more moduli are present, the formula is still invariant under
S-duality.

As we discussed in chapter \ref{ChapterFourDimensionalStringsAndM}
we can describe a
subsector of the ${\cal N}=4$ heterotic compactification
in terms of the tree level ${\cal N}=2$ theory. The full
${\cal N}=4$ theory contains four additional gauge fields and 
therefore there are four electric and four magnetic 
charges that one has to switch off. The full ${\cal N}=4$
entropy formula \cite{CveYou:1995,CveTse:1995}
is obtained from our tree level ${\cal N}=2$
formula by replacing the 
$O(2,N_V-1,\mathbb{Z})$ invariants by the corresponding
$O(6,24,\mathbb{Z})$ invariants.

Furthermore
for specific choices
of the charges,  like purely electric or magnetic black holes,
and, more generally, for 
black holes where $M^I$ and $N_I$ are parallel as vectors,
one finds that one gets a vanishing entropy. The cases of purely
electric or magnetic charges correspond to singular solutions of
the stabilization equations, where the moduli take singular values
either on the horizon or at infinity. The associated
black hole solutions are degenerate at the horizon, which signals
that the solution does not make sense as a four-dimensional geometry.
This can be understood
from the ${\cal N}=4$ point of view as follows: In ${\cal N}=4$
theories there are two types of BPS states, short and intermediate
ones. But only black holes that are members of
intermediate multiplets and which have
4 Killing spinors correspond to regular solutions of the 
stabilization equations, whereas black holes that
are in short ${\cal N}=4$ BPS multiplets and which have 8 Killing spinors 
necessarily are degenerate. One can show
that charge configurations with parallel magnetic and electric
charge vectors correspond to short multiplets whereas
the more generic charge configurations with non-parallel 
charge vectors correspond to intermediate multiplets.

\subsubsection{Perturbative Heterotic Black Holes}

Let us now consider black hole entropy on the basis of the
full perturbative heterotic prepotential 
\cite{BehCardWKalLueMoh:1996,Rey:1996}. 
Remember that the
perturbative prepotential has the form
\be
{\cal F} = - \frac{Y^1 Y^a \eta_{ab} Y^b}{Y^0} + (Y^0)^2 h(T^a) \;,
\ee
where the one-loop correction $h(T^a)$ does not depend on the 
dilaton S.

The action of the T-duality group $O(2,N_V - 1,\mathbb{Z})$
on the heterotic section now takes the form
\be
\Gamma_{\mscr{pert}} = \left( \begin{array}{cc}
{\bf U} & \mathbb{O} \\
{\bf U}^{T,-1} \Lambda & {\bf U}^{T,-1} \\
\end{array} \right) \;,
\label{PertMonodr1}
\ee
where the symmetric matrix $\Lambda$ accounts for the perturbative
modification. The charges now transform in the following way under
T-duality:
\be
(N^I) \rightarrow (N^I) {\bf U}^T \mbox{   and  }
(M_J) \rightarrow (M_J) {\bf U}^{-1} + (N^I) \Lambda^T {\bf U}^{-1} \;.
\ee
Note that the transformation law of the electric charges 
is modified whereas the magnetic charges $N^I$ still transform
as at tree level. Therefore $\la N, N\ra$ is still an
invariant, whereas $\la M,M \ra$ and $M \cdot N$ are not.

Proceeding as in the tree level case one can show that the
entropy is given by
\be
{\cal S} = \la N, N \ra \left. \frac{4 \pi^2}{g_{\mscr{pert}}^2}
\right|_{\mscr{Horizon}} \;,
\label{EntropyCoupling}
\ee
where $g_{\mscr{pert}}$ is the perturbative string coupling
defined by
\be
\frac{4 \pi}{g^2_{\mscr{pert}}} = \ft12 ( S + \ov{S} + V_{GS}(T^a,
\ov{T}^a ) ) \;.
\ee
Since the perturbative coupling is by construction 
T-duality invariant, we have shown that
the entropy is T-duality invariant, irrespective of
the precise form of $h(T^a)$.

Having derived this important general result we now turn to 
explicit examples. Beyond tree level we can only get explicit results for
the entropy in terms of the charges
if the
function $h(T^a)$ is sufficiently simple and if we restrict the
charges. 
Let us first consider the $STU$-model and include the cubic part of the
one loop correction, which is tree level in $\alpha'$.
Then the prepotential is
\be
F = - \frac{Y^1 Y^2 Y^3 + \ft13 (Y^3)^3}{Y^0} \;.
\ee
The coefficients of this cubic polynomial are now proportional to the
triple intersection form of the dual Calabi-Yau space.
The solution 
for the entropy is a special case of the discussion
given for IIA black holes, where we discussed three different
configurations of charges that allow explicit solutions. 
Here we only display the formula for 
non-axionic black holes with $p^0=q_A=0$.
One finds
\be
{\cal S} = 2 \pi \sqrt{ |q_0| (p^1 p^2 p^3 + \ft13 (p^3)^3) } \;.
\ee
Note that the $(p^3)^3$ term is due to the one-loop contribution
to the heterotic prepotential, i.e. to the two-derivative part
of the action. This is the most simple example of a
quantum effect that modifies the entropy of a black hole.
From the IIA perspective the prepotential is
purely classical, whereas the higher derivative couplings
are loop effects. 

With some more effort one can treat the $\alpha'$-loop correction
to the $STU$ prepotential exactly. More generally we can consider 
axion-free black holes in a theory with prepotential
\be
F = D_{ABC} \frac{Y^A Y^B Y^C}{Y^0} + i c (Y^0)^2 \;,
\ee
where the coefficients $D_{ABC}$ are arbitrary and need not
correspond to a perturbative heterotic model. The constant
$c$ is chosen to be real because its imaginary part could be
removed by a symplectic transformation. In IIA models
$D_{ABC}$ is proportional to the triple intersection numbers whereas
$c$ is proportional to the Euler number. The $STU$ model has
\be
D_{ABC} = - \ft16 \ve_{ABC} - \ft13 \d_{A3} \d_{B3} \d_{C3}
\mbox{   and   }
c = \frac{\chi \zeta(3)}{16 \pi^3}  \;,
\ee
where the Euler number of the dual Calabi-Yau space is 
$2(h_{1,1} - h_{2,1}) = 2 \cdot ( 3 - 243 ) = - 480$.

Using the stabilization equations and proceeding like in
the previous IIA discussion of non-axionic black holes we find that 
the entropy is given by
\be
{\cal S} = - 2 \pi (q_0 - 2 c \lambda) \left[
\lambda + \frac{(p^0)^2}{\lambda} \right]\;,
\ee
where $\lambda$ is now a solution of the cubic equation
\be
D_{ABC}p^A p^B p^C + 2 c \lambda^3 = q_0 \lambda^2 \;.
\ee
This equation can be solved explicitly, but the resulting
expressions are not very illuminating, and therefore we will not
write them down. One can explicitly verify that the corrections
due to the constant terms are small in the large volume limit,
by expanding the solution of the full cubic equation in terms
of the approximate solution with $c=0$. In the general non-axionic
case only the charges $q_0,p^0,p^A$ are independent whereas
the $q_A$ are determined by
\be
q_A = - 9\frac{p^0}{\lambda^2} D_{ABC}p^B p^C \;.
\ee
This can be understood in terms of the underlying reality
constraint that restricts the moduli to purely real values
(in the heterotic parametrization by $S$, $T^a$). Since there are
$2N_V + 2$ charges but only $2N_V$ real moduli, one has $N_V$
real constraints, leaving $N_V+2$ independent charges
$q_0,p^0,p^A$. 

If one imposes in addition $p^0=0$ it follows that 
$q_A=0$ and the nonvanishing changes are $q_0,p^A$.\footnote{When
solving the stabilization equations for the explicit form of the 
entropy and of the moduli one has to discuss this case separately, because
one cannot divide by $p^0=0$.}

One can also discuss black hole solutions based on the same 
prepotential with the constraint $\mbox{Re} \; Y^0$. In this case
there is no dependence on $c$ and we get nothing new compared to
the previous discussion.

\subsubsection{World-Sheet Instanton Corrections to the Heterotic
Black Hole Entropy}

One might wonder whether it is possible to derive explicit results
when the world sheet instanton corrections are taken into account.
It turns out that one can get concrete results when expanding
the prepotential around one of the special loci, where 
a particular contribution to $h(T^a)$ is dominant. 
This was discussed in \cite{BehGai:1997} for the $STU$ and in
\cite{BehCarGai:1997} for the $ST$ model with the result that one gets
corrections to the entropy that depend logarithmically on the
charges, but in such a way that the entropy is invariant under
monodromy transformations. One sees explicitly that the entropy,
in the IIA language, depends on the rational instanton numbers.

One other point about higher $\alpha'$-corrections is noteworthy.
The higher $\alpha'$-corrections have the particular feature that
they introduce transcendental numbers such as $\zeta(3)$ into the
entropy formula. It is difficult to imagine how such 
numbers could be understood in terms of counting internal excitations.
In \cite{BehGai:1997,BehCarGai:1997} a microscopic 
model for the $\alpha'$-loop contribution
was proposed in terms of a gas of membranes. But
it was remarked in \cite{Moo:1998} that special values of the 
$\zeta$-function such as $\zeta(3)$ are related to the world sheet
instanton coefficients. Thus it might be that
one has to take into account all such terms simultanously and 
that the full result then yields a rational contribution.
In conclusion the issue of $\alpha'$-corrections is not yet properly
understood and our later discussion of the microscopic entropy will
exclusively concern contributions at $\alpha'$-tree level.

\subsubsection{Higher Curvature Contributions to the Heterotic Black 
Hole Entropy}

So far our discussion of heterotic black holes was 
based on the prepotential. We will now consider the effect
of higher curvature corrections \cite{CardWMoh:1999/06}. 
Again, this is technically
a special case of our discussion of IIA black holes, but
interesting regarding the physical content. In particular
the heterotic higher coupling functions are modular 
forms and we expect that the entropy is T-duality invariant.

Let us first consider a general heterotic model at tree level 
in the string coupling, which is defined by the function
\bea
F(Y,\Upsilon) = - \frac{Y^1 Y^a \eta_{ab} 
Y^b}{Y^0} + c_1 \; \frac{Y^1}{Y^0} \; \Upsilon \;\;.
\label{hetprep}
\eea
Here 
\bea
 Y^a \eta_{ab} Y^b = Y^2 Y^3 - \sum_{a=4}^{n} (Y^a)^2 \;,\qquad 
a=2, \ldots, n \;,
\eea
with real constants $\eta_{ab}$ and $c_1$. 
The calculation 
goes through as in the case $\Upsilon=0$,
with minor modifications. Using the
stabilization equations we first derive
\bea
{\cal S} = \ft{1}{2}\, \pi \, (S + {\bar S})
\, \Big(\langle N,N \rangle - 512 \, c_1 \,   \Big)
\;. 
\label{entros}
\eea
The value of the dilaton at the horizon is found to be
\bea
S  &=&  
\sqrt{ {\langle M,M \rangle \langle N,N \rangle - ( M \cdot N )^2}\over
{\langle N,N \rangle\,(\langle N,N \rangle  - 512 \,c_1) } } \; + \; i \;
\frac{ M \cdot N }{ \langle N,N \rangle}
\;\;\;, \nonumber\\
\eea
and the entropy therefore is
\bea
{\cal S} &=& \pi \; \sqrt{ \langle M,M \rangle 
 \langle N,N \rangle  - ( M \cdot N )^2}
\;  \sqrt{1 - \frac{512 \,c_1 \,}{\langle N,N \rangle}}
\;\;\;,
\label{entrofin}
\eea
where we used the classical T-duality invariants.
These formulae reduce properly to the ones derived from the tree level
prepotential. Moreover they are manifestly T-duality invariant.
As in the case of the heterotic tree level black holes 
we can regard this as the truncation of an ${\cal N}=4$ model.
Conversely we can lift our ${\cal N}=2$ result to the full
${\cal N}=4$ theory by replacing $O(2,N_V-1,\mathbb{Z})$
by the T-duality group $O(6,22,\mathbb{Z})$ of   
a heterotic ${\cal N}=4$ compactification. We will return to 
${\cal N}=4$ compactifications in the next section.

The tree level result can now be extended to more general models
with higher curvature corrections
in the same way as we did in the IIA case. Let us then continue
the discussion of black holes with $\mbox{Re}\;Y^0=0$, that we 
started when considering IIA black holes. 
We will  restrict ourselves
to type-II models with a dual heterotic description, 
\be
D_{ABC}\, Y^A Y^B Y^C = - Y^1 Y^a \eta_{ab} Y^b \;\;, \qquad 
D_{A} Y^A = c_1 Y^1 \,,
\ee
where $D_{1ab} = -\ft13 \eta_{ab}$. 
The first equation in (\ref{stabrmy}) yields 
\bea
Y^a&=& {1\over Y^1+\bar Y^1} \,\Big[ \ft14 p^0\,\eta^{ab}q_b + i \bar
Y^1\,p^a\Big]\,, \nonumber\\
4c_1\,\Upsilon + p^0q_1 + p^a\eta_{ab}p^b &=& {1\over (Y^1+\bar Y^1)^2}
\, \Big[\ft14 (p^0)^2 \,q_a\eta^{ab}q_b + (p^1)^2 p^a\eta_{ab}p^b + p^0
p^1\,p^aq_a \Big] \,.\quad~
\eea
Substituting this into the above entropy formula, we obtain
 \bea
{\cal S} &=& -{2\pi\over p^0(Y^1+\bar Y^1)} \,\Big[\ft14 (p^0)^2
\,q_a\eta^{ab}q_b + (p^1)^2 p^a\eta_{ab}p^b + p^0 p^1\,p^aq_a \Big]
-2i\pi (G-\bar G) \nonumber\\ 
&=&  -{\pi\over p^0} 
\sqrt{\Big(
(p^0)^2 q_a \eta^{ab} q_b + 4 (p^1)^2 p^a \eta_{ab} p^b + 4 p^0 p^1 q_a p^a 
\Big)
\Big( 
q_1 p^0 + p^a \eta_{ab} p^b + 4 c_1  \Upsilon
\Big)}  \nonumber\\
&& -2i\pi (G-\bar G) \,.
\eea
We thus see that we have to choose $p^0 < 0$.
This expression can be rewritten as follows 
in terms of the heterotic electric and magnetic
charges $M_I$ and $N^I$,
\begin{eqnarray}
{\cal S} &=& -{\pi\over p^0}  \sqrt{ \left( (p^0)^2 \la M, M \ra + (p^1)^2
\la N,N \ra + 2 p^0 {p^1} M \cdot N \right) 
\left( \la N,N \ra + 8 c_1 \Upsilon \right) } \nonumber \\
 & & - 2 i \pi (G - \bar{G} ) \,. \label{hetentg}
\end{eqnarray}
Observe that the charges are subject to the constraint
(\ref{charconstr}), which in the  
case at hand reads 
$p^0 M \cdot N + p^1 \langle N,N \rangle = 
-i (p^0)^2 \,(G_0-\bar G_0)$. 
Substituting this into (\ref{hetentg}) yields
\begin{eqnarray}
{\cal S} &=& \pi \sqrt{ \la M, M \ra \la N,N \ra - (M\cdot N)^2
- (p^0)^2 (G_0 - {\bar G}_0)^2 } \;
\sqrt{ 1 -  \frac{512 \, c_1 }{\la N,N \ra} } \nonumber \\ 
 & & - 2 i \pi (G - {\bar G}) \;, 
\end{eqnarray}
where we also used $\Upsilon = -64$. This expression reduces 
to (\ref{entrofin}) in the case of $G=0$.

\subsubsection{Non-Perturbative and Non-Holomorphic Contributions to the
Heterotic Black Hole Entropy}

We already remarked that tree level
${\cal N}=2$ heterotic black holes can be considered as special cases of
tree level ${\cal N}=4$ black holes. We now return to this subject
and discuss non-perturbative aspects of such black 
holes \cite{CardWMoh:1999/06}. This 
provides an example where the entropy is modified through non-holomorphic
contributions to the gravitational couplings.

First recall that in ${\cal N}=4$ models (and in (${\cal N}=4$)- like
${\cal N}=2$ models such as the FHSV model)
the minimal terms in the Lagrangian are exact, 
whereas the $C^2$-coupling $F^{(1)}(Y)$ is modified
by non-perturbative corrections. The reason is that the
dilaton dependence of $F^1(Y)$ is given by
\be
{\cal F}^{(1)}(S) = 24 S \,,
\ee
which is not invariant under S-duality. Therefore the tree level
result has to be modified by non-perturbative contributions such
that the full function $F^{(1)}(S)$ is compatible with 
S-duality.

Let us discuss this in the ${\cal N}=2$ language by considering
a function of the form
\bea
F(Y,\Upsilon) = - \frac{Y^1 Y^a \eta_{ab} 
Y^b}{Y^0} + \; F^{(1)} (S) \;  \Upsilon \;.
\label{hetprepnp}
\eea
The discussion is most easily done in the heterotic basis
$(Y^0, F_1, Y^2, \ldots, F_0, - Y^1, F_2, \ldots)$.
As a first step one can verify that the invariance under the T-duality
group $O(2,N_V-1,\mathbb{Z})$ holds irrespective of the 
form of $F^{(1)}(S)$. (See \cite{CardWMoh:1999/06} for a 
more detailed account.)

Next one studies S-duality transformations
\be
S \rightarrow  \frac{aS -ib}{icS +d} \;.
\label{SdualityOfS}
\ee
Such a transformation is induced by the following symplectic transformation
of the section:
\be
\begin{array}{rcl}
Y^0 &\to& \widetilde Y^0= {d}\, Y^0 +{c} \,Y^1 \,, \\
Y^1 &\to& \widetilde Y^1 = {a}\, Y^1 +  {b}\, Y^0 \,, \\ 
\qquad Y^a &\to& \widetilde Y^a = {d} \,Y^a  -\ft12 \, {c}\, 
\eta^{ab}\,F_b \,, 
\end{array} 
\qquad
\begin{array}{rcl}  
F_0 &\to& \widetilde F_0= {a}\, F_0 -{b} \,F_1 \,, \\
F_1 &\to& \widetilde F_1 = {d}\, F_1 -{c}\, F_0 \,, \\
F_a &\to& \widetilde F_a = {a}\, F_a - 2 \, {b}\, \eta_{ab}\,Y^b\,.
\end{array}
\label{SdualityOfSection}
\ee
The corresponding action on the charges is
\be
M_I \to \widetilde M_I ={a}\, M_I - 2\, 
{b}\,\eta_{IJ} N^J \,,\qquad N^I \to
\widetilde N^I =  { d}\, N^I -\ft12 \, {c}\, \eta^{IJ} M_J \, ,
\label{mntilde}
\ee
implying that the T-duality invariant combinations of charges transform
under S-duality according to
\bea
\langle M,M \rangle & \to &  { a^2} \,\langle M,M \rangle 
+ { b^2} \, \langle N,N
\rangle - 2 \,{ab} \, M\cdot N \,, \nonumber \\ 
\langle N,N \rangle & \to & {c^2} \,\langle M,M \rangle 
+ {d^2} \, \langle N,N
\rangle - 2 \,{ cd} \, M\cdot N \,, \nonumber \\
M\cdot N & \to & - {ac}\,\langle M,M \rangle - {bd} \, \langle N,N
\rangle  + ({ ad + bc}) \, M\cdot N \,.
\label{mtilde}
\eea
We have now to take into account that the $F_I$ are not independent objects,
but functions of $S$ and the moduli. Therefore the transformation
behaviour (\ref{SdualityOfSection}) 
of the $F_I$ has to be induced by the transformations
of the $Y^I$.
The crucial observation is that the S-duality 
transformation (\ref{SdualityOfS})
does in general not induce the correct transformation behaviour
of the $F_I$, but that the correct transformation follows if
$f(S) = -i \der F^{(1)}/ \der S$ is a modular function of weight 2,
\be
f(S) \rightarrow (icS +d)^2 f(S) .
\label{fSweight2}
\ee
It is well known that the tree level function $F^{(1)}(S)=ic_1S$ indeed
receives corrections that are positive powers of $e^{-S}$ that are
not visible in perturbation theory and can complete $F^{(1)}(S)$ 
to a covariant object \cite{HarMoo:1996}. 
In ${\cal N}=2$ models (including the FHSV model)
there are also
moduli dependent corrections \cite{HarMoo:1996a}, but our focus 
here is S-duality 
in ${\cal N}=4$ models, where such corrections are absent. Therefore 
the function $F^{(1)}(S)$ does not depend on geometric moduli.
The full function $F^{(1)}(S)$ cannot be 
{\em holomorphic}  and satisfy property 
(\ref{fSweight2}) at the same time,
because there are no (holomorphic) modular forms of weight 2.
There are, however, non-holomorphic objects with the desired transformation
behaviour. We arrive at the same conclusion as in the discussion of
the T-duality properties of higher couplings: The physical 
couplings, which are invariant under all symmetries of the theory differ
by non-holomorphic terms from the Wilsonian couplings, which are 
holomorphic, but do not necessarily have all symmetries.

Therefore we have to replace the holomorphic function $F^{(1)}(S)$ by
a non-holomorphic function $F^{(1)}(S, \ov{S})$. A concrete example
of a non-holomorphic, but covariant function is provided by
\be
F^{(1)}(S,\ov{S}) = - i c_1 \ft6{\pi} ( \log \eta^2(S) +
\log(S + \ov{S}) )\;, 
\ee
where $\eta(S)$ is the Dedekind $\eta$-function. This function reduces
to $ic_1S$ in the limit $S\rightarrow \infty$, whereas
\be
f(S,\ov{S}) = -i \der_S F^{(1)}(S,\ov{S}) = c_1 \ft3{\pi^2}
G_2(S, \ov{S})
\ee
is a non-holomorphic modular form of weight two. Here
\be
G_2(S, \ov{S}) = G_2(S) - \frac{2 \pi}{ S + \ov{S}}
\ee
is the non-holomorphic, but modular covariant second Eisenstein 
series, whereas
\be
G_2(S) = - 4 \pi \der_S \log \eta(S)
\ee
is the holomorphic but non-covariant second Eisenstein series.
Note that the function $F^{(1)}(S,\ov{S})$ is not strictly invariant
but transforms as
\be
F^{(1)}(S,\ov{S})  \rightarrow F^{(1)}(S,\ov{S})
+ i c_1 \ft{6}{\pi} \log( - i c \ov{S} +d ) \;.
\ee
This does not change the gravitational coupling but acts as
a graviational $\theta$-shift. Also note that above we ignored
possible anti-holomorphic contributions, i.e. terms holomorphic
in $\ov{S}$, which vanish in the perturbative limit. In comparison
to our discussion in chapter \ref{ChapterFourDimensionalStringsAndM}
we have identified the
non-holomorphic part of $F^{(1)}(S,\overline{S})$ 
by symmetry arguments.

One can now analyse the stabilization equations based on the function
\bea
F(Y,\Upsilon) = - \frac{Y^1 Y^a \eta_{ab} 
Y^b}{Y^0} + \; F^{(1)} (S,\ov{S}) \;  \Upsilon \;.
\eea
The resulting 
\be
|Z|^2 =  {(M_I + 2 i \, S \, \eta_{IK} \,N^K) \, \eta^{IJ} \, 
(M_J - 2 i \, \bar S \,\eta_{JL} N^L)\over 2(S + \bar S)} 
\ee
is manifestly S- and T-duality invariant.
When computing the entropy one finds that it is not invariant under
S-duality. This indicates that besides the non-holomorphic contribution
to the prepotential there must be an additional non-holomorphic
contribution to the physical $C^2$-coupling. S-duality is restored if the
physical $C^2$ coupling is given by $F^{(1)}(S,\ov{S}) + i c_1 \ft3{\pi}
\log( S + \ov{S})$. Then the manifestly S- and T-duality invariant
formula for the entropy is
\bea
{\cal  S} &=& \pi\, \Big[ \; |Z|^2 
 +4 \, {\rm Im} \,\Big( \Upsilon\, F^{(1)} (S, {\bar S}) + i \, c_1 \,
 \ft{3 }{\pi} \, \Upsilon  \log (S + {\bar S}) 
 \Big)\; \Big] \nonumber \\
&=& { \pi\over 2} \, {(M_I + 2 i \, S \, \eta_{IK} \,N^K) \, \eta^{IJ} \, 
(M_J - 2 i \, \bar S \,\eta_{JL} N^L)\over S + \bar S} + 768\,c_1 
\, \log \Big[ (S+\bar S)\,|\eta(S)|^4\Big]  \,.\quad~ 
\label{entropiaynonholo}
\eea
This formula depends implicitly on the function $f(S,\ov{S})$ through
the equation
\be
f(S,\bar S)\,\Upsilon = \ft{1}{4}\, \frac{1}{(S + \bar S)^2}\, (M_I - 2 i \,
\bar S\, \eta_{IK} \,N^K) \, \eta^{IJ} \, 
(M_J - 2 i \,
\bar S \,\eta_{JL} N^L) \,, \label{sn2}
\ee
which determines $S$ in terms of the charges.
This equation can be solved iteratively in $\Upsilon$. In the case
of a purely imaginary dilaton, the tree level solution
$S = i M \cdot N / \la N, N \ra$ is exact.

\section{The Microscopic Black Hole Entropy\label{SectionMicroEntropy}}

\subsection{Computation of the Microscopic Black Hole Entropy}

In this section we will review the derivation of the microscopic
black hole entropy. The microscopic picture of a four-dimensional
black hole is obtained by embedding it into ten-dimensional string
theory or eleven-dimensional M-theory. These higher dimensional
theories have $p$-brane solitons, which can upon compactification
reduce to four-dimensional black holes. Type II $p$-branes with
Ramond-Ramond charge provide the low energy description of the
type II D-branes that one can introduce in string perturbation 
theory \cite{Pol:1995}. Therefore $p$-branes with Ramond-Ramond charges
are themselves called D-branes or
Dp-branes for simplicity. The p-brane solitons of eleven-dimensional
supergravity are called Mp-branes. The correspondence between
D-branes and Ramond-Ramond charged p-branes can be used to 
compute the microscopic entropy of five- and four-dimensional
extremal black holes. Starting from the work of \cite{StrVaf:1996},
which introduced the method and for the first time achieved 
quantitative agreement between macroscopic and microscopic
black hole entropy, this has been extended to 
five- and four-dimensional black holes in compactifications 
with ${\cal N}=8,4,2$ supersymmetry. The D-brane model
also allows to study the entropy of near-extremal black holes,
of Hawking radiation and of greybody factors. We refer to 
\cite{Mal:1996} for a review and references. The microscopic
entropy of four-dimensional ${\cal N}=2$ black holes was computed
in \cite{MalStrWit:1997,Vaf:1997}.\footnote{The relevant brane
configurations and the microscopic interpretation of the leading
term were already discussed in \cite{BehMoh:1996,Mal:1996}}

A $p$-brane is a solution of the effective supergravity equations of
motion, which is translationally invariant along $p$ space directions
and behaves in the transversal directions like a charged black hole.
We refer to \cite{Ste:1997} for an extensive review.
There is an extremal limit where the tension of a $p$-brane equals
its charge(density), and in this limit the solution has Killing spinors
and is a BPS state of the underlying supersymmetry algebra. Such extremal
$p$-branes are higher-dimensional generalizations of the extremal
Reissner-Nordstr{\o}m black hole. As usual for BPS states one can have
multi-centered solutions, which describe static configurations of 
extremal $p$-branes located at arbitrary positions. 

In order to describe black holes of ${\cal N}=2$ supergravity in terms
of $p$-branes of type II string theory or M-theory, one has to 
compactify $p$-branes on Calabi-Yau manifolds. Since this is complicated
we first illustrate the structure of the configurations in the context
of toroidal compactifications of type II or M-theory. 
It turns out that the toroidal
compactification of
a single $p$-brane yields a four-dimensional black hole with vanishing
event horizon. This singularity is a compactification artefact which
signals that close to the horizon the solution does not make sense
as a four-dimensional geometry. In order to describe four-dimensional
black holes with a finite event horizon one has to use more complicated
$p$-brane bound states. Here one makes use of the fact that specific
combinations of $p$-branes (or other BPS solitons) still have Killing
spinors and therefore are themselves BPS states. Four-dimensional
black holes are obtained by combining four different species of 
BPS solitons. Here $p$-branes are refered to as different species if
they yield different charges by compactification. This is the case 
when they have different values of $p$, or if they are wraped on
homologically different cycles. In order to have a finite horizon
the configuration must have precisely four Killing spinors, 
i.e. it breaks $7/8$ of the
32 supersymmetries of the higher-dimensional vacuum.

We will now give two explicit examples which have generalizations
to Calabi-Yau compactifications. The first one is a IIA configuration
of three D4-branes with charges $p^1, p^2, p^3$ and $q_0$ D0-branes.
The geometry is such that the $D4$ branes are wraped on the three
different four-cycles of the internal six-torus, such that they
mutually intersect transversely 
on two-cycles and triple-intersect over a zero-cycle.
The corresponding ten-dimensional string frame metric is
\cite{Tse:1996a}
\bea
ds^2_{10} &=& \frac{-1}{\sqrt{H_0 H^1 H^2 H^3}} dt^2 + 
\sqrt{H_0 H^1 H^2 H^3}  d{\bf x}^2 \nonumber \\
& & + \sqrt{ \frac{H_0 H^1}{H^2 H^3}} ( dy_1^2 + dy_2^2)
+ \sqrt{ \frac{H_0 H^2}{H^1 H^3}} ( dy_3^2 + dy_4^2) 
+ \sqrt{ \frac{H_0 H^3}{H^1 H^2}} ( dy_5^2 + dy_6^2) \;,\\
\nonumber
\eea
where $y_i$ are coordinates along the torus directions and 
$x_m$ are coordinates along the three space directions.
The functions $H_0, H^A$ only depend on the radius $r = \sqrt{x_1^2 +
x_2^2 + x_3^3}$ and are harmonic functions. By toroidal compactification
one obtains an extremal  four-dimensional 
black hole with charges $q_0, p^A$ and metric
\be
ds^2_{4} = \frac{-1}{\sqrt{H_0 H^1 H^2 H^3}} dt^2 + 
\sqrt{H_0 H^1 H^2 H^3} d{\bf x}^2 \,.
\label{MetricToroidalBH}
\ee
The charges are related to the harmonic function as in
chapter \ref{ChapterN=2BlackHoles}. String and Einstein frame
coincide because the four-dimensional IIA dilaton is constant
for this solution.
We refer to \cite{BehGaiLueMahMoh:1997} for a more detailed 
discussion.

From the M-theory point of view $D0$ branes are Kaluza-Klein modes
of the eleven-dimensional supergravity multiplet along the M-circle,
whereas D4 branes result from wraping one dimension of the 
M5-brane on the M-circle. This explains the structure of the 
M-theory configuration that describes the same four-dimensional black hole
from the eleven-dimensional point of view. One takes a configuration of
three M5-branes with charges $p^A$
which mutually intersect transversely on 
three cycles and triple-intersect over a string (one-cycle).
One now has to compactify on the M-circle and on a six-torus.
The configuration is wraped such that the string wraps on the M-circle.
If the radius of the M-circle is taken to be 
much larger then the six-torus, 
the configuration describes a five-dimensional black string. In order
to account for the D0 branes of the corresponding IIA configuration one
has to put $|q_0|$ quanta of lightlike, left-moving (for definiteness)
momentum on the string. In four dimensions one obtains the same black hole
as before with charges $q_0,p^A$. The corresponding eleven-dimensional
metric is \cite{Tse:1996}
\bea
ds_{11}^2 &=& \frac{1}{ ( H^1 H^2 H^3)^{1/3} } 
\big( du dv + H_0 du^2 + H^1 H^2 H^3 d{\bf x}^2  \nonumber \\
 & & + H^1 ( dy_1^2 + dy_2^2 ) 
+ H^2 ( dy_3^2 + dy_4^2 ) 
+ H^3 ( dy_5^2 + dy_6^2 ) \big) \;, \\
\nonumber
\eea
where $u,v$ are light cone coordinates involving time and the 
M-circle direction.

The above pattern of compactification generalizes to compactifications
on $K3 \times T^2$ and on Calabi-Yau threefolds, and then describes
extremal black holes of ${\cal N }=4$ and ${\cal N}=2$ supergravity, 
respectively. Like the six-torus $T^6$ these manifolds posess non-trivial
four-cycles on which the above brane configurations can be wraped,
and the four-cycles intersect over two-cycles and triple-intersect
over zero-cycles. A configuration with charges $p^A$ corresponds to
wraping the D4 or M5 branes on a four-cycle ${\cal P}$ in the homology
class $p^A [\Sigma_A]$, where $\Sigma_A$ is a basis of $H_4(X,\mathbb{Z})$.
Note that $A$ now runs over the number of homology generators, which
equals the Betti-number $b_2(X) = b_4(X)$, where $X$ is either a
Calabi-Yau threefold, $K3 \times T^2$ or $T^6$. The four-cycle
has to be holomorphic in order that the configuration is a
BPS state \cite{BecBecStr:1995}. The validity of both the macroscopic
black hole solutions and the state counting
require that we are working in a limit where both $\alpha'$-corrections
and space-time loops are suppressed. This implies that one has to take all 
the charges to be large, and $|q_0|$ must be much 
larger than all the other charges.
Thus, one has to impose the same hierarchy of charges both from the
macroscopic and the microscopic point of view:
\be
|q_0| >> p^A >> 0 \;.
\ee
The fact that the $p^A$ are large implies that ${\cal P}$ is,
in the language of algebraic geometry, a very ample divisor of $X$.
Technically this means that $X$ can be embedded into a higher dimensional
projective space, such that ${\cal P}$ is a hyperplane section.
Very ample divisors have various nice properties that have to be
used for the state counting.

The divisor ${\cal P}$ is not a rigid object but can be holomorphically
deformed inside $X$. For very ample divisors
a generically chosen representative of such a family
is always smooth and therefore is a compact complex K\"ahler surface.  
The state counting will be done for such smooth choices of ${\cal P}$.
The concrete cycle that we described in the context of toroidal
compactification is not smooth, because it is a sum of
cycles which mutually intersect. At the intersections
the cycle is not a manifold.
But if the charges $p^A$ are sufficiently big, it is guaranteed that
every generic deformation of this configuration is smooth. 
The adaequate picture for the state counting is that one wraps
a single D4- or M5-brane on a smooth complex K\"ahler surface of
the homology type specified by the charges.

The state counting has been performed both in the M-theory and in the
type IIA picture \cite{MalStrWit:1997,Vaf:1997}. 
We will review the M-theory analysis of \cite{MalStrWit:1997}
in the following and we treat the cases of $T^6$, $K3 \times T^2$ and
Calabi-Yau threefolds in parallel. A more detailed presentation can be
found in \cite{CardWMoh:1999/06}. The M-theory analysis starts with the 
world volume theory of the M5-brane. The zero modes of the 
M5-brane fit into a tensor multiplet of six-dimensional $(0,2)$
supersymmetry. One takes the radius of the M-circle to be much larger
than the radius of the Calabi-Yau manifold and compactifies the
M5-brane theory on the divisor ${\cal P}$. The resulting two-dimensional
effective theory is a superconformal theory with $(0,4)$ supersymmetry.
As discussed above the black hole does not correspond to the ground
state of this theory but to an excited BPS state where one has
switched on $|q_0|$ quanta of left-moving momentum. In the limit of
very large $|q_0|$ the asymptotic density of states $d$ and the
corresponding microscopic entropy are given by Cardy's formula
\be
d \simeq \exp({\cal S}_{\mscr{micro}}) \simeq 
\exp \left( 2 \pi \sqrt{\ft16 |q_0| c_L } \right) \;,
\ee
where $c_L$ is the left-moving central charge of the two-dimensional
effective M5-brane theory. Thus the entropy is determined by 
counting in how many ways the $|q_0|$ quanta of momentum can be
distributed among the zero mode 
excitations of the M5-brane. The central charge
$c_L$ encodes how many distinct degrees of freedom exist. This
number has to be determined by counting the zero modes of 
the compactification of the M5-brane theory on ${\cal P}$.

The six-dimensional $(0,2)$ tensor multiplet contains a selfdual
antisymmetric tensor, five scalars, which describe the motion of
the M5 brane along the five transverse directions and two
Weyl fermions. Since ${\cal P}$ is a K\"ahler manifold one 
can express the numbers of the corresponding two-dimensional zero modes
in terms of the independent Hodge numbers of ${\cal P}$, which are
$h_{1,0}({\cal P}), h_{2,0}({\cal P}) $ and $h_{1,1}({\cal P})$.
The numbers $b^{\pm}_2({\cal P})$ 
of selfdual and antiselfdual harmonic two-forms are given in terms
of Hodge numbers by
$b_2^+({\cal P}) = 2 h_{2,0}({\cal P}) + 1$ and
$b_2^-({\cal P}) = h_{1,1}({\cal P}) -1$.

By dimensional reduction of the antisymmetric tensor one gets
$b_2^-({\cal P})$ left-moving and $b_2^+({\cal P})$ right-moving 
scalars. Three of the five directions transverse to the
M5-brane are non-compact, whereas the other two directions
lie inside the Calabi-Yau space. The zero modes associated
with these directions correspond to deformations of ${\cal P}$
inside $X$, i.e. they are sections of the normal bundle of
${\cal P}$. For a very ample divisor the number of these sections
can be counted by a Riemann-Roch theorem, with the result that
one gets $2h_{2,0}({\cal P}) - 2 h_{1,0}({\cal P})$ left- and
right-moving scalars. Finally the compactification of the 
2 six-dimensional Weyl spinors yields left- and right-moving 
two-dimensional fermions that are counted by the sums of the 
odd and even Betti numbers, respectively. 
In summary one gets the following
numbers of massless two-dimensional fields:
\bea
N^{\mscr{left}}_{\mscr{bosonic}} &=& 2 h_{2,0}({\cal P}) + h_{1,1}({\cal P})
+2 - 2 h_{1,0}({\cal P}) \nonumber \;,\\
N^{\mscr{right}}_{\mscr{bosonic}} &=&
4 h_{2,0}({\cal P}) + 4 - 2 h_{1,0}({\cal P}) \nonumber \;,\\
N^{\mscr{left}}_{\mscr{fermionic}} &=& 4 h_{1,0}({\cal P}) \,, \nonumber\\
N^{\mscr{right}}_{\mscr{fermionic}} &=& 4 h_{2,0}({\cal P})
+ 4 \,,  \\
\nonumber
\eea
where we used $h_{0,0}({\cal P}) = 1$.

The next step is to express the Hodge numbers of ${\cal P}$ in terms
of topological data of $X$ and the homology of ${\cal P}$. 
The topological data are the triple intersection form $C_{ABC}$,
the components $c_{2A}$ of the second Chern class of $X$ and
the Hodge number $h_{1,0}(X)$. The homology class of ${\cal P}$
is given by the charges $p^A$. Then, the Hodge numbers of ${\cal P}$
are given by
\bea
h_{2,0}({\cal P}) &=& \ft16 C_{ABC} p^A p^B p^C + \ft1{12}
c_{2A}p^A + h_{1,0}({\cal P}) -1 \;,\nonumber \\
h_{1,1}({\cal P}) &=& \ft23 C_{ABC} p^A p^B p^C + \ft56 c_{2A}p^A
+ 2 h_{1,0}({\cal P}) \nonumber \;,\\ 
h_{1,0}({\cal P}) &=& h_{1,0}(X) \;.  \\
\nonumber
\eea
The first two lines follow from the index theorems for the Euler number
and Hirzebruch signature, whereas the last line is the Lefshetz hyperplane
theorem applied to a very ample divisor.

We can now write down the left- and right-moving central charges
$c_{L/R} = N^{\mscr{left/right}}_{\mscr{bosonic}} + \ft12 
N^{\mscr{left/right}}_{\mscr{fermionic}}$:
\bea
c_L &=& C_{ABC} p^A p^B p^C + c_{2A} p^A + 4 h_{1,0}(X)\;, \nonumber \\
c_R &=& C_{ABC} p^A p^B p^C + \ft12 c_{2A} p^A 
+ 4 h_{1,0}(X) \;.  \\
\nonumber
\eea
The corresponding entropy is
\be
{\cal S}_{\mscr{micro}} = 2 \pi \sqrt{\ft16 | q_0| \left( C_{ABC} p^A p^B
p^C + c_{2A} p^A + 4 h_{1,0}(X) \right)} \;.
\label{MicroWrong}
\ee
This is not yet our final result, due to a subtlety that concerns the
cases $K3 \times T^2$ and $T^6$, but is irrelevant for Calabi-Yau threefolds.
As we explained above the two-dimensional conformal field theory 
has to be a $(0,4)$ supersymmetric sigma-model, 
so that our configuration describes
a BPS state. This requires that the number of right-moving bosons and
fermions has to match. Moreover the right-moving scalars must 
parametrize a quaternionic manifold and therefore the number of 
right-moving real scalars should be a multiple of four.
If $X$ is a Calabi-Yau threefold, then $h_{1,0}(X) = 0$ and the
above constraints are satisfied. This is, however, not the case
for $K3 \times T^2$, which has $h_{1,0}(X) = 1$ and for $T^6$
which has $h_{1,0}(X) = 3$. Therefore our mode counting is not
consistent with supersymmetry. Note however that these two cases
differ from the case of a Calabi-Yau threefold in that one also
gets $h_{1,0}(X)$ gauge fields from the dimensional reduction of the
six-dimensional tensor field. These two-dimensional gauge fields are
not dynamical and therefore do not contribute to the entropy as
degrees of freedom. But they might modify the counting of the
other modes, provided that these modes are charged. If we assume
this, then the number of left- and right-moving scalars is reduced 
by $2h_{1,0}({\cal P})$ due to gauge symmetry. 
In order to get a supersymmetric right-moving spectrum 
the same mechanism must remove $4 h_{1,0}({\cal P})$ right-moving 
fermions. If one furthermore assumes that this mechanism is left-right
symmetric, then all left-moving $4 h_{1,0}({\cal P})$ fermions
are removed. As a result the entropy formula takes the form
\be
{\cal S}_{\mscr{micro}} = 2 \pi \sqrt{\ft16 | q_0| \left( C_{ABC} p^A p^B
p^C + c_{2A} p^A   \right)} \;.
\ee
Note that this formula also applies 
to Calabi-Yau threefolds, where $h_{1,0}({\cal P})=0$.

Besides two-dimensional supersymmetry we can invoke two further independent
arguments in favour of the modified state counting for $K3 \times T^2$ and
$T^6$. The first is microscopic in nature. Anomaly inflow arguments 
can be used to determine the numbers $c_L$ and $c_R$
and they yield the same values as our modified state counting 
\cite{FreHarMinMoo:1998}.
The second argument comes from comparison with the macroscopic
black hole entropy. As we showed in the beginning of this chapter
the entropy
must be an even function of the charges, implying that $c_L$ must
be odd in the charges. An entropy formula of the type
(\ref{MicroWrong}) with $h_{1,0}(X) \not=0$ is therefore not compatible
with a macroscopic entropy formual based on low energy supergravity.

Finally we note that one can generalize the above discussion from
black holes with charges $q_0,p^A$ to the more general case 
where only $p^0=0$, while the charges $q_A$ are non-vanishing.
This can be done by adding an M2-brane to the configuration, which
is wraped on a two-cycle in the homology class $q_A \Sigma^A$, where
the $\Sigma^A$ form a basis of $H_2(X,\mathbb{Z})$, \cite{Mal:1996}.
As discussed in \cite{Mal:1996,MalStrWit:1997} 
this has the effect that $q_0$ is replaced
by
\be
\widehat{q_0} = q_0 + \ft1{12} D^{AB} q_A q_B
\ee
in the entropy formula, where $D^{AB}$ is the inverse
of $D_{AB} = D_{ABC} p^C$. 
Thus the final entropy formula, which is valid for black holes
with $p^0=0$ at tree level in $\alpha'$ is
\be
{\cal S}_{\mscr{micro}} = 2 \pi \sqrt{\ft16 | \widehat{q}_0| 
\left( C_{ABC} p^A p^B
p^C + c_{2A} p^A   \right)} \;.
\label{MicroEntropy}
\ee

\subsection{Comparison of Macroscopic and Microscopic 
Results \label{Comparison}}

Comparing the formulae (\ref{EntropyCYmacro}) and (\ref{MicroEntropy})
we find that the macroscopic and microscopic
entropies agree. This is a highly non-trivial test of the microscopic
picture of black holes provided by string theory in view of the complications
we had to go through in order to derive these formulae.
On the macroscopic side we had in particular to include
the higher curvature terms using an elaborate formalism
and we had to take into account the modifications of the area law.
It is illustrative to look at certain points in more detail. The first
subtle point is that the model-independent entropy formula 
(\ref{ModIndEntropyFormula})
which was 
derived by substituting our near-horizon solution into Wald's formula
turned out to be symplectically covariant. Despite the fact that the
Lagrangian contains several terms involving the Riemann tensor we only
got one correction term to the area law, and this term involved
the function 
$F_{\widehat{A}}$, which is a symplectic function, in contradistinction
to almost all other expressions that one can form out of $F(X,\widehat{A})$.
All non-covariant contributions to Wald's formula either cancel or
vanish for the near-horizon solution. This illustrates that the 
generalized entropy
formula makes perfect sense in supersymmetric theories.

Our second remark concerns the explicit form of the two terms
in the entropy formula. It is very crucial that the higher curvature
terms enter in two different ways: First they explicitly modify the near
horizon solution and in particular the area $A$ of the event horizon
and second there is the modification of the area law itself. It is
instructive to write out both terms separately:
\bea
\pi |Z|^2 &=& 2 \pi \frac{ |\widehat{q}_0| \left(
\ft16 C_{ABC}p^A p^B p^C + \ft1{12} c_{2A}p^A \right)}{
\sqrt{ |\widehat{q}_0| \left( \ft16 C_{ABC} p^A p^B p^C + \ft16 c_{2A}p^A \right)
}} \label{AreaContribution} \\
\left. 4 \pi \mbox{Im} ( \Upsilon F_{\Upsilon} ) \right|_{\Upsilon=-64} &=&
2 \pi \frac{ \ft1{12} |\widehat{q}_0| c_{2A}p^A }{
\sqrt{ |\widehat{q}_0| \left( \ft16 C_{ABC} p^A p^B p^C + \ft16 c_{2A}p^A \right)
}} \label{WaldContribution} \\
\nonumber
\eea
Note that both single terms are more complicated than the full answer.
The fact that they combine into one single square root depends on the
precise prefactor of the term $T^{ij ab}D_a D^c T_{cbij}$ in the
highest component of the Weyl multiplet (\ref{WeylMultComp}). As we saw
in section \ref{SectionEntropyFormula} this is the only term besides
the Einstein-Hilbert term that actually contributes to the entropy.

Moreover when expanding the area term in $c_{2A}p^A$ and comparing
with the corresponding expansion of the microscopic entropy formula
we find a mismatch already in the leading correction term. In the
expansion of the area term the leading 
correction term even vanishes:
\be
\frac{A}{4} = \pi |Z|^2 = 2 \pi \sqrt{
\ft16 |\widehat{q}_0| C_{ABC} p^A p^B p^C} + 2 \pi \left( 
\ft1{12} - \ft1{12} \right) c_{2A} p^A \sqrt{ \frac{ 6 |\widehat{q}_0|}{
C_{ABC} p^A p^B p^C }} + \cdots \;,
\ee
whereas the expansion of (\ref{MicroEntropy}) yields
\be
{\cal S}_{\mscr{micro}} = 2 \pi \sqrt{
\ft16 |\widehat{q}_0| C_{ABC} p^A p^B p^C} +
2 \pi \cdot \ft1{12} \cdot c_{2A} p^A \sqrt{ \frac{
6 |\widehat{q}_0| }{C_{ABC} p^A p^B p^C }} + \cdots \;.
\label{ExpandMicro}
\ee
The observation that the area law under very general assumptions
cannot account for the microscopic entropy (\ref{MicroEntropy})
was made
in \cite{BehCardWLueMohSab:1998}. This then motivated the detailed
investigation \cite{CardWMoh:1998/12} which used
Wald's entropy formula and the off-shell formalism
of ${\cal N}=2$ supergravity to properly include the effect of
higher curvature terms.

Next we would like to point out that the correction is strictly
speaking not a curvature squared correction. 
As we saw in section \ref{SectionPoincareGaugeFixing}
the curvature squared term in the highest component of the Weyl multiplet 
involves the square of the Weyl tensor (see formula
\ref{WeylPowersInAction}). Since the near-horizon geometry is
conformally flat
this term vanishes at the event horizon and does not contribute
to the entropy. The only correction term to the area law 
instead comes from the term $T^{ij ab}D_a D^c T_{cbij}$
in the highest component of the Weyl multiplet which contains
terms linear in the Riemann tensor as we discussed in section
\ref{SectionEntropyFormula}.
The original proposal for a macroscopic origin of the $c_{2A}p^A$ term
in the entropy formula \cite{MalStrWit:1997}
was based on the observation that a 
term $c_{2A}z^A (R_{abcd})^2$ in the Lagrangian gives the leading order
term in the expansion (\ref{ExpandMicro})
of the microscopic entropy.\footnote{In \cite{MalStrWit:1997}
the entropy was computed using Euclidean methods. Such methods
are capable of treating deviations from the area law, and they
agree with Wald's formula in all cases where both methods can be
applied \cite{IyeWal:1995}.} Now the full supergravity analysis has shown
that this term combines with other terms and does not contribute
to the entropy at all, whereas the term 
$T^{ij ab}D_a D^c T_{cbij}$, which sits in the same component
$\widehat{C}$ of the Weyl multiplet does not only give the leading
correction but the full result.

Finally we comment on the case of black holes 
in ${\cal N}=4$ and ${\cal N}=8$ supergravity. Above 
we argued that the microscopic entropy formula (\ref{MicroEntropy})
applies to these cases as well. On the supergravity side we worked
with the off-shell formulation of ${\cal N}=2$ supergravity
coupled to vector multiplets, which
does not cover the full ${\cal N}=4,8$ theories. One can, however,
describe a subsector of these theories, and for black holes 
which only depend on fields of this subsector it is reasonable
to expect that our results apply. In particular we expect that
the entropy is even in the charges and that the macroscopic
entropy formula (\ref{EntropyCYmacro}) is valid.
With the modified state counting of the last section the
macroscopic and microscopic entropies agree in 
these cases as well.

Since the torus $T^6$ is flat the second Chern class vanishes and
the entropy formula reduces to the first term. The intersection
form of $T^6$ is $C_{ABC} = \ve_{ABC}$, where $A,B,C=1,2,3$, and we 
get the formula
\be
{\cal S}_{\mscr{micro}} = 2 \pi \sqrt{q_0 p^1 p^2 p^3} \;,
\label{MicroN=8}
\ee
which coincides with the Bekenstein-Hawking entropy extracted from
the metric (\ref{MetricToroidalBH}). Since higher curvature terms do not
contribute, the use of the area law is justified. A microscopic explanation
for this entropy formula was first given in 
\cite{LarWil:1995,CveTse:1995,Tse:1996}, based on the relation
between chiral null models and black hole solutions in toroidal 
compactifications. Note that this derivation does not use D-branes,
but gives the same result, as required by U-duality.

In the case of $K3 \times T^2$ the second Chern class is non-vanishing
and therefore there is a subleading term in the microscopic
entropy which corresponds to a higher curvature correction on the
supergravity side. We refer to \cite{CardWMoh:1999/06} 
for a more detailed account. By heterotic - type II duality the
same entropy formula applies to heterotic string compactifications
on $T^6$. The leading part of the entropy has of course 
the same structure (\ref{MicroN=8}) as a type II on $T^6$. But in 
addition there is a higher curvature correction.\footnote{Remember that for
heterotic models the coefficient of the leading term in the
function $F^{(1)}$ is not related to the second Chern class of the
internal manifold, but is a universal constant.}

In conclusion we see that the matching of the macroscopic and
microscopic entropy depends on many subtle details. Any small
mistake in one of the above points would result in a completely
different and much more complicated macroscopic entropy formula.
Thus the matching with the microscopic entropy is a strong 
argument in favour of the microscopic picture that string theory
provides for black holes. Complicated as it was, our investigation
was limited on the microscopic side to terms which are tree level
in $\alpha'$. A better microscopic understanding of stringy
$\alpha'$-corrections remains to be found.

\chapter{Summary, Discussion and Outlook \label{ChapterLast}}

At the end of this expedition through black hole entropy in supergravity
and string theory it might be useful to recall and list the key
results.

\begin{enumerate}
\item
Consider ${\cal N}=2$ supergravity with an arbitrary number
of vector multiplets and with a general function  $F(X,\widehat{A})$.
Then the most general static and spherically symmetric 
field configuration with full ${\cal N}=2$ supersymmetry has
the geometry $AdS^2 \times S^2$ and can be fully specified in 
terms of the field $Z$, (\ref{BRvacuum}, \ref{BRvacuum1}).
\item
By substituting this field configuration, which describes the
horizon of a BPS black hole, into the generalized entropy formula
(\ref{WaldEntropy})
one obtains the model-independent expression (\ref{ModIndEntropyFormula})
for the entropy, which specifies
the entropy in terms of the field $Z$. By model-independent we mean that
the formula holds for all possible prepotentials.
The formula is covariant with respect to 
symplectic transformations. 
\item
As a consequence of symplectic invariance the values of the 
scalar fields at the horizon are related 
to the charges by the stabilization equations (\ref{StabEquatWithR2}). 
By solving the
stabilization equations one can obtain an expression for
the entropy as a function of the charges.
\item 
The entropy is a series in even powers of the charges and
the $g$-th contribution is due to the $g$-th coupling function
$F^{(g)}(X)$, see (\ref{EntropyEvenPowers}).
\item
For black holes in type II compactifications the stabilization
equations can be solved in the large volume limit  in the three cases $p^0=0$ 
(\ref{EntropyPNull=Null}), axion-free
black holes (\ref{EntropyAxionFree}) and 
$\mbox{Re}\;Y^0=0$ (\ref{EntropyRe=0}). We can also include a certain
class of higher order terms, described by the function $G(Y^0,\Upsilon)$.
\item 
For black holes in heterotic compactifications the entropy is
T-duality invariant. Neglecting higher curvature terms
but taking into account all perturbative quantum corrections
the entropy is related to the perturbative string coupling,
evaluated at the horizon (\ref{EntropyCoupling}). 
Taking into account higher curvature 
corrections but working at tree level in $\alpha'$
one can derive an
explicit formula for the entropy in terms of the charges which
is manifestly invariant under T-duality (\ref{entrofin}).
\item
For black holes in ${\cal N}=4$ compactifications 
we have found the manifestly S- and T-duality invariant entropy 
formula (\ref{entropiaynonholo}).
This requires to take into account non-perturbative and non-holomorphic
contributions to the gravitational $C^2$-couplings. 
\item
For ${\cal N}=8,4,2$ compactifications on $T^6, K3 \times T^2$ and  
Calabi-Yau threefolds one can derive the microscopic entropy 
formula (\ref{MicroEntropy}) \cite{MalStrWit:1997,Vaf:1997}
by counting, in the M-theory picture,  the collective modes
of an M5-brane wrapped on a very ample divisor. The formula agrees
with the corresponding macroscopic entropy formula.
In the cases
${\cal N}=2$ and ${\cal N}=4$ one has a subleading term
which on the macroscopic side is a higher curvature correction.
One has to deviate from the Bekenstein-Hawking formula and to
use the generalized formula (\ref{WaldEntropy}) in order to
agree with the microscopic entropy formula.
\end{enumerate}

In conclusion we have shown that the macroscopic and microscopic
black hole entropies are equal for ${\cal N} \geq 2$ compactifications,
to leading order in $\alpha'$.
The understanding
of higher $\alpha'$-corrections remains to be achieved.
Moreover the state counting in the ${\cal N}=4$ case involves
indirect arguments, based on supersymmetry or anomaly considerations.
A direct derivation based on the effective M5-brane Lagrangian is
desirable. Moreover we have shown that the entropy of heterotic
black holes in T-duality invariant, and in the case of ${\cal N}=4$
compactifications, S-duality invariant, even in presence of non-trivial
$C^2$- and $\alpha'$-corrections.

An intriguing feature of the entropy is its close relation
to couplings in the Lagrangian. We have seen that there is a
general relation between the entropy and the series $F^{(g)}(X)$ of
gravitational couplings. In one example, where the higher couplings
were neglected, we could prove a relation which is valid to all
orders in $\alpha'$. The relation between black hole entropy and
the couplings is a natural one, once it is appreciated that the
entropy in general is not just the area of the event horizon, but
is given by a variation of the Lagrangian with respect to the Riemann
tensor, evaluated at the horizon. Our observation also fits 
with the fact that the black hole attractor mechanism and the 
corresponding flow of the scalar fields closely resemble fixed points
of the $\beta$-function and renormalization group flows. One
of the important developments of the last years was the renewed 
interest in the relation between string theory and gauge theories.
This has proceded in various steps from the discovery of the role
of D-branes \cite{Pol:1995}, through the matrix formulation of 
M-theory \cite{BanFisSheSus:1996}
to the AdS - CFT correspondence \cite{Mal:1997}. We think that a
deeper understanding of the modified entropy formula and of the
black hole attractor mechanism will help to improve our understanding
of both string theory and gauge theories.

It is important to realize that most of the results discussed in 
this work are robust: Though we have
no concrete doubts concerning specific details of string theory that
we mentioned or used in the paper, we are also aware that our picture of
string theory might change significantly in the next years. The 
discovery of string dualities is certainly a big step forward
but we have become aware at the same time 
that we do not understand well what string 
theory really is. Before the second string revolution it was clear
how to define string theory:
it was given by the perturbative 
quantization of the Polyakov action in a given background geometry
(and other classical background fields). Nowadays we believe in
one single underlying theory, but it is not clear what are
the truly fundamental objects, or more radically, whether there
are truly fundamental objects at all. Therefore it is important to
reflect on the assumptions that we need to make in deriving the
above results. Somewhat ironically in view of the fact that we
strongly advocate string theory, we find that most of the results
and the most interesting results,
do not depend on string theory in detail. All one has to assume is 
that there is a consistent quantum theory behind supergravity.
At time string theory is unique as a candidate.

In the above list, the results one to five 
are derived on the basis of four-dimensional
${\cal N}=2$ supergravity, whereas the entropy counting in M-theory
works by a collective mode analysis for solitons in eleven-dimensional
supergravity and does not use string theory either.\footnote{This
was emphasized in \cite{MalStrWit:1997}.}
However the other results concern the T- or S-duality invariance
of entropy formulae and therefore they belong into the context of
string theory and are not motivated by supergravity alone.

Let us finally indicate what are the natural concrete steps
in extending the results listed above.
We focussed on black hole entropy and the 
near horizon geometry of black holes in the presence of higher 
curvature terms. One should now construct the 
full black hole solutions. This is currently under investigation.
Next one can generalize this to multi-centered solutions and to adiabatic
motion in the moduli space.
Such solutions can be described by 
superconformal quantum mechanics (see for example 
\cite{GibTow:1998,MicStr:1999,MalSprStr:1999}),
and since this is a specific case of the AdS-CFT correspondence,
we can make contact with the gravity and string theory -
gauge theory correspondence discussed above. We also think
that a better understanding of Wald's construction of surface
charges is desirable and might be helpful to explore the relation between
the supersymmetric attractor mechanism and renormalization group flows.
Note that Wald's construction should have an interpretation
in terms of cohomology, and that it resembles the descent equations
that are familar from the analysis of anomalies in field theory.

Concerning black hole physics the intrinsic limitation of an
approach based on BPS states is of course that one can only
treat supersymmetric states. The next level of understanding
will be reached when non-supersymmetric states can be brought
under quantitative control. This includes in particular the
Schwarzschild black hole. But as soon as one understands
far non-extreme charged black holes this can be treated by switching
off the charges. 

Clearly the understanding of non-extreme black holes requires
a major conceptual step forward. The first indication that 
this might be possible is provided by what one might call
(in the spirit of Wigners remark concerning the relation of
mathematics to physics) the unreasonable effectiveness
of D-branes. Though the extrapolation to the perturbative
regime is strictly speaking only justified for BPS states,
it 'just works' more generally. In particular one can relate
the entropy of near-extreme black holes to state counting
and the resulting formula has a suggestive (though formal)
interpretation in terms of branes and antibranes, see
\cite{Mal:1996}. Hawking radiation
can be computed by considering 
interactions of open strings living on non-extremal
D-brane configurations.
Another way of going away from the extreme limit is to 
slightly perturb static multiblack hole solution. The dynamics of
the resulting 
system of slowly moving, interacting black holes is determined
by the metric on the moduli space of the multi-centered solution.
This metric has been computed for five- and four-dimensional
${\cal N}=2$ supergravity coupled to vector multiplets 
\cite{GutPap:1999,MalSprStr:1999,GutPap:2000}. In the 
near-coincident limit the system can be described in terms
of superconformal quantum mechanics on the moduli space,
see \cite{StrEtAl:1999} for a recent review. This surprising and
fascinating structure might give further insight into the entropy
and dynamics of black holes. The impact of $R^2$-terms on the 
metric on moduli space is currently under investigation
\cite{CardWKaeMoh2}. It seems that Wald's entropy formula
appears in the metric, giving further evidence that the study of 
multiblack hole system will lead to a deeper understanding of
black hole entropy.

Ultimately one has to understand the entropy of generic, non-extreme
black holes of Schwarzschild and Kerr type. As already mentioned in
the introduction there are general arguments which allow to explain
the entropy of general black holes up to one order of magnitude in
terms of string (or brane) states \cite{Sus:1993,HorPol:1996}. Moreover
the Schwarzschild black hole has been
analysed in the matrix formulation of M-theory 
\cite{BanFisKleSus:1997,KleSus:1997,BanFisKleSus:1997a}. 
The study of stable non-BPS
brane configurations (see \cite{Sch:1999} for a review)
and the renewed interest in non-supersymmetric string 
theories \cite{Pol:1998,KleTse:1998} also
suggest that non-supersymmetric situations are tractable.
We think that all these activities encourage the belief that the 
gravity / string theory - gauge theory correspondence is the
right strategy for a full understanding 
of the entropy of non-extreme black holes.

These ideas may rise questions about the ultimate role
of supersymmetry. Above we argued that our results are robust,
because they can be derived to a large extent
on the basis of supersymmetry,
now we suggest that the we should try to do 'without supersymmetry'.
But this is not necessarily a contradiction. We believe that 
supersymmetry plays a role at the fundamental level, and there
are plenty of arguments in favour of this, which we do not need
to review here. At the same time we know that our universe is 
in a non-supersymmetric state. In this situation it is natural to
start with objects which share many of the symmetries of the vacuum,
such as BPS states and then to improve on it by studying less
symmetric states.

\section*{Acknowledgements}

First of all I would like to thank Gabriel Lopes Cardoso and Bernard
de Wit for the fruitful and pleasent collaboration over the
last years. The more recent results, which are briefly summarized
in section \ref{AddedSection}, were obtained with J\"urg K\"appeli
as a new member of this collaboration.

I am also indepted to my other coworkers 
Klaus Behrndt, Gottfried Curio, 
Ingo Gaida, Renata Kallosh, Dieter L\"ust, Swapna Mahapatra,
Soo-Jong Rey and Wafic Sabra for sharing their insights and
knowledge with me. The quantum field theory group
in Halle provided a stimulating working environment for me over the
last years. Many thanks for this to Robert B\"ohm,
Holger G\"unther, Michael
Haack, Carl Herrmann, Matthias Klein, Jan Louis and Monika Marquart. 
I would also like to thank Bruce Hunt for several illuminating
discussions and remarks on Calabi-Yau manifolds.
Part of this work was written
during a stay at Cern and I would like to thank the theory division
for hospitality and financial support.

\begin{appendix}

\chapter{Space-Time Geometry \label{AppSpaceTime}}

Part of the material in this appendix was taken from 
\cite{Kleijn:1998,Clausetal:1998,BakBorCar:1997}.

\section{Tensors}

\subsection{Metric and Vielbein}

Lorentz indices: $\mu, \nu, \ldots$ are curved indices, 
$a,b,\ldots$ are flat indices (local Lorentz indices, tangent space
indices). They take values $0,1,2,3$.

We work with Minkowski signature $(-+++)$. The flat metric
is denoted $\eta_{ab}$.

The vielbein is $e_{\mu}^{\;\;a}$ and the inverse vielbein
is $e_a^{\;\;\mu}$. 
Thus $e_{\mu}^{\;\;a} e_a^{\;\;\nu} = \delta_{\mu}^{\nu}$.
The curved and flat metrics are related by
\be
g_{\mu \nu} = e_{\mu}^{\;\;a} e_{\nu}^{\;\;b} \eta_{ab}\;, \;\;\;
\eta_{ab} = e_a^{\;\;\mu} e_b^{\;\;\nu} g_{\mu \nu} \;.
\ee
Curved and flat indices are converted by
\be
V_a = e_a^{\;\;\mu} V_{\mu} \;, \;\;\;
V_{\mu} = e_{\mu}^{\;\;a} V_a \;.
\ee
Curved and flat indices are moved up and down with $g_{\mu \nu}$ and
$\eta_{ab}$ and their inverses $g^{\mu\nu}$ and $\eta^{ab}$,
respectively. We define
\be
e = \sqrt{ |\det(g_{\mu \nu})| } \;.
\ee

\subsection{Symmetrization and Antisymmetrization}

Symmetrization and antisymmetrization are
done with the following normalization:
\be
V_{(a_1 \cdots a_p)} = \frac{1}{p!} \sum_{\sigma} V_{ \sigma a_1 \cdots
\sigma a_p}\;, \;\;\;
V_{[a_1 \cdots a_p]} = \frac{1}{p!} \sum_{\sigma} (-1)^{\mscr{sgn}(\sigma)}
V_{ \sigma a_1 \cdots \sigma a_p} \;,
\ee
where $\sigma$ runs over all permutations of $p$ objects and
$\mbox{sgn}(\sigma)$ is the signum of $\sigma$.
The normalization is such that $S_{(ab \cdots )}=S_{ab \cdots}$
for symmetric tensors $S_{ab \cdots}$ and
$T_{[ab \cdots ]}=T_{ab \cdots}$
for antisymmetric tensors  $T_{ab \cdots}$.

The standard normalization of the
completely antisymmetric four-index symbol with tangent space indices is  
\be
\epsilon^{0123} = 1\;.
\ee
When dealing with antisymmetric tensors in Minkowski signature it
is convenient to use the modified symbol
\be
\ve^{0123} := i \epsilon^{0123} = i \;.
\ee
When contracting the $\ve$-tensor with itself, one has to
apply the Euclidean rather then the Minkowskian formula, because
the signature dependent factor $(-1)^s$ has been absorbed by
the explicit $i$:
\be \ve_{k_1 \cdots k_p,i_1 \cdots i_q}
\ve^{k_1 \cdots k_p,j_1 \cdots j_q}   = p!
\delta^{j_1 \cdots j_q}_{i_1 \cdots i_q}
\label{vecontraction}
\ee
where
\be  \delta^{i_1 \cdots i_n}_{j_1 \cdots j_n} := \det
\left( \begin{array}{ccc}
\delta^{i_1}_{j_1} &\cdots& \delta^{i_n}_{j_1}\\
\vdots & &\vdots \\
\delta^{i_1}_{j_n} & \cdots & \delta^{i_n}_{j_n} \\
\end{array} \right)
\ee

The fully antisymmetric tensor density (with world indices) is
\be
\ve^{\mu \nu \rho \sigma} = i \epsilon^{\mu \nu \rho \sigma}
= e \;e_a^{\;\;\mu}e_b^{\;\;\nu}e_c^{\;\;\rho}e_d^{\;\;\sigma}
\ve^{abcd} \;.
\ee

In the main text various kinds of $\ve$-symbols appear, 
like the $SU(2)$ $\ve$ symbol $\ve_{ij}$, $i=1,2$ and the binormal
$\ve_{ab}$, $a,b=0,1$. We use the letter $\ve$ rather then $\epsilon$ to
avoid confusion with the $Q$-supersymmetry parameters
$\epsilon^i, \epsilon_i$. Note however that all the other 
$\ve$-symbols are conventionally normalized, i.e. they are real.

\subsection{Selfdual and Antiselfdual Antisymmetric Tensors}

The dual of an antisymmetric Lorentz tensor is
\be
\tilde{F}_{ab} = \ft12 \ve_{abcd} F^{cd} \;.
\ee
The selfdual and antiselfdual parts of $F_{ab}$ are
\be
F^{\pm}_{ab} = \ft12 ( F_{ab} \pm \tilde{F}_{ab} ) \;.
\ee
Note that $F^{\pm}_{ab}$ are complex conjugated quantities
in Minkowski space:
$(F^+_{ab})^{*} = F^-_{ab}$.

The projection operator onto the (anti-) selfdual part is
\be
\Pi_{ab}^{\pm \;\;cd} =  \frac{1}{2}  \left(
\d_{[a}^c d_{b]}^c \pm \frac{1}{2}  \ve_{ab}^{\;\;\;\;cd} \right) \;.
\ee

The following identities for (anti-) selfdual tensors are
useful:
\bea
 G^\pm_{[a[c}\, H_{d]b]}^\pm &=& \pm\ft18 G^\pm_{ef} \,H^{\pm ef} 
\,\varepsilon_{abcd} -\ft14( G^\pm_{ab}\, H^\pm_{cd} +G^\pm_{cd}\, 
H^\pm_{ab}) \,,\nonumber \\ 
G^{\pm}_{ab} \, H^{\mp cd} + G^{\pm cd} \, H^{\mp}_{ab} &=& 4 
\d^{[c}_{[a} G^{\pm}_{b]e} \, H^{\mp d]e} \,,\nonumber \\
\ft12 \varepsilon^{abcd} \,G^{\pm}_{[c}{}^e \, H^\pm_{d]e} &=& \pm  
G^{\pm [a}{}_{e} \, H^{\pm b]e}\,,\nonumber \\ 
G^{\pm ac}\,H^\pm_c{}^b + G^{\pm bc}\,H^\pm_c{}^a &=& -\ft12 \eta^{ab}\, 
G^{\pm cd}\,H^\pm_{cd} \,,\nonumber \\ 
G^{\pm ac}\,H^\mp_c{}^b &=&G^{\pm bc}\,H^\mp_c{}^a\,, \qquad 
G^{\pm ab}\,H^\mp_{ab} =0 \,. 
\label{IdASDT}
\eea

\subsection{Spin Connection and Riemann Tensor}

The anholonomicity coefficients of the vielbein $e_\m^{\;\;a}$
are:
\be
\Omega_{\m \n a} := 2 \der_{[\m} e_{\n] a} \;.
\ee
The spin connection is
\be
\omega_{\m ab} := \ft12 e_a^{\;\;\rho} \Omega_{\mu  \rho b} -
\ft12 e_{b}^{\;\;\sigma} \Omega_{\mu \sigma a} -
\ft12 e_a^{\;\;\rho} e_b^{\;\;\sigma} e_{\m}^{\;\;c}
\Omega_{\rho \sigma c} \;
\ee
and the Riemann tensor is
\be
R_{\m \n}^{\;\;\;\;ab} := 2 \der_{[\m} \omega_{\n]}^{ab}
- 2 \omega_{[\m}^{ac} \omega_{\n]}^{db} \eta_{cd} \;.
\ee
By contraction one gets the Ricci tensor
\be
R_\m^{\;\;a} = R_{\m\n}^{\;\;\;\;ab} e_b^{\;\;\n}
\ee
and the Ricci scalar
\be
R = R_\m^{\;\;a} e_a^{\;\;\m} \;.
\ee
The Weyl tensor is the trace-free part of the Riemann tensor,
\be
C_{\m\n}^{\;\;\;\;ab} = R_{\m\n}^{\;\;\;\;ab}
- 2 \d_{[\m}^{[\;\;a} R_{\n]}^{\;\;b]}
+ \ft13 R \d_{[\m}^{\;\;[a} \d_{\n]}^{\;\;b]} \;.
\ee

\subsection{Consequences of Rotational Invariance}

A rotationally invariant metric can be brought to the form
\be
ds^2 = - e^{2g} dt^2 + e^{2f} ( dr^2 + r^2 \sin^2 \theta d\phi^2
+ r^2 d \theta^2 ) \;,
\ee
with two arbitrary functions $f=f(t,r)$ and $g=g(t,r)$. If the metric
is in addition static, then one can achieve $f=f(r)$ and
$g=g(r)$.

In a spherically symmetric background curved indices take values
$\mu, \nu, \ldots  = t,r,\phi,\theta$ whereas flat indices take values
$a,b, \ldots = 0,1,2,3$. The vielbein and its inverse can be
brought to the form
\be
e_{\mu}^{\;\;a} = \left( \begin{array}{cccc}
e^g & 0 & 0 & 0 \\
0 & e^f & 0 & 0 \\
0 & 0 & e^f r \sin \theta & 0 \\
0 & 0 & 0 & e^f r \\
\end{array} \right) \;, \;\;\;
e_{a}^{\;\;\mu} = \left( \begin{array}{cccc}
e^{-g} & 0 & 0 & 0 \\
0 & e^{-f} & 0 & 0 \\
0 & 0 & e^{-f} r^{-1} \sin^{-1} \theta & 0 \\
0 & 0 & 0 & e^{-f} r^{-1} \\
\end{array} \right)\;.
\ee

The following components of the spin connection are non-vanishing:
\be
\omega_t^{01}= -g^\prime \,e^{g-f}\,, \quad \omega_\phi^{12}= (1 + r \,
f^\prime) \sin\theta\,,\quad \omega_\theta^{13}= 1 + r \,
f^\prime\,,\quad \omega_\phi^{23}= -\cos\theta\,.
\label{SphericConnection}
\ee
The corresponding curvature tensor $R_{\m\n}^{ab}$ has the components
\be
R_{tr}^{01} = \Big[g'' + g'(g'-f')\Big] e^{g-f}\,,\quad R_{t\phi}^{02} =
g^\prime(1+r\,f') \,e^{g-f} \,\sin\theta\,,\quad
\quad R_{t\theta}^{03} = 
g^\prime(1+r\,f') \,e^{g-f} \,,
\ee
\be
R_{r\phi}^{12} = (r\,f'' +f')\sin\theta\,,  \quad R_{r\theta}^{13} =
(r\,f'' +f') \,,  \quad
R_{\phi\theta}^{23} = -\sin\theta\Big[ 1 - (1+r\,f')^2 \Big] \,.
\ee
With tangent-space indices, these curvatures 
read
\bea
R_{01}^{01} &=& \Big[g'' + g'(g'-f')\Big] e^{-2f}\,, \nonumber \\
R_{02}^{02} &=&R_{03}^{03} = \Big[ f'\,g' + {1\over r} \,
g^\prime\Big] \,e^{-2f}\,,\nonumber \\  
R_{12}^{12} &=& R_{13}^{13}= \Big[ f'' + {1\over r}\,f'\Big]\,
e^{-2f} \,, \nonumber \\ 
R_{23}^{23} &=& \Big[f'^2 + {2\over r}\,f'  \Big] e^{-2f} \,. 
\eea
The Ricci tensor ($R_{\m\n}^{ab} \,e^\n_b\, e^\m_c$) and the Ricci 
scalar are:
\bea
R_0^0 &=& \Big[ g'' + g'(g'+f') + {2\over r} g'\Big] {\rm 
e}^{-2f}\,,\nonumber\\
R_1^1 &=& \Big[2 f'' +  g'' + g'(g'-f') + {2\over r} f'\Big] {\rm 
e}^{-2f}\,,\nonumber\\
R_2^2 &=& R_3^3= \Big[ f'' + f'(g'+f') + {1\over r} (3f' +g') 
\Big] {\rm e}^{-2f}\,,\nonumber\\
R&=& \Big[ 2(2 f''+g'')  + 2(f'^2+ g'^2 +f'\,g')+ {4\over r} (2f' 
+ g') \Big] e^{-2f} \,. 
\eea 
Finally the components of the Weyl tensor are:
\bea
C_{01}^{01} = C_{23}^{23} = -2C_{02}^{02} =-2C_{03}^{03} = 
-2C_{12}^{12} = -2C_{13}^{13}= && \nonumber\\
 \ft13 \Big[-f'' + g'' + (g'-f')^2 +{1\over 
r}(f'-g') \Big] e^{-2f}\,. &&
\eea

A rotationally invariant antisymmetric tensor takes the form
\be
F_{tr} = F_E(t,r)\;, \;\;
F_{\phi \theta} = F_M(t,r) \sin \theta\;,
\ee
with all other independent components vanishing.
If $F_{\mu \nu}$ is a field strength satisfying the 
field equations and Bianchi identities
\be
\der_{\mu} ( e F^{\mu \nu} ) = 0\;, \;\;\;
\ve^{\mu \nu \rho \sigma} \der_{\nu} F_{\rho \sigma} = 0 \;,
\ee
then
\be
F_E (t,r) = e^{g-f} \frac{q}{r^2} \;, \;\;\;
F_M (t,r) = p \;,
\ee
with constants $q,p$ that are proportional to the electric and
magnetic charge, respectively.
Converting to flat indices one finds
\bea
F_{01} &=& e^{-(f+g)} F_E(t,r) = e^{-2f} \frac{q}{r^2} \;,\\
F_{23} &=& e^{-2f} r^{-2} \sin^{-1} \theta F_M(t,r) = 
e^{-2f} \frac{p}{r^2} \;. \\
\nonumber
\eea
The components of the dual tensor are:
\be
\tilde{F}_{01} = -i F_{23} \;, \;\;\;
\tilde{F}_{23} = i F_{01} \;.
\ee
The components of the selfdual and antiselfdual parts are:
\bea
F^{\pm}_{01} &=& \ft12 ( F_{01} \mp i F_{23} ) = \mp i F_{23}^{\pm} \;, \\
F^{\pm}_{01} &=& \mp \ft{i}{2} 
e^{-2f} r^{-2} ( F_M \pm i e^{f-g} r^2 F_E ) \;. \\
\nonumber
\eea

\section{Spinors}

\subsection{$\gamma$-Matrices}

The $\gamma$-matrices satisfy
\be
\gamma_a \gamma_b = \eta_{ab} + 2 \sigma_{ab}\;, \;\;\;
\gamma_5 = i \gamma_0 \gamma_1 \gamma_2 \gamma_3 \;.
\ee
The symmetric and antisymmetric part of the product are
\be
\begin{array}{lclcl}
\gamma_{(a} \gamma_{b)} &=& \ft12 \{ \gamma_{a}, \gamma_{b} \}&=& \eta_{ab}\;, \\
\gamma_{[a} \gamma_{b]} &=& \ft12 [ \gamma_{a} ,
\gamma_{b} ]&=& 2\sigma_{ab}\;. \\
\end{array}
\ee
The following identities for $\g$-matrices are useful:
\be 
\begin{array}{rclrcl}
\s_{ab} &=& -\ft12\ve_{abcd}\s^{cd}\g_5 \,, \hspace{2cm} 
&\g^b\g_a\g_b &=& -2\g_a \,,\\ 
\s^{ab}\s_{ab} &=& -3 \,,&\s^{cd}\s_{ab}\s_{cd} &=& \s_{ab}\,,\\
\g^c \s_{ab}\g_c &=& 0 \,,&\s^{bc}\g_a\s_{bc} &=& 0 \,,\\
{[\g^c,\s_{ab}]} &=& 2 \d_{[a}^c \,\g_{b]} \,,
&\{\g^c,\s_{ab}\} &=& \ve_{ab}{}^{cd}\g_5\g_d \,,\\
{[\s_{ab},\s^{cd}]} &=& -4\d_{[a}{}^{[c}\s_{b]}{}^{d]}\,,
&\{\s_{ab},\s^{cd}\} &=& -\d_{[a}^c\,\d_{b]}^d
+\ft12\ve_{ab}{}^{cd}\g_5\,.
\end{array}
\ee
Also note that if
$T^{ab}\sigma_{ab}$ acts on a spinor of positive chirality,
the tensor is projected onto its antiselfdual part:
\be 
T^{ab} \sigma_{ab} \e^i = T^{-ab} \sigma_{ab} \e^i \;.
\ee

\subsection{Charge Conjugation, Dirac-, Majorana- and Weyl-Spinors}

The charge conjugation matrix $C$ is defined by:
\be
- \gamma_a^T = C \gamma_a C^{-1} \;, \;\;\;
\gamma_5^T = C \gamma_5 C^{-1} \;, \;\;\;
C^T = -C \;.
\ee

We work with four-component spinors. The Dirac conjugate is defined 
by
\be
\ov{\psi} = \psi^{+} \gamma_0 \;,
\ee
whereas the Majorana conjugate is 
\be
\psi^{M} = \psi^T C \;.
\ee
Majorana spinors are subject to the reality constraint
\be
\psi^{M} = \ov{\psi} \;,
\ee
whereas Weyl spinors are subject to
the chirality constraint
\be
\gamma_5 \psi_{\pm} = \pm \psi_{\pm} \;.
\ee
Thus, the chiral projections of a spinor are
\be
\psi_{\pm} = \ft12 ( \mathbb{I} \pm \gamma_5 ) \psi\;.
\ee
In four-dimensional Minkowski space the Majorana and Weyl conditions
are not compatible, i.e. they cannot be imposed simultaneously.
The chiral projections of a Majorana spinor (which are not
Majorana-Weyl spinors as explained in the last sentence)
are not independent,
as is the case for a Dirac spinor, but are related by
\be
\psi_{\pm} = C^{-1, T} \gamma_0^T \psi_{\mp}^{*} \;.
\ee
This follows by chiral decomposition of the Majorana constraint 
into equations for its left- and righthanded part.
(We use a convention where $\gamma_0^+ = - \gamma_0$ and 
$C^* C = - \mathbb{I}$.)

\subsection{Spinor Bilinears and Fierz Rearrangements}

Complex conjugation of spinor bilinears:
\be
(\overline{\psi} \phi)^{\star} = \overline{\phi} \psi \;, \;\;\;
(\overline{\psi} \g_a \phi)^{\star} = - \overline{\phi} \g_a \psi \;.
\ee
Transpositions of bilinears of Majorana spinors:
\be
\overline{\psi} \phi = \overline{\phi} \psi \;, \;\;\;
\overline{\psi} \g_a \phi = - \overline{\phi} \g_a \psi \;.
\ee
Fierz rearrangement formula:
\be
\phi \overline{\psi} = - \ft14 ( \overline{\psi} \phi) \mathbb{I}
- \ft14 ( \overline{\psi} \g^a \phi ) \g_a
- \ft14 ( \overline{\psi} \g_5 \phi ) \g_5
+ \ft14 ( \overline{\psi} \g^a \g_5 \phi ) \g_a \g_5
+ \ft12 ( \overline{\psi} \sigma^{ab} \phi) \sigma_{ab}
\ee
These formulae are needed to work out the components of the
Weyl multiplet.

\chapter{Abelian Gauge Fields \label{AppAbelianGaugeFields}}

\section{Maxwell and Einstein-Maxwell Theory}

In this section we collect a few useful facts about
abelian gauge fields, in particular various forms of the
Lagrangian,
equations of motion and Bianchi identities and about their
solution in static curved backgrounds. 

The Lagrangian for a single abelian gauge field is:
\be
{\cal L} = \frac{1}{2} F \wedge {\star}F 
= \frac{1}{4} e F_{\mu \nu} F^{\mu \nu} \;.
\label{MaxwellLagrangian}
\ee

The Maxwell equations in a curved background are, in the absence of
sources,
\be
\nabla_{\m} F^{\m\n} = 0 \mbox{   and   }
\e^{\mu \nu \rho \sigma}\der_{\n} F_{\rho \sigma} = 0 \;.
\ee
where $\nabla_\m$ is the Christoffel connection.
The first equation is the Euler Lagrange equation obtained 
by varying the Lagrangian (\ref{MaxwellLagrangian}) with respect
to the vector potential $A_\m$, whereas the second 
is a Bianchi identity, i.e. the integrability condition for the
existence of the vector potential
\be
F_{\m\n} = \der_\m A_\n - \der_\n A_\m =\nabla_\m A_\n - \nabla_\n A_\m \;.
\ee

The equations can be rewritten in various ways. Using the formula
for the covariant derivative of an antisymmetric tensor the
field equation becomes
\be
\frac{1}{\sqrt{-g}} \der_{\mu} \left( \sqrt{-g} F^{\mu \nu} \right)= 0\;.
\ee
The second equation is a Bianchi identity and therefore 
does not depend on the metric. 
Equivalent forms are
\be
\e^{\mu \nu \rho \sigma}\nabla_{\n} F_{\rho \sigma} = 0
\mbox{   or   } 
\nabla_{\n}( \e^{\mu \nu \rho \sigma}F_{\rho \sigma}) = 0
\mbox{   or   } 
\der_{\n}( \e^{\mu \nu \rho \sigma}F_{\rho \sigma}) = 0 \;.
\ee

Here and in other equations we use that the Christoffel connection
drops out because of the antisymmetrization. The
curved space $\e$-tensor is covariantly constant by the vielbein
postulate $\nabla_\m e_\n^{\;\;a}=0$.

A more interesting way of rewriting the field equations is to 
go to the Hodge dual fields
\be
{}^{\star} F_{\m \n}=\ft12 \e_{\m\n \rho \sigma} F^{\rho \sigma} \;.
\ee
In Minkowski space the Hodge-$\star$-operator on 
two-forms satisfies $\star^{2}=-1$ and therefore
\be
F_{\m \n}= - \ft12 \e_{\m\n \rho \sigma}{}^{\star} F^{\rho \sigma} \;.
\ee
In terms of the dual field the Maxwell equations take the form
\be
\e^{\mu \nu \rho \sigma}\der_{\n} {}^{\star} F_{\rho \sigma} = 0
\mbox{   and   }
\nabla_{\m} {}^{\star}F^{\m\n} = 0 \;.
\ee
Thus the role of Euler-Lagrange equations and Bianchi identities
is reversed, provided one introduces a dual gauge potential
\be
{}^{\star} F_{\m \n}  = \der_\m \breve{A}_{\n} - \der_\n \breve{A}_\m \;.
\ee
Note that the original and the dual gauge field are not
related by a local field redefinition. Moreover the duality is only valid
in the absence of electric and magnetic sources.

The duality is manifest when writing both equations in the same form
either as
\be
\nabla_{\m} F^{\m\n} = 0 \mbox{   and   }
\nabla_{\m} {}^{\star} F^{\m\n} = 0 
\ee
or as
\be
\e^{\m \n \rho \sigma} \der_{\n} F_{\rho \sigma}= 0
\mbox{   and   }
\e^{\m \n \rho \sigma} \der_{\n} {}^{\star} F_{\rho \sigma}= 0 \;.
\ee
In this formulation it is obvious that one cannot only 
exchange $F$ with ${}^{\star}F$, but that one more generally has
the freedom of a uniform $Gl(2,\mathbb{R})$ rotation
of the vector $(F_{\m\n}, {}^{\star}F_{\m\n})$, i.e. one can
take arbitrary linear combinations.

We can decompose the field
strength into a selfdual and an antiselfdual part.
Since $\star^2=-1$ in spaces with Minkowski signature selfdual and 
antiselfdual tensors are complex:
\be
{}^{\star} F^{+}_{\mu \nu} = -i F^{+}_{\mu \nu}\;,\;\;\;
{}^{\star} F^{-}_{\mu \nu} = i F^{-}_{\mu \nu}\;.
\ee
The field strength can be decomposed as
\be
F_{\mu \nu} = F^{+}_{\mu \nu} + F^{-}_{\mu \nu}\;,\;\;\;
{}^{\star} F_{\mu \nu} = -i F^{+}_{\mu \nu} 
+ i F^{-}_{\mu \nu}\;
\ee
and conversely the self and antiselfdual parts are obtained by
projection:
\be
F^{+}_{\mu \nu} = \frac{1}{2}(F_{\mu \nu} 
+ i \; {}^{\star}F_{\mu \nu}) \;,\;\;\;
F^{-}_{\mu \nu} = \frac{1}{2}(F_{\mu \nu} 
- i \; {}^{\star}F_{\mu \nu}) \;.
\ee
Note that the selfdual and antiselfdual parts are  complex conjugated,
$(F^+_{\mu \nu})^* = F^-_{\mu \nu}$ for Minkowski signature.

We also note that the other Lorentz scalar that one can form out of the
field strength $F$ is a topological term:
\be
{\cal L} = \frac{1}{2} F \wedge F =
\frac{1}{2} (F,{\star}F) =
\frac{1}{4} e F_{\mu \nu} {}^{\star}F^{\mu \nu} =
\frac{1}{8} \epsilon^{\mu \nu \rho \sigma}
F_{\mu \nu } F_{\rho \sigma} = \ft14 \der_\m
\e^{\m \n \rho \sigma}  A_\n F_{\rho \sigma}  \,.
\ee
Nevertheless such terms can play a role for example for field
configurations with a non-trivial behaviour at infinity. In effective
$U(1)$ field theories such terms are generated with a field depedent
coefficient, and then they are relevant for the local dynamics as well.

In the main part of the paper we use a non-standard definition
for the $\e$-tensor with the effect that Minkowski-signature
formulae look like Euclidean signature formulae. Then we use
a modified definition for the dual tensor, denoted by
\be
\tilde{F}_{\m\n} = i\; {}^{\star}F_{\m\n} \;.
\ee

Consider now finding static solutions, $\der_t F_{\mu \nu}=0$,
of the Maxwell equations  in a 'conformastatic' curved background.
Such a metric can be brought to the isotropic form
\be
ds^2 = -e^{2g(\vec{x})} dt^2 + e^{2f(\vec{x})} d\vec{x}^2 \;.
\label{staticmetricApp}
\ee
The corresponding world indices are denoted by 
$\mu =t,m=x,y,z$.

First look for electric solutions, $F_{mn}=0$. Then the Bianchi
identity reduces to
\be
\der_m F_{tn} = \der_{n} F_{tm} \;,
\ee
which is solved by introducing an electrostatic potential,
\be
F_{tm} = \der_m \frac{1}{H} \;.
\ee
The field equation then implies
\be
\sum_{m=x,y,z} \der_m \left( e^{2f} \frac{1}{H^2} \der_m H
\right) = 0 \;,
\ee
which is solved by 
\be
e^{f} = H \mbox{   and   } \sum_{m=x,y,z} \der_m \der_m H =
\Delta H = 0 \;,
\ee
where $\Delta$ is the (flat) Laplace operator. The Einstein equation
then implies $g=-f$. 

Consider now a magnetic solution, $F_{tm}=0$. This time the
Bianchi identities read
\be
\e^{tmnp} \der_{m} F_{np}= 0
\ee
and are solved by
\be
F_{np} = \sum_{q=x,y,z} \e_{npq} \der_q H \;,
\ee
where $\e_{npq}$ is the flat $\e$-symbol, $\e_{xyz}=1$ and
$H$ must be a harmonic function,
\be
\sum_{q=x,y,z} \der_q \der_q H = 0 \;.
\ee
Then the field equation is satisfied if one takes $f=-g$ and
finally the  Einstein equation implies $e^f =H$.

Both solutions are related by a electric-magnetic duality
rotation and belong to a class of solutions of Einstein-Maxwell
theory called the Majumdar Papapetrou solutions. The general form
of the solution is
\be
ds^2 = - H^{-2} dt^2 + H^2 d \vec{x}^2 \;,
\ee
\be
F_{\mu \nu} = \cos \theta F^{(0)}_{\mu \nu} 
+ \sin \theta \; {}^{\star}F^{(0)}_{\mu \nu} \;,
\ee
\be
F^{(0)} := \ft12  F^{(0)}_{\m \n} dx^\m \wedge  dx^\n
= \vec{\nabla} \frac{1}{H} d\vec{x} \wedge dt \;,
\ee
where $H$ is a harmonic function,
\be
\Delta H = \sum_{m=x,y,z} \der_m \der_m H =0 \;.
\ee
The parameter $\theta$ parametrizes an electric-magnetic
duality rotation.

The most simple solution, which is given by a spherically symmetric
choice
\be
H = 1 + \frac{M}{r}
\ee
is the dyonic extreme Reissner-Nordstrom black hole with mass M and 
electric and magnetic charges $p,q$ given by $q= \cos \theta M$ and
$p=\sin  \theta M$.
Thus the mass is related to the charges by $M^2 = p^2 
+ q^2$, and the values $\theta=0,\frac{\pi}{2},\pi, \frac{3\pi}{2}$
of the parameter $\theta$
correspond to the cases of positive electric, postive magnetic, 
negative electric and negative magnetic charge, respectively.
Our duality rotation is an $SO(2)$ transformation rather then a
general $Gl(2,\mathbb{R})$ transformation in order to keep the
normalization of the electric and magnetic charge fixed.

A generalization of this is provided by the multi-centered solution, where
\be
H = 1 + \sum_{i=1}^{N} \frac{M_i}{ | \vec{x} - \vec{x_i} | } \;.
\label{MC}
\ee
The $\vec{x}_i$ are arbitrary points in space. When approaching
any of these points the metric becomes asymptotic to the
Bertotti-Robinson metric. The 'points' $\vec{x_i}$ represent
the postions of 
the finite size horizons of Reissner-Nordstrom black holes in
(asymptotically) isotropic coordinates. The electric and magnetic
charges and the mass of the $i$-th black hole are given by
$M_i^2 = p_i^2 + q_i^2$. Note that this configuration has the
astonishing property that it is static, i.e. the gravitational attraction
and the electro- and magneto-static repulsion cancel for all choices
of the locations $\vec{x}_i$ of the black holes. In particular the 
mass is additive and is given by the total charge.
Note that to achieve a static solution all the black holes must have charges
with the same 'sign' in the sense that all the complex charges
$q_i + i p_i$ must lay on the same ray in the complex plane, i.e. one
does not have the freedom to make electric-magnetic duality rotations
on individual black holes. In order to avoid naked singularities,
all the parameters $M_i$ in (\ref{MC}) have to be positive. The 
field equations fix the relation between $M_i$ and $q_i + i p_i$ up
to a uniform duality rotation \cite{Cha:1992}.

One can consider more general choices for the harmonic functions,
but it was shown in \cite{HarHaw:1972}
that all other choices lead to naked singularities.

\section{Abelian Gauge Fields coupled to ${\cal N}=2$ Supergravity}

We now generalize to the case of $N$ abelian gauge fields
with Lagrangian 
\be
{\cal L} = e \frac{1}{2} (- \gamma_{IJ} F^{I}_{\m\n}F^{J\m\n}
+ \theta_{IJ} F^{I}_{\m\n}{}^{\star}F^{J \m\n}) \;.
\ee
$\gamma_{IJ}$ generalizes the gauge 
coupling $1/g^{2}$ that we surpressed above. 
The gauge kinetic term of abelian vector multiplets coupled to 
${\cal N}=2$ supergravity takes precisely this form with
couplings $\gamma_{IJ}$ 
and 'theta-angles' $\theta_{IJ}$ that are field-dependent. 

Now 
introduce a complex symmetric coupling matrix:
\be
{\cal N}_{IJ} = -i \gamma_{IJ} + \theta_{IJ}
\ee
and rewrite the Lagrangian:
\be
\begin{array}{lcl}
{\cal L}/e &=& \frac{1}{2} (-F^{I} \gamma_{IJ} F^{J}
+ F^{I} \theta_{IJ} {}^{\star} F^{J})
= \frac{i}{2} ( F^{+I} {\cal N}_{IJ} F^{+J} - 
\mbox{c.c.} ) \\
 & & = - \mbox{Im}( F^{+I}{\cal N}_{IJ} F^{+J})
= \mbox{Im}( F^{-I} \overline{\cal N}_{IJ} F^{-J}) \;.\\
\end{array}
\ee

The Euler-Lagrange equations for the vector fields
are
\be
\nabla_{\mu} ( {\cal N}_{IJ} F^{+J \mu \nu} 
- \mbox{c.c.} ) = 0\;
\ee
and the Bianchi identities 
\be
\epsilon^{\mu \nu \rho \sigma} \der_{\nu} F_{\rho \sigma}^I =0
\ee
can be rewritten as
\be
\nabla_{\mu}( F^{+I \mu \nu} - F^{-I \mu \nu})=0 \;.
\ee

As is the case in Maxwell theory one can cast 
the Euler-Lagrange equations and Bianchi identities
into a symmetric form, which displays a continuous
electric-magnetic duality symmetry (in the absence of
sources).
To do so in this more general situation one 
defines a set of so-called dual gauge fields by
\be
G^{+}_{I \mu \nu} = {\cal N}_{IJ} F^{+ J}_{\mu \nu} \;.
\ee
Then the Euler-Lagrange equations read
\be
\nabla_{\mu} ( G^{+\mu \nu} _{I} - G^{- \mu \nu}_I ) = 0
\ee

When considering the full ${\cal N}=2$ Lagrangian the 
situation is slightly more general because 
additional moment couplings
$O^{+}_{I \mu \nu} F^{+I\mu \nu} + \mbox{c.c.}$
are present. These describe fermion couplings, and in the 
off-shell formulation the coupling to a bosonic auxiliary field.
Then, the dual gauge fields are defined by
\be
{\cal L}/e = \frac{i}{2} (F^{+I}_{\mu \nu} G^{+\mu \nu}_{I}
- \mbox{h.c.} )
\ee
or equivalently by
\be
G^{+\mu \nu}_{I} = e^{-1} 
\frac{2}{i} \frac{\der {\cal L}}{\der F^{I+}_{\mu \nu}} \;.
\ee
With this definition
the Euler Lagrange equations and Bianchi identities still have the
form given above.

The combined system of Euler-Lagrange equations and
Bianchi identities is manifestly invariant under
duality rotations:
\be
\left( \begin{array}{c}
F^{\pm I} \\ G^{\pm}_{J} 
\end{array} \right)
\longrightarrow
\left( \begin{array}{cc}
U^I_{\;\;K} &  Z^{IL} \\ W_{JK} & V_J^{\;\;L} \\
\end{array} \right)
\left( \begin{array}{c}
F^{\pm K} \\ G^{\pm}_L \\
\end{array} \right) = 
\left( \begin{array}{c}
\breve{F}^{\pm I} \\ \breve{G}^{\pm}_J\\
\end{array} \right) \;.
\ee
A detailed discussion of duality in this case was given in the main text.

Due to the presence of field dependent couplings the field equations
are now harder to solve. In a static background one can, however,
still express the magnetic parts of the field strength and its
dual in terms of harmonic functions. This is due to the fact that
the equations take the form of Bianchi identities, which do not
involve the metric:
\bea
\e^{\m \nu \rho \sigma} \der_{\n}  F^I_{\rho \sigma} &=&0 \;,\nonumber \\
\e^{\m \nu \rho \sigma} \der_{\n}  G_{J\rho \sigma} &=&0 \;.  \\
\nonumber
\eea
Looking for static solutions in a static metric background
(\ref{staticmetricApp}) this is solved by
\bea
F^I_{mn} &=&\sum_p  \e_{mnp} H^I \;, \nonumber \\
G_{Jmn} &=&\sum_p  \e_{mnp} H_J \;,  \\
\nonumber
\eea
where $H^I,H_J$ are harmonic functions,
\be
\Delta H^I = 0 = \Delta H_J \;.
\ee
Since the electric parts $F^I_{tm}$, $G_{Jtm}$ are determined
by the magnetic parts, this already fixes the gauge fields.
But in order to explicitly specify the electric parts one
needs to know the functions $f,g$ and the metric and the
field dependent matrix ${\cal N}_{IJ}$. This requires 
to solve the gravitational and scalar field equations.

This has been done explicitly for various cases (but typically 
in cases where half of the electric and magnetic charges have
been set to zero). The resulting metrics generalize the 
Majumdar-Papapetrou solutions and describe static configurations
of extreme black holes carrying electric and magnetic charges with
respect to several gauge fields. Generically theses solutions 
also contain nontrivial space-dependent scalar fields. The functions
$f$ and $g$ are still related by $f=-g$ as a consequence of
supersymmetry. They are complicated functions of the harmonic 
functions $H^I,H_J$ that parametrize the solution.

In the main text we need the explicit solution for a metric
which is spherically symmetric in addition to static. Introducing 
spheric coordinates in the space part the metric takes the form
\be
ds^2 = - e^{2g} dt^2 + e^{2f} ( dr^2 + r^2 ( \sin \theta d\phi^2 +
d\theta^2)) \;.
\ee
We denote world indices by $t,r,\phi,\theta$. Then we can take the magnetic
gauge fields to have the form
\be
F^I_{\phi \theta}(r) = -\der_r H^I(r), \;\;\;
G_{J \phi \theta}(r) = -\der_r H_J(r) \;.
\ee
Spherical symmetry requires to take the harmonic functions to be of 
single-centered type,
\be
H^I = h^I + \frac{p^I}{r} \;,\;\;\;
H_J = h_J + \frac{q_J}{r} \;,\;\;\; 
\ee
with the result
\be
F^I_{\phi\theta} = \frac{p^I}{r^2} \;,\;\;\;
G_{J \phi\theta} = \frac{q_J}{r^2} \;.\;\;\;
\ee
Switching to flat indices we find
\be
F^I_{23} = \frac{e^{-2f}}{r^2} p^I \;,\;\;\;
G_{J 23} = \frac{e^{-2f}}{r^2} q_J \;,\;\;\;
\ee
which is what we use in the main text.
For a single gauge field with no coupling to scalar fields 
we have $G_{23} = F_{01}$ and we have rederived 
the result we found in appendix \ref{AppSpaceTime}.

\chapter{Covariant Derivatives  \label{AppSpecialGeometry}}

We review the definition of covariant quantities and covariant
derivatives given in \cite{vP:1983,Kleijn:1998}: 
a quantity is covariant with respect
to a local transformation 
if the transformation law does not involve derivatives of the
transformation parameter. A derivative is called covariant if 
its application to a covariant quantity gives another covariant quantity.
A given derivative can be made covariant by adding an object that 
transforms appropriately (namely the connection related to the transformation
under consideration). This process we call covariantization, and
we illustrate it in a schematic example. 

Let $\phi$ be a field that transforms as
\be
\delta \phi(x) = \epsilon(x) \psi(x) \;,
\label{transf}
\ee
where $\epsilon(x)$ is the transformation parameter. (In general
the fields $\phi,\psi$ and the parameter $\epsilon$
can be tensors or spinors with respect to
several transformation groups. Note also that $\psi$ could be 
identical to $\phi$ as is the case for abelian gauge transformations
or dilatations for instance.
The relevant fact we want to focus on
in our example is that the transformation is linear in a space-time
dependent parameter.) Now consider how the derivative transforms:
\be
\delta \der_{\mu} \phi = \epsilon \der \psi + \der \epsilon \psi \;.
\ee
This is not covariant according to our definition because of the 
second term. To covariantize we add a second term to the derivative
\be
D_{\mu}  \phi = \der_{\mu}  \phi  - h_{\mu} \psi \;,
\label{covariantization}
\ee
where the newly introduced connection $h_{\mu}$ has to transform as
\be
\delta h_{\mu}   \psi =  \der_ {\mu}  \epsilon \psi + \cdots \;.
\ee
As indicated there might be further terms 
which do not contain $\der_{\mu} \epsilon$ and
are linear in $\epsilon$. Such additional terms are in fact present in many
cases, in particular in ${\cal N}=2$ supertransformations.
Note that the second term contains the transformed field $\psi$ rather
than the original field $\phi$. Thus,
as an operator, the covariant derivative has the form
\be
D_{\mu} = \der_{\mu} - \delta(h_{\mu}) \;,
\ee
where the operator $\delta(h_{\mu})$ generates a transformation
with parameter $h_{\mu}$ on the object to its right (compare to 
(\ref{transf},\ref{covariantization})). 
The new covariant derivative transforms covariantly:
\be
\delta D_{\mu} \phi  = \epsilon ( \der_{\mu} \psi + \cdots  ) \;.
\ee

When considering several kinds of transformations, one has to add 
several connections,
\be
D_{\mu} = \der_{\mu} - \sum_T \delta(h_{\mu}(T)) \;.
\ee   
In the superconformal context $D_{\mu}$ denotes the covariant
derivative with respect to all superconformal transformations,
whereas ${\cal D}_{\mu}$ is the convariant derivative with
respect to Lorentz transformations, dilatations, $SU(2)_R \times U(1)_R$
and gauge transformations. ${\cal D}_{\mu}$ is useful when going
to the Poincar\'e gauge.

As a concrete example let us work out the covariant derivative
of $X^I$, which has Weyl and chiral weights $w=1,c=-1$. 
First covariantize with respect to
$U(1)$ transformations and denote the corresponding 
covariant derivative by ${\cal D}'_{\m}$:
\be
\delta X^I = - i \alpha X^I \;,\;\;\;
{\cal D}'_{\mu} X^I = \der_{\mu} X^I - h_{\mu}(A) X^I\;,\;\;\;
\delta h_{\mu} = -i \der_{\mu} \alpha \;.
\ee
Since $h_{\mu}(A) = -i A_{\mu}$ this means 
$\delta A_{\mu} = \der_{\mu} \alpha$.

After covariantization with respect to dilatations we get the
full ${\cal D}_{\mu}$ since $X^I$ is neutral with respect to the
other relevant transformations:
\be
{\cal D}_{\mu} X^I = ( \der_{\mu} - b_{\mu} + i A_{\mu}) X^I
= \der_{\mu} X^I - w  b_{\mu} X^I - i c A_{\mu} X^I \;.
\ee

To obtain the superconformal derivative $D_\m$ we have to
take into account that $X^I$
transform under $Q$-supertransformations:
\be
\delta X^I = \ov{\epsilon}^i \Omega_i^I \;.
\ee
Therefore further covariantization is needed:
\be
D_{\mu} X^I = {\cal D}_{\mu} X^I - \ft12 \ov{\psi}_{\mu}^i \Omega_i^I \;,
\ee
where we used $h_{\mu}^i(Q) = \ft12 \psi_{\mu}^i$ and that 
$\psi_{\mu}^i$ transforms into ${\cal D}_{\mu} \ov{\epsilon}^i + \cdots$.

\chapter{Modular Geometry \label{AppModularGeometry}}

In this appendix we illustrate the geometry of vector multiplet
moduli spaces using the most simple examples.

The vector multiplet moduli space of a four-dimensional 
${\cal N}=2$ supergravity theory or string compactification
is a special K\"ahler manifold. In the most simple 
case this is a symmetric space. This happens 
for example in tree level heterotic compactifications,
where the moduli space is locally
\be 
\frac{SU(1,1)}{U(1)} \times \frac{SO(2,N_V-1)}{SO(2) \times SO(N_V-1) } \;.
\label{FamMS}
\ee
In the following we will discuss some local and global 
properties of the moduli space. The fact that it is special
K\"ahler will not play an explicit role. Using the local
isomorphisms
\be
\frac{SU(1,1)}{U(1)} \simeq \frac{SO(2,1)}{SO(2)} \simeq
\frac{SL(2,\mathbb{R})}{SO(2)} \mbox{  and   }
\frac{SO(2,2)}{SO(2) \times SO(2)} \simeq \frac{SO(2,1)}{SO(2)} \times
\frac{SO(1,2)}{SO(2)} 
\ee
we realize that for $N_V\leq 3$ the moduli space locally is a product
of $SL(2,\mathbb{R})/SO(2)$ cosets. This is the space we will consider
in the following. The standard realization is given by the 
complex upper half plane 
\be
{\cal H} = \{ \tau \in \mathbb{C} | \mbox{Im} \; \tau > 0 \}
\ee
equipped with the Poincar\'e metric 
\be
ds^2 = ( \mbox{Im} \; \tau)^{-2}  d\tau d\ov{\tau} \;,
\ee
which is a K\"ahler metric with K\"ahler potential
\be
K = - \log ( i(\tau - \ov{\tau})) \;.
\ee
The group $SL(2,\mathbb{R})$ acts from the left on the coset
by fractional linear transformations,
\be
\tau \rightarrow \frac{a \tau + b}{c \tau +d},\mbox{   where   }
\left( \begin{array}{cc} 
a&b\\c&d\\ \end{array} \right)  \in SL(2,\mathbb{R})\;.
\label{SL2action}
\ee
Since the matrix $-\mathbb{I}$ acts trivially, the group which 
properly acts on ${\cal H}$ is the projective group
$PSL(2,\mathbb{R})$.

A trivial reparametrization is obtained by replacing the upper
half plane ${\cal H}$ by the right half plane
$\{ t \in \mathbb{C} | t  = - i \tau, \; \tau \in {\cal H} \}$. 
This is the standard parametrization for
heterotic moduli, whereas for IIA moduli one conventionally
prefers the upper half plane. A less trivial reparametrization
is provided by a conformal transformation that maps the upper
halfplane onto the interior of the unit disc. This is the natural
realization of $SU(1,1)/U(1)$. Realizations of non-compact 
symmetric spaces by bounded open domains are somewhat distinguished.
In particular they are rare. The spaces (\ref{FamMS}) have
several realizations by unbounded but only one realization by a 
bounded domain.
Explicit parametrizations of the higher dimensional members of the
family (\ref{FamMS}) can be found in \cite{CarLueMoh:1995,HarMoo:1995}.

In string theory 
the global structure of the moduli space is determined by the discrete
duality group.
For moduli spaces of local structure 
$G/H$ (where $G$ is non-compact, $H$ is a maximal compact subgroup
and $G/H$ is symmetric, as is the case for supergravity models with 
sufficiently many supercharges, and toroidal string compactifications)
the discrete group is a discrete
subgroup $G(\mathbb{Z})$ of $G$. The true moduli space is then
the left-right coset $G(\mathbb{Z}) \backslash G / H$, whereas the covering
space $G/H$ is sometimes called the Teichm\"uller space ( borrowing
terminology from Riemann surfaces). Depending on the context $G(\mathbb{Z})$
is called the T, S or U-duality group. The relevant case for our
purposes is $G=SL(2,\mathbb{R})$, $H=SO(2)$ and 
$G(\mathbb{Z})=SL(2,\mathbb{Z})$. Depending on the actual case 
$SL(2,\mathbb{Z})$ is either T-duality or S-duality.

The presence of the discrete duality group implies that points
related by the action (\ref{SL2action}), with integer values
of $a,b,c,d$ are to be identified. Since $SL(2,\mathbb{Z})$ 
is generated by the elements $\tau \rightarrow \tau +1$ and
$\tau \rightarrow \ft{-1}{\tau}$, a realization of the 
moduli space is provided by the fundamental domain 
\be
{\cal F} = \{ \tau \in {\cal H} | - \ft12 < \mbox{Re}\;\tau
<  \ft12 \mbox{   and   } |\tau| > 1 \}
\ee
when supplemented by adding boundary points such that all
inequivalent points occur once. The boundary contains
the fixed points $\infty, i, \rho=\exp(2 \pi i/3)$ 
of the modular group,
which are relevant for the discussion of gauge symmetry enhancement.

Since physical quantities have to be duality invariant, 
automorphic forms and automorphic
functions of the group $G(\mathbb{Z})$ make their
appearence. We discuss such objects
for the most simple case $SL(2,\mathbb{Z})$. Part of the following
material is taken from \cite{CveFonIbaLueQue:1991}.

A function has weight $k$ under $SL(2,\mathbb{Z})$, iff
\be
f(\gamma \cdot z) = (cz + d)^k f(z)
\ee
for $\gamma \in SL(2,\mathbb{Z})$. 
Holomorphic weight $k$ functions on $H$ are classified according
to their behaviour at $\infty$: They are called modular functions,
modular forms, or cusp forms, iff they are meromorphic, holomorphic
or vanishing at $\infty$, respectively.

We now turn to explicit examples.
For $k \geq 4$ and $k$ even 
\be
G_k(z) =  \sum_{(m,n) \not= (0,0)} \frac{1}{ (mz + n)^k }
\ee
is 
the $k$-th Eisenstein series. It is absolutely convergent and defines
a modular form of degree $k$. $G_k$ is finite at $\infty$:
\be
G_k(\infty)= \sum_{n \not=0} n^{-k} = 2 \zeta(k) \;.
\ee
The values of the $\zeta$ function at even $k$ are related to the
Bernoulli numbers:
\be
\zeta(k) = - \frac{ (2 \pi i)^k}{2} \frac{B_k}{k!} \;,
\ee
where 
the Bernoulli numbers are defined by
\be
\frac{x}{e^x -1} = \sum_{k=0}^{\infty} B_k \frac{x^k}{k!} \;.
\ee

The normalized Eisenstein series are
\be
E_k(z) = \frac{1}{2 \zeta(k)} G_k(z) 
= 1 - \frac{2k}{B_k} \sum_{n=1}^{\infty} \sigma_{k-1}(n) q^n \;.
\ee
They have rational coefficients in the
$q= e^{2 \pi i z}$ expansion. 
The coefficents of the $q$ expansion are arithmetic:
\be
\sigma_{k-1}(n) = \sum_{d|n} d^{k-1} \;,
\ee
where the sum is over all divisors of $n$.

The second Eisenstein series is only conditionally convergent. Choosing
\be
E_2(z) = 1 + \frac{6}{\pi^2} \sum_{m=1}^{\infty} 
\sum_{n = - \infty, \not=0 }^{\infty} \frac{1}{(mz+n)^2} = 1 -
24 \sum_{n=1}^{\infty} \sigma_1(n) q^n
\ee
gives a holomorphic function which has anomalous, inhomogenous
behaviour under modular transformations: 
\be
z^{-2} E_2 (- \frac{1}{z} ) = E_2(z)  + \frac{12}{2 \pi i z} \;.
\ee
The non--holomorphic but modular covariant (weight 2) Eisenstein
series is defined by
\be
\wh{G}_2 (\tau, \ov{\tau}) = G_2 (\tau) - \frac{2 \pi i}{\tau 
- \ov{\tau}} \;.
\ee

The absolute modular invariant $j$ function
\be
j(z) = 
1728 \frac{E_4^3(z)}{E_4^3(z) - E_6^2(z) }
\ee
is a modular function of weight 0, and all modular functions
of weight 0 are rational functions of $j$. $j$ has a simple pole
with residue 1 at $\infty$ and is holomorphic in $H$. 
Some special values are
$j(\rho) = 0, j'(\rho) = 0, j''(\rho) = 0, j'''(\rho) \not= 0$,
where $\rho=e^{2 \pi i/3}$ and  
$j(i) = 1728, j'(i) = 0, j''(i) \not= 0.$ The order of these zeros
plays a crucial role in the discussion of gauge symmetry enhancement.

The $\eta$ function
\be
\eta(z) = e^{2 \pi i z /24} \prod_{n=1}^{\infty} ( 1 - e^{2 \pi i n z})
\ee
is a modular form of weight $1/2$ up to phase:
\be
\eta( - \frac{1}{z} ) = \sqrt{\frac{z}{i}} \eta(z) \;,
\ee
where the root has nonnegative real  part.\\
The $\eta$ function is related  to $E_2(z)$ by:
\be
\frac{\eta'(z)}{\eta(z)} = \frac{2 \pi i}{24} E_2(z) \;.
\ee

The derivative of a modular form is in general not a modular form.
A covariant derivative can be defined, using the inhomogenous
transformation property of $E_2$ to cancel the 'anomaly':
\be
D f(z) = f'(z) - 2 \pi i \frac{k}{12} E_2(z) f(z) \;.
\ee
This covariant derivative relates modular forms to modular forms.

Any modular form of weight $k$ is a polynomial in $E_4$ and $E_6$:
\be
f(z) = \sum_{4i + 6j =k} c_{ij} E_4^i(z) E_6^j(z) \;.
\ee
A modular form of weight $k$ vanishes at $i$, iff $k \not= 0$
modulo 4 and vanishes at $\rho$, iff $k \not= 0$ modulo 3.

A modular form of weight $k$ with multiplyer system
is by definition an object that transforms as
\be
f(\gamma \cdot z) = e^{i \Phi(\gamma)} (cz + d)^k f(z) \;.
\ee
$\eta$ is a form of weight $1/2$ with non--trivial multiplyer system.
Any such generalized modular form with weight $k$ which is regular
in the fundamental domain can be written as
\be
f(z) = \eta^{2k} \left( \frac{G_6}{\eta^{12}} \right)^m
\left( \frac{G_4}{\eta^{8}} \right)^n P(j) \;,
\ee
where $m,n \in \mathbb{N}$ and $P$ is a polynomial.

Note that one can also consider modular forms of negative
weight which then have poles at $i$ and $\rho$ instead of zeros.

The theory of automorphic or modular forms of more complicated
groups is less developed. There seems to be a deep relation to
infinite dimensional Lie algebras such as Borcherds algebras
\cite{HarMoo:1995}.
Both structures are intimately related in string theory.

Since we mentioned the T-duality group $O(2,N_V-1,\mathbb{Z})$ 
several times in the paper, we illustrate the complications involved
in properly defining and describing this group using the example
$O(2,2,\mathbb{Z})$ which is the T-duality group of a 
two-torus.\footnote{The structure of the group $O(d,d,\mathbb{Z})$
is discussed in \cite{GivPorRab:1994}.}
Whereas the corresponding continuous group locally factorizes
as $O(2,2) \simeq SL(2,\mathbb{R}) \times SL(2,\mathbb{R})$ it
is by no means true that $O(2,2,\mathbb{Z})$ is just the direct
product of two $SL(2,\mathbb{Z})$ groups, where one acts on the
complex structure modulus whereas the other acts on the
(complexified) K\"ahler modulus. The full T-duality group 
contains two more $\mathbb{Z}_2$  transformations. One of them
is mirror symmetry, which exchanges the K\"ahler and the complex
structure modulus and, hence, the two $SL(2,\mathbb{Z})$ groups.
The second (which we ignored in the main part) is world sheet parity.
A detailed description of this group and its fundamental
domain is found in \cite{Erl:1992,GivPorRab:1994,Asp:2000}.

\chapter{The Polylogarithmic Functions \label{AppPolylog}}

We quote the definition and some basic properties of 
the polylogarithmic functions, see for example
\cite{HarMoo:1995,BehCarGai:1997,MarMoo:1998}.

The series
\be
Li_{k}(x) = \sum_{n=1}^{\infty} \frac{x^{k}}{n^{k}},\;\;\;
\mbox{for} \;\;\;
0<x<1
\ee
can be extended to a multivalued analytic function on $\mathbb{C}$,
called the $k$-th polylogarithmic function, or $k$-th polylog
for short. 

In particular we have $Li_{1}(x)=-\log(1-x)$ 
and $x \frac{d}{dx} Li_{k}(x)=
Li_{k-1}(x)$.
Special values are
$Li_{k}(x\rightarrow 0+) =0$
and $Li_{k}(1) = \zeta(k)$. Some integral representations:
\begin{eqnarray}
Li_{2}(x) &=& \int_{0}^{1} \frac{dt}{t} \frac{1}{1-xt} \\
Li_{3}(x) &=& - \int_{0}^{1} \frac{dt}{t} \int_{0}^{1}
\frac{ds}{s} \log (1-xts)
\end{eqnarray}

In non-perturbative world-sheet or space-time contributions in
string theory one encounters expressions of the type
$Li_k(e^{-z})$, where $z\rightarrow \infty$ corresponds to the
classical or large volume limit and 
$z \rightarrow 0$ corresponds to a special locus, for example
the boundary of the K\"ahler cone or the boundary of a Weyl chamber.
In the classical limit this expression goes exponentially to zero, which
shows that the polylogarithmic term is non-perturbative either in
$\alpha'$ or in $g_S$, since $z$ is linear in the moduli.
Concerning the behaviour on special loci we note that
\begin{eqnarray}
Li_{3}(e^{-z}) &=& \zeta(3) - \frac{\pi^{2}}{6} z +
\left( \frac{3}{4} - \frac{\log (z)}{2}
\right) z^{2}  + {\cal O}(z^{3})\,, \\
Li_{2}(e^{-z}) &=& \frac{\pi^{2}}{6} + (\log (z) -1) z
+ \frac{z^{2}}{2} + {\cal O}(z^{3})
\end{eqnarray}
where we used $\zeta(2) = \frac{\pi^{2}}{6}$. \\ 

For $k\leq0$ the polylog is an elementary function:
\be
Li_0(x) = \frac{x}{1-x} \;,
\ee
\be
Li_k(x) = \left( x \frac{d}{dx} \right)^{|k|} \frac{1}{1-x}, \;\;\;
k<0 \;.
\ee
These functions appear in the higher gravitational couplings
${\cal F}^{(g>1)}$.

\end{appendix}


\bibliographystyle{h-physrev}

\bibliography{Habil}

\end{document}